\documentclass{lanl-report}
\usepackage[utf8]{inputenc}

\usepackage[utf8]{inputenc}
\usepackage{svg}
\usepackage{graphicx}
\usepackage{chemformula}
\usepackage{bm}
\usepackage{epsfig,graphicx,amssymb,color,amssymb,amsmath,mathrsfs,bm}
\usepackage[listings,skins,breakable,most]{tcolorbox}
\usepackage{float}
\usepackage{aas_macros}
\usepackage{url}
\usepackage{makecell}
\usepackage{braket}
\usepackage{hyperref}
\usepackage{xcolor}
\usepackage[normalem]{ulem}
\usepackage[version=4]{mhchem}
\usepackage{multirow}

\newcommand{\calL}{\mathcal{L}}

\newcommand{\br}{\mathbf{r}}
\newcommand{\bR}{\mathbf{R}}
\newcommand{\hH}{\hat{H}}
\newcommand{\hT}{\hat{T}}
\newcommand{\hV}{\hat{V}}
\newcommand{\dg}{^\dagger}

\tcbset{enhanced,width=\columnwidth,
                  colback=orange!48!white,breakable}

\usepackage{bbold}

\usepackage{wrapfig}

\usepackage[justification=centering]{caption}
\RequirePackage{subfiles}
\RequirePackage{subcaption}
\usepackage{biblatex}

\usepackage{lipsum}

\title{Potential Applications of Quantum Computing at Los Alamos National Laboratory}
\subtitle{v0.3.0}
\author{
Andreas B\"artschi, 
Francesco Caravelli,
Carleton Coffrin\footnote{Corresponding author cjc@lanl.gov},\hspace{5em}
Jonhas Colina,
Stephan Eidenbenz,
Abhijith Jayakumar,\hspace{5em}
Ammar A. Kirmani,
Scott Lawrence,
Minseong Lee,\hspace{8em}
Andrey Y. Lokhov,
Avanish Mishra,
Sidhant Misra,\hspace{8em}
Zachary Morrell,
Zain Mughal,
Duff Neill,
Andrei Piryatinski,
Allen Scheie,
Marc Vuffray,
Yu Zhang
}

\LAUR{LA-UR-26-24592}
\date{March 2026} 

\preparer{Name}
\preparertitle{Title}
\preparerorg{Org}

\preparedfordept{Department}
\preparedforoffice{Office}

\preparedforname{Name}
\preparedfortitle{Title}
\preparedfororg{Org}

\OUOname{OUO Name}
\OUOorg{OUO org}
\OUOdate{OUO date}
\OUOguidance{OUO guidance}

\DCname{DC name}
\DCZ{DC Z\#}
\DCposition{DC position}
\DCorg{DC org}
\DCderived{DC derived}

\UCNIname{UCNI Name}
\UCNIorg{UCNI Org}
\UCNIdate{UCNI Date}
\UCNIguidance{UCNI Guidance}

\begin{document}

\LANLTitlePage{}{Distribution Statement A – Approved for Public Release, Distribution Unlimited}


\tableofcontents




\begin{chapter}{Overview} \label{ch:overview}

\begin{section}{Introduction} 

Since the scientific revolution in the 16th and 17th centuries, the process of scientific discovery has followed an iterative feedback process of observation, hypothesis development and testing with physical experiments, which is widely referred to as the scientific method.
This process remained largely unchanged until the middle of the 20th century, when the emergence of digital computers empowered scientist to build and inspect detailed simulations of physical phenomena.
Over the last century, computational tools have transformed modern approaches to scientific discovery by enabling fast and affordable hypothesis testing before physical experiments are conducted, shown in Figure \ref{fig:science-workflow}.
Some notable examples include: global climate forecasts to understand how the environment may change over decades \cite{e3smHomePage}; modeling the behavior of plasma to design fusion reactors \cite{osti_1995209}; and understanding the behavior of molecules in biological processes \cite{Girodat2023,10.1002_jcc.25840}.

\begin{wrapfigure}{r}{0.45\textwidth}
    \centering
    \includegraphics[width=0.33\textwidth]{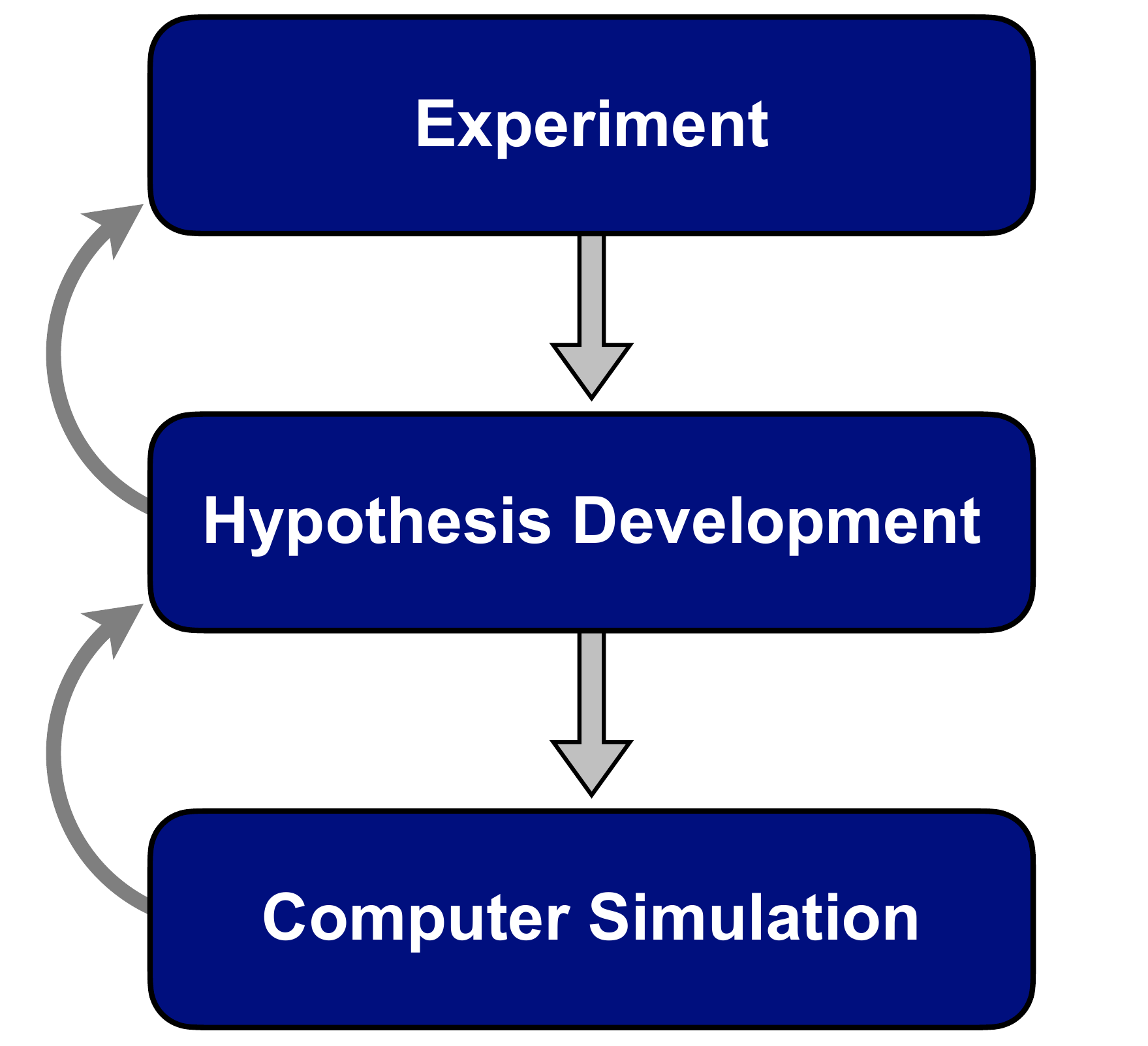}
    \caption{For the majority of modern history the scientific discovery process has been characterized by a feedback loop between experimental observation and hypothesis development. The emergence of physics simulation on digital computers in the 20th century introduced a new layer in the scientific discovery process, which has dramatically increased research productivity.}
    \label{fig:science-workflow}
\vspace{-0.3cm}
\end{wrapfigure}
Due to the transformational impact of computing technology for accelerating scientific discovery, the United States National Laboratories were some of the first developers of High-Performance Computing (HPC) facilities going back as far as the 1950's \cite{lanlComputingMesa}.
These facilities leverage so-called supercomputers to solve large and complex physics simulation tasks.
Modern supercomputers often take the form of large clusters of general purpose commodity computers with specifications that are tailored to the requirements for physics simulations \cite{energySupercomputingExascale}.
However, for most of the 20th century, it was common for National Laboratories to develop computing hardware that was specialized to physics simulation, such as the \textit{vector machines} developed by Cray Research \cite{lanlComputingMesa,osti_7081495}.
These large-scale computing facilities continue to be an essential component of the modern scientific discovery process with the Department of Energy investing over one billion dollars annually to provision, operate and maintain a range of cutting edge computing facilities across more than ten national laboratories \cite{doe-hpc}.

There is no doubt that physics simulation with computing technology is a key component of the scientific discovery process for many important subject areas.
However, physics regimes that exhibit quantum mechanical effects largely have not benefited from this computational transformation in scientific discovery.
The foundational challenge is that each quantum mechanical particle that is added to the simulation roughly doubles the amount of traditional (i.e., classical) computing resources that are required to perform this simulation. 
Consequently, the amount of classical computing resources needed to simulate quantum physics grows exponentially with the size of the quantum system, dramatically limiting the types of quantum systems that can be studied through computational simulation.
Over the years many classical algorithm designers have worked to reduce these requirements and have been successful in a variety of specific cases, such as simulations where limited entanglement is required.
However, a breakthrough method for \textit{general purpose} quantum physics simulation on classical computers has resisted over fifty years of exploration.

The lack of flexible, scaleable and accurate simulation tools for quantum mechanical physics has been holding back the understanding of important foundational topics such as super conductivity, chemistry and magnetism, for at least four decades.
This challenge motivated Feynman's proposal to develop a quantum mechanical computer for the study of quantum physics in the 1980's \cite{Feynman1982}.
The last decade has seen the emergence of the first generation of such devices, widely referred to as Noisy Intermediate Scale Quantum (NISQ) computers \cite{Preskill2018quantumcomputingin}, to recognize the challenges that noise and limited scalability present to conducting computations of interest.
It is anticipated that the next generation of quantum computers will leverage error correction protocols to provide a fault tolerant quantum computation, which addresses many of the challenges faced by current NISQ devices \cite{Preskill2018quantumcomputingin,Cerezo2021,Wang2021,Cerezo2022,2312.09121}.
Regardless of the specific trajectory of quantum technological development, the emergence of this technology suggests the possibility for science in quantum mechanical systems to experience the same computer-accelerated discovery that classical physics has experienced over the last century.

The development of quantum computing technology naturally leads to the following questions:
\begin{enumerate}
    \item \textit{If a sufficiently large and performant quantum computer existed, what would National Laboratories (and other large research intuitions) use it for?}
    \item \textit{What are the capability requirements for a quantum computer to meet the simulation needs of the application?}
    \item \textit{How much would these types of institutions be willing and able to pay for improved quantum simulation capabilities?}
\end{enumerate}
The primary objective of this report to develop initial answers to these questions and provide details of how quantum computing technologies could impact scientific discovery.
This is accomplished by conducting a broad investigation of quantum physics research activities at Los Alamos National Laboratory (LANL), one of the Department of Energy's seventeen laboratories.
This report includes both a survey of how LANL currently uses classical computers for scientific discovery and develops a collection of detailed use cases for the kinds of quantum simulations that LANL scientist would like to conduct, if a sufficient computing technology was available.
While the report strives to provide a broad portfolio of possible use cases, our investigation also suggests that many additional use cases could be documented with sufficient time and effort.

\subsection{Report Structure}

The remainder of this report is structured as follows.
It begins with a brief overview of the current methods for conducting quantum physics research at LANL (Section \ref{sec:current-methods}).
This section highlights the need for quantum simulation capabilities at this type of intuition and introduces how improved quantum simulation capabilities would support the scientific discovery process, in broad terms.
The next section (Section \ref{sec:ue-methods}) introduces the methodical foundations for estimating the value of quantum simulation capabilities.
These methods are used throughout the document for developing concrete value estimates for specific science applications.
The last in component of the overview chapter (Section \ref{sec:qc-methods}) provides a broad introduction to the well-established algorithmic tools for quantum computers that could benefit quantum simulation applications.
Through these sections, this overview chapter provides a foundation for understanding the remaining \textit{application instance} chapters.

Each subsequent chapter conducts a deep-dive on a specific application topic.
These chapters focus on documentation of the application requirements needed for a quantum simulation capability to have a transformational impact on the subject area.
These application examples tend to focus on grand challenge tasks that can serve as guiding stars as quantum computational capabilities develop over time.
A subset of application chapters also include references to external resources that provide implementations of the proposed quantum computational tasks using established quantum algorithms.
Such transformational workloads are too challenging for existing NISQ devices but are valuable as a tool for conducting quantum computer \textit{resource estimates}, which quantify the computational requirements of future quantum computers to address these applications.
As the field of quantum computer algorithms is nascent, it is expected that the implementation details in these external resources will see consistent improvements over the next several years.

This document follows a three digit version numbering the convention \texttt{vX.Y.Z}, which is inspired by the Semantic Versioning standard \cite{semver}.
A leading zero in this version number system indicates that this document is still under active development.
Improved subsequent versions are expected until \texttt{v1.0.0} is released.

\end{section}

\begin{section}{Quantum Physics Research at Los Alamos National Laboratory}
\label{sec:current-methods}

Los Alamos National Laboratory's (LANL) interest in quantum physics dates back to the inception of laboratory, with a wide variety of research continuing on to present day in areas such as nuclear physics, high energy physics, condensed matter and chemistry.
All of these research areas include theoretical, experimental and computational components, which work together to produce scientific insights and discoveries.
This section highlights a few examples of how computation and experimental capabilities support this kind of research.
This is not a comprehensive list of LANL's capabilities in quantum physics research.

\subsection{Quantum Physics Simulation}

LANL hosts a variety supercomputers that support a wide collection of computational needs.
Based on the performance specifications of the Trinity supercomputer \cite{lanlTrinityPlatforms},
one can estimate that LANL's total classical computational capability is on the order of 10 billion core hours per year.
Around 1 billion of these annual core hours are provided by a system called Chicoma, which is dedicated to supporting basic scientific research needs.
The Chicoma supercomputer includes around 1800 compute nodes, each with two AMD EPYC 7H12 processors (64 cores at 2.6 GHz clock) and 512 GB RAM.
Throughout this document the value of computations that are executed on Chicoma are estimated using the a value of \$4.00 to \$8.00 USD per node hour.
This is based on the costs required to provision similar resources on established cloud computing providers.
The large capital investments in specialized HPC hardware at National Laboratories dramatically reduces the cost of these computations, however detailed information about the costs of specific computational capabilities at National Laboratories is not readily available.

To better understand the needs for quantum physics simulation at LANL, this project analysed the workloads conducted on Chicoma in 2022 to understand how much of this compute might benefit from quantum computation.
In 2022 the Chicoma computer supported around 500 million core-hours of research spanning more than 100 individual research projects.
However, the bulk of total capacity (75\%) of the system was consumed by approximately 30 flagship projects.
These 30 flagship projects spanned a range of subject areas including: Plasma Physics, Chemistry, Biology, Materials Science, Earth Science, Astrophysics, and Particle Physics.
The relative breakdown of the total compute into these subject areas is shown in Figure \ref{fig:lanl-hpc-fy22}.
Plasma physics and Material Science were the clear leading workloads followed by significant contributions from Astrophysics, Chemistry and Earth Science models.

\begin{wrapfigure}{r}{0.45\textwidth}
    \centering
    \includegraphics[width=0.25\textwidth]{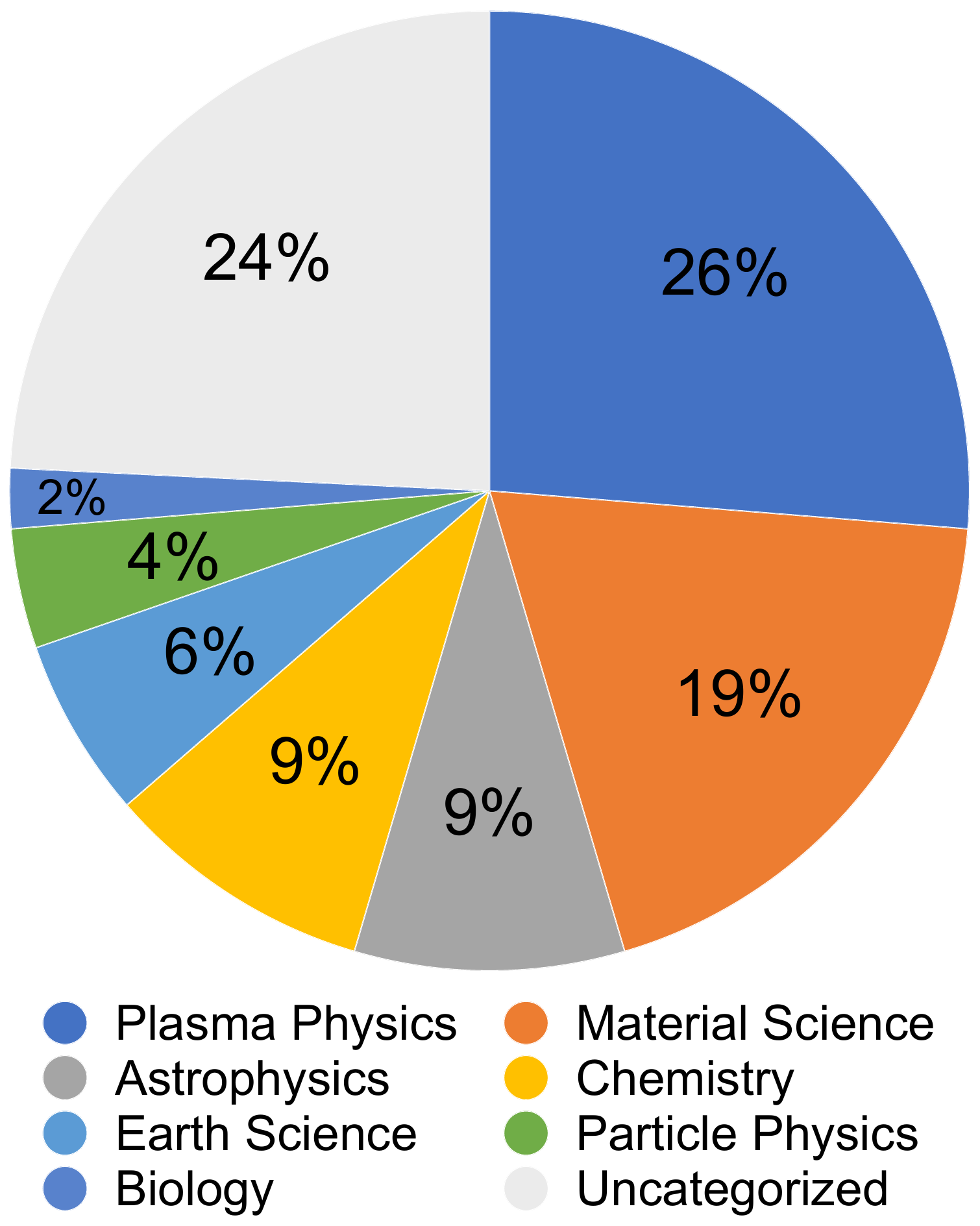}
    \caption{A summary of the application topics that were studied using 500 million core-hours of computing workload on LANL's Chicoma supercomputer in 2022. The workloads in Materials Science, Chemistry and Particle Physics focused on simulations of quantum physics, indicating that around 33\% of this computing has the potential to benefit from quantum computation.}
    \label{fig:lanl-hpc-fy22}
\end{wrapfigure}
Given this project's interest in quantum computation, the specific types of physics simulation that were conducted in each of these 30 flagship projects was also investigated.
This investigation identified that the majority of computations conducted in the Chemistry, Materials Science and Particle Physics subjects were simulations of quantum physics.
These workloads included both exact methods for quantum simulation (e.g., diagonalization) and approximate methods such as Density Functional Theory (DFT), Tensor Netowrk Methods (e.g., DMRG) and Quantum Monte Carlo methods; all of which have notable limitations in scalability or accuracy.  
Overall, these results indicate that 33\% of Chicoma's workload in 2022 were quantum physics simulations, which are natural candidates to benefit from quantum computation.
There is an ongoing debate in the research community whether quantum computers will be able to improve classical physics simulation, such as the Plasma Physics, Biology, Earth Science and Astrophysics workloads identified in this study.
If this turns out to be the case, then it reasonable to assume that a quantum computer might benefit more than $75\%$ of Chicoma's workloads. 

Based on this analysis, it is clear that there is a significant demand at LANL for quantum physics simulations, which would likely benefit from quantum computing technology.
However, note that computing capability of Chicoma is less than $1\%$ of the total compute resource in the Department of Energy.
If similar demand for quantum simulation exists broadly across the entire department, that would represent a large demand for computing technology that specializes in quantum simulation tasks.

\subsection{Quantum Physics Experimentation}

In scientific research topics where computer simulation is inadequate for hypothesis testing, laboratory experiments are the sole method for testing theories and making new discoveries.
In the context of quantum physics, the Large Hadron Collider (LHC) stands out as a widely known example. 
This experimental facility was provisioned to investigate the properties of subatomic particles, including but not limited to confirming the existence of the Higgs-Boson particle.
This experimental facility cost more than \$4 Billion USD and has been producing scientific insights in particle physics since 2010.

The LHC is one example of an experimental user facility, where capital investments in extreme technologies enable experimental scientists to make unique observations of physical systems.
However, a wide range of smaller scale facilities exist to provide insights into a variety different aspects of physics.
Two examples of such facilities at LANL that are particularly interested in quantum physics are the National High Magnetic Field Laboratory (MAGLAB) \cite{nationalmaglabNationalMagLab}, which specializes in exposing materials to extremely strong magnetic fields (up to 100 Tesla) to probe the quantum limits of materials and the Center for Integrated Nanotechnologies (CINT) \cite{osti_1764870}, which specializes in the fabrication and characterization of materials at nanoscales, including those exhibiting quantum mechanical properties.
The productivity of both of these facilities would be increased by improved computer simulation capabilities of quantum mechanical systems.
Indeed, these experimental physics facilities are a significant source for the quantum simulations workloads that are run on LANL's HPC facilities.

In the context of experimental physics, computer simulations provide valuable insights in several ways.
First, using a computer simulation to reproduce observations that have been made in a experimental facility confirms an understanding of the specific physics that are driving the observed phenomena.
Second, once a simulation can reproduce experimental data it can be inspected in a variety of ways that are not feasible in the laboratory, providing deeper insights into underlying mechanisms.
Third, computer simulations can be used to automate the search through a large collection of physical systems, to identify the most promising candidates for experimental validation.
This report includes several examples of how a quantum computer could be used to provide these kinds of insights to experimental quantum research being conducted at LANL.

\end{section}

\begin{section}{Utility Estimation Methods} 
\label{sec:ue-methods}

In this section, we provide an introduction to the various approaches to estimating the utility of a given set of computational tasks. Each computational task of value is termed an \emph{application} in this document and are presented in the subsequent area-specific chapters. Broadly, the set of applications can be divided into two categories; (i) scientific utility, and (ii) commercial utility. The scientific utility category contains applications that are of value to researchers in a research facility such as national laboratories, universities and other research organizations. The commercial utility category contains applications with direct economic value to various industries such as the chemical manufacturing and drug-design, among others. All the applications presented in this document belong to the scientific utility category. Although many of them are expected to have commercial utility, we do not pursue commercial utility estimation. Regardless of the category, the added value provided by a fault-tolerant quantum computer with appropriate specifications can be attributed to one or more of its following capabilities.

\begin{itemize}
\item \emph{Speed:} Solve a computational task much faster than current  methods. 
\item \emph{Accuracy:} Find a solution with higher accuracy than current methods. 
\item \emph{Scalability:} Solve a problem at a larger scale than is possible with current methods. 
\end{itemize}
Although scalability and speed are both a consequence of faster computational capability, it is natural from an application perspective to make a distinction between them by comparison to current (classical computing) approaches to the problem. For an application where a faster solution time is desired, for example to solve a larger number of instances of the problem, we attribute it to the \emph{speed} feature. If instead, a larger problem instance is desired, for example to eliminate finite-size effects, then we attribute it to the \emph{scalability} feature. 

Two main approaches are employed in this document for scientific utility estimation. Our goal is to use them to provide a numerical monetary value to each application instance (orange highlighted boxes in subsequent chapters) while maintaining a bias towards under-estimating these values. These approaches are described in the following sections, along with the estimation methods, and the caveats and uncertainties inherent to them.

\subsection{Method 1: Offsetting current computational investments}
In many research projects, it is necessary to obtain solutions to highly complex models. Consequently, a significant fraction of the research budget is dedicated to affording computational capabilities ranging from individual computing nodes to high performance computing (HPC) platforms. Suppose that in a given research project, an average of $D$ dollars per application instance is spent on computing capabilities. We determine that a quantum computer can provide utility for this research project because it can solve these instances \emph{at least as well as the currently utilized computing capabilities}, because of one or more of its \emph{speed, accuracy} and \emph{scalability} features. Then we can conclude that the value of a quantum computation $U$ for this application is at least $D$ per application instance on average, i.e.
\begin{align}
    U \geq D.
\end{align}
Method $1$ is employed to estimate the numerical utility value of the application instance provided in Chapter~\ref{ch:fermi-hubbard} for studying phase diagrams of Fermi-Hubbard models in the context of high-temperature superconductivity, in Chapter~\ref{ch:q_chem} for quantum chemistry applications in catalyst design.

\subsection{Method 2: Enabling successful completion of a research project}
A research project is funded with the goal of producing answers to a specific set of project-defining research questions. Some examples of such project-defining questions are: ``determine if a given material has a spin liquid phase", or ``find the optimal reaction pathway for a given chemical reaction". If the project has a total budget of $B$ dollars, we can then assign a value of $B$ dollars for answering the set of research questions. Depending on the project area, achieving this goal can require a combination of experimental, theoretical and computational capabilities and efforts. Ideally, the fraction  $\alpha_{comp}$ of the budget required for computational efforts is accurately estimated and allocated. Suppose that a set of $N_Q$ application instances cover the computational part of the project-defining research questions, then each application instance has an average utility of 
\begin{align}
    U = \frac{\alpha_{comp} B}{N_Q}.
\end{align}

While the number of required application instances $N_Q$ can be accurately determined, the total budget $B$ and the fraction dedicated to computational tasks $\alpha_{comp}$ may or may not be available. In certain cases, a project is part of an ongoing effort and is expected to continue for several more years in the future with the project-defining research questions remaining unchanged. In this case, we determine the total budget of the project by multiplying the expenditure of the project thus far by an estimated constant greater than one to account for the total duration of the project. 

In many projects the fraction $\alpha_{comp}$ is not specified or allocated and is determined on-the-fly using the best judgement of the performers in the project. Therefore, $\alpha_{comp}$ must be estimated and this estimate is subject to significant uncertainty. A particularly challenging scenario occurs when the application instances are well beyond the capabilities of classical computing hardware often due to their scaling requirements. In this case, researchers use alternative approaches such as approximation algorithms, heuristics, reduced models and experimental protocols, thus leading to $\alpha_{comp}$ being excessively underestimated when using historical data.  However, although these alternative approaches provide scientific progress, a quantum computer with the appropriate specifications will be the most efficient and therefore the preferred choice owing to its \emph{scalability} feature. In such scenarios, we provide estimates of $\alpha_{comp}$ by direct communication with project performers and by comparing to other projects where this data is available.

Method $2$ is used to provide numerical utility value of application instances in Chapter~\ref{ch:maglab} for the the study of magnetic materials in the MAGLAB user facility, Chapter~\ref{ch:neutron_scattering} for the study of exotic phases in spin systems and Chapter~\ref{ch:optical_cavity} for analyzing the ultra-strong coupling regime in optical cavities.

\end{section}

\begin{section}{Quantum Computational Kernels} 
\label{sec:qc-methods}

The promise of quantum computing is to develop algorithms and methods for solving problems more efficiently than classical computers. Among the multiple quantum algorithms developed over the past decades, we have selected those that solve tasks that we believe have the highest chances to be implemented in the near future. In agreement with Feynman's insight~\cite{feynman1982simulating}, the tasks that we have identified as the most promising for quantum computers to solve pertain to scientific computation. The general concept is that quantum computers are efficient tools for finding properties of quantum systems. Four general tasks, which we call Quantum Computational Kernels, were found to be central for speeding-up current scientific computations. These four tasks are \emph{Hamiltonian Simulation}, \emph{Open Quantum System Simulation}, \emph{Ground-state Preparation} and \emph{Thermal State Preparation}. The goal of the first two kernels is to simulate the time evolution of an isolated and open quantum system respectively, while the goal of the last two is to prepare the lowest energy state or a thermal mixed state of a quantum system respectively. 

In what follows, we expand in more detail about the nature of these quantum kernels and provide a list of developed quantum algorithms that achieve these tasks based on our current literature survey, quantum algorithms reviews~\cite{Dalzell:2023ywa,lin2022lecture,au2023quantum} and~\cite{Nielsen_Chuang}. The main strengths and weaknesses of each method are briefly reviewed as well as their computational complexity.
Note that in addition to these algorithms, there are two main hidden computational costs that need to be considered based on specific applications. Namely, \emph{initial state preparation} and  \emph{information extraction}. While preparing the initial state of the register is usually considered a separate task in literature, it can come with a prohibitive cost, see for instance~\cite{aaronson2015read}. Similarly, computing observables can be a challenging task as reconstructing the full final state of a quantum system is an inherently inefficient procedure \cite{haah2016sample}. However, several strategies exist for constructing highly efficient procedures to obtain estimations of a large numbers of few-qubit observables~\cite{cotler2020quantum, yang2023experimental, huang2020predicting}.


\subsection{Hamiltonian Simulation}\label{Hamiltonian}
Hamiltonian simulation is the fundamental computational primitive that can be performed efficiently on a quantum computer. The general task on a $n-$qubit system can be stated as follows; given a Hamiltonian operator $H \in \mathbb{C}^{2^n \times 2^n}$ and a time $t$ apply the unitary time evolution operator, $\exp(-iHt)$ to the $n-$qubit register.

The relevance of this task comes from the fact that this time evolution operator is the solution to the Schr\"{o}dinger equation that governs how a  quantum system with a Hamiltonian $H(t)$ evolves in time,
\begin{equation}\label{eq:schrodinger}
    \dfrac{\partial \ket{\psi(t)} }{\partial t} = -i H(t) \ket{\psi(t)}.
\end{equation}

The solution to this equation for a given initial state  tells us how the quantum state of a system evolves in time. If the Hamiltonian is not a function of time (i.e. $H(t) = H$), it can be shown that the state of a quantum system, initially in a state $\ket{\psi}$, evolves to the state $\exp(-iHt) \ket{\psi}$ after time $t$ \cite{Nielsen_Chuang}. 
This time independence assumption  on the Hamiltonian does not hold for some systems, and relaxing this assumption leads to time-dependent Hamiltonian simulation problem. Simulating time-dependent systems is a more complex algorithmic task. 

 The Hamiltonian for an $n-$qubit system is a $2^n \times 2^n$ matrix. The task in time-independent simulation is simply to apply the operator $e^{-iHt}$ to a quantum register. For physically relevant systems, the Hamiltonian often has several features that makes this task feasible in time that scales at worst as a polynomial in the number of qubits of the system and the evolution time $t$. For Hamiltonians of locally interacting systems, there is strong theoretical evidence that the complexity of this task in the worst case cannot scale sub-linearly i.e. scaling slower than $nt$ \cite{berry2007efficient,childs2010limitations}. However for a certain subsets of Hamiltonians this evolution can be simulated faster \cite{gu2021fast}.

\begin{itemize}
    
    \item \textbf{Common features of the Hamiltonian.} A commonly assumed feature of the Hamiltonian is \emph{locality}, i.e. it only contains terms that describe interactions between qubits/particles that are close to each other in space. The most favorable assumption that one can make in this context is strict geometric locality, i.e. qubits/particles only interact with their nearest neighbors in a $D-$dimensional lattice. This assumption is true for many Hamiltonians that come up frequently in condensed matter or high-energy physics. Another assumed feature of Hamiltonians is \emph{sparsity}, i.e. the number of non-zero elements in each row of the matrix is significantly smaller than $2^n$. Locality can often be shown to imply sparsity.
    
   Physical systems can also have interactions that whose strength decay as an inverse polynomial of distance between particles (e.g. Coulomb interactions). Hamiltonians of such systems are harder to simulate compared to the strictly local systems.
    
    \item \textbf{Specification of Hamiltonian.} The Hamiltonian is an exponentially large matrix, so for efficient quantum simulation it is necessary that there exists a way to efficiently compute the elements of this matrix. The most common way to represent Hamiltonians in this way is to write them as a sum of Pauli terms, where each term contains only Pauli-matrices on only a few qubits,
    \begin{equation} \label{eq:pauli_model}
        H = \sum_{l = 1}^L H_l.
    \end{equation}
For instance, in the well studied Heisenberg model Hamiltonian in 1D, $H_l$ would be a term of the form $X_lX_{l+1} + Y_l Y_{l+1} + Z_l Z_{l+1}$ which is supported on a nearest neighbour pair of qubits.

Some quantum simulation algorithms assume a different access model based on an oracle (i.e. a black box function) that can compute in coherent superposition the matrix entries of the Hamiltonian given the corresponding indices \cite{childs2010relationship}. Such an oracle can be constructed with polynomial overhead from the representation in Equation \eqref{eq:pauli_model}.

    \item \textbf{Error definitions.} The runtime complexity of Hamiltonian simulation algorithms is often specified in terms of an error guarantee $\epsilon.$ This error is measured in terms of operator norm, which is defined as the largest eigenvalue of an operator in absolute value. Simulation algorithms give a method to efficiently implement an operator $U$ such that $|| U - e^{-iHt}|| \leq \epsilon.$

\end{itemize}

\subsubsection{Algorithms for time-independent Hamiltonians}
An extensive amount of work has been done on the problem of time-independent Hamiltonian simulation. Often, complexity bounds in literature are stated in terms of the operator norm $||H||.$ Keep in mind that in many physical systems this quantity is linear (or of higher-order) in the number of particles in the system.
\begin{itemize}
    \item \textbf{Trotter product formulas.}
Product formulas give simple ways to simulate the time evolution of $H$ using time evolution of few-body interaction terms (i.e. the $H_l$ in Equation \eqref{eq:pauli_model}), which make up the full Hamiltonian. For the simplest case of simulation $H = H_1 + H_2$ for time $t$, the second order Trotter product formula gives the following recipe for simulation,
\begin{equation}
    e^{-i H t} =  (e^{-i (H_1 + H_2) \Delta t})^{N} \approx  (e^{-i H_1 \Delta t/2} e^{-i H_2 \Delta t} e^{-i H_1 \Delta t/2} )^{N},~~~~~ 
\end{equation}

Here the evolution is broken up into $N$ time steps of time $\Delta t$ and the evolution in each time step is approximated using the second-order Trotter formula. Higher-order product formulas are more accurate but a single step requires more overhead to implement. It can be shown that achieving a final error of $\epsilon$ in the simulation using  a $(2k)$-th order formula for a physical Hamiltonian  requires a circuit with a  gate-count that scales as $O(\alpha^{1/{2k}}_{k,comm}5^{2k} nL \frac{t^{1 + 1/{2k}}}{\epsilon^{1/2k}}).$ Here $\alpha_{k,comm}$ is a quantity that captures the commutativity of the terms in the Hamiltonian \cite{childs2021theory,Dalzell:2023ywa}. In theory and in practice, the performance of these algorithms can be improved if the Hamiltonian has many terms that commute with each other.

Trotter product formulas are simple to implement and work well in practice. These methods do not require ancilla qubits or extra control operations. These features make them ideal for implementation in near-term quantum computers.
However, there are  two main drawbacks: higher order product formulas lead to higher circuit depth and they do not scale well in terms of the error $\epsilon$. 

\item \textbf{Randomization and qDRIFT.}
It has been observed that randomizing the order in which terms are applied in a Trotter formula can improve its performance \cite{childs2019faster} . A popular algorithm that uses randomization is \emph{qDRIFT}. This method has a scaling that is independent of the number of terms in the Hamiltonian $(L)$ \cite{campbell2019random}. The downside is that the complexity scales quadratically in $t$. This might be useful for cases where the Hamiltonian is dominated by a few terms, refer to \cite{chen2021concentration} for details.

\item \textbf{HHKL algorithm for simulation on lattices.} 
For Hamiltonians defined on a $D$-dimensional lattice with geometric locality, the algorithm described in \cite{haah2018QAlgSimLatticeHam} can achieve $\Tilde{O}(nt \log(\frac{1}{\epsilon}))$ scaling. The algorithm is similar to a product formula and exploits locality of quantum evolution to optimally break it up into smaller pieces. This gives a circuit complexity that is asymptotically optimal with much improved dependence on error compared to Trotter product formulas. However, the algorithm is not applicable to more complex geometries.

\item \textbf{LCU series expansion method.}
For general sparse Hamiltonians, a method based on series expansion can give better asymptotic scaling compared to product formulas. These algorithms exploit various expansions (Taylor series, Dyson series etc..) to write $e^{-iHt}$ as a weighted sum of easily implementable unitary operations \cite{kothari2014PhDThesis,berry2014HamSimTaylor}. Such a series can be implemented on a quantum computer using a technique known as Linear Combination of Unitaries (LCU). The complexity of this method for simulating an $s-$sparse Hamiltonian is $O(s^2 nt ||H||_{max} \log(1/\epsilon))$ given Oracle access to such a Hamiltonian ($||H||_{max}$ scales with $n$ for physical systems). This algorithm only succeeds with a certain probability and hence requires an amplitude amplification step at the end to boost this success probability to an acceptable level. The main drawback of this method is identified as the potentially large number of ancillas required for implementation \cite{childs2018toward}.

\item \textbf{Qubitization and QSVT.}
These are related techniques that work by manipulating the so-called \emph{block encoding} of the Hamiltonian, which is a larger unitary matrix that has a matrix proportional to $H$ as a submatrix. Such a block encoding necessarily requires ancillary qubits but fewer ancillas are required compared to LCU. The complexity of these methods depend heavily on the various input models assumed  \cite{low2019hamiltonian}. For simulation on lattices the complexity is seen to scale with $n^2t$ \cite{haah2018QAlgSimLatticeHam}. Quantum Singular Value Transformation (QSVT) is a generalization of qubitization and some related techniques \cite{low2017optimal} These allow for implementing any function $f(H)$ of the Hamiltonian, given an appropriate block encoding of $H$ \cite{gilyen2018QSingValTransfThesis,gilyen2018QSingValTransfArXiv}. These techniques are not as simple to implement as product formulas, but they have asymptotically better complexity in terms of the error parameter.
\end{itemize}

\subsubsection{Time-dependent Hamiltonians}
Time-dependent Hamiltonian simulation is a strictly harder task than time-independent simulation. However, many ideas from the time-independent case carry over. Using ideas from Trotterization, product formulas have been developed that can be used in the time dependent case \cite{wiebe2010TimeDepTrotter}. The main issue with this approach is the strong dependence of the complexity on the rate of change of the Hamiltonian. This issue can be mitigated by using a randomization approach \cite{poulin2011TimeDepTrotterRandomized}. The HHKL algorithm \cite{haah2018QAlgSimLatticeHam} also works natively for time-dependent geometrically local time-dependent Hamiltonians. 

Time-dependent evolution can also be approximated using a Dyson series and simulated using the LCU technique discussed above \cite{low2018hamiltonian,kieferova2019DysonSeriesSimulation}. These methods lead to improved dependence in the rate of change of the Hamiltonian but are more difficult to implement. Developing a time-dependent equivalent of QSVT type techniques is an open problem \cite{Dalzell:2023ywa}.

\subsubsection{Simulating Fermionic Hamiltonians}
A large fraction of Hamiltonians seen in areas such as quantum chemistry or condensed matter physics  describe the physics of fermions. Fermionic Hamiltonians require some overhead to be mapped to  Hamiltonians that describe interactions between qubits. The main reason for this overhead is the following fact: Fermionic operators that have disjoint supports anti-commute and this constraint is highly non-trivial to satisfy for a system of qubits. Following are some techniques used in literature to map fermionic problems to qubits.

\begin{itemize}
    \item \textbf{Jordan-Wigner}: This is the standard conversion that is used between fermions and qubits. However for Hamiltonians that are not one-dimensional, this will turn a local Fermionic Hamiltonian into a highly non-local qubit Hamiltonian. This produces high-weight Pauli terms in the qubit Hamiltonian that can be challenging for simulation algorithms \cite{tranter2018comparison}.
    \item \textbf{Bravyi-Kitaev}: This is similar to Jordan-Wigner and produces a non-local Hamiltonian. But the mapping itself is more efficient in theory. Both Jordan-Wigner and Bravyi-Kiratev does not require ancillary qubits to do the mapping \cite{bravyi2002fermionic}.
    \item \textbf{Techniques using auxiliary fermions}: These preserve locality of the Hamiltonian but require ancillary qubits. Key idea is to use extra fermions to mediate interactions so as to preserve locality. Refer \cite{verstraete2005mapping} and the corresponding section in \cite{haah2018QAlgSimLatticeHam} for one such technique.
    \item \textbf{First quantization}: For electronic Hamiltonians the issue of anti-commutation can be avoided if the simulation is performed in the first-quantized representation \cite{su2021fault,low2018hamiltonian}. These approaches require that the initial state be prepared to respect the exchange symmetry of electronic wave functions. While this is a promising approach for quantum chemistry problems, it is under-explored in other areas like QCD.
\end{itemize}
\subsection{Open quantum systems}
Consider a more general version of Hamiltonian simulation that applies to an open quantum system that is interacting with an environment. The dynamics of an open quantum system under Markovian assumptions is governed by the Lindblad master equation \cite{manzano2020short}. The Lindblad equation describes a  non-unitary dynamics and this makes it more challenging to simulate on a closed quantum computer.
The dynamics of the system qubits alone (now given as a density matrix $\rho$), can be described by a \emph{Lindblad master equation}:
\begin{align}
    \label{eq:lindbladian}
    \frac{\partial \rho(t)}{\partial t} 
       =  \calL \rho(t) := \underbrace{-i \left[H,\rho\right]}_{\substack{\text{\large coherent part}\\ \text{\small (Schrödinger equation)}}}
        + ~~\underbrace{\sum_{j=1}^{m} \overbrace{\left(L_j \rho L_j^{\dagger}\right)}^{\text{transition}} - \overbrace{\tfrac{1}{2}\left( L_j^{\dagger}L_j \rho + \rho L_j^{\dagger}L_j \right)}^{\text{decay}}}_{\text{\large dissipative part}}
\end{align}
The operators $L_i$ define the interaction between the system and the environment and have to specified as part of the input to the algorithm. Many ideas from Hamiltoninan simulation can be generalized to the case of open quantum systems.
\begin{itemize}
\item \textbf{Simulation of environment}: Simulation of open quantum systems can be mapped to the Hamiltonian simulation problem by using ancilla qubits to represent the environment and then simulating the system and ancilla qubits together as the larger closed system. Then any property of the system can be estimated by measuring the system qubits alone \cite{terhal2000problem}. This technique has a large overhead in terms of ancillas. Recent work by Ding et.al gives another such prescription to map open system dynamics to a unitary evolution of a system with ancillary qubits. The open nature of the evolution in this setting is imposed by measuring the ancillary system periodically. This will require the introduction of fresh ancillas after every step of the evolution algorithm \cite{ding2024simulating}. 

    \item \textbf{Trotterization for Lindbladians}: 
    The solution of the Lindblad master equation can be given by the super operator (i.e. a linear operator acting on the space of operators) $e^{t \calL}.$ Trotter product formulas can be developed for this operator.
    Reference \cite{childs2016SparseLindbladianSim}   gives a Trotterization framework for such open evolutions improving on \cite{kliesch2011dissipative}. An experimental implementation can be found in \cite{han2021experimental}. Going beyond second-order trotter formulas for such evolutions is an open problem.
    \item \textbf{Series expansion}:  A generalized LCU technique can be used to simulate the Lindblad equation \cite{cleve2016EffLindbladianSim}. Just like for closed quantum system, this technique can improve asymptotic scaling of the algorithm but requires ancillary control to be implemented.
    
\end{itemize}

\subsection{Finding ground states}
The ground state of a system is the quantum state with the lowest energy. Given the system Hamiltonian, the ground state is also the eigenstate with the lowest eigenvalue.  In terms of worst-case complexity, finding the ground state is thought of as a harder problem than simulating the Hamiltonian \cite{kempe2006complexity}. But it is far less clear if the Hamiltonians found in nature correspond to  worst-case instances for this problem.  Quantum simulation is often used as a subroutine in algorithms that aim to find the ground state. The main quantum algorithms for finding ground states are the following:
\begin{enumerate}
    \item \textbf{Quantum Phase Estimation (QPE)}:  Given an $n-$qubit Hamiltonian with the spectral decomposition, $H = \sum_i E_i \ket{\lambda_i} \bra{\lambda_i}$ with $E_i \in [0,1]$ and an initial state $\ket{\psi} = \sum_i \psi_i \ket{\lambda_i}$, the QPE algorithm essentially implements the following transformation,
    \begin{equation}\label{eq:QPE}
        \sum_i \psi_i \underbrace{\ket{0}}_{t\text{ ancillas }}\underbrace{\ket{\lambda_i} }_{n  \text{ qubit system}}  \xrightarrow[]{QPE }~~~\sum_i \psi_i \ket{\Tilde{E_i}}\ket{\lambda_i}
    \end{equation}
    Here $\tilde{E}_i$  is a $t$ bit approximation of the energy eigenvalue $E_i$. For physical systems, the constraint $E_i \in [0,1]$ can be satisfied by scaling the Hamiltonian. The QPE circuit consists of a sequence of quantum simulation steps controlled on the ancilla register followed by quantum Fourier transform. Measuring the ancilla register then gives us the state $\ket{\Tilde{E_i}}\ket{\lambda_i}$ with probability $|\psi_i|^2$. Thus, if the initial state $\ket{\psi}$ is prepared such that it has a significant overlap with the ground state, then this measurement will have a correspondingly large probability to prepare the ground state. The smaller this overlap the more number of times the QPE circuit will have to be run  to guarantee the preparation of the ground state. Also notice that the number of ancillary qubits ($t$) sets the ability of the algorithm to resolve different eigenstates, hence the number of ancillas required will depend on the energy gap between the ground state and the first excited states. The number of ancillas required to have $\epsilon$ accuracy in eigenvalues with a  success probability of $ 1 -\delta$ is  $t = \lceil \log(\frac{1}{\epsilon}) \rceil + \lceil \log(2 + \frac{1}{2\delta}) \rceil$ \cite{Nielsen_Chuang,lin2022lecture}. 

    Additional errors can be incurred in QPE due to errors coming from the Hamiltonian simulation steps in QPE. For Hamiltonian simulation using Trotterization, this error is studied in Ref. \cite{kivlichan2020improved}.    The cost of Hamiltonian simulation can be offset by using QPE directly with the qubitized version of the Hamiltonian \cite{poulin2018SpectralQubitization}. A recent efficient implementation of QPE can be found in \cite{ding2023even}. The key issue in using QPE is finding a good initial state which has a significant overlap with the ground state. Existing classical approaches like Hartree-Fock can help here and QPE can be used to further refine them \cite{tubman2018postponing}.
   
    \item \textbf{Adiabatic theorem.} The adiabatic theorem is a result that converts the problem of ground state preparation to one of time-dependent Hamiltonian simulation \cite{ambainis2004elementary,albash2018adiabatic}. For $t \in [0,T]$, consider the following time dependent Hamiltonian, $H(t) = (1 - \frac{t}{T})H_i + \frac{t}{T}H_f$ and a system whose dynamics is governed by this Hamiltonian via \eqref{eq:schrodinger}. If the initial state of this system is prepared to be ground state of $H_i$, then for large enough $T$,  the state at time $T$ can be shown to have a large overlap with the ground state of $H_f.$ $H_i$ is often taken to be Hamiltonian whose ground state is easy to prepare and $H_f$  is taken to the Hamiltonian whose ground state we are interested in preparing. The interpolation between these two Hamiltonians can be a general non-linear function and in some cases this is advantageous \cite{roland2002quantum}.  There is a theoretical characterization of how large $T$ should be based on some properties of the interpolation used and hence cannot be generally bounded.  Ref. \cite{wecker2015solving} gives an end-to-end description of an adiabatic approach for the Hubbard model.
    The adiabatic theorem can also be used in conjunction with QPE to prepare ground states \cite{boixo2010fast}. 

    \item \textbf{Variational approaches \cite{cerezo2021variational, tilly2022variational}.} Variational approaches work constructing by a quantum circuit with variational parameters which are then optimized to minimize the energy of the system. These techniques usually have a hybrid quantum-classical nature. The energy of the ansatz is first estimated by measuring an observable by repeatedly applying this circuit to a register and then measuring it. Then the variational parameters of the circuit are updated using an optimization algorithm running on a classical computer. The promise of these methods come from the fact that variational quantum circuits have enough expressibility to represent highly entangled states efficiently. It is impossible to give theoretical guarantees for these types of methods as the quality of their solution depends on the the choice of the variational circuit. Just like for classical variational methods, an effective ansatz would be one which can represent the ground state with only a few variational parameters.
\end{enumerate}

\subsection{Preparation of thermal states}
Preparation of thermal states is an important algorithmic task that can be seen as a generalization of finding ground states. The state of a quantum system with Hamiltonian $H$ at thermal equilibrium with a bath of inverse temperature $\beta$ is given by the mixed state $\rho = \exp(-\beta H)/Z.$ Here $Z$ is known as partition function which is an important thermodynamic quantity that also fixes the normalization of the state. The ability to prepare copies of such states or compute observable from them is crucial in studying quantum systems in thermal equilibrium. We can see that in the $\beta \rightarrow \infty$ limit the thermal state reduces to the ground state of $H$. In practice, this problem is often harder for higher values of $\beta$ than for lower values. The exact algorithmic task here is to compute observables from the state $\rho \propto \exp(-\beta H)$ given access to a Hamiltonian $H$. A more technical overview of the algorithms known for this task can be found in Reference \cite{chen2023quantum}.

\begin{itemize}
    \item \textbf{Series expansion}: Reference  \cite{chowdhury2016QGibbsSampling} shows the application of the LCU technique  to the problem of thermal state preparation. The key innovation is a cleverly constructed series approximation (a discretized Hubbard-Stratonovich transform) for the exponential function. The complexity of this algorithm scales as $O(\sqrt{ \frac{2^{n/2}\beta}{Z}} poly(\log(\frac{1}{\epsilon}) ).$ This is very inefficient in the worst case.
    \item \textbf{Using Quantum Singular Value Transforms (QSVT)}: QSVT is meta-algorithm that can be seen as a generalization of the qubitization procedure. For a given function $f$ and a Hamiltonian $H$, this technique allows us to construct operators proportional to $f(H).$  QSVT techniques can also be used  with a polynomial approximation for the exponential function to achieve this task (see Section 5 in Reference \cite{gilyen2018QSingValTransfArXiv} for detailed dsicussion on complexities).
\item \textbf{Quantum Gibbs sampling}: Classically thermal state distributions are most easily prepared by Monte-Carlo-Markov-Chain (MCMC) samplers (Glauber dynamics, Metropolis-Hastings etc.). At a high level these methods exploit the physical fact that a carefully constructed open system evolution eventually converges to a thermal state (i.e by cooling/heating by a thermal bath). This opens the possibility of using open quantum system evolution to prepare quantum thermal states. These algorithms are most closely related to how thermal states are prepared by natural processes  \cite{chen2023quantum, chen2023efficient, rall2023thermal}.  Just like for classical MCMC samplers, the time taken for such algorithms to converge to the thermal states cannot be tightly upper bounded in general. However, each step of the algorithm can be implemented efficiently. We should note that these class of algorithms are very recent and as of writing this it is not clear which proposal will be the most useful one in practice.
\item \textbf{QPE-based methods}: 
Another way to leverage classical MCMC to prepare quantum thermal states is to run MCMC directly in the energy eigenbasis of the Hamiltonian  \cite{temme2011quantumMetropolis}. These methods use the QPE transform to achieve this task. The disadvantage of this method  compared to quantum Gibbs sampling is the large overhead of the QPE, induced by very high precision requirements in the energy estimation task \cite{chen2023quantum}. Initial attempts to fix this inefficiency using a boosted version of QPE were later shown to be incorrect \cite{chen2023efficient,chen2023quantum}.
\end{itemize}
\end{section}
\newpage
\end{chapter}

\begin{chapter}{Experimental analysis of magnetic materials at the MAGLAB user facility} \label{ch:maglab}

This chapter presents several applications relevant to the MAGLAB user facility. The MAGLAB is an experimental facility that analyzes several magnetic materials of interest. These include Kitaev quantum spin liquids (KQSLs), multi-ferroic materials for memory and high temperature superconductivity. We focus on the applications of quantum computing in the study of KQSLs, where the broad research goal is to identify the effective spin Hamiltonian and search for the existence of the KQSL phase in the phase space. The computational capabilities required are methods for quantum Hamiltonian simulation and computation of ground states for spin Hamiltonians on given regular two-dimensional lattices. Current classical approaches cannot be scaled to sufficiently large system sizes ($>10000$ sites) that are required to avoid significant errors from finite size effects. Consequently, the primary potential benefit of a quantum computer is the ability to perform these computations at a scale where the results accurately represent experimental observables and properties of the material. The following table summarizes the computational requirements that must be met to be of value to research at the MAGLAB. \\

\noindent \textbf{Hamiltonian Type}:  Spin Kitaev/Heisenberg Hamiltonian\\
\textbf{Quantum Computational Kernels}: Ground State Preparation, Hamiltonian simulation.



\section{Application area overview}

\subsection{High-level description of the application area}

Materials that exhibit strong spin-spin interactions typically manifest long-range magnetic order. When subjected to a magnetic field, these materials undergo a spin structure evolution that is reflected in distinct magnetization curves. Analyzing these curves enables the deduction of the spin Hamiltonian governing the magnetic systems. The recent identification of potential quantum spin liquid (QSL) candidates has generated considerable excitement in the condensed matter community \cite{zhou2017quantum, kitaev2006anyons,trebst2022kitaev,savary2016quantum,broholm2020quantum}. However, the intricate magnetization behavior, stemming from purely quantum mechanical origins, poses challenges in interpreting experimentally measured magnetization and thermodynamic properties \cite{loidl2021proximate}. Consequently, the advent of scalable quantum computers holds great promise for efficiently deciphering complex spin Hamiltonians, facilitating a deeper understanding of quantum effects.

The quest for materials featuring a robust coupling between spin and electrical degrees of freedom is pivotal for realizing ultra-low-power devices compatible with existing industrial technology \cite{gao2021review,cheong2007multiferroics}. This is because current-based electronic devices inevitably dissipate heat, whereas controlling spin-based devices does not. The coupling between spin and electrical degrees of freedom is achieved through a robust spin-orbit coupling, which induces spin-dependent charge motion and vice versa. While current studies often rely on symmetry arguments \cite{eerenstein2006multiferroic}, the absence of simulations considering both spin and charge degrees of freedom hinders the discovery of ideal magnetoelectric materials capable of operating at room temperature with minimal magnetic and electric fields. A crucial imperative lies in formulating a Hamiltonian that encompasses both charge and spin degrees of freedom, utilizing extensive magnetization and electric polarization data accumulated over the years. Identifying key parameters that enhance operational temperatures to room temperature is paramount for guiding materials design in this pursuit.

Aforementioned studies are actively pursued in The National High Magnetic Field Laboratory (NHMFL), which operates seven user facilities spread across three campuses in the United States \cite{maglab1}. The headquarters is situated in Tallahassee, Florida, and specializes in ultra-high 'dc' magnetic fields. It boasts numerous Guinness World Records, including the most potent dc magnet capable of reaching up to 45 T \cite{maglab2}. Another branch is located in Gainesville, Florida, where ultra-low temperatures down to 300 $\mu$K can be combined with a 20 T dc magnet. The final branch is situated at Los Alamos National Laboratory, which operates pulsed magnets reaching up to 100.75 T \cite{maglab2}, making it the most powerful pulsed magnet globally. Annually, more than 1,800 researchers utilize these facilities, with most incurring no cost, and collectively, they produce over 400 peer-reviewed publications studying magnetic materials. NHMFL supports a diverse range of experimental techniques, including various magnetometry methods, torque magnetometry, electrical property measurements, and thermodynamic measurements.



\subsection{Utility estimation}
\subsubsection{Overview of the value of the application area}
The accurate identification of a spin Hamiltonian for a Quantum Spin Liquid (QSL) phase is a critical task, as it provides insights into the quasi-particle excitations that the QSL phase can accommodate. Some quasi-particle excitations within certain QSL phases exhibit valuable physical properties, paving the way for fault-tolerant topological quantum computations \cite{kitaev2003fault}. These computations hold the promise of mitigating the noise issues inherent in quantum computations \cite{lahtinen2017short}.

Furthermore, the realization and identification of devices with ultra-low energy consumption, leveraging magnetoelectric coupling, are indispensable for sustaining Moore's Law \cite{klimov2017magnetoelectric}. The conventional trajectory of Moore's Law is impeded by heat generation \cite{shalf2020future}, and the pursuit of reduced energy consumption without compromising computational capabilities is paramount.

Hence, a fundamental goal is to comprehend the magnetic and electric properties of materials, propelling our society toward smaller, faster, smarter, and more robust technology. This endeavor not only addresses the challenges posed by issues we are facing but also has the potential to usher in innovative technologies that will reshape the way we perceive and experience the world. 

\subsubsection{Concrete utility estimation}
The MAGLAB receives a budget of $\$40M$ per year \cite{maglab2022funding} which is utilized to cover labor costs, the cost of experimental equipment and materials. On average a total of $20$ materials are studied within one fiscal year \cite{maglab_budget}, thus resulting in a budget of $\$2M$ per year per material. A given material, especially the more promising ones, are studied for multiple years in a row with the goal of answering a given set of questions. The research goals for the specific material will be complete once these set of questions are answered. Following this argument, if a material is studied for $N_Y$ years with the goal of answering $N_Q$ questions, then the allocated budget for answering each question on average is $\$2M \times \frac{N_Y}{N_Q}$.

In Section~\ref{sec:KQSL}, we consider the Kitaev Quantum Spin Liquid candidate materials using $\alpha$-RuCl$_{3}$ as the concrete example. A set of four applications, of which two are presented in the chapter, cover the set of questions that constitute the current research goals. These set of questions are currently unanswered due to the computational bottlenecks of classical approaches, and the applications presented will be able to resolve this problem thus successfully achieving the research goals for $\alpha$-RuCl$_{3}$. Consequently, we assign a cumulative value for the four applications equal to the total budget for the current research goals. The material $\alpha$-RuCl$_{3}$ has been studied in the MAGLAB for at least $5$ years. Therefore, we estimate the value of each of the four applications presented to be $\$2M \times 5/4 = \$2.5M$.


\section{Problem and computational workflows: Kitaev QSLs}  \label{sec:KQSL}

\subsection{Detailed background for Kitaev QSLs}

We first define some notation that we will use for the rest of the chapter regarding the bonds and coordinates. We use $(\mathcal{X,Y,Z})$ to denote the set of edges in the lattice in Figure~\ref{fig:honeycomb lattice} corresponding to the $X, Y$ and $Z$ bonds respectively. Each site $i$ has a magnetic spin $S_i$ which can be decomposed in the Pauli basis as $S_i  = (S_i^x, S_i^y, S_i^z)$ denoting the component of the spin along the orthogonal $(\hat x,\hat y,\hat z)$ directions shown in Figure~\ref{fig:groundstate}.
Note this is different from the capitalized letters $(X,Y,Z)$ used to denote bond types. An alternative orthogonal coordinate system $(\hat a,\hat b,\hat c)$ that is relevant to the experimental protocols and observables is shown in Figure~\ref{fig:groundstate}, where $\hat c$ is the out of place component. The relationship between the two orthogonal coordinate systems is given as
\begin{align}   \label{eq:coordinate_transform}
    \begin{bmatrix}
        \hat a \\
        \hat b\\
        \hat c
    \end{bmatrix} = 
    D
    \begin{bmatrix}
        \hat x \\
        \hat y \\
        \hat z
    \end{bmatrix}, \qquad \mbox{where } D = \begin{bmatrix} 
        1/\sqrt{6} & 1/\sqrt{6} & -2/\sqrt{6} \\
        1/\sqrt{2} & -1/\sqrt{2} & 0 \\
        1/\sqrt{3} & 1/\sqrt{3} & 1/\sqrt{3}
    \end{bmatrix} 
\end{align}
Therefore, the decomposition of the spin $S_i = (S_i^a, S_i^b, S_i^c)$ along the $(\hat a,\hat b,\hat c)$ coordinates and along the $(\hat x, \hat y, \hat z)$ coordinates are given by 
\begin{align} \label{eq:spin_coordinate_transform}
    (S_i^x, S_i^y, S_i^z) = (S_i^a, S_i^b, S_i^c) \  D, \qquad (S_i^a, S_i^b, S_i^c) = (S_i^x, S_i^y, S_i^z) \ D'.
\end{align}

The quantum spin liquid phase is a magnetic state that lacks magnetic long-range ordering even at the lowest temperatures, despite strong spin-spin interactions \cite{savary2016quantum}. Although the ground state appears genuinely chaotic without any discernible local structure, the long-range quantum entanglement among spins enables the emergence of exotic excitations \cite{wen2002quantum}. These excitations are topologically protected, providing a foundation for fault-tolerant quantum computations. One prominent example of a quantum spin liquid phase predicted to host such topologically protected excitations is the Kitaev quantum spin liquid phase, proposed by Alexey Kitaev in 2003 \cite{kitaev2003fault, kitaev2006anyons}.
He examined the honeycomb lattice, where each vortex with a spin-1/2 is linked to its nearest neighbor spin through a bond-dependent spin interaction as shown in \ref{fig:honeycomb lattice}. 

 \begin{figure}
    \centering
     \includegraphics[width=0.8\textwidth]{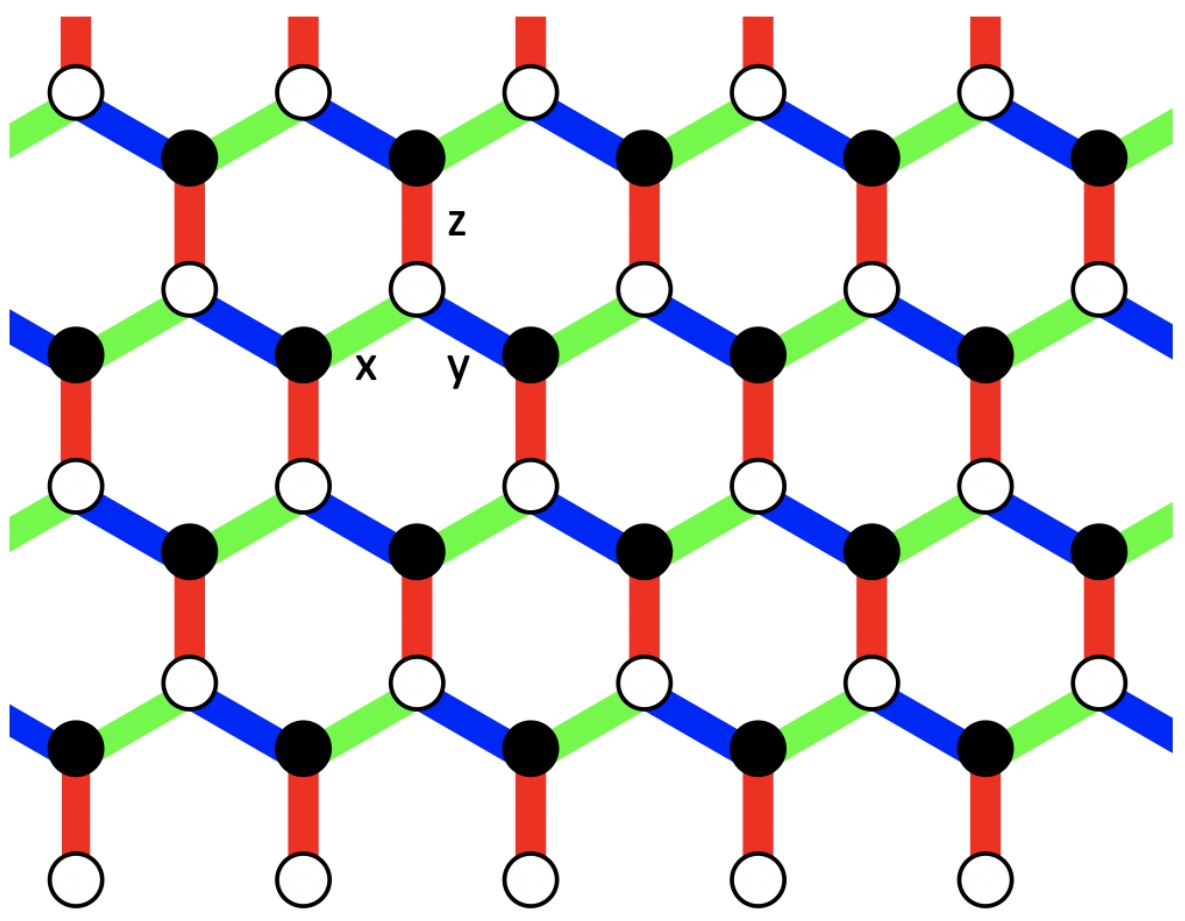}
     \caption{Each color indicates $K_{x}$, $K_{y}$, $K_{z}$ bonds \cite{roberts2022fidelity}}
     \label{fig:honeycomb lattice}
 \end{figure}
 
The spin Hamiltonian is expressed as follows.
\begin{equation}
    \mathcal{H} = K_{x}\sum_{ij \in \mathcal{X}}S_{i}^{x}S_{j}^{x} + K_{y}\sum_{ij \in \mathcal{Y}}S_{i}^{y}S_{j}^{y} + K_{z}\sum_{ij \in \mathcal{Z}}S_{i}^{z}S_{j}^{z},
    \label{eq:KHamiltonian}
\end{equation}
This Hamiltonian emerged as a rare instance of an exactly solvable model for a two-dimensional system. In its ground state, it was established that only very short-range spin-spin correlations exist at zero temperature, indicating the absence of magnetic long-range ordering \cite{kitaev2006anyons}. Additionally, two types of Majorana Fermions were identified as excitations: 
local flux excitations and itinerant Majorana Fermions. Initially serving as a useful toy model for physics exploration, subsequent theoretical investigations \cite{jackeli2009mott,rau2016spin,kim2015kitaev} demonstrated that $\alpha$-RuCl$_{3}$ could realize this Hamiltonian. However, further investigations revealed that while $\alpha$-RuCl$_{3}$ 
indeed captures the bond-dependent interactions, other unwanted terms such as Heisenberg exchange interactions up to third nearest neighbors and symmetric off-diagonal terms are also permissible \cite{rau2014generic}. The presence of these terms in the spin Hamiltonian leads to undesirable magnetic long-range ordering \cite{lee2014heisenberg}. 

It was discovered that the magnetic field destabilizes the magnetic long-range ordering and gives rise to an intermediate phase before saturating the system \cite{kasahara2018majorana}. In this intermediate state, a disordered configuration is observed, marked by the presence of half-quantized thermal Hall conductivity—a characteristic feature of a topologically nontrivial state and chiral spin liquid states \cite{kasahara2018majorana}. Nevertheless, the debate surrounding the nature of the intermediate phase persists, and the features of this state remain considerably uncertain, despite extensive experimental studies \cite{loidl2021proximate}, including magnetization measurements in various configurations. The absence of an appropriate spin Hamiltonian capable of describing the magnetic properties of the compound adds to the complexity in comprehending the nature of the intermediate state.



\subsection{Application category 1: Hamiltonian search for spin liquids}
In this category of applications, we pose the problem of determining the accurate spin Hamiltonian of candidate Kitaev QSLs using various experimental data such as net magnetization, specific heat, etc. The computational workflow will be different for the various applications in this category depending on the type of experimental data being utilized. 

We will use the compound $\alpha$-RuCl$_{3}$ as an example for this section. A similar description is relevant for several other candidate spin liquid materials. The permitted terms in the spin Hamiltonian for $\alpha$-RuCl$_{3}$ are determined by its local and global symmetry. Through symmetry arguments, researchers have derived the most general form of the spin Hamiltonian for $\alpha$-RuCl$_{3}$ \cite{rau2014generic}. This includes: 1) Heisenberg-type exchange interactions up to the third nearest neighborhood, 2) Kitaev exchange interactions up to the third nearest neighborhood, 3) Off-diagonal symmetric interactions between nearest neighbor exchange interactions, and 4) Single-ion anisotropy with the Zeeman term. 
The most generic spin Hamiltonian possesses a number of parameters as shown in Eq. (\ref{eq:genericHamiltonian}).
\begin{eqnarray}
    \nonumber\mathcal{H} &=& K_{x}\sum_{ij \in \mathcal{X}}S_{i}^{x}S_{j}^{x} + K_{y}\sum_{ij \in \mathcal{Y}}S_{i}^{y}S_{j}^{y} + K_{z}\sum_{ij \in \mathcal{Z}}S_{i}^{z}S_{j}^{z}\\
    \nonumber &+& J\sum_{ij}{\bf S}_{i}\cdot{\bf S}_{j}\\
    \nonumber &+& \Gamma_{z}\sum_{ij \in \mathcal{Z}}\left(S_{i}^{x}S_{j}^{y} + S_{i}^{y}S_{j}^{x}\right) + \Gamma_{y}\sum_{ij \in \mathcal{Y}}\left(S_{i}^{z}S_{j}^{x} +  S_{i}^{x}S_{j}^{z}\right) + \Gamma_{x}\sum_{ij \in \mathcal{X}}\left(S_{i}^{y}S_{j}^{z} + S_{i}^{z}S_{j}^{y}\right)\\
    &+& A\sum_{i}(S_{i}^{c})^{2} + \sum_{i}{\bf S}_{i}\cdot{\bf H}
    \label{eq:genericHamiltonian}
\end{eqnarray}
\noindent The term $\bf H$ denotes the external magnetic field applied during experimental protocols. A tabulated list of potential spin Hamiltonian terms has been compiled based on these considerations and is given in Figure~\ref{fig:table}. 

To elucidate the properties of Kitaev quantum spin liquid candidates such as $\alpha$-RuCl$_{3}$, we conduct a series of experiments. The initial investigation involves magnetization measurements, wherein we assess magnetization as a function of temperature with magnetic fields at various angles, primarily along and perpendicular to the bond within the plane and out-of-plane directions. We explore potential distinctions between zero-field-cooled conditions—where we cool the compound to its lowest temperature without a magnetic field—and field-cooled magnetization, where we cool the compound with a magnetic field before conducting measurements. Given that a QSL is anticipated to lack long-range ordering, we scrutinize magnetization versus temperature curves for any anomalies. Additionally, we examine magnetic anisotropy by comparing the magnitude of magnetization along different directions as a function of magnetic field, aiming to deduce interaction types embedded in the spin Hamiltonian. Sweeping the magnetic field allows us to observe the evolution of the spin structure. In the case of $\alpha$-RuCl$_{3}$, the role of the magnetic field is pivotal, as it destabilizes long-range magnetic ordering, leading the system into the putative Kitaev QSL phase.

Another commonly conducted measurement involves magnetic specific heat measurement for the Kitaev QSL phase. The magnetic specific heat is anticipated to exhibit two distinctive broad anomalies at certain temperatures. These anomalies are attributed to the release of entropy from itinerant Majorana fermions and Z$_{2}$ flux excitations. The precise locations of these two anomalies serve to pinpoint the energy scale of the Kitaev exchange interactions \cite{nasu2015thermal,yamaji2016clues}.

Furthermore, we explore how the magnetic specific heat evolves with varying magnetic fields to gain insights into the stabilization of the long-range ordered phase and the scaling of the spin gap with magnetic field strength. In the Kitaev QSL phase, the magnetic gap is expected to exhibit a cubic scaling relationship with the magnetic field \cite{tanaka2022thermodynamic}.

A crucial final measurement involves thermal conductivity assessments. The Kitaev Quantum Spin Liquid (KQSL) phase is topologically distinct from the vacuum, and the presence of charge-neutral particles at the edge state is imperative. Given that these carriers lack charge, their detection is possible solely through the flow of thermal energy. The KQSL phase under a magnetic field is expected to exhibit half-integer quantization of the thermal Hall conductance \cite{kasahara2018majorana}.



\begin{tcolorbox}[enhanced,width=\columnwidth,
                  colback=white,breakable]
                  
\subsection*{{Application 1.1} - Seeking the accurate spin Hamiltonian for $\alpha$-RuCl$_{3}$ using experimental data on magnetization.}

\subsubsection*{Specific background of the application - magnetization experimental protocol}

The measurable parameter is the net magnetization denoted as $g\mu_{B}\sum_{i}\left<\psi\left|{\bf S}_{i}\right|\psi\right>$. For dc 
field experiment, we employ vibrating sample magnetometry for its determination. This technique involves oscillating a sample around a coil and measuring the induced voltage across the coil, directly proportional to the net magnetization of the sample. Through meticulous calibration against a standard sample, we can obtain the absolute value of magnetization. Our typical temperature scan ranges from 0.5 K to 400 K under a small magnetic field of up to 14 T. An example data is shown in Fig. \ref{fig:magnetization_1}. During the temperature scan, we discern whether the sample exhibits spontaneous magnetic ordering.

Furthermore, we perform a field scan to detect spin flip or spin flop phase transitions, providing insights into the energy scale of magnetic anisotropy. Moreover, the saturation field aids in determining the energy scale of spin interactions. For pulsed field measurements, we utilize extraction magnetometry, measuring the change in magnetization over time (dM/dt). Subsequently, we integrate the signal with time to derive the magnetization. The pulsed field profile as a function of time is depicted in Figure~\ref{fig:pulsed_field_profile} and typical data of magnetization are shown in Fig. \ref{fig:magnetization_2}.


\subsubsection*{Objective} 

The objective is to discern which candidate Hamiltonians are the most accurate predictors of existing magnetization data as a function of external magnetic fields. 




\subsubsection*{End-to-end computational workflow}
\begin{itemize}
\item \textbf{Inputs}:
\begin{enumerate}
    \item List of candidate Hamiltonians $\{H_i, \ i \in [m]\}$.
    \item External magnetic field settings $\{B_k, \ k\in[K]\}$ and simulation times $\{t_{sim}^k, \ k \in [K]\}$.
    \item For each tuple of candidate Hamiltonian $H_i$, external field setting $B_k$ and simulation time $t_{sim}^k$, the full description of the resulting time-varying Hamiltonian in the magnetic field sweep experimental protocol  $\{H_i^k(t), \ t \in [0,t_{sim}^k]\}$.
    \item Initial state $\rho_{0}$.
    \item Experimental data on magnetization $\{M_k^d, \ k \in [K]\}$.
\end{enumerate}

\item \textbf{Outputs}
\begin{enumerate}
    \item Net in-plane magnetization perpendicular to the bond direction : $g_{a}\mu_{B}\sum_i {\bf S}_{i}\cdot {\hat a}$, where ${\hat a}$ is perpendicular to one of the bond directions shown in Fig. \ref{fig:honeycomb lattice}, where $g_{a}$ is $g$-factor anisotropy that is close to 2.3 and $\mu_{B}$ is Bohr magneton. 
    \item Net in-plane magnetization along the bond direction:  $g_{b}\mu_{B}\sum_i {\bf S}_{i}\cdot {\hat b}$, where ${\hat b}$ is  one of the bond directions shown in Fig. \ref{fig:honeycomb lattice}, where $g_{b}$ is $g$-factor anisotropy that is close to 2.3.
    \item Net out-of-plane magnetization $g_{c}\mu_{B}\sum_i {\bf S}_{i}\cdot {\hat c}$, where ${\hat c}$ is perpendicular direction to the honeycomb plane, where $g_{c}$ is $g$-factor anisotropy that is close to 1.3.
\end{enumerate}

\item \textbf{Workflow}: The flowchart showing the \emph{end to end workflow} is given in Figure~\ref{fig:net_mag_workflow}. The \emph{hard computational module} is the \emph{Hamiltonian simulation}. For the the lattice sizes relevant to study this problem, the exponential dimension of the Hilbert space makes the simulation classically intractable. 
\end{itemize}

\subsubsection*{Why classical methods are not sufficient to perform the hard computational module}
Classical methods to solve the hard computational module of Hamiltonian simulation include \emph{exact diagonalization} and heuristics such as \emph{density matrix renormalization group (DMRG)} and \emph{quantum monte carlo (QMC)}. These methods are either too limited in size, not sufficiently accurate or not applicable in some situations. The details and limitations of these methods are given in the supplementary material Section~\ref{sec:KQSL_suppmat}.

\subsubsection*{Concrete problem instantiations}
Our focus is on $\alpha$-RuCl$_{3}$, a key compound anticipated to exhibit a quantum spin liquid phase under a magnetic field. Researchers have compiled potential spin Hamiltonians through simulations on classical computers.
\begin{itemize}
\renewcommand\labelitemi{}
    \item \emph{Input 1}: The list of candidate Hamiltonians is given in the table in Figure~\ref{fig:table}. The size of the Hamiltonian is 100 $\times$ 100 and is chosen to avoid finite size effects. 
\end{itemize}

In experiments, the sample is cooled to an extremely low temperature of approximately 2 K, applying a magnetic field to establish a single magnetic domain. We anticipate observing the antiferromagnetic phase transition around 6.5 K. During this process, we gradually sweep the magnetic field between $0$ and $60$ Tesla while measuring the magnetization. A suitable spin Hamiltonian should be capable of capturing: 1) the magnitude of the magnetization within 5\% of the experimentally measured magnetization, 2) the critical magnetic field between magnetically long-range ordering and the intermediate phase, as well as between the intermediate phase and the saturated phase, within 5\%, 3) magnetic anisotropy between in-plane and out-of-plane magnetization, and 4) the magnetic anisotropy within the plane, specifically the subtle magnetization difference between the magnetic field along the bond and perpendicular to the bond. In the magnetization data shown in Figure~\ref{fig:magnetization_2}, quantum critical fields are evident, marking the transitions from the antiferromagnetic phase to the intermediate phase and from the intermediate phase to the fully saturated phase. It is imperative that these quantum critical fields are reproduced within a 5\% accuracy. Additionally, the ratio between in-plane and out-of-plane magnetizations should be faithfully reproduced, as it provides insight into the energy scale of off-diagonal terms and single ion anisotropy. \\

\noindent Based on the above discussion, a resolution of $0.5$ Tesla is chosen for the external magnetic field settings, with a finer resolution of $0.1$ Tesla in the region believed to contain the phase transitions. This specifies the list of external field set points.
\begin{itemize}
\renewcommand\labelitemi{}
    \item \emph{Input 2}: The external field set points are $B_k \in 0:0.5:60$ Tesla, with an increased resolution in the critical range $B_k \in 7:0.1:9$ Tesla. The The simulation times $t_{sim}^k$ are given by the x-axis values in Figure~\ref{fig:pulsed_field_profile} corresponding to the $B_k$ values specified above. 
\end{itemize}
The time varying form of the Hamiltonian can be obtained by incorporating the external field with the chosen candidate Hamiltonian $H_i$ among the list given in Input 1.
\begin{itemize}
\renewcommand\labelitemi{}
    \item \emph{Input 3}: The time varying Hamiltonian is given by  
    \begin{align}
        H_i^k(t) = H_i  + f(t) \sum_j S_j^a, \quad \text{for} \ t \in [0,t_{sim}^k],
    \end{align}
    The time dependent magnetic field profiles $f(t)$ of both dc and pulsed magnets are shown in Figure~\ref{fig:pulsed_field_profile}.
\end{itemize}
\noindent The initial state of the sample can be inferred from experimental data. See supplementary material section for more details.
\begin{itemize}
\renewcommand\labelitemi{}
    \item \emph{Input 4}: The initial state $\rho_0$ is given in Section~\ref{sec:RuCl_GS}.
    \item \emph{Input 5}: The experimental data on net magnetization is given in Figure~\ref{fig:magnetization_2}. For $\alpha$-RuCl$_{3}$ it is known that the time scales of the material are much faster than the time scale at which the external magnetic field in Figure~\ref{fig:magnetization_2} changes. Consequently, it is sufficient to perform a time evolution for $1000 J_{max}T$ normalized time units starting from the inital state defined in Input 4 for each value of the magnetic field values in the x-axis of Figure~\ref{fig:magnetization_2}. Here $J_{max}$ refers to the maximum absolute coupling value.
\end{itemize}

\begin{itemize}
\renewcommand\labelitemi{}
    \item \emph{Output precision requirement}: The magnetization measurements $M_k^i$ from the Hamiltonian simulation must be computed to $1\%$ relative accuracy. Given that a candidate Hamiltonian is considered suitable if it can predict experimental observables $M_k^d$ within $5\%$ accuracy, and the experimental measurement precision is less than $1\%$.
\end{itemize}
\begin{itemize}
\renewcommand\labelitemi{}
    \item \emph{Output wall-clock time limit}: Finding a suitable Hamiltonian by comparing with the existing experimentally measured magnetization data requires lots of iteration and we need to scan many possible form of Hamiltonian. To this end, therefore, we need to simulate a spin Hamiltonian with a couple minutes. For the entire workflow, less than two weeks time scale is acceptable. 
\end{itemize}


\subsubsection*{List of candidate systems where similar process is relevant?}
Numerous quantum spin liquid candidates face a common challenge: determining their spin Hamiltonians is challenging with current computational power and methods. This set of candidates includes iridates compounds such as Na$_{2}$IrO$_{3}$, $\alpha$-Li$_{2}$IrO$_{3}$, and $A'_{3}$LiIr$_{2}$O$_{6}$ ($A'$ = H, Cu, and Ag) \cite{rau2016spin}. Recently, certain cobaltites have garnered interest from researchers, including Na$_2$Co$_{2}$Te$_{2}$O$_{6}$ \cite{zhang2023electronic} and BaCo$_{2}$(AsO$_{4}$)$_{2}$ \cite{zhang2023magnetic}.

\end{tcolorbox}

\begin{figure}[!ht]
    \centering
    \includegraphics[width=0.6\textwidth]{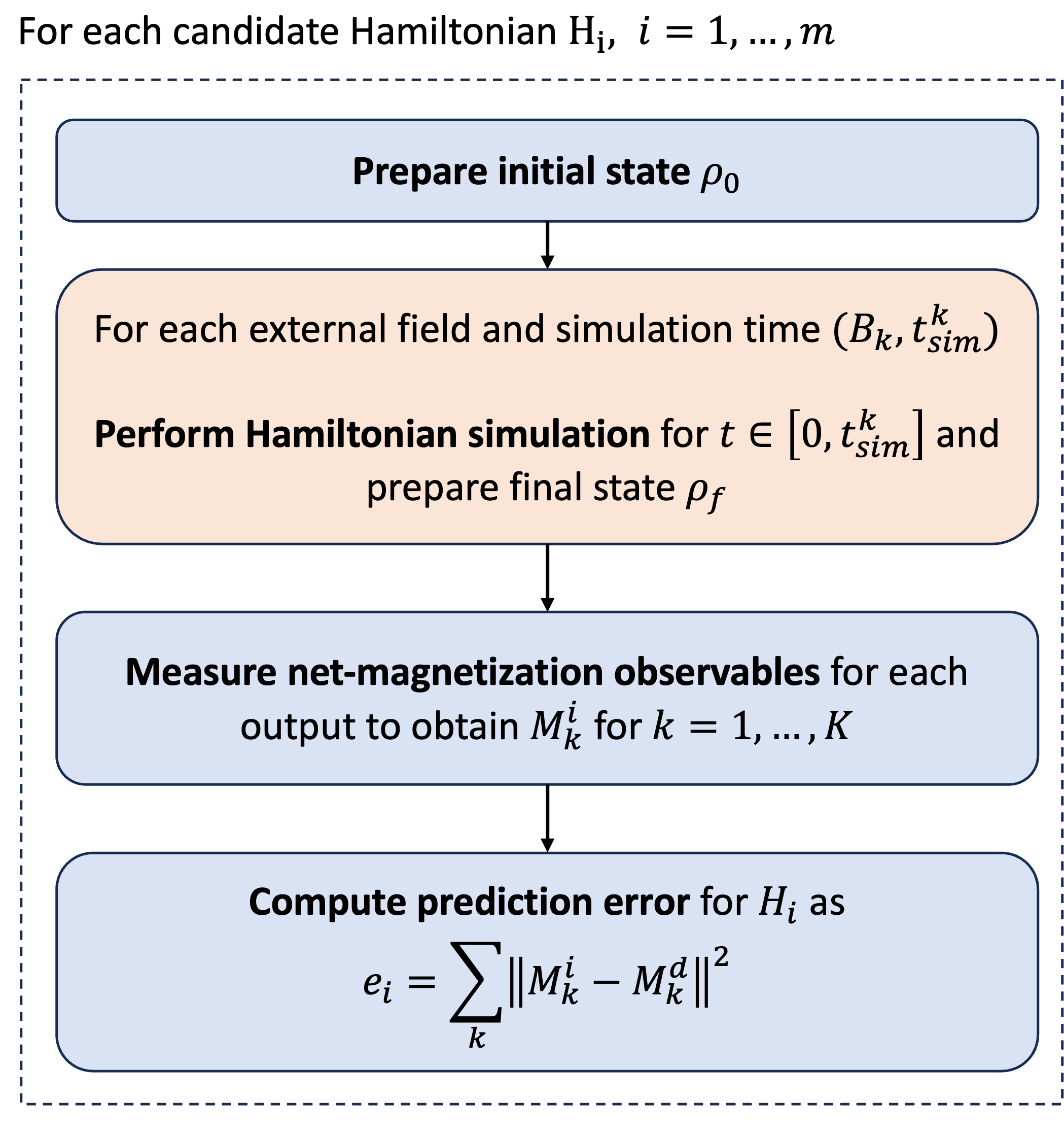}
    \caption{Workflow for Hamiltonian search using magnetization experimental data. The red box (Hamiltonian Simulation) contains the hard computational task/kernel.
    }
    \label{fig:net_mag_workflow}
\end{figure}

\begin{figure}[!ht]
    \centering
    \includegraphics[width=0.5\textwidth]{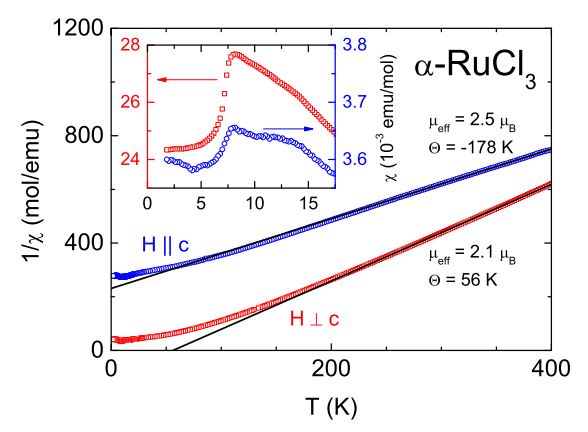}
    \caption{The inverse of magnetization divided by external magnetic field ({$1/\chi$}) as a function of temperature with magnetic field along two different directions are shown \cite{reschke2018sub}.
    \label{fig:specific_heat}
    }
    \label{fig:magnetization_1}
\end{figure}

\begin{figure}[!ht]
    \centering
    \includegraphics[width=0.7\textwidth]{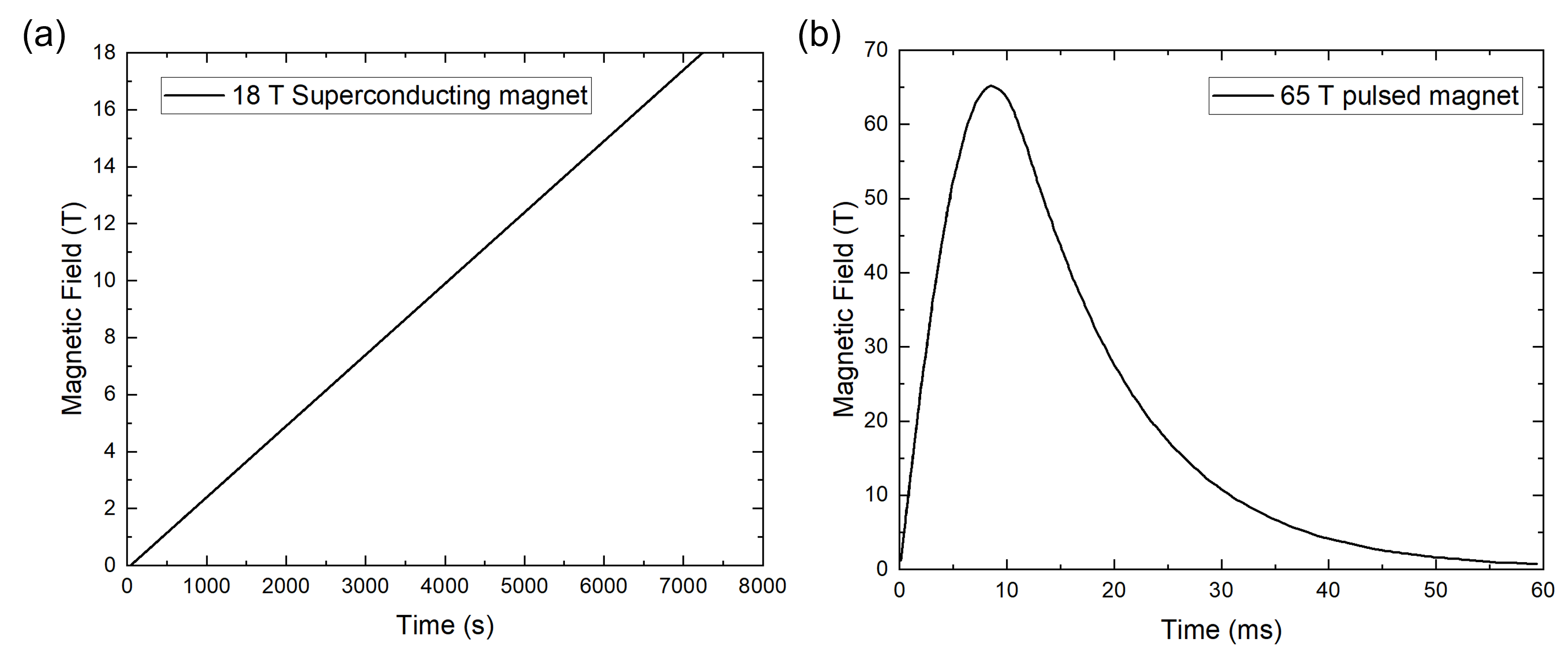}
    \caption{Field vs. time profiles of superconducting dc magnet (a) and 65 T short pulse magnet (b) \cite{maglab1}. Note that the time scale difference between dc magnet and pulsed field magnet.
    }
    \label{fig:pulsed_field_profile}
\end{figure}

\begin{figure}[!ht]
    \centering
    \includegraphics[width=0.7\textwidth]{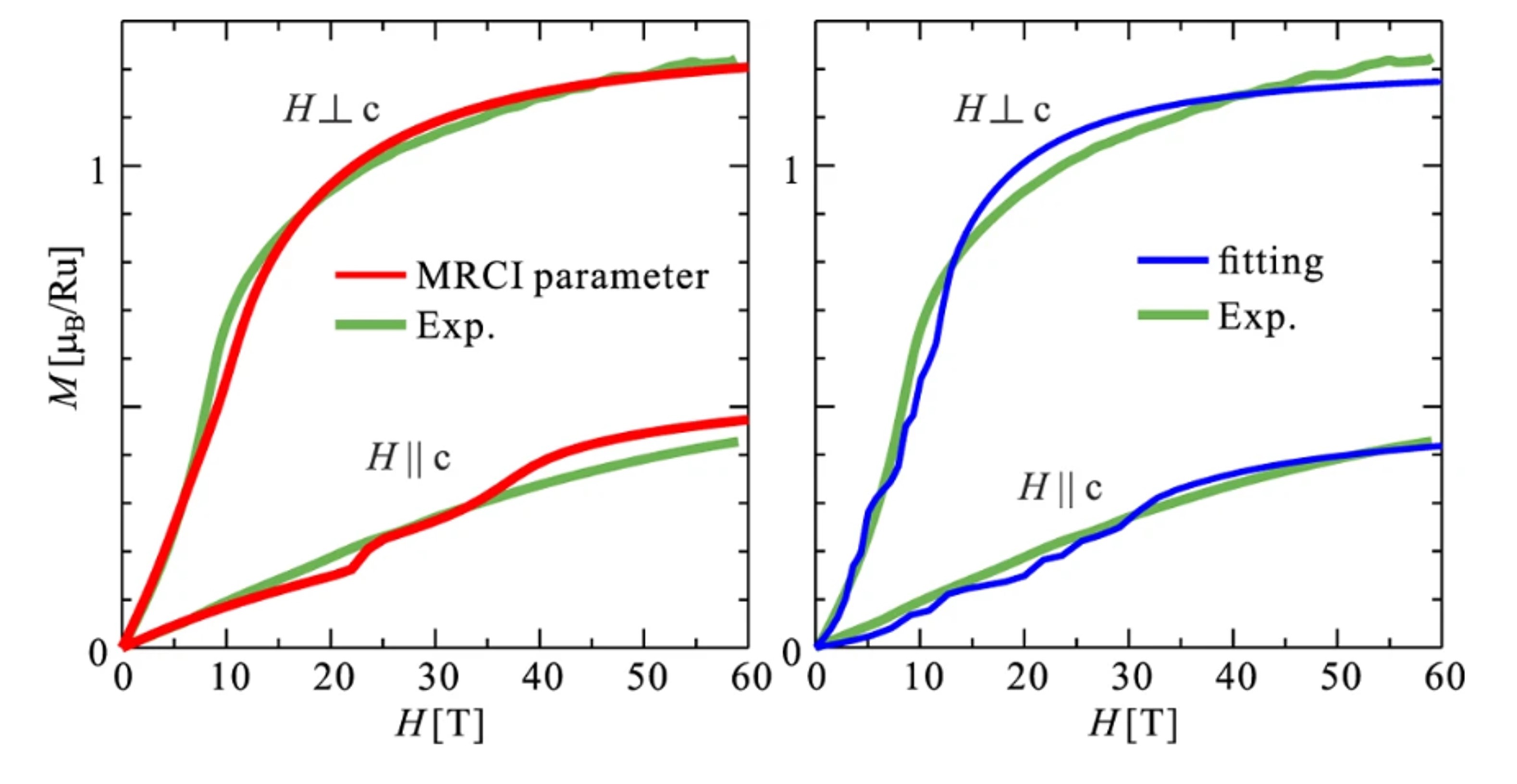}
    \caption{The magnetization measurements (solid green lines) with field along in-plane with larger magnetization and out-of-plane with smaller magnetization \cite{johnson2015monoclinic}. The red and blue solids lines are from exact diagonalization simulations \cite{yadav2016kitaev}.}
    \label{fig:magnetization_2}
\end{figure}


\newpage

\begin{tcolorbox}[enhanced,width=\columnwidth,
                  colback=white,breakable]

\subsection*{{Application 1.2} - Testing candidate Hamiltonians for $\alpha$-RuCl$_{3}$ using ground state prediction.}

\subsubsection*{Specific background of the application - Experimental protocol to determine the ground state}

The spin structure of the ground state at \emph{zero external magnetic field} can be inferred from magnetization-temperature and magnetic field measurements. Initially, we categorize whether a system exhibits magnetic long-range ordering, discernible through singularities such as divergences or kinks in the magnetization measurement, indicative of a symmetry-breaking phase transition.

Subsequently, we ascertain if the magnetic ordering possesses a net magnetization. In the presence of a net magnetic component, particularly a ferromagnetic one, distinctions often arise between zero field cooled and field cooled states owing to magnetic domain effects. The direction of the ferromagnetic component can be deduced from the magnitude of magnetization.

In the case of antiferromagnetism, understanding how spins counterbalance is achieved by analyzing magnetization against temperature and magnetic field in various orientations. When the magnetic field aligns with spin directions, the net magnetization approaches zero in the limit of zero temperature; however, when perpendicular to the spins, a temperature-independent finite magnetization is observed. The magnetization-field curve illustrates spin flip or flop transitions, marked by sudden changes when the field aligns with spin directions and gradual increases when perpendicular.

Symmetry-sensitive techniques, such as electric polarization, are also employed. Depending on the spin structure's symmetry, linear magnetoelectric coupling or second-order magnetoelectric coupling may be allowed.

Finally, we validate the symmetry through neutron diffraction experiments. In the presence of magnetic ordering, magnetic Bragg peaks emerge in momentum space, aligning with the wave vector of the spin ordering. Analyzing these magnetic Bragg peaks allows us to infer the spin arrangement and magnetic interactions within the material. The intensity of the peaks serves as an indicator of the magnetic order's strength, while the positions of the peaks unveil the spatial distribution of magnetic moments.


\subsubsection*{Objective} 
The objective is to identify which of the candidate Hamiltonians in Figure~\ref{fig:table} is able to predict the experimentally observed ground state in Figure~\ref{fig:groundstate} at \emph{zero external magnetic field}. This procedure serves as a screening process for the proposed candidate Hamiltonians.

\subsubsection*{End-to-end computational workflow}
\begin{itemize}
\item \textbf{Inputs}:
\begin{enumerate}
    \item List of candidate Hamiltonians $\{H_i, \ i \in [m]\}$.
    \item A set $\mathcal{P}$ of $m$ points $\bf p_1, \ldots, p_m$ in the two dimensional momentum space for which the Fourier transform is desired.
    \item The experimentally observed intensities $y^d$ of the Fourier transform of the ground state $\rho_{0}$ at zero (near-zero) external field.
    \item (Optional) Use classical methods such as DMRG for initialization.
\end{enumerate}

\item \textbf{Outputs}
\begin{enumerate}
    \item The Fourier transform intensities of the spin ground state at the $m$ points specified in $\mathcal{P}$. This can be obtained by computing the quantities for each $k = 1, \ldots, m$,
    \begin{align}
        F(\mathbf{p}_k)  = \frac{1}{N}\sum_j \exp(i \mathbf{r}_j.\mathbf{p}_k) S_j,
    \end{align}
    and the intensities can be computed as $I({\bf p_k}) = F^{*}({\bf p_k})F({\bf p_k})$. In the above $N$ is the total number of magnetic moments,  ${\bf r}_j$ is the position vector in real space of spin vector $S_j$.

    
    
\end{enumerate}


\item \textbf{Workflow}: The flowchart showing the \emph{end to end workflow} is given in Figure~\ref{fig:groundstate_workflow}. The \emph{hard computational module} is the \emph{ground state preparation}.

\end{itemize}

\subsubsection*{Why classical methods are not sufficient to perform the hard computational module}
Same as Application 1.1.

\subsubsection*{Concrete problem instantiations}
The compond in consideration is $\alpha$-RuCl$_{3}$.
\begin{itemize}
\renewcommand\labelitemi{}
    \item \emph{Input 1}: The list of candidate Hamiltonians is given in the table in Figure~\ref{fig:table}. The size of the Hamiltonian is 100 $\times$ 100 and is chosen to avoid finite size effects. 
\end{itemize}
\begin{itemize}
\renewcommand\labelitemi{}
    \item \emph{Input 2}: We select 100 $\times$ 100 uniformly spaced data points in two dimensional reciprocal space within the first Brillouin zone. 
    \item \emph{Input3 }: The experimentally observed ground state is given in Figure~\ref{fig:groundstate}.
\end{itemize}

\begin{itemize}
\renewcommand\labelitemi{}
    \item \emph{Output precision requirement}: The output precision requirement is 0.01 for the observables $I({\bf p_k})$ at each specified point ${\bf p_k}$ in the momentum space.
    \item \emph{Output wall-clock time limit}: Finding a suitable Hamiltonian by comparing with the existing experimentally measured magnetization data requires lots of iteration and we need to scan many possible form of Hamiltonian. To this end, therefore, we need to simulate a spin Hamiltonian with a couple minutes. For the entire workflow, finding the proper ground state within two weeks is acceptable. 
\end{itemize}

\subsubsection*{List of candidate systems where similar process is relevant?}
Same as Application 1.1.
\end{tcolorbox}
\begin{figure}[!ht]
    \centering
    \includegraphics[width=0.6\textwidth]{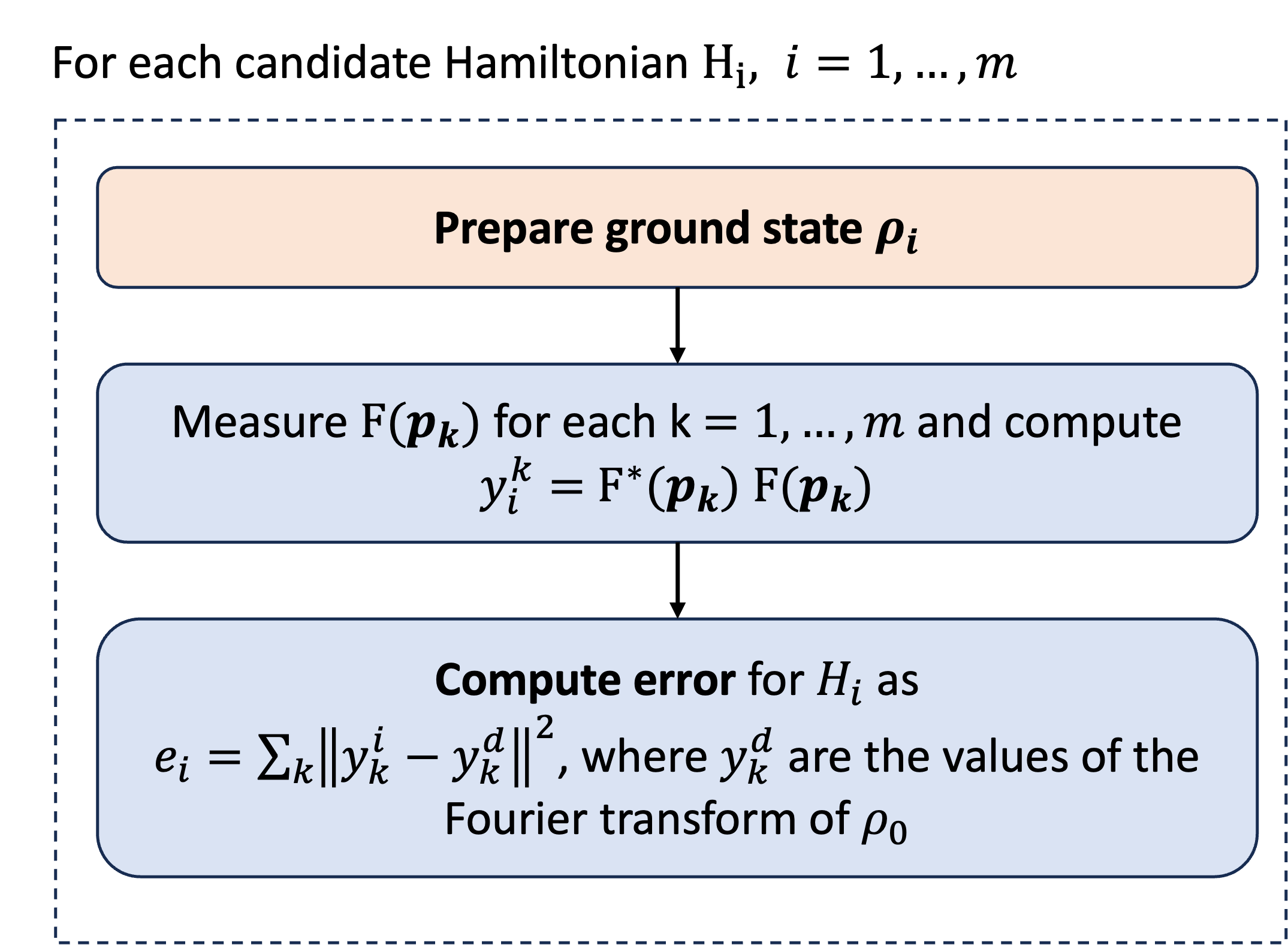}
    \caption{Workflow for Hamiltonian search using experimental ground state. The red box contains the hard computational task/kernel.}
    \label{fig:groundstate_workflow}
\end{figure}



\newpage
\section{Requirements summary}
This section summarizes the application requirements for all the applications described in this chapter.
\subsection*{Application 1.1: Seeking the accurate spin Hamiltonian for $\alpha$-RuCl$_{3}$ using experimental data on magnetization}


\begin{tabular}{ |l l l| }
    \hline
        & & \\
     & Total time limit & 2 weeks - 1 month \\ [1ex]
    \textbf{Workload:} & Number of subroutine calls required & $140\times15$ \\ [1ex]
     & Maximum subroutine time limit & 2 weeks - 1 month \\ [1ex]
     & & \\[1ex]
     & Model type & Kitaev/Heisenberg spin hamiltoninan \\[1ex]
     & Size & minimum  $30 \times 30$ sites \\
     &  & target $100 \times 100$ sites \\[1ex]
     \textbf{Problem specifications:} &  Interaction Structure & Sparse regular (2-D Hexagonal latice) \\[1ex]
     & Computational target & Net magnetization after time evolution \\[1ex]
     & Accuracy requirement & $1\%$ relative on net magnetization \\
     & & \\
     \hline
\end{tabular}

\subsection*{Application 1.2: Testing candidate Hamiltonians for $\alpha$-RuCl$_{3}$ using ground state prediction}

\begin{tabular}{ |l l l| }
    \hline
        & & \\
     & Total time limit & 2 weeks - 1 month \\ [1ex]
    \textbf{Workload:} & Number of subroutine calls required & $100\times 100 \times 15$  \\ [1ex]
     & Maximum subroutine time limit & 2 weeks - 1 month \\ [1ex]
     & & \\[1ex]
     & Model type & Kitaev/Hiesenberg spin hamiltoninan \\[1ex]
     & Size & $100 \times 100$ sites\\ [1ex]
     \textbf{Problem specifications:} &  Interaction Structure & Sparse regular (2-D Hexagonal latice) \\[1ex]
     & Computational target & Fourier transform of ground state \\
     & & on a specified lattice \\[1ex]
     & Accuracy requirement & $0.01*(2*pi/a)$ where $a$ is lattice constant \\
     & & \\
     \hline
\end{tabular}


\newpage
\subsection{Supplementary material for Kitaev quantum spin liquids} \label{sec:KQSL_suppmat}

\subsubsection{Classical methods} \label{sec:classical_methods_KQSL}
\begin{enumerate}
    \item Exact diagonalization (ED) faces computational challenges beyond a certain lattice size. While ED accurately captures quantum effects, the spin Hamiltonian matrix grows exponentially with the number of spin sites, making it impractical for lattices exceeding 50-60 sites. The need for a $2^{60}\times 2^{60}$ matrix, even with high-performance computing (HPC), renders ED time-consuming and  impossible for the sizes of problems presented in this chapter. Our objectives involve exploring numerous parameter combinations, making ED unfeasible despite its insights into quantum effects.
    
    The resulting finite size effects can make the predictions too inaccurate to be relevant. The intermediate phase in spin liquids that is topologically distinct from the vacuum entails edge states. To model edge states effectively, a large lattice is essential, as closely situated edge states can interact, obscuring their properties. ED, limited to small lattices, is unsuitable for detecting critical properties like edge states in the topologically distinct quantum spin liquid phase. The small lattice size also fails to capture long-range quantum entanglement, another distinctive feature of the quantum spin liquid phase.
    
    \item Heuristic methods like density matrix renormalization group (DMRG) and quantum Monte Carlo (QMC) simulations have been devised for simulating quantum magnetism. Regrettably, these methods are not readily applicable to the quantum spin liquid phase. The Kitaev quantum spin liquid, characterized by magnetic frustration resulting from bond frustration due to competing bond interactions, presents challenges for both QMC and DMRG. In QMC simulations, the presence of magnetic frustrations gives rise to an unsolvable sign problem. Additionally, when a system exhibits significant magnetic anisotropy, as evidenced by a large magnetic gap in the magnon band structure, studying the impact of quantum fluctuations becomes challenging, limiting the analysis to the classical ground state. While DMRG method is a highly effective tool for one-dimensional systems, it is not suitable for investigating quantum long-range entanglement.
\end{enumerate}

\subsubsection{List of Hamiltonians}    \label{sec:Rucl_candidate_Hs}
The  table in  Figure~\ref{fig:table} is a list of candidate Hamiltonians proposed for $\alpha$-RuCl$_{3}$ based on symmetry and other theoretical arguments.
\begin{figure}[!ht]
    \centering
    \includegraphics[width=0.8\textwidth]{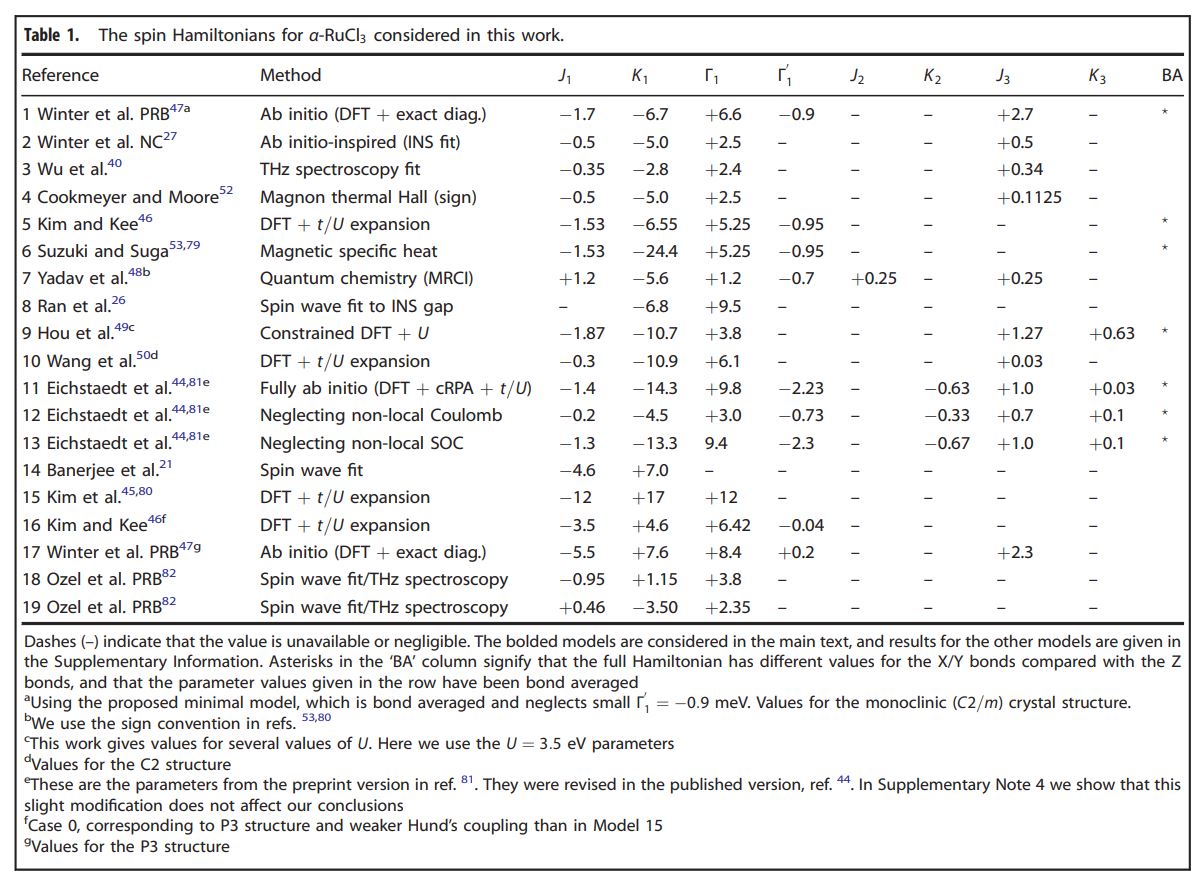}
    \caption{List of parameters for $\alpha$-RuCl$_{3}$ suggested by different theoretical approaches.}
    \label{fig:table}
\end{figure}

\subsubsection{Experimental ground state} \label{sec:RuCl_GS}
Symmetries inferred from experimental observations in $\alpha$-RuCl$_{3}$ can be used to conclude that its ground state can one of three equivalent states. This symmetry can be broken and one among the three is chosen in the presence of an external magnetic field. The ground state of $\alpha$-RuCl$_{3}$ is called a zigzag spin structure and it is given in Fig. \ref{fig:groundstate}. From the figure, we see that there are two category of nodes distinguished by arrows colored blue and red. The spin configuration of the ground states are given by
\begin{align}
    S_{blue}^a = \sqrt{1-\epsilon^2},  \ S_{blue}^b = 0,  \ S_{blue}^c = \epsilon, \qquad \mbox{and} \qquad S_{red} = -S_{blue}.
\end{align}
Here $\epsilon = 0.1$. There is some uncertainty about the value of the out-of-plane component quantified by $\epsilon$ based on currently available experimental observaions. However, we deem it will be consequential to the result of the Hamiltonian simulation.

\begin{figure}
    \centering
    \includegraphics[width=0.8\textwidth]{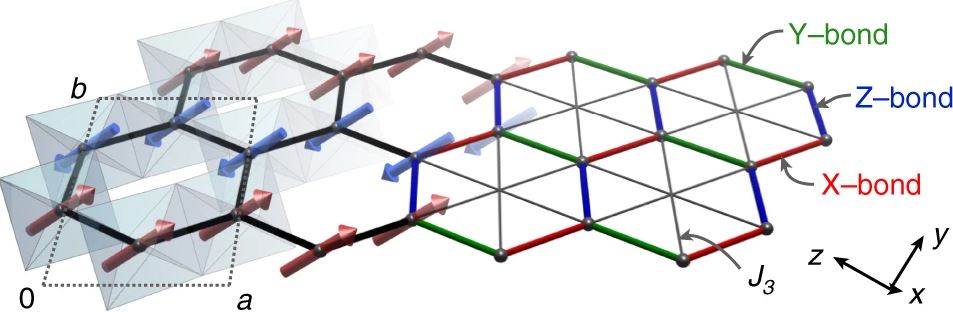}
    \caption{The spin structure of the magnetic ground state of $\alpha$-RuCl$_{3}$}
    \label{fig:groundstate}
\end{figure}


\section{Quantum implementation}
The fully specified Hamiltonians in Figure~\ref{fig:table} and quantum implementation and resource estimation can be found in \cite{qc-applications-notebooks}.

\newpage

\end{chapter}

\begin{chapter}{Exploring exotic phases of magnetic materials near instabilities} \label{ch:neutron_scattering}
    

This chapter presents applications related to the study of exotic phases in magnetic materials. For many Hamiltonians, either of theoretical importance or those proclaimed to be effective models for certain compounds, there are regions in the parameter space of the exchange interactions that defy classical or semi-classical description. This phenomenon often arises from the presence of long range entanglement that prevents understanding and prediction of properties of such phases. The computational capabilities required to enable the study of these exotic phases is the ability to compute specified observables either from the ground state or after simulating simulate the time-dynamics for a spin Hamiltonian defined on a 
lattice. Existing classical methods cannot be scaled to large lattice sizes, where a fault tolerant quantum computer can provide utility by computing these observables at the desired scale and accuracy.
The following table provides a condensed summary of the computational requirements of the applications. \\

\noindent
\textbf{Hamiltonian Type}:  Heisenberg spin Hamiltonian.\\
\textbf{Quantum Computational Kernels}: Ground State Preparation, Hamiltonian simulation. 

\clearpage

\section{Application area overview}

\subsection{High-level description of the application area}

In 1964, Richard Feynman wrote, ``given a lot of electron spins in a regular lattice, interacting with such and such a law, what do they do? It is simply stated, but it has defied complete analysis for years.''\cite{feynmanLectures} 
Now, in \the\year{}, this basic physics problem remains unsolved. Despite tremendous efforts over the decades, we have no universal way to calculate and predict  interacting magnetic atoms in a crystal lattice. 

By and large, physics can predict and explain magnetic behavior if materials ``behave'' classically or semiclassically \cite{fishman2018spin}, but these methods fail when the system is tuned to an instability and becomes ``highly quantum.'' And, unfortunately, such regions are where the most exciting properties exist. 
For all but a few such cases, we cannot actually calculate or simulate their behavior, because the state involves such a large number of entangled spins. 
Thus, their study is riddled with speculation, uncertainty, and controversy. 

Quantum spin liquids are a high-profile example of a magnetic state beyond the reach of classical simulation, but there are in fact a large number of unstable regions where current algorithms fail. This has three consequences: (i) Many exotic magnetic  states have been proposed to exist, but there is often no way to calculate or verify whether these predictions are correct. (ii) Many quantum materials have been found which act strange, but their underlying physics is unknown because their behavior cannot be compared with simulations. (iii) Many unstable regions are known to exist in theory, but there is no robust way to calculate what actually happens at these unstable points, and whether there is an exotic phase. 
As a result, the study of quantum magnets is awash in speculation and unanswered questions. 
In our opinion, this is a prime application for quantum computers.

\subsection{Utility estimation}

\subsubsection{Overview of the value of the application area}

The motivation for this research is both for fundamental physics and technology. On the fundamental side, interacting quantum spin systems are some of the simplest models in many-body quantum physics. The problem of interacting quantum spins dates back to the origin of quantum mechanics, attracting such brilliant minds as Hans Bethe \cite{bethe1931theorie}, Lars Onsager \cite{onsager1944crystal}, Phil Anderson \cite{anderson1952approximate}, and Duncan Haldane \cite{haldane1983continuum}; and it has been the subject of four Nobel prizes in physics (1970, 1977, 2007, 2016). Their study has produced foundational physics concepts like spontaneous symmetry breaking \cite{anderson1952approximate} and quantum criticality \cite{sachdev2011quantum}.
Meanwhile, on the technology side, real materials with interacting magnetic atoms can give unique and useful properties like giant magnetoresistance (used in magnetic sensing and computer memory) \cite{grunberg2008nobel}, atomically thin electronic switches \cite{dieny2020opportunities}, or---looking to the future---topological qubits \cite{Nayak_2008}. On a basic level, quantum magnetic materials offer the ability to harness quantum mechanics at higher temperatures and longer length-scales than is possible with other quantum devices. 
Indeed, if physicists could reliably predict the behavior of materials, this could yield a veritable technology revolution \cite{tokura2017emergent}. 

\subsubsection{Concrete utility estimation}
The three applications presented in this chapter provide a comprehensive list of computational tasks that will enable completion of the short-term research goals for a given material and set of models. We calculate the total budget used for the study of the models and materials presented in this chapter. The Department of Energy's budget request for running the neutron facilities for FY24 was $\$373$M which includes a total of $33$ instruments. For the neutron scattering experiments for a given material (KYbSe$_2$) presented in this chapter, 
2.5 weeks of beamtime were utilized
out of a total annual up-time of $8$ months. From this we compute the experimental expenditure to be $\frac{373}{33}\times \frac{2.5}{0.66\times52} \approx 1$M dollars. The time and labor costs are funded by the quantum science center and is estimated to be  $\approx \$0.5$M for four years for a total of $\$2$M. We assign a equal value to each of the applications presented resulting in  per application resulting in a value of $\$0.75$M per application.

\section{Problem and computational workflows}

\subsection{Detailed background of the application area}

The basic scientific problem is simple: we assume a lattice of interacting spins with the overall Hamiltonian
\begin{equation}
    \mathcal{H} = \sum_{i,j} {\bf S}_i \mathbb{J}_{ij} {\bf S}_j
    \label{eq:GeneralHamiltonian}
\end{equation}
where ${\bf S}_i$ are quantum spin operators and $\mathbb{J}_{ij}$ is a second rank tensor governing the interactions between spins. 
In most cases of interest, the interactions are nonzero only for the nearest few neighbors on the lattice. 
Also, for many classic theoretical models, $\mathbb{J}_{ij}$ is isotropic, meaning it is replaced by a scalar $J_{ij}$ and the spins interact via a dot product:
\begin{equation}
    \mathcal{H} = \sum_{i,j} J_{ij} {\bf S}_i \cdot  {\bf S}_j
\end{equation}
which is commonly called ``Heisenberg exchange''.

After the Hamiltonian is defined, the computational challenge is to calculate the ground state and excitation spectrum. 
In trivial cases, such as long range magnetic order, the quantum ground state is the same as the classical ground state. However, there are many cases where the quantum ground state involves nontrivial superpositions which are impossible classically, and the full quantum ground state must be computed. 
This is trivial for a small number ($< 10$) of spins; but the real problems of interest are in the thermodynamic limit, where the lattice has Avogadro's number of spins. 
Because this is impractical, the usual approach is to assume a ``large enough'' lattice (often $\sim 10^2$ sites along each dimension) and calculate the properties based on this. Thus for a two-dimensional lattice, $\sim 10^4$ spins would be required to simulate the thermodynamic limit. The size of the Hamiltonian matrix scales as $2^n$ (where $n$ is the number of spins), so such a lattice would have a $10^{300} \times 10^{300}$ Hamiltonian matrix, far beyond the reach of a classical computer!

Because the general solution of Eq. \ref{eq:GeneralHamiltonian} is computationally hard, classical or semiclassical approximations are usually employed. 
If we choose to completely ignore quantum effects, the spin dynamics and ground state can be simulated with classical Monte Carlo and Landau Lifschitz dynamics \cite{Zhang_2021_SUN}. Also, semiclassical dynamics can be calculated with spin wave theory \cite{fishman2018spin} which assumes the excitations can be described by a boson quasiparticle effective field theory. These approaches work well when the system is ``well-behaved,'' which means quantum effects are small and the system can indeed be described by boson quasiparticles. However, there are many models (and correspondingly many materials) where these assumptions break down (see Table \ref{tab:exampleModels} for models and Table \ref{tab:exampleCompounds} for materials), and the full quantum solution is necessary.

\begin{table}
    \centering
    \begin{tabular}{p{3.5in}p{2.7in}}
        Model & References \\ \hline
        
         2D square lattice antiferromagnet, with second neighbor $J_2$ over first neighbor $J_1$ tuned to $J_2/J_1 = \frac{1}{2}$ & \cite{LIU20221034,PhysRevB.38.9335,PhysRevLett.63.2148,PhysRevB.38.9335,PhysRevB.41.4619,PhysRevB.41.9323,PhysRevB.43.10970,PhysRevB.44.12050,schulz1992finite,PhysRevB.46.8206,PhysRevB.51.6151,schulz1996magnetic,PhysRevB.54.9007,PhysRevB.60.7278,PhysRevLett.84.3173,PhysRevLett.87.097201,PhysRevLett.91.067201,PhysRevLett.91.197202,PhysRevB.73.184420,PhysRevLett.97.157201,PhysRevB.74.144422,PhysRevB.78.214415,PhysRevB.78.224415,PhysRevB.79.024409,PhysRevB.79.195119,PhysRevB.79.224431,richter2010spin,PhysRevB.85.094407,PhysRevB.86.024424,PhysRevB.86.045115,PhysRevLett.111.037202,PhysRevB.88.060402,PhysRevB.89.104415,PhysRevB.89.235122,PhysRevLett.113.027201,PhysRevB.90.041106,morita2015quantum,richter2015spin,PhysRevB.94.075143,PhysRevB.96.014414,PhysRevLett.121.107202,PhysRevB.97.174408,PhysRevB.98.241109,poilblanc2019critical,PhysRevB.102.014417,PhysRevX.11.031034}\\ \hline
         2D triangular lattice antiferromagnet, with second neighbor $J_2$ over first neighbor $J_1$ tuned to $J_2/J_1 = \frac{1}{8}$ & \cite{PhysRevB.92.041105,PhysRevB.92.140403,PhysRevB.93.144411,PhysRevB.94.121111,PhysRevB.95.035141,PhysRevB.96.075116,PhysRevLett.123.207203}\\ \hline
         3D pyrochlore lattice with anisotropic nearest neighbor exchange & \cite{benton16-pinchline,Yan_2017} \\ \hline
         2D Kagome lattice antiferromagnet & \cite{benton16-pinchline,Yan_2017} \\ \hline
         3D pyrochlore Heisenberg antiferromagnet & \cite{PhysRevX.9.011005,PhysRevLett.126.117204,PhysRevX.11.041021,PhysRevLett.131.096702,pohle2023ground} \\ \hline
         2D honeycomb anisotropic $\Gamma$ antiferromagnet & \cite{Luo2021} \\ \hline
         2D Shastry Sutherland lattice & \cite{liu2023deconfined} \\ \hline
         Hyperkagome antiferromagnet & \cite{PhysRevLett.101.197201,PhysRevB.94.235138} \\ \hline
         Hyper-hyperkagome antiferromagnet & \cite{PhysRevB.104.094413} \\ \hline
         $S=1$ spins with biquadratic exchange on various lattices &  \cite{WenJun_2019,Pohle_2023} \\
         \hline
    \end{tabular}
    \caption{Well-studied examples of interacting quantum spin systems where classical methods fail, and quantum simulation is necessary.}
    \label{tab:exampleModels}
\end{table}

\begin{table}
    \centering
    \begin{tabular}{p{1.2in}p{4.8in}}
        Compound & Description \\ \hline
        
    Herbertsmithite & A kagome quantum spin liquid candidate \cite{han2012fractionalized,RevModPhys.88.041002} \\
    $A$YbSe$_2$ & A family of triangular quantum spin liquid candidates, where $A$ is an Alkalai metal \cite{Dai_2021,Xie2023,Scheie2024,Scheie_2024_KYS_PRB} \\
    NaCaNi$_2$F$_7$ & A nearly isotropic pyrochlore magnet with exchange disorder \cite{plumb2019continuum,Zhang_2019} \\
    Yb$_2$Ti$_2$O$_7$ & A pyrochlore a well-studied Hamiltonian \cite{PhysRevLett.119.057203,Yan_2017} but a puzzling excitation spectrum \cite{scheie2020multiphase} \\
    YbMgGaO$_4$ & An intrinsically disordered triangular lattice quantum magnet \cite{li2015gapless,paddison2017continuous,shen2016evidence}, with great controversy over the excitations \cite{PhysRevX.8.031028,rao2021survival}. \\
    RuCl$_3$ & A candidate Kitaev spin liquid candidate with a well-measured neutron spectrum \cite{Banerjee2017Science,banerjee2016proximate} \\
         \hline 
    \end{tabular}
    \caption{Examples of nontrivial quantum magnetic materials with well-measured properties but poorly understood quantum ground states.}
    \label{tab:exampleCompounds}
\end{table}

\subsection{Classical methods} \label{sec:classical}
In the case of the full quantum treatment, there are some quantum methods for classical computers that work with limited success.  Some of the most popular are:
\begin{itemize}
    \item Exact Diagonalization (ED), wherein a system of interacting spins is diagonalized using Lanczos techniques \cite{PhysRevB.49.5065}. Current state of the art is around 32 spins \cite{Wu_2021}. This is a far cry from an extended system, and finite size effects can severely affect the calculations (especially for non-local long-range entangled states). 
    \item Matrix Product methods, most notably density matrix renormalization group (DMRG). This method is numerically accurate in one dimension \cite{Schollwock_2005}, able to simulate lattices of $\approx 100$ spins \cite{scheie2021witnessing}. Recent advances have applied this to simulate two-dimensional systems by wrapping the spins along a cylinder \cite{drescher2022dynamical,Sherman_2023}. However the limited dimensionality around the cylinder ($\sim 6$ sites) means that this method too is severely limited by finite size effects in two dimensions, and cannot be used for three-dimensional lattices. 
    \item Quantum Monte Carlo  is a stochastic variational wave-function approach to calculating quantum systems \cite{RevModPhys.73.33}. This technique is extremely powerful, easily able to simulate $\sim 10,000$ spins, but it suffers from a so-called ``sign problem'' which prevents numerical convergence for frustrated lattices \cite{PhysRevB.92.045110} --- which is precisely where some of the most interesting physics lies. 
\end{itemize}
None of these methods are able to calculate the ground state of the most interesting problems. 
Going beyond these methods requires a radically new approach, such as a quantum computer.




\subsection*{Application 1: Ground state of systems near instabilities}

Computing the magnetic ground state of interacting magnetic spins where entanglement becomes long ranged and semiclassical methods fail. 

\subsubsection*{Specific background of the application}

Spin ground states can be calculated by classical and semiclassical methods, but these methods fail when the exchange interactions are tuned to an instability: either an exotic magnetic phase which defies semiclassical description, or a phase boundary between two types of magnetic order. In such cases, it is often unclear what the ground state is and where the phase boundaries lie. 

\subsubsection*{Objectives}

There are two objectives: (1) To compute the magnetic ground state of spins at or near an instability, and determine whether an additional non-trivial magnetic phase exists at this point; and (2) compute the extent in parameter space that this ground state is stabilized (i.e., where the phase boundaries lie).

\subsubsection*{End-to-end computational workflow}
For a Hamiltonian of the form in Eq. \eqref{eq:GeneralHamiltonian}  with values defined from regions where semiclassical methods fail or a proposed exotic quantum phase, the goal is to compute the spin correlation functions of the ground state. The most commonly used is the lattice-averaged two-spin correlation $\langle S_i \cdot S_j \rangle$, but depending on the geometry of the lattice and the postulated order, this can include three, four, or six spin correlators as a means of identifying distinct phases among the ground states.  In many cases, the unstable region is already known from where semi-classical methods break down, and the range will be well-defined. 
\begin{itemize}
\item \textbf{Inputs}:
\begin{itemize}
    \item A Hamiltonian of the form in Eq. \eqref{eq:GeneralHamiltonian}  with values defined from regions where semiclassical methods fail or a proposed exotic quantum phase (Fig. \ref{fig:workflow}) exists.
\end{itemize}

\item \textbf{Outputs}
\begin{itemize}
    \item Spin correlations of the ground state.
\end{itemize}

\item \textbf{Workflow}: The flowchart showing the \emph{end to end workflow} is given in Figure~\ref{fig:workflow}. The \emph{hard computational module} is the \emph{ground state preparation}.

\end{itemize}


\subsubsection*{Why classical methods are not sufficient to perform the hard computational module}

As explained in Section~\ref{sec:classical}, approximating the behavior of a thermodynamic quantum spin system requires diagonalizing huge matrices, often greater than $10^{300} \times 10^{300}$. 

\subsubsection*{Concrete problem instantiations}

For the input, the workflow is applicable to all the listed systems in Table \ref{tab:exampleModels}. We choose two especially long-standing problems.
\begin{itemize}
\renewcommand\labelitemi{}
    \item \emph{Input Option 1}: 2D square lattice tuned to instability, with second neighbor $J_2$ over first neighbor $J_1$ tuned to instability $J_2/J_1 = \frac{1}{2}$ (see Table \ref{tab:exampleModels} for references). The key question to answer for this model is: what is the nature of the intermediate phase? Is it a valence bond solid, or quantum spin liquid \cite{qian2023absence,LIU20221034}?
    \item \emph{Input Option 2}: 2D triangular lattice antiferromagnet, with second neighbor $J_2$ over first neighbor $J_1$ near $J_2/J_1 = \frac{1}{8}$. In the quantum limit, there is a quantum spin liquid phase between approximately $0.07 < J_2/J_1 < 0.16$~\cite{PhysRevB.92.041105,PhysRevB.92.140403,PhysRevB.93.144411,PhysRevB.94.121111,PhysRevB.95.035141,PhysRevB.96.075116,PhysRevLett.123.207203}. The key questions for this model are: what is the exact range of this phase in $J_2/J_1$? Is it a resonating valence bond liquid, or another type of spin liquid?
\end{itemize}

\begin{itemize}
\renewcommand\labelitemi{}
    \item \emph{Discretization of parameter space}: The range of the parameter values $J_1$ is disretized into $\approx 100$ points to provide sufficient accuracy for mapping out the local phase diagram around an instability or quantum phase.
    \item \emph{Output precision}: The ground state must be evaluated to an energy accuracy $\approx |J_{1}|/1000$ where $J_1$ is the largest exchange in the Hamiltonian.
\end{itemize}


\begin{figure}[!ht]
	\centering
	\includegraphics[width=0.5\textwidth]{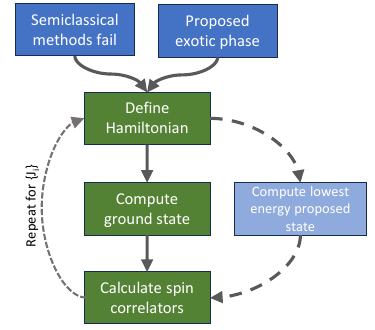}
	\caption{\small Computational workflow for ground state calculation.}
	\label{fig:workflow}%
\end{figure}



\subsection*{Application 2: Validating the spin Hamiltonian of a compound and predicting the dynamics of an exotic quantum phase}

Computing the magnetic dynamics of model spin systems for comparison with experimental data from neutron scattering experiments. 

\subsubsection*{Specific background of the application}

For quantum magnetic materials, a standard measurement to study the magnetic dynamics is inelastic neutron scattering. This measures the lattice-averaged two-point spin correlation function $\langle S_i \cdot S_j \rangle$, Fourier transformed in time and space \cite{boothroyd2020principles}. For a given magnetic Hamiltonian, in the case where the magnetic excitations act like boson quasi-particles, the inelastic neutron spectrum can be modeled by spin wave theory \cite{fishman2018spin} and can be efficiently predicted using current computing hardware. This prediction can be used to fit the parameters of the magnetic Hamiltonian by comparing with exprerimental data. 
However, for more exotic quantum materials, such analysis tools break down. 

A central goal in experimental condensed matter is understanding why certain materials act the way they do, and this requires understanding both the interactions between spins and the nature of the excitations. 
However without the ability to compare with a theoretical model, the underlying magnetic interactions (the spin exchange Hamiltonian) and magnetic dynamics (the type of quasiparticles) are extremely hard to determine.

\subsubsection*{Objectives}

The objective is to calculate the time-evolved two-point quantum spin correlation function of a particular lattice Hamiltonian in eq.~\eqref{eq:GeneralHamiltonian} after a local spin flip corresponding to a scattered neutron. The purpose of this is to compare with experimental neutron scattering data to verify or rule out a particular Hamiltonian, a parameter range in the Hamiltoninan or {type of dynamics to describe a particular material} (e.g., whether the boson quasiparticle description matches the full quantum simulation). 

\subsubsection*{End-to-end computational workflow}
We begin with a specific Hamiltonian, initially obtained by an educated guess at a compound's interactions (for an example in KYbSe$_2$, see below). From this, there are two possible routes in the computational workflow with the same goal: compute 
the two-point spin correlation $\langle S_i \cdot S_j \rangle$ as a function of time/energy. The first route is to calculate the spin correlations as a function of time by simulating the time dynamics corresponding to a neturon scattering experiment. The second is to calculate the spin correlations as a function of energy by calculating the correlations corresponding to the low energy wavefunctions of the Hamiltonian. For the first route, we begin from a sufficiently low energy state, randomly flip a spin corresponding to a scattered neutron in experiment, and allow the system to time-evolve with quantum interactions. After a specified time, an observation is made correlating the flipped spin to the rest of the lattice. The whole process is repeated for many different flipped spins and many different times to gain a complete, lattice-averaged picture.  The second route involves obtaining a low energy state and computing the corresponding spin correlation and repeating the process sufficiently many times until a complete picture of the energy dependence of the correlation is obtained. In the following we only consider the first route and present its inputs, outputs and specifications. 

\begin{itemize}
\item \textbf{Inputs}:
\begin{enumerate}
    \item A list of candidate Hamiltonians $\mathcal{H}$. This is usually obtained by an educated guess at a compound's interactions.
    \item Time length of the simulation $t_{sim}$ and time resolution $dt$.
    \item Total number of samples $M$ required to obtain a good lattice averaged picture of the time evolution. 
\end{enumerate}

\item \textbf{Outputs}
\begin{enumerate}
    \item For each of the $M$ samples, the lattice averaged two point spin correlation $\langle S_i \cdot S_j \rangle$ 
  as a function of time.
\end{enumerate}

\item \textbf{Workflow}: The flowchart showing the \emph{end to end workflow} is given in Figure~\ref{fig:neutron_scattering}. The \emph{hard computational modules} are the \emph{ground state preparation} and \emph{Hamiltonian simulation} as highlighted by the red boxes.

\end{itemize}

\subsubsection*{Why classical methods are not sufficient to perform the hard computational module}

Simulating quantum dynamics on large system sizes requires matrices beyond the reach of classical methods, see above. 

\subsubsection*{Concrete problem instantiations}
This workflow is applicable to many systems,  see Table \ref{tab:exampleCompounds}. We present specifications for two specific compounds
KYbSe$_2$ \cite{Scheie2024,Scheie_2024_KYS_PRB} and a sister material NaYbSe$_2$ \cite{Dai_2021,Scheie_2024_KYS_PRB}. \vspace{0.8em}

The material KYbSe$_2$ has a triangular lattice of magnetic Yb atoms, and behaves in a way that suggests it is close to a quantum instability, specifically a quantum spin liquid phase \cite{Scheie2024}. The exchange Hamiltonian is the 2D Heisenberg triangular model, with empirically determined $J_2/J_1 = 0.044 \pm 0.005$ \cite{Scheie_2024_KYS_PRB}. However, NaYbSe$_2$, has much larger uncertainty with $J_2/J_1 = 0.071 \pm 0.015$ \cite{Scheie_2024_KYS_PRB}, and appears to be a quantum spin liquid at low temperatures \cite{Dai_2021}.
\begin{itemize}
\renewcommand\labelitemi{}
    \item \emph{Input 1}: Two dimensional triangular lattice with a range of $J_2/J_1$ values between 0.035 and 0.09 in steps of 0.001  chosen to be able to address both the compounds.
    The size of the lattice to avoid finite size effects is chosen to be $100 \times 100$.
\end{itemize}

The experimental neutron scattering spectra for KYbSe$_2$ are shown in Fig. \ref{fig:kys}. The data were collected at very low temperatures with $k_B T/J_1 = 0.06$, 0.20, and 0.39, where { $J_1 = 0.438 
\pm 0.008$~meV}. The total timescale over which the inelastic spectra are probed is $\approx 10^{-10}$~s. The energy resolution used in the experiment is 0.04~meV (corresponding to $10^{-10}$~s) compared to a bandwidth of 1.5~meV (corresponding to $2.8 \times 10^{-12}$~s). Therefore, the experiment is equivalent to a quantum system simulation with a time-resolution of $2.8 \times 10^{-12}$~s and a total time of $10^{-10}$~s corresponding to $37$ time steps. This suggests that the system is being probed in a phase far from thermal equillibrium. For the application specification we conservatively suggest a higher resolution than the experiment. 
\begin{itemize}
\renewcommand\labelitemi{}
    \item \emph{Input 2}: Total simulation time $t_{sim} = 10^{-10}$~s and time resolution $dt = 2 \times 10^{-12}$~s.
    \item \emph{Input 3}: We choose the total number of samples $M = 100$, chosen to be the same as the lattice extent along one dimension to capture the motif of the ground state.

\renewcommand\labelitemi{}
    \item \emph{Output precision requirement}: The energy precision of the final state of the Hamiltonian simulation is chosen to be $\frac{|J_1|}{100}$ and a $1\%$ precision is chosen on the two-point correlation.
    \item \emph{Output wall-clock time limit}: One month  
\end{itemize}

 Constraining the exchange Hamiltonian in NaYbSe$_2$ (and other members of the chemical series) would be an important role of simulating the inelastic neutron spectrum. What is more, an accurate simulation of $S(q,\omega)$ could be compared with different quasiparticle simulations \cite{PhysRevB.105.224404}, and would give great confidence that the observed behavior is indeed from this non-trivial quantum liquid phase. 

\begin{figure}[!ht]
	\centering
	\includegraphics[width=0.8\textwidth]{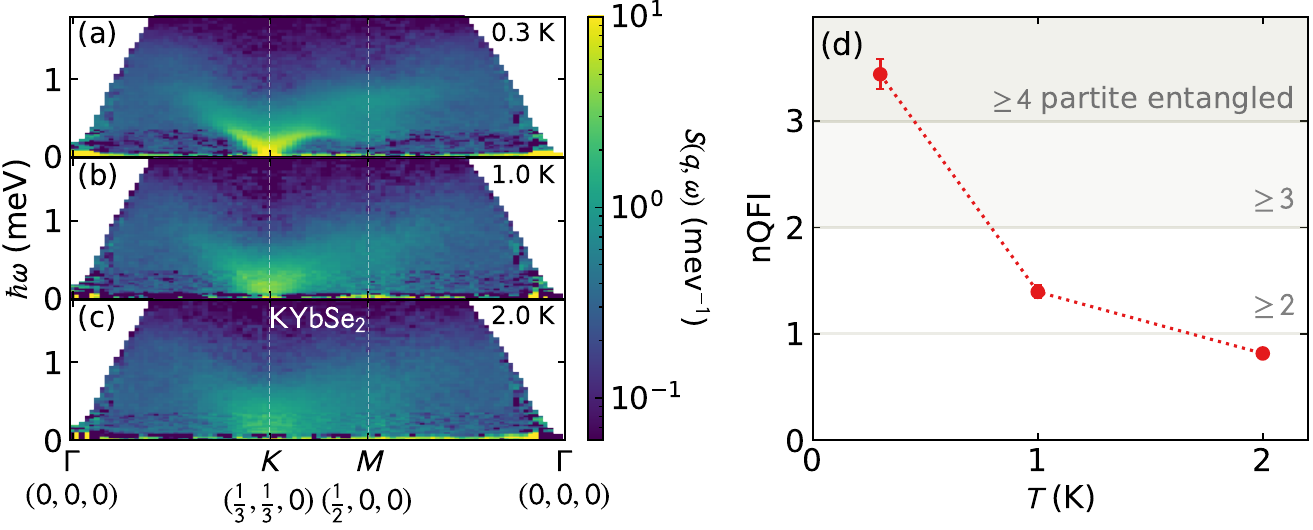}
	\caption{\small Experimental neutron scattering on KYbSe$_2$ at three different temperatures (a)-(c). The $y$ axis gives energy transfer, the $x$-axis gives different scattering directions in reciprocal space units, and the color scale indicates the scattered neutron intensity. Panel (d) shows the Quantum Fisher Information (d) computed from the spectra in panels (a)-(c) via Eq. \ref{eq:QFI:Hauke}. Here ${\rm nQFI} = f_\mathcal{Q}/(12 S^2)$ gives a lower bound on entanglement depth. 
 }

	\label{fig:kys}
	\vspace{-0.2cm}
\end{figure}

\begin{figure}[!ht]
	\centering
	\includegraphics[width=0.8\textwidth]{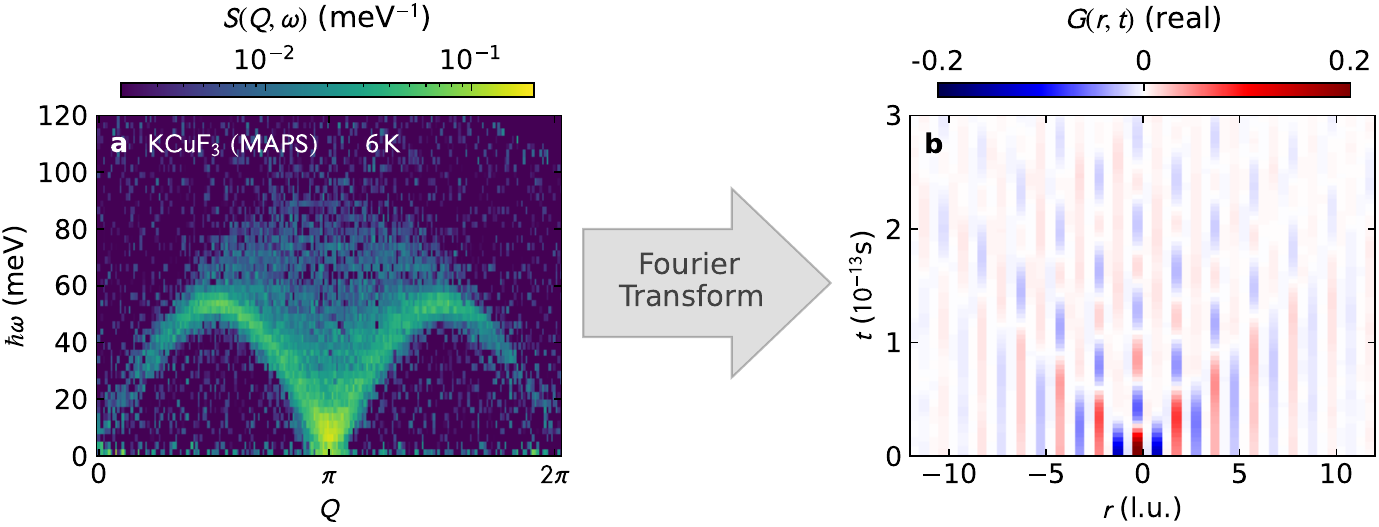}
	\caption{\small Fourier-transforming neutron scattering from frequency and reciprocal space to time and distance. Panel (a) shows the experimental neutron spectrum of KCuF$_3$. Panel (b) shows the spectra transformed into real space on the $x$ axis and time on the $y$ axis \cite{scheie2022quantum}, which corresponds to a lattice-averaged time-evolved spin correlation.}
	\label{fig:FourierTransform}
	\vspace{-0.2cm}
\end{figure}

\begin{figure}[!ht]
    \centering
    \includegraphics[width=0.5\textwidth]{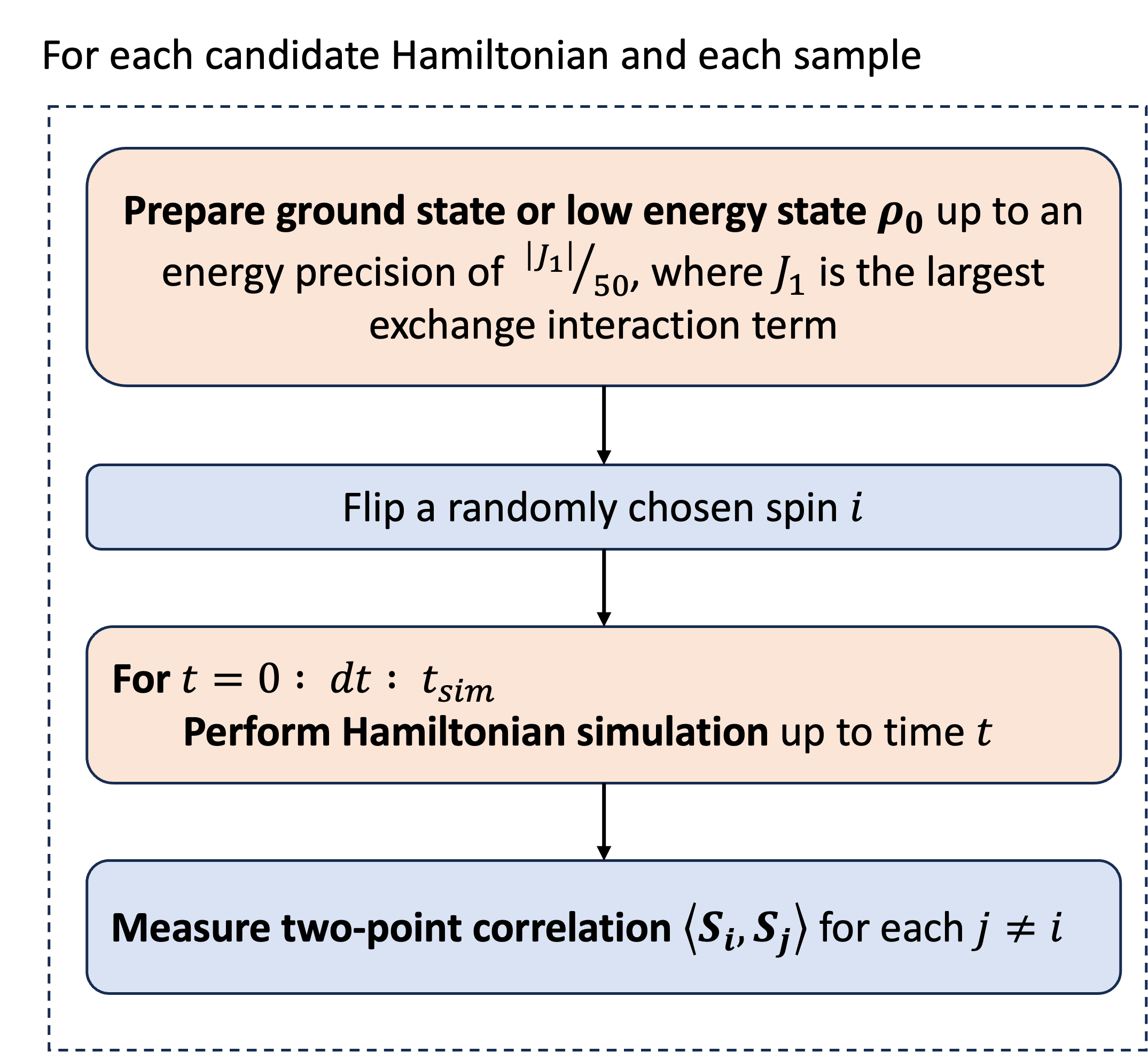}
    \caption{Workflow for Hamiltonian dynamics to compare with neutron scattering data. The red boxes contain the hard computational task/kernel.}
    \label{fig:neutron_scattering}
\end{figure}


\subsection*{Application 3: Using quantum correlations and quantum information to certify the existence and extent of exotic phases}

Calculate quantum correlation functions for given spin Hamiltonians to understand their quantum ground state and compare with theory.

\subsubsection*{Specific background of the application}

Although computing total correlation functions like $\langle S_i \cdot S_j \rangle$ can reveal much about a system, even more can be revealed by {\bf quantum correlation functions}. For example, quantum Fisher information, which for a pure state is expressed as 
\begin{equation}
    F_Q = 4(\langle \psi| \hat{S}(-\vec{Q}) \hat{S}(\vec{Q}) |\psi \rangle - \langle \psi| \hat{S}(\vec{Q})  |\psi \rangle ^2 )
    \label{eq:QFI}
\end{equation}
\cite{Hauke2016} gives a lower bound on entanglement depth \cite{Hyllus_2012}. Here $\vec{Q}$ is the translation vector in the reciprocal space and $\hat{S}(\vec{Q}) = \sum_i  \hat{S}_{i} e^{i \vec{r}_i \cdot \vec{Q}}$ where $\vec{r}_i$ refers to the position of the spin in the 2-D lattice. In fact, there are a large number of experimentally accessible quantum correlation functions \cite{Roscilde_2004,PhysRevB.94.075121,Frerot2019}, which have distinct meaning in terms of entanglement witnesses (akin to Bell's theorem), and can be used to diagnose quantum entanglement transitions \cite{laurell2020dynamics}.  
Simulating highly entangled spin Hamiltonians using a quantum computers offers the possibility to study the entanglement structure and dynamics in far larger systems than is possible to study with classical computers.

In the last few years it has been demonstrated that quantum correlations, and correspondingly quantum entanglement, can be extracted from the spin correlations $S({\bf Q},\omega)$ measured by neutron scattering via the equation
\begin{equation}
	f_\mathcal{Q} \left[ {\bf Q} \right]  = 4 \int_{0}^\infty \mathrm{d}(\hbar \omega) \tanh \left( \frac{\hbar \omega}{2 k_BT }\right) 
 \left( 1-e^{-\hbar\omega/ k_B T} \right)
 S({\bf Q},\omega)	,	\label{eq:QFI:Hauke}
\end{equation} 
\cite{scheie2021witnessing,laurell2020dynamics,scheie2022quantum,PhysRevB.106.085110,PhysRevB.107.054422,scheie2023reconstructing}, also see Fig. \ref{fig:kys}(d) where ${\rm nQFI} = f_\mathcal{Q}/(12 S^2)$. This allows simulations to be compared to experiment in new ways, and to establish the entanglement structure of experimental quantum materials---which are not guaranteed to be the same as a simplified theoretical model \cite{syzranov2022eminuscent}.

\subsubsection*{Objectives}

The first objective is to understand the fundamental quantum properties of exotic quantum phases of matter. We do this by calculating quantum correlation functions of exotic quantum spin phases (see Table \ref{tab:exampleModels})  to distinguish between their quantum properties \cite{li2024multipartite}. 
The second objective is to compare the entanglement structure of simulated lattices to real materials. This is important to discern, e.g., whether infinitesimal amounts of spin exchange disorder (which can be theoretically controlled but not experimentally controlled) destroy a given quantum phase \cite{syzranov2022eminuscent}.

\subsubsection*{End-to-end computational workflow}

This workflow is identical to Application 1, but instead of simplistic spin correlation functions, we calculate quantum correlation functions. The quantum correlation function most straightforward to calculate and interpret is Quantum Fisher Information (QFI, eq. \ref{eq:QFI}), which can be extracted from experimental data \cite{scheie2021witnessing}. This is then used to analyze the entanglement structure of a given phase (where QFI is analyzed in real space, looking for the pattern of highly entangled bonds), or for comparison with experimental data (where QFI is analyzed in Fourier space). 
The input and output of the workflow are as follows: 

\begin{itemize}
\item \textbf{Inputs}:
\begin{enumerate}
    \item A list of Hamiltonians $\mathcal{H}$ around a quantum phase or quantum instability (see Table \ref{tab:exampleModels}), with sufficient granularity in the parameters to distinguish phase boundaries. 
    \item A specified set of points for $\vec{Q}$ in the two-dimensional reciprocal space. 
\end{enumerate}

\item \textbf{Outputs}
\begin{enumerate}
    \item Quantum Fisher Information of the ground state, calculated via Eq. \ref{eq:QFI} for pure states. 
    \item Optionally, other quantum correlation functions as defined in Ref. \cite{scheie2023reconstructing}. 
\end{enumerate}

\item \textbf{Workflow}: The flowchart showing the \emph{end to end workflow} is given in Figure~\ref{fig:workflow}, but with ``calculate spin correlators'' indicating the quantum correlation functions discussed above. The \emph{hard computational module} is the \emph{ground state preparation}. 

\end{itemize}

\subsubsection*{Why classical methods are not sufficient to perform the hard computational module}

Simulating quantum dynamics on large system sizes requires matrices beyond the reach of classical methods, see above. 

\subsubsection*{Concrete problem instantiations}

Any of the phases in Table \ref{tab:exampleModels} would be relevant to this application, as would the [Na,K]YbSe$_2$ compound described above. 

For a concrete example, let us take the 2D square lattice antiferromagnet with $J_1$ nearest neighbor exchange (along square edges) and $J_2$ second neighbor exchange (along square diagonals). In this case, the region of interest would be between 
\begin{itemize}
\renewcommand\labelitemi{}
    \item \emph{Input 1}: Two dimensional triangular lattice with a range of $J_2/J_1$ values between 0.5 and 0.65 in steps of 0.002. 
    The size of the lattice to avoid finite size effects is chosen to be $100 \times 100$.
    \item \emph{Input 2}: A grid of $100 \times 100$ uniformly spaced points in the region $\vec{Q} \in [0,1/2]\times[0,1/2]$. 
    \item \emph{Input 3} (Optional): random Gaussian noise to the $J_1$ exchange interactions, with FWHM $\delta_J$ between 0 and $J_1$ in steps of $J_1/50$.
\end{itemize}

\begin{itemize}
\renewcommand\labelitemi{}
    \item \emph{Output- precision requirement}: The energy precision of the ground state is chosen to be $\frac{|J_1|}{1000}$. 
    \item \emph{Output wall-clock time limit}: One month. 
\end{itemize}

Obtaining quantum correlation/entanglement as the lattice is tuned through nontrivial quantum points will allow the underlying quantum state to be rigorously characterized, and studying the quantum phase as a function of exchange disorder will give a picture of disorder in real materials might affect the stability of the quantum phase. 


% ...




\section{Requirements summary}
This section summarizes the application requirements for all the applications described in this chapter.

\begin{small}
\subsection*{Application 1: Ground state of systems near instabilities}
\begin{tabular}{ |l l l| }
    \hline
        & & \\
     & Total time limit & 1 month \\ [1ex]
    \textbf{Workload:} & Number of subroutine calls required & $100$ \\ [1ex]
     & Maximum subroutine time limit & 1 month \\ [1ex]
     & & \\[1ex]
     & Model type & Hiesenberg spin hamiltoninan \\[1ex]
    & Size & minimum  $30 \times 30$ sites \\
     &  & target $100 \times 100$ sites \\[1ex]   
     \textbf{Problem specifications:} &  Interaction Structure & Sparse regular (2-D square or triangular lattice) \\[1ex]
     & Computational target & $2,4,6$-point spin correlations of ground state \\[1ex]
     & Accuracy requirement & $|J_{1}|/1000$ for ground state energy  \\
     & & and $1\%$ relative on correlations \\
     & & \\
     \hline
\end{tabular}
\end{small}

\subsection*{Application 2: Validating the spin Hamiltonian of a compound and predicting the dynamics of an exotic quantum phase}
\begin{small}

\begin{tabular}{ |l l l| }
    \hline
        & & \\
     & Total time limit & 1 month \\ [1ex]
    \textbf{Workload:} & Number of subroutine calls required & \#spin-flips$\times$paramter resolution$\times$time-steps \\
    & & $= 100 \times 100 \times 55 \times 50$ \\ [1ex]
     & Maximum subroutine time limit & 1 month \\ [1ex]
     & & \\[1ex]
     & Model type & Hiesenberg spin hamiltoninan \\[1ex]
     & Size & minimum  $30 \times 30$ sites \\
     &  & target $100 \times 100$ sites \\[1ex]    
     \textbf{Problem specifications:} &  Interaction Structure & Sparse regular (2-D triangular lattice) \\[1ex]
     & Computational target & $2$-point correlations after time evolution \\[1ex]
     & Accuracy requirement & $|J_{1}|/100$ for energy of the final state  \\
     & & and $1\%$ relative on correlations \\
     & & \\
     \hline
\end{tabular}
\end{small}

\subsection*{Application 3: Using quantum correlations and quantum information to certify the existence and extent of exotic phases} 
\begin{small}
\begin{tabular}{ |l l l| }
    \hline
        & & \\
     & Total time limit & 1 month \\ [1ex]
    \textbf{Workload:} & Number of subroutine calls required & \# parameters $ = 75$ \\ [1ex]
     & Maximum subroutine time limit & 1 month \\ [1ex]
     & & \\[1ex]
     & Model type & Hiesenberg spin hamiltoninan \\[1ex]
      & Size & minimum  $30 \times 30$ sites \\
     &  & target $100 \times 100$ sites \\[1ex]     
     \textbf{Problem specifications:} &  Interaction Structure & Sparse regular (2-D square or triangular lattice) \\[1ex]
     & Computational target & Quantum correlation functions (Fischer \\
     & & information) computed from the ground state \\[1ex]
     & Accuracy requirement & $|J_{1}|/1000$ for ground state energy and \\
     & & $1\%$ relative on quantum correlation functions \\
     & & \\
     \hline
\end{tabular}
\end{small}

\newpage
\end{chapter}

\begin{chapter}{Driven-dissipative Dicke model in the ultrastrong coupling regime} \label{ch:optical_cavity}
    

The interaction between light and matter is at the core of most our nanotechnologies. Using atoms to produce highly coherent sources of light , i.e., lasers, that in turns can be used to control particles at the individual level is one of the most remarkable success of quantum optics. Even though the nature of the interactions between light and matter are well-understood, the consequences of these interactions are not, in particular when matter and light are highly entangled. The Dicke model is one of the simplest example of light-matter coupling, realizable experimentally, that exhibits such exotic behavior in what is called the ustrastrong coupling regime. While being proposed more than fifty years ago, the analysis of this model is still elusive due to its difficult highly quantum nature. In this chapter, we introduce the Dicke model and its connection to quantum optic. We then describe the computational workflow needed to understand the propriety of this model and how quantum computing resources can help in portraying its phase diagram.\\

\noindent \textbf{Hamiltonian Type}:  Spins $+$ multi-level state Hamiltonian.\\
\textbf{Quantum Computational Kernels}: Ground State Preparation, Open Quantum System Simulation.

\clearpage

\section{Application area overview}

\subsection{High-level description of the application area}

Recent advances in the development of optical functional materials and the fabrication of complex nanostructures incorporating these materials have unlocked broad opportunities for designing on-chip photonic circuits. These circuits hold vast potential for applications, including tunable coherent light sources, optoelectronics, light harvesting and energy conversion devices, and the development of solid state devices for quantum communication and computation. The functionality of these nanostructures relies on our capacity to control the interaction between light (photons) and optically active nanoparticles (quantum emitters). For the sake of technological advances to achieving this control, fundamental physical aspects governing various regimes of coupled photons and quantum emitter excitations should be understood. The coupling strength, which quantifies the rate of energy exchange between photons and quantum emitters is a key parameter governing the emergence of novel states in the composite light-matter excitations known as polaritons. When the coupling is weaker than any rate of energy losses in the quantum emitters and the confining optical cavity, this regime is identified as the weak coupling regime. Conversely, the strong coupling regime characterizes the opposite situation with distinct features in polariton physics. The ultra-strong coupling (USC) regime is defined by a coupling rate that constitutes a significant fraction (0.1 or larger) of the photon energy that the quantum emitters can absorb or emit. This regime manifests distinctly different quantum  properties than those observed in the weak and strong coupling regimes. Theoretical analysis and modeling of such properties encounter considerable challenges, even with substantially simplified light-matter interaction models. 
      
\subsection{Utility estimation}
\subsubsection{Overview of the value of the application area}
Addressing the challenges in USC theory promises to yield new fundamental insights into emergent phenomena, laying the groundwork for experimental investigations. Together, these efforts will expand our fundamental understanding of the behavior of light and matter. Moreover, exploring the USC regime in various nanostructures holds the potential for technological advancements, this includes the design of new photonic circuits for the use in telecommunication and optoelectronics. As detailed below, USC has been demonstrated not only in the optical spectrum but also in microwave superconducting circuits and magnetic materials, unlocking technological possibilities for designing novel information storage and processing devices, potentially including ultrafast quantum computers. 

\subsubsection{Concrete utility estimation}
Three projects were identified at LANL that are connected with the study USC regime. These projects were or are funded internally, each for a duration of three years, and include both an experimental and a theoretical component.  The first research project entitled ``Hybrid Photonic-Plasmonic Materials: Toward Ultimate Control Over the Generation and Fate of Photons" took place in FY 2017-2019 with a total budget of \$4.8M, for which $30\%=\$0.5$M per year was allocated in labour cost to theoretical investigations of the USC regime. The two other projects, ``Complementary Metal–oxide–semiconductor (CMOS)-Compatible Exploitation of Emergent Nanoplasmonic Phenomena in the Near-Infrared" and ``Paired Exciton-Polariton Condensates and Bloch Surface Waves as Neurons and Synapses in Photonic Neuromorphic Computers" are currently undergoing from FY 2023 to FY 2025 and have each a total budget of approximately $\$750$K, The part of this budget allocated in labour cost to the study of USC regime represents $\$95$K per year and $\$90$K per year respectively.
Therefore, the total investment made by these projects on answering this particular scientific question is approximately $\$2$M. It is worth recalling that the approximation methods used in these projects for producing phase diagrams, such as depicted latter in this chapter, are insufficient to accurately describe and predict the physics of the problem.

\section{Problem and computational workflows}

\subsection{Detailed background of the application area}

Over the last decade, the USC regime has undergone extensive theoretical and experimental investigations. Comprehensive reviews of this topic can be found in Refs.~\cite{friskNRP:2019,fornRMP:2019}. According to these reports,  the USC has been experimentally realized across diverse quantum systems, including semiconductor quantum wells, molecules within optical and plasmonic cavities, microcavity exciton-polaritons, magnons in microwave cavities, optomechanical systems, and superconducting circuits. It is noteworthy that some of the experimental implementations have served as a hardware platform for quantum simulations, validating theoretical predictions of various USC models.~\cite{langfordNatComm:2017,lvPRX:2018,petersonPRL:2019} Consequently, the applications of quantum systems in the USC regime span from low-threshold solid-state lasers and efficient quantum photon sources to quantum spectroscopies, control of chemical reactions,  ultrafast quantum computing, and optomechanics.~\cite{friskNRP:2019,fornRMP:2019} 

Understanding the complex cavity quantum electrodynamics in the USC regime relies on theoretical methods developed in the past decade as reviewed in Ref.~\cite{leBoiteAQT:2020}. The review outlines efforts to investigate the spectral properties (eigenstates and eigenenergies) of coupled quantum emitters and a cavity mode forming closed quantum system under USC.  Using the Rabi model describing a single two-level quantum emitter interacting with the cavity mode, the importance of accounting for the effect of counter-rotating terms breaking standard rotating-wave approximation (RWA) has been demonstrated. Yet, the interpretation of the experimentally studied quantum systems calls for treating both the quantum emitters and the photonic cavity as an open quantum system, accounting for dissipation and energy supply through various driving mechanisms. Ref.~\cite{leBoiteAQT:2020} underscores the challenges posed by the USC when applying two key theoretical approaches to open quantum systems—the reduced density matrix master equation and quantum Langevin equations in conjunction with the input-output formalism. Among these considerations, it is necessary to introduce common bath degrees of freedom for the coupled quantum emitters and cavity photons. The dissipation rates should be treated as time-dependent, necessitating the inclusion of non-Markovian effects. Finally, ensuring the fulfillment of gauge invariance is crucial for models describing quantum systems in the USC regime~\cite{leBoiteAQT:2020}. These constraints impose restrictions on conventional quantum optics approximations. For example, the use of a two-level approximation for quantum emitters in certain cases yields inconsistent outcomes in the Coulomb and multipolar gauges. Additionally, the diamagnetic contribution of the photon fields cannot be neglected.
 
Theoretical challenges outlined above impose significant limitations on the analytical and computational techniques developed for studying ensembles of quantum emitters coupled to cavity photons in the strong coupling regime while extending them to the USC. Within this framework, we explore a representative generalization of the Rabi model to the driven-dissipative Dicke model. In the strong coupling regime the driven-dissipative Dicke model predicts different non-equilibrium phases of polariton state~\cite{kirtonAQT:2019}. However, the extension of such phase diagrams to the USC is yet to be done. We argue that the utilization of noiseless quantum computers should enhance our modeling capabilities and offer insights into non-equilibrium polariton states under the USC regime beyond the constraints imposed by classical computing.


\subsection{Application 1: Low energy spectrum of Dicke and Tavis-Cummings models}

\begin{figure}[b]
\centering
\includegraphics[width=0.4\textwidth]{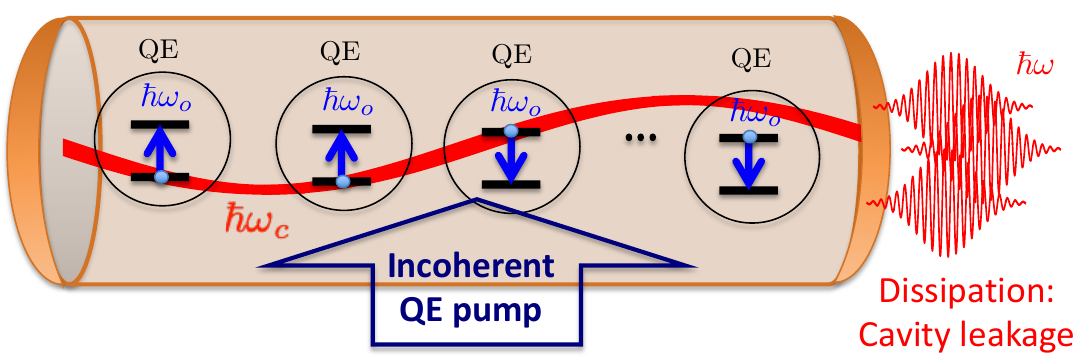}
\caption{Schematic of driven-dissipative Dicke model: An ensemble of ${\cal N}_o$ two-level quantum emitters with transition transition energy $\hbar\omega_o$ and cavity photon mode of energy $\hbar\omega_c$ interact via coherent quantum exchange rate $\lambda$. The quantum emitters are subject to the incoherent pump and internal relaxation processes whereas the cavity leakage results in the photon emission outside the cavity.
}
\label{Fig:SchDicke}
\end{figure}

\subsubsection{Specific background of the application}

The Dicke model's schematic, illustrated in Fig.~\ref{Fig:SchDicke}, shows the photonic cavity mode of energy $\hbar\omega_c$, described by the Bose annihilation and creation operators $\hat a$ and $\hat a^\dag$, respectively. Identical two-level quantum emitters with energy $\hbar\omega_o$ positioned at sites $n=\overline{1,{\cal N}_o}$ and characterized by the Pauli SU(2) operators $\hat \sigma_n^\pm=\hat \sigma_n^x \pm i\hat \sigma^y_n, \sigma_n^z$, interact with the cavity mode. The strength of this interaction is defined by the coherent quantum exchange rate, $\lambda$. Consequently, we describe the system through the following Dicke Hamiltonian
\begin{eqnarray}
\label{H-Dk}
\hat H_D &=& \hbar\omega_c\hat a^\dag\hat a 
		+\frac{\hbar\omega_o}{2}\sum\limits_{n=1}^{{\cal N}_o}\left(\hat \sigma_n^z+1\right) 
		+\hbar\lambda\left(\hat a+\hat a^\dag\right)\sum\limits_{n=1}^{{\cal N}_o}\left(\hat \sigma_n^{-}+\hat \sigma_n^{+}\right),
\end{eqnarray}
which is invariant under $\mathbb{Z}_2$ symmetry transformations.

The Dicke Hamiltonian gives the most general account of the quantum-emitter-cavity-mode interactions including the rotating terms $\hat a^\dag\hat\sigma^{-}_n+\hat\sigma^+_n \hat a$ and the counter rotating terms $\hat a^\dag\hat\sigma^{+}_n+\hat\sigma^{-}_n \hat a$ standing for simultaneous (de)excitation of the quantum emitters and cavity mode. In the Rotating Wave Approximation (RWA), neglecting the later terms, the Dicke model becomes the Tavis-Cummings model with the Hamiltonian
\begin{eqnarray}
\label{H-TC}
\hat H_{TC} &=& \hbar\omega_c\hat a^\dag\hat a 
		+\frac{\hbar\omega_o}{2}\sum\limits_{n=1}^{{\cal N}_o}\left(\hat \sigma_n^z+1\right) 
		+\hbar\lambda\sum\limits_{n=1}^{{\cal N}_o}\left(\hat a^\dag\hat\sigma_n^{-}+\hat \sigma_n^{+}\hat a\right).
\end{eqnarray}
Notice, that dropping the counter-rotating terms lifts symmetry of the Tavis-Cummings model to U(1). Finally, the singe quantum emitter (n=1) limit of the Dicke and Tavis-Cummings models is refereed to as the Rabi and Jaynes-Cummings models, respectively.

\subsubsection{Objectives}

Our initial goal is to conduct a spectral analysis of the Dicke Hamiltonian~(\ref{H-Dk}) within the USC regime to specifically identify the parameter range where the inclusion of the counter-rotating terms becomes important. Achieving this will need a comparative analysis involving the solution of the spectral problem for the Tavis-Cummings Hamiltonian~(\ref{H-TC}). Additionally, we will explore variations in the number of quantum emitters, ranging from $N=1$ to  anticipated order of $10^2-10^3$. This investigation aims to examine the evolution of the eigenspectra comparing with the  well-established spectra of the Rabi and Jaynes-Cummings models. 

\subsubsection{End-to-end computational workflow}

The input to the workflow are the Hamiltonians in (\ref{H-Dk}) and (\ref{H-TC}) and the parameter ranges. The trial ground state should be prepared in the form of squeezed vacuum state.
\begin{small}
\begin{lstlisting}
set N0 in range 100 to 1000
    for omega_0:  0.7 omega_c to 1.3 omega_c, step domg = 0.05 omega_c
        for lambda: 0.5 omega_0 to 1.5 omega_0, step dlmb = 0.01 omega_0
             # Find first 100 eigenenergies of the Hamiltonians (1) and (2)
             # with precision in range 0.01 omega_0 to 0.001 omega_0
        end
    end
output: array of eigenergies vs corresponding omega_0 and lambda values   
\end{lstlisting}
\end{small}

\subsubsection{Why classical methods are not sufficient to perform the hard computational module}

The capacity of classical hardware introduces constraints on the number of quantum emitters and bosonic degrees of freedom that can be accommodated in the simulations. The bottleneck in this computational process is the limitation on the number of bosonic quanta of the cavity mode. For instanse, simulations using the QuTiP package~\cite{johanssonCPC:2012} allow for numerically exact handling of approximately 10 quantum emitters coupled to about 50 bosonic quanta. Classical methods can be employed to diagonalize the Hamiltonians~(\ref{H-Dk}) and (\ref{H-TC}) using the Lanczos algorithm that enables the extraction of a few, 1 to 10, lowest energy eigenstates. For larger number of eigenstates quantum algorithms must be employed. Pushing the boundaries to encorporate 1000 quantum emitters and hundreds of bosonic quanta becomes crucial when dealing with USC regime, however, these scenarios become challenging to realize using the classical hardware.

\subsubsection{Concrete problem instantiations}

The eigenenergies obtained in the workflow outlined above will be plotted against coupling strength for a fixed number of quantum emitters and various values of detuning $\omega_c-\omega_o$. This graphical analysis should facilitate the identification of coupling parameters beyond which the Dicke spectrum diverges from that of the Tavis-Cummings model due to the effect of the counter-rotating terms. In addition, by comparison with the lowest eigenstates obtained using the classical hardware, one will identify the limitations of this approach. Conversely, the case of a single quantum emitter will be compared with the outcomes of the Rabi and Jaynes-Cummings models found in the existing literature\cite{friskNRP:2019} and the evolution of the spectra for the increasing number of quantum emitters analyzed.  

\subsubsection{List of other candidate instances}

The Dicke Hamiltonians~(\ref{H-Dk}) of interest have identical coupling rates $\lambda$ for quantum emitters with both rotating and counter-rotating terms. Nevertheless, these parameters can be varied, leading to what is known as the asymmetric Dicke model~\cite{kirtonAQT:2019,leBoiteAQT:2020}. The numerical examination of the interplay between these parameters and their impact on spectral properties is a worthwhile investigation.

\subsection{Application 2: Phase diagram of the driven-dissipative Dicke model}

\subsubsection{Specific background of the application}

By taking into account that the cavity and quantum emitters form an open quantum system, we introduce the associated reduced density operator $\hat\rho$ whose time evolution is described by the Liouville-Lindblad master equation
\begin{eqnarray}
\label{Liov-Lind-eom}
\partial_t\hat\rho = -\frac{i}{\hbar}\left[\hat H,\hat\rho\right] 
    + \frac{\gamma_c}{2}\hat{\cal D}_{\hat a}[\hat\rho]
    + \frac{\gamma_\uparrow}{2}\sum\limits_{n=1}^{{\cal N}_o}\hat{\cal D}_{\hat \sigma_n^{+}}[\hat\rho]
	+ \frac{\gamma_\downarrow}{2}\sum\limits_{n=1}^{{\cal N}_o}\hat{\cal D}_{\hat \sigma_n^{-}}[\hat\rho]
	+ \gamma_\phi\sum\limits_{n=1}^{{\cal N}_o}\hat{\cal D}_{\hat \sigma_n^z}[\hat\rho]
    +\eta\sum\limits_{n=1}^{{\cal N}_o}\hat{\cal D}_{\hat a,\hat s_n^{+}}[\hat\rho].
\end{eqnarray}
Here, the first term in the right-hand side accounts for coherent dynamics governed by the Hamiltonian $\hat H$ which is given by either Eq.~(\ref{H-Dk}) or (\ref{H-TC}) and the rest of the terms account for the incoherent energy pump and the dissipation  processes using the Lindblad superoperator $\hat{\cal D}_{\hat{\cal O}}[\hat\rho] = 2\hat{\cal O}\hat\rho\hat{\cal O}^\dag-\hat{\cal O}^\dag\hat{\cal O}\hat\rho-\hat\rho\hat{\cal O}^\dag\hat{\cal O}$. Specifically, we care about the cavity leakage with the rate $\gamma_c$, the quantum emitter population (pump) decay with the rate ($\gamma_\uparrow$) $\gamma_\downarrow$, and the quantum emitter pure dephasing processes with the rate $\gamma_\phi$. The last term in Eq.~(\ref{Liov-Lind-eom}) represents the dissipative interaction via common bath degrees of freedom between the cavity mode and quantum emitters characterized by the rate $\eta$. The associated Lindblad operator is 
$\hat{\cal D}_{\hat a,\hat s_n^{+}}[\hat\rho]= 2\hat a\hat\rho\hat s_n^{+}-\hat s_n^{+}\hat a\hat\rho-\hat\rho\hat s_n^{+}\hat a +2\hat s_n^{-}\hat\rho\hat a^\dag-\hat a^\dag\hat s_n^{-}\hat\rho-\hat\rho\hat a^\dag\hat s_n^{-}$. We introduced this term into a generalized version of the Dicke model in Ref.~\cite{piryatinskiPRR:2020}. 

By tuning the quantum emitter and cavity mode frequencies in resonance, $\omega_o\approx\omega_c$, distinct coupling regimes can be distinguished. The strong coupling regime occurs when the coupling strength, scaling as $\sqrt{{\cal N}_o}\lambda$,  exceeds any of the relaxation rates listed above. In this regime, excitations of quantum emitters can form hybrid states with the cavity mode known as polariton states. These states exhibit emergence of spontaneous coherence between quantum emitters and cavity mode beyond the critical coupling rate, i.e., for $\lambda>\lambda_c$. In the framework of the Dicke model, spontaneous coherence can be associated with the nonequilibrium phase transition to the superradiant state, breaking $\mathbb{Z}_2$ symmetry. In contrast, the Tavis-Cummings model allows for a lasing phase transition associated with U(1) symmetry breaking. Conventional approach to explore the phase diagram involves a two-step process: initially, stability analysis is conducted in the mean-field approximation, which holds in the thermodynamic limit. Subsequently, long-range correlations are incorporated, e.g., via the second cumulant approximation which works well for relatively large ${\cal N}_o\gtrsim 10^2$ finite-size ensembles of quantum emitters.\cite{kirtonAQT:2019} 

An example of the phase diagram, depicted in Fig.~\ref{Fig:PHD}, shows results obtained using both the mean-field (a-c) and the second cumulant (d-i) approximations.\cite{piryatinskiPRR:2020} In the left and central columns of Fig.~\ref{Fig:PHD}, corresponding to the quantum emitter populations below inversion, we predict the coexistence of superradiant (SR) and lasing without inversion (LWI) polariton phases. In the right column of Fig.~\ref{Fig:PHD}, presenting the population inversion regime, only the regular lasing (RL) phase is observed. Of particular interest is the scenario where the coupling rates are ${\cal N}_o\lambda^2/\omega^2_o\gtrsim 0.3$ and $\eta/\omega_o\gtrsim 0.3$, indicative of the USC. In this instance, our phase diagram shows the overlap of the SR and LWI phases. This ``gray" region warrants further in-depth investigation, holding the potential for identifying novel polariton phases. Investigating the USC phase diagram, we should consider the higher-order cumulants, given that the second-order approximation may break down. Additionally, incorporating common bath conditions and accounting for potential non-Markovian effects in the calculations, as outlined earlier, is important. While other examples of Dicke model phase diagrams exist~\cite{kirtonAQT:2019}, extending these diagrams to the USC regime remains an open research problem, awaiting resolution. We foresee the utilization of noiseless quantum computers as a reliable tool for overcoming aforementioned problems and gaining deeper insights into this intriguing class of problems.

\begin{figure}[t]
\centering
\includegraphics[width=0.8\textwidth]{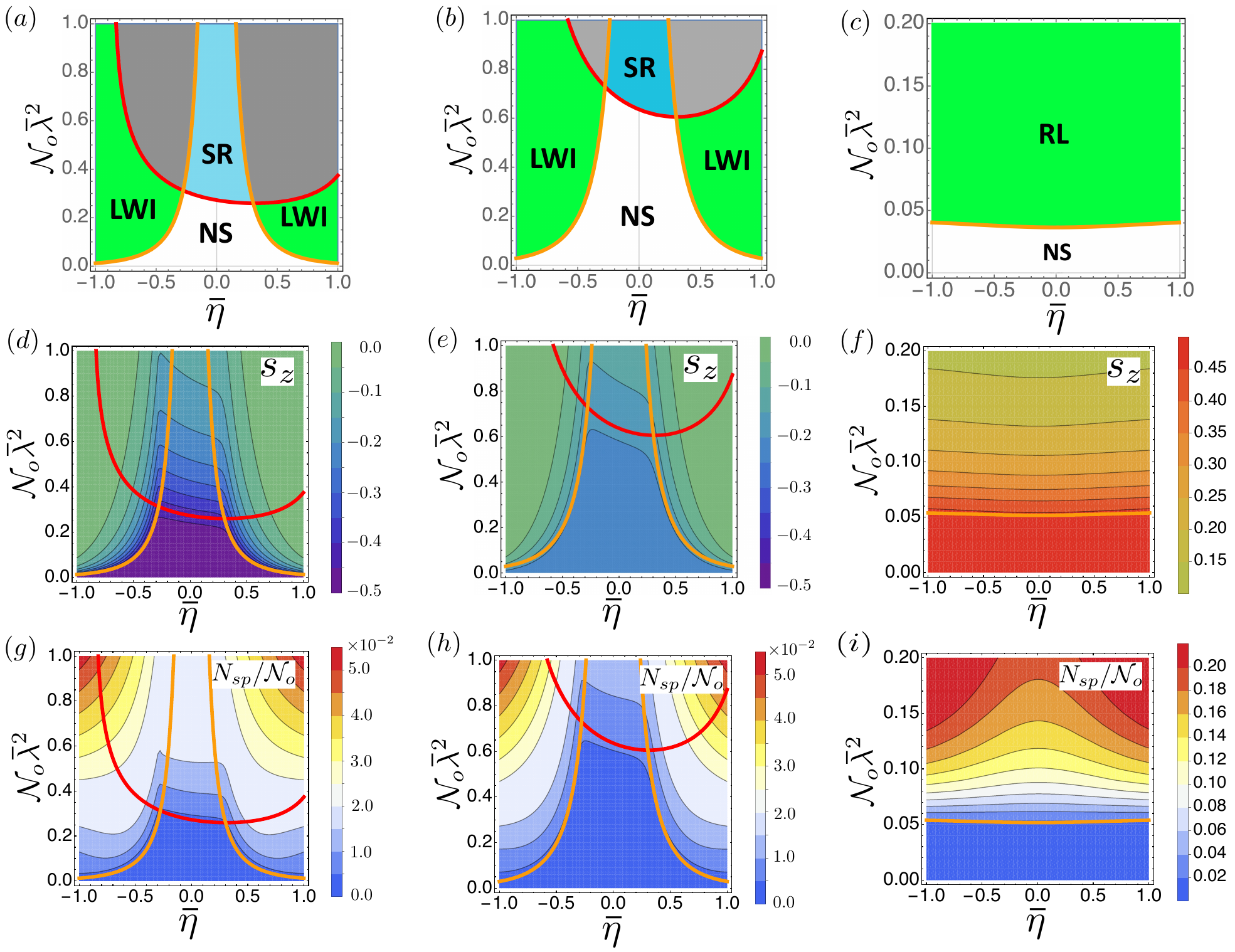}
\caption{Phase diagram of generalized Dicke model accounting for the coherent and dissipative coupling.~\cite{piryatinskiPRR:2020}
(a)-(c): Mean-field phase diagrams marking the normal (NS), superradiant (SR), lasing without inversion (LWI), and regular lasing (RL) states. (d)-(f) The steady state projection on the quantum emitter population inversion $s_z=\sum_{n=1}^{{\cal N}_o}\langle\hat\sigma^z_n\rangle/2{\cal N}_o$ and (g)-(i) reduced cavity mode population (the order parameter) ${\cal N}_{sp}=\langle\hat a^\dag \hat a\rangle$ normalized per total number of quantum emitters ${\cal N}_o$  calculated for the driven dissipative Dicke model with the coherent $\bar\lambda=\lambda/\omega_o$ and dissipative $\bar\eta=\eta/\omega_o$ couplings. The left and central columns represent quantum emitters below population inversion and the right one above. The diagram segments for $\sqrt{{\cal N}_o}\bar\lambda >0$ and $\bar\eta>0$ are attributed to the USC regime.}
\label{Fig:PHD}
\end{figure}

\subsubsection{Objectives}

According to the discussion above, our objective is to investigate the phase diagram of the generalized driven-dissipative Dicke model, as defined by Eqs.~(\ref{H-Dk}) and (\ref{Liov-Lind-eom}) where the size of the ensemble of quantum emitter and the coherent and dissipative coupling parameters are varied. To understand the influence of the counter-rotating terms (reflecting different spontaneous symmetry breaking) in the emergence of phases, we aim to conduct a comparative analysis with the driven-dissipative Tavis-Cummings model described by Eqs.~(\ref{H-TC})-(\ref{Liov-Lind-eom}) within the same parameter range. Furthermore, exploring variations in the number of quantum emitters will be helpful in identifying emergent signatures of phase transitions and their deviations from the thermodynamic limit.

\subsubsection{End-to-end computational workflow}

Solution of Lindblad Eq. (3) with the Hamiltonians (1) and (2)
\begin{small}
\begin{lstlisting}
set N0 in range 100 to 1000
set omega_0 = delta omega_c with delta in range 0.7 to 1.3
set dissipative parameters [gamma_c, gamma_uo, gamma_down, gamma_phi] 
    in Eq. (3) in range 0.01 omega_c to 0.1 omega_c
    for lambda:  0.5 omega_c to 1.5 omega_c, step dlmbd = 0.01 omega_c
        for eta: 0.5 omega_c to 1.5 omega_c, step deta = 0.01 omega_c
             # Initialize reduced density operator as direct product of 
             # density operator for photon vacuum and quantum emitters 
             # in certain population state rho(0) = rho_photon_vacuum rho_QE 
             # Run time dependant quantum trajectories  with timestep  
             # dt = 0.01/omega_c for rho until reaching steady state 
             # with precision 0.01
             # Measure order parameters such as QE population inversion sigma_z 
             # and cavity photon population a^dag a
             # Store measured values in array vs corresponding lambda 
             # and eta values
        end
    end
output: array of order parameter values vs corresponding lambda and eta values   
\end{lstlisting}
\end{small}

\subsubsection{Why classical methods are not sufficient to perform the hard computational module}

 State of the art numerically exact simulations meant to explore the phase diagram of the Dicke model in the strong coupling regime have been reported to handle ensembles of ${\cal N}_o=25$ quantum emitters.\cite{kirtonAQT:2019} This truncation of the interacting quantum emitter chain defines the scope of the physics we can accurately model. Although phase transitions are well-defined in the thermodynamic limit, we can detect them for relatively large but finite ensembles of quantum emitters by adopting the second-order cumulant approximation. Classical computations can handle scenarios with about $10^3$ interacting quantum emitters using the second cumulant approximation.\cite{piryatinskiPRR:2020} However, extending these simulations to incorporate higher-order cumulants becomes impractical due to memory scaling limitations and the inherent complexity in deriving analytical expressions. The utilization of QPUs holds promise in overcoming these challenges, allowing for the natural incorporation of higher-order cumulants in the numerically exact modeling of the extended ensembles of quantum emitters entering the USC regime.

\subsubsection{Concrete problem instantiations}

Using computational workflow outlined above explore the phase diagram similar to that depicted in Fig.~\ref{Fig:PHD} for an ensemble of quantum emitters of the size of $N\sim 10^2-10^3$ where we suggest using $N=1$, $N=50$, $N=100$, $N=500$, and $N = 1000$, which will be helpful in identifying emergent signatures of phase transitions and their deviations from the thermodynamic limit. To understand the influence of the counter-rotating terms (reflecting different spontaneous symmetry breaking) in the emergence of phases, we aim to conduct a comparative analysis with the driven-dissipative Tavis-Cummings model described by Eqs.~(\ref{H-TC})-(\ref{Liov-Lind-eom}) within the same parameter range. Comparison to the second cumulant classical calculation in Fig.~\ref{Fig:PHD} will be performed to identify the limitations of this method.

\subsubsection{List of other candidate instances}

Expanding the phase diagrams for different generalizations of the Dicke model \cite{kirtonAQT:2019,liNatMat:2023}, for instance, the asymmetric variant, obtained in the strong coupling regime,  into the USC range is of particular interest. Furthermore understanding the impact of non-Markovian effects requires further studies involving the modification of the Lindblad master equation~(\ref{Liov-Lind-eom}) by incorporating the appropriate relaxation memory kernel, along with the implementation of suitable quantum algorithms.
To ensure the fulfillment of the gauge invariance requirement, an augmentation of the Dicke Hamiltonian~(\ref{H-Dk}) to multi-level quantum emitters is necessary. Furthermore, a broadening of the Dicke Hamiltonian introducing multiple cavity modes could be relevant.

\section{Requirements summary}
This section summarizes the application requirements for all the applications described in this chapter.
\subsection*{Application 1: Low energy spectrum of Dicke and Tavis-Cummings models}
\begin{small}
\begin{tabular}{ |l l l| }
    \hline
        & & \\
     & Total time limit & 2 months \\ [1ex]
    \textbf{Workload:} & Number of subroutine calls required & $13\times 100 \times 100$ \\ [1ex]
     & Maximum subroutine time limit & NA \\ [1ex]
     & & \\[1ex]
     & Model type & Multiple-spins with one multi-level state\\[1ex]
     & Size & minimum $100$ spins $+$ $100$ levels, \\
     & & target $1000$ spins $+$ $500$ levels\\[1ex]
     \textbf{Problem specifications:} &  Interaction Structure & Star Graph (the multi-level state is the center)\\[1ex]
     & Computational target &100 lowest energy values\\[1ex]
     & Accuracy requirement & Relative with respect to multi-level coupling,\\ &&minimum $10^{-2}$, target $10^{-3}$\\[1ex]
     & & \\
     \hline
\end{tabular}

\subsection*{Application 2: Phase diagram of the driven-dissipative Dicke model}

\begin{tabular}{ |l l l| }
    \hline
        & & \\
     & Total time limit & 2 months \\ [1ex]
    \textbf{Workload:} & Number of subroutine calls required & $13\times 100 \times 100$ \\ [1ex]
     & Maximum subroutine time limit & NA \\ [1ex]
     & & \\[1ex]
     & Model type & Multiple-spins with one multi-level state\\[1ex]
     & Size & minimum $100$ spins $+$ $100$ levels, \\
     & & target $1000$ spins $+$ $500$ levels\\[1ex]
     \textbf{Problem specifications:} &  Interaction Structure & Star Graph (the multi-level state is the center)\\[1ex]
     & Computational target &Average quantum emitter population\\
     &&Average cavity population\\
     &&(after reaching dynamical steady-state)\\[1ex]
     & Accuracy requirement & For averages: Relative to number\\
     &&of quantum emitters, $10^{-2}$\\[1ex]
     && For considering dynamical steady-state:\\ 
     &&relative changes over time, $10^{-2}$\\[1ex]
     & & \\
     \hline
\end{tabular}
\end{small}
\newpage 

\section{Quantum implementation}
A quantum implementation and resource estimation for computing the ground state energy for the Dicke model Eq.~\eqref{H-Dk} and the Tavis-Cummings model Eq.~\eqref{H-TC} can be found in \cite{qc-applications-notebooks}.



\newpage
\end{chapter}

\begin{chapter}{High-temperature superconductivity and exotic properties of Fermi-Hubbard models} \label{ch:fermi-hubbard}
    

The design of new crystals and alloys is currently hindered by the lack of theoretical guidance and predictions for the so-called strongly correlated materials. The principal limitations are not believed to reside in the quality of the models but in the inherent classical constraints to simulate them. In this chapter, we describe applications of the Fermi-Hubbard model(s) for predicting high-temperature superconductivity and other exotic properties in materials. We briefly introduce how the Fermi-Hubbard model arises as an effective description of the behavior of strongly correlated electrons in crystals. We depict what the expected phases, or material's behavior, of the Fermi-Hubbard model look like and what physical observables are identified as key markers of these different phases. Finally, we show how these different phases can be computed using quantum computing resources and we provide a series of concrete use cases that are of relevant and important scientific value.\\
\noindent
\textbf{Hamiltonian Type}:  Fermionic Hamiltonians.\\
\textbf{Quantum Computational Kernels}: Ground State Preparation, Thermal State Preparation.

\clearpage

\begin{figure}
    \centering
    \includegraphics[scale=0.6]{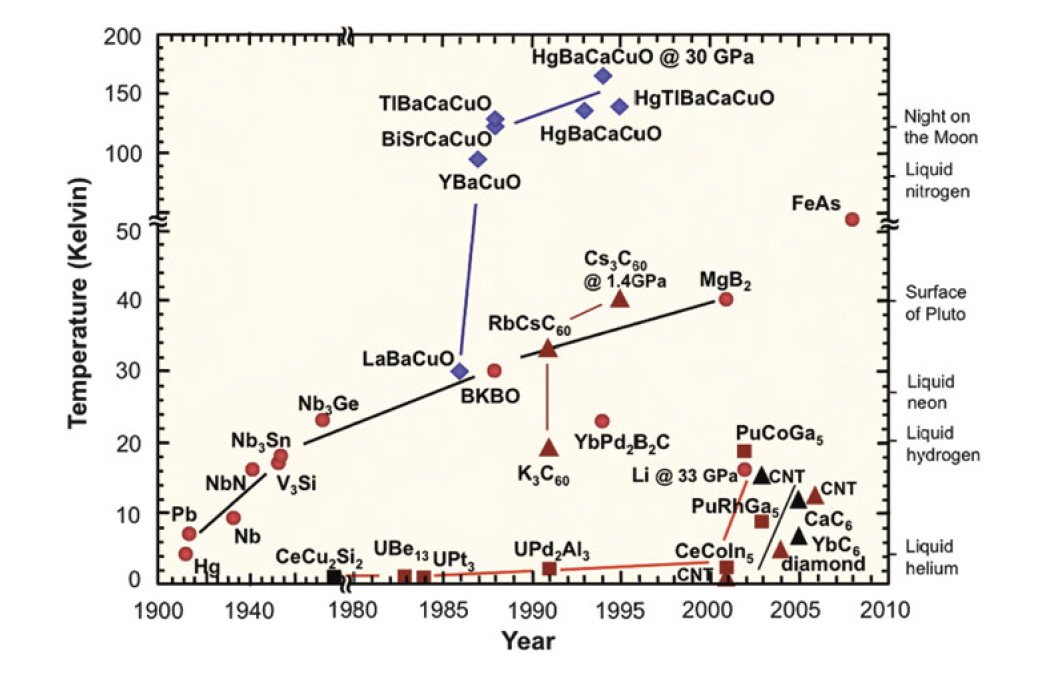}
    \caption{Evolution of the superconductivity critical temperature over the years for various materials. The gap between low-temperature superconductors and HTC is over a hundred of Kelvin.}
    \label{fig:htcs}
\end{figure}

\section{Application area overview}

\subsection{High-level description of the application area}
Superconductivity manifests itself physically as virtually zero electric resistance and the expulsion of magnetic fields from the bulk of certain materials (Meissner effect). Overall, there is a good understanding of low-temperature superconductivity, which is due to the interaction between electrons and lattice phonons (quantized vibrations), leading to the Bardeen-Cooper-Schrieffer theory (BCS) of late 1950s. Above a certain critical temperature $T_c$, conventional superconductors behave as ``regular" metals, and below this temperature their resistivity drops to zero by the current testing methods. The basic insights that BCS theory provides is that electrons, naturally repelling each other because of their charge, become attracted as a result of the mediation of phonons. In contrast to the BCS theory of conventional superconductors, high-temperature superconductors (HTC) requires grasping a variety of concomitant phenomena \cite{Dagotto1994-jj} and its understanding remains one of the biggest open scientific and technological challenge. The goal of this review is to point out to some models that currently require intensive numerical methods to study and for which quantum computer may prove useful at removing computational barriers.

In Fig.~\ref{fig:htcs}, we see the evolution of superconductivity critical temperatures over the years. In 1986 HTC materials branched off from the pattern (Blue line in Fig.~\ref{fig:htcs}), giving us a new direction of achieving superconductivity at the room temperature. In materials based on Lanthanide (La) compounds, pressure displaces lanthanum and induces high-temperature superconductivity at higher critical temperatures~\cite{Wang2017}. For example, LaBaCuO superconductor is a HTC superconductor where increasing the pressure can increase its critical temperature. This shows how the critical temperature is sensitive to the material structure and physical conditions. In what follows, we provide a review of some basic methods to probe the physics of various materials:
\begin{itemize}
    \item Mercury Barium Calcium copper Oxide (HgBa$_2$Ca$_2$Cu$_3$O$_8$ or sometimes simply Hg-1223) is an important cuprate (based on copper oxides) superconductor with a superconducting transition temperature that can exceed 138 Kelvin. Its complex structure, particularly the interactions within the CuO$_2$ planes as we will see below, contributes to its high-temperature superconducting properties. The reduction of this compound to layers of 2-dimensional lattices of copper and oxygen can be studied using the Fermi-Hubbard model. The cuprates are founded on two-dimensional planes comprising a square lattice of Cu ions connected by O ions at the links \cite{Dagotto1994-tz}. 
\item Iron pnictides are a class of high-temperature iron-based superconductors with a layered crystal structure \cite{Si2016}. For instance, in Iron-Pnictide LaFeAsO (e.g., LaFeAsO$_{1-x}$F$_x$), superconductivity is observed with critical temperatures (T$_c$) up to 26 K when doped with fluorine and is one of the materials believed to require a multi-band Fermi-Hubbard model. Various compounds are reported to exhibits up to 150 Kelvin transitions. Iron-based superconductors are a class of high-temperature superconductors discovered in the late 2000s \cite{Kamihara2006-nx} in LaOFeP. Unlike traditional HTC, which often contain copper, these compounds feature iron atoms at their core. They exhibit superconducting behavior at relatively higher temperatures compared to conventional superconductors. LaOFeP represents a specific composition within this class. LaOFeP's structure consists of an intricate arrangement: it comprises alternating layers of lanthanum oxide (composed of $\text{La}_3^+$ and $\text{O}_2^-$ ions) and iron pnictide (made up of $\text{Fe}_{2}^+$ and $\text{P}_3^-$ ions) \cite{Johannes2008-qc}.  An illustrative instance of the family of Fe-based superconducting compounds is exemplified by La$O_{1-x}$F$_x$FeAs, which is among the extensively studied pnictides. Notably, a remarkable superconducting critical temperature of approximately 55 K \cite{renzhi} has been achieved in a related compound, SmO$_{1-x}$ F$_x$FeAs. These critical temperature values are surpassed only by those observed in the Cu-oxide family of high critical temperature superconductors. The physics of the new Fe-based superconductors exhibits several resemblances to the properties of cuprates, which are however much more studied in the literature.

\item In some types of compounds based on lanthanides, a stripe phase is observed, e.g. a modulation of the charge density on top of antiferromagnetism  \cite{Emery1999}. This suggests the presence of ordered stripes or charge density waves within the material, that occur as a modulation of the strength of the magnetic field on top of the antiferromagnetic state. These stripes are believed to arise from a delicate interplay between charge, spin, and lattice degrees of freedom and are mostly observed in materials of the La$_2$CuO$_4$ family \cite{Tranquada1995-uw,Wu2011-ws,Fujita2012-qa}. An interesting fact is that in nickelates (which we do not discuss here at length) a charge density wave leads to insulators, while in cuprates it leads to superconductors \cite{Kivelson1998-jg};
    
    \item Graphene-based superconductors are an old family of conventional superconductors that have been proposed to have a high-temperature superconductive phase when properly doped. These types of compounds are currently under study, predicted but never observed experimentally \cite{Hai2023,Nandkishore2012-kx}.  In addition to the physics of cuprate-based superconductors, a fresh area of research has emerged in the form of graphene-based superconductivity and iron-based superconductivity. For graphene based superconductivity, there have been a few proposals of superconductivity on honeycomb lattice systems \cite{Nandkishore2012-kx, Nandkishore2014-kx}. These are typically either hexagonal or triangular plaquette materials, inspired by the physics of graphene. Recently, a study introduced a single orbital Fermi-Hubbard model on a octagraphene lattice, which is a square-octagon lattice.
\end{itemize}
Below we will provide more background on the origin of superconductivity in Fermi-Hubbard models.
\subsection{Role of Fermi-Hubbard physics in the investigations of HTC}
As we discuss below, the physics of HTC is deeply intertwined with the theory of Mott insulators that was introduced to explain why some materials that would be conducting electricity according to band theory, are effectively insulators \cite{Carlson2002-dw}. The Fermi-Hubbard model was central in predicting the physics of Mott insulators and the Mott transition \cite{Mott1937-ki,Arovas2022-mt,Zhou2021-av,Lee2006-bz}.

From an explanatory point of view, there are two global pictures in HTC physics, both connected to a Mott-Insulator transition but near two phases that, from a material perspective, are considered different: the stripe phase and the d-wave picture. The conventional d-wave superconductive picture refers to the symmetry of the superconducting order parameter in momentum space near a Mott transition, in the presence of an antiferromagnetic state. This picture emphasizes the d-wave nature of the superconducting pairing, where the order parameter changes sign along specific directions on the Fermi surface, on top of the antiferromagnetic state. 

One of the reasons why there is so much effort and numerical studies in the study of the Fermi-Hubbard model to explain HTC is that it exhibits a large number of strongly competing orders observed in the cuprates, including both d-wave superconducting and charge/spin stripe order. 
Another physical picture which was very popular among HTC explanations, i.e. Anderson's resonating valence bond state developed for La$_2$CuO$_4$ \cite{Anderson1987-rt} and topological order, while having an important role in physics as a theory of non-Landau symmetry breaking, is now considered an independent area of research which does not apply to ``conventional" high-temperature superconductors or does not completely explain it. The physics of HTC is among the most important aspects of the more general field of strongly correlated materials. In general, it is also observed experimentally that the superconductive critical temperature can be increased at high pressure \cite{Gorkov2018-sx}.

In the survey of applications below, we describe four models. In particular, we  suggest instances based on cuprates  and those based on iron-based materials for superconductors (with square plaquette), those based on graphene (honeycomb plaquette), and a particular choice of parameters for a 2D square lattice which is of academic interest.
The model used to describe the hopping of electrons is the Fermi-Hubbard model, which describes electrons as particles (of spin up and down) hopping on a certain lattice or crystal structure. The two key quantities of interests are the onsite repulsion (due to the Coulomb energy in the Hamiltonian, and Pauli's principle at the wavefunction level) and the hopping energy. The hopping energy requires a deep knowledge of the crystal structure, the orbitals involved, and the overlaps between the orbitals, which all contribute to the hopping energy (later called more generically $t_{ij}$ between sites $i,j$). 
While the model is predominantly studied on square lattices or decorations thereof, the study of real materials thus requires a careful characterization of the crystal structure. However, models derived on certain sublattices of the original one are typically used to explain some aspects of the nature of the HTC phase. For instance, in Fig. \ref{fig:reduction}, La$_2$CuO$_4$ is reduced from a full crystal structure to three bands involving only O and  Cu atoms, and subsequently reduced to a single-band model with the O atoms integrated out\footnote{In the literature, bands and orbitals are used interchangeably.}. Cuprates are the most studied materials from the point of view of conventional HTC. Generically, the important quantities which affect the physical behavior of the system are the number of electrons on the lattice (at most two per sites, later referred as electron filling), the temperature and the intrinsic parameters of the Hamiltonian (hopping energies and onsite repulsion, or ratios thereof). These three sets of parameter inputs that characterizes the Fermi-Hubbard physics from a computational perspective \cite{Zhou2021-av}.

\begin{figure}
    \centering
    \includegraphics[width=0.6\linewidth]{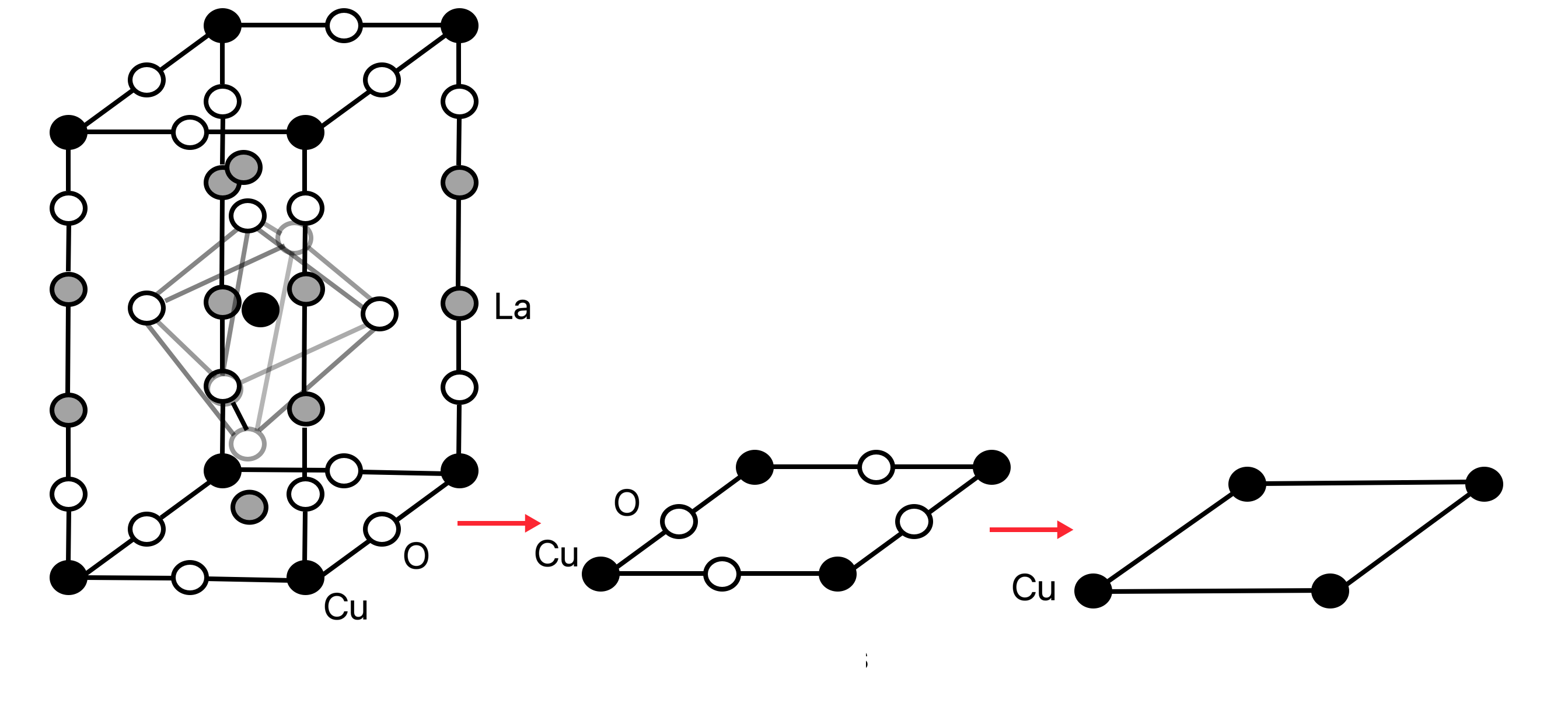}
    \caption{The crystal structure of La$_2$CuO$_4$ (or Sr$_2$CuO$_4$), a high temperature superconductor, and its reduction to a single band effective model by the progressive elimination of degrees of freedom until the square lattice of Fig. \ref{fig:hoppingf} is reached. In the two-dimensional sublattice, Cu has four nearby O atoms.}
    \label{fig:reduction}
\end{figure}


High-temperature superconductivity can be studied, to some degree, using the Fermi-Hubbard model. The material's electron filling then can be modeled by the number of electrons in the model. The FH model incorporates the most important electronic interactions in a lattice structure, considering the hopping of electrons between lattice sites and their repulsive interactions. The cuprate-based single-band models primarily focus on the behavior of doped Mott insulators. Upon electron filling with charge carriers (holes or electrons), these materials undergo a transition from an antiferromagnetic insulating phase to a superconducting phase at relatively high temperatures compared to conventional superconductors \cite{Keimer2015-qi}. 


As we will discuss below, from a numerical perspective evaluating the physical properties of these systems becomes challenging because of the necessity of analyzing the Fermi-Hubbard model numerically: this implies calculating ground and thermal states of a Hamiltonian with a large number of states. The phases of the Fermi-Hubbard model (FH) is rather complicated. It involves various phases and multiple order parameters that need to be evaluated. Classical methods can be pushed, with supercomputers of various generations, to systems with 20-25 sites at most. 

In the following Section~\ref{sec:utes}, we provide an overview of the utility of the applications. Section~\ref{sec:der} contains a brief introduction of the derivation of the model from localized orbitals, with further basic background provided in Appendix.~\ref{sec:methodsder}, while in Appendix~\ref{sec:FH_classical_methods} we survey a few classical methods used to analyze the FH model. In Section~\ref{sec:observ}, we characterizes the main observables used to describe each phase of the model, based on a consensus from a broad literature survey. In Section~\ref{sec:application}, we provide a pseudo-algorithmic approach to the study of superconductivity, with concrete examples drawn from the literature in Section~\ref{sec:problem_instances}.

\begin{figure}
    \centering
    \includegraphics[scale=0.5]{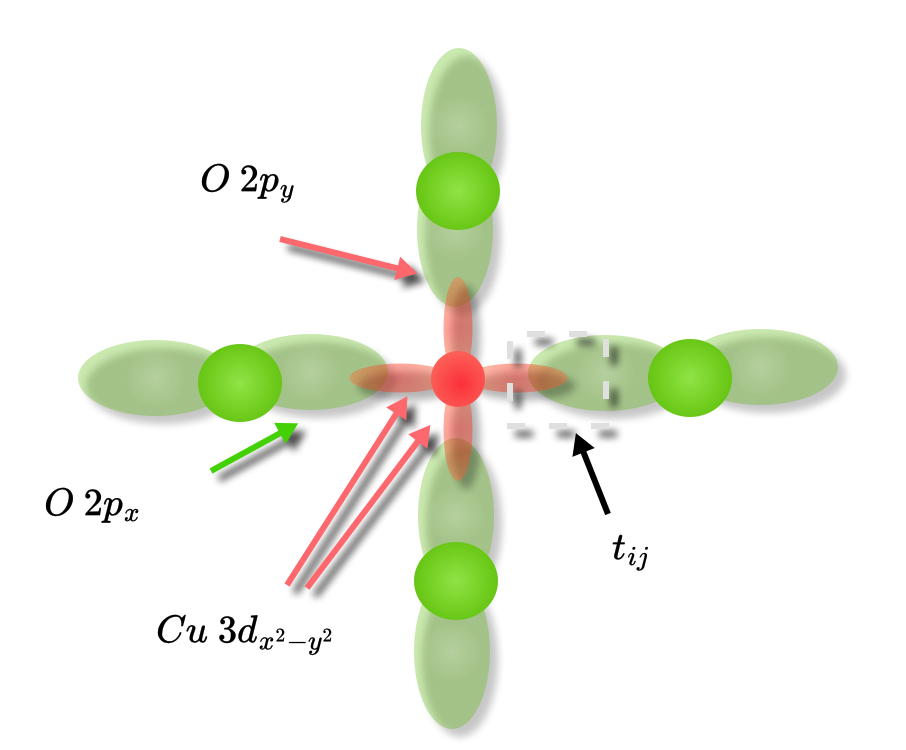}
    \caption{The Fermi-Hubbard model originates from the hopping of electrons between the orbitals of the crystal's atoms in a copper oxide. In the case of the two dimensional Fermi-Hubbard model on the square lattice, the hopping parameters $t_{ij}$ originate from the overlap of the orbitals of oxygen and copper.}
    \label{fig:cup}
\end{figure}

\subsection{Utility estimation}\label{sec:utes}
\subsubsection{Overview of the value of the application area}
Predictions of the phase diagram of Fermi-Hubbard models is an important scientific endeavor as it would provide guidance for the exploration and development of new strongly correlated materials. The exotic phases of these crystals are currently elusive to classical simulations precisely because they require to take into account the full quantum nature of a large multitude of interacting electrons. Having a better understanding of how the lattice geometry, interaction intensities and electron filling impact superconductivity at higher-temperature constitutes the flagship application and primary scientific motivation for doing Fermi-Hubbard simulations.
The promise of high-temperature superconductors is the emergence of stronger, more compact and more sensitive magnets with applications in all sectors of our society: Transportation with smaller electric drives, Industry with kaolin/clay separation, Medical with stronger MEG coils, Communication with more sensitive antennas, Electric power with resilient auxiliary control equipment and Defense \& Space with more sensitive satellite and submarine sensors~\cite{congress_1990_HTC, CCAS_2014_Brochure}.
Therefore, it is not surprising that a considerable amount of effort and compute time is dedicated every year by the scientific community to understand and simulate Fermi-Hubbard models.

\subsubsection{Concrete utility estimation}
In FY22, the total computing power at LANL is on the order of 10~billions core-hours. Open-science computations amount for $10\%$ of the total load which represents 1~billion core-hours. The fraction of the open-science compute time dedicated to solving material science problem is around $20\%$, see Section~\ref{sec:current-methods}. Approximately $50\%$ of this later computational burden is focused on simulating model like Fermi-Hubbard alone\footnote{his number has been obtained via private communication with LANL's SMEs.}. In summary, on the order of 100~million core-hours are spent on solving Fermi-Hubbard models at LANL using different classical algorithms. Moreover, these simulations require access to large amount of RAM with the nodes in LANL Chicoma supercomputer possessing $128$ cores and $512$Gb of memory. For these specifications, the typical cloud computing cost is around \$$0.03$ per core-hour. Therefore, the open science Fermi-Hubbard simulations performed at LANL in 2022 would cost approximately \$$3$~million per year at a commercial cloud computing vendor.
Unfortunately, the exact number and nature of the Fermi-Hubbard models that were studied is unknown. However, based on the 2022 list of publications involving these computations, we estimated that approximately 10 different models were studied. This would bring the cost of characterizing part of the phase diagram of small Fermi-Hubbard models with classical algorithms to \$$300$K per model within a time-frame of 1-2 months.

\section{Problem formulation and computational workflows}
As mentioned earlier, the Fermi-Hubbard model is a simplified, effective, theoretical model that describes the behavior of electrons in a strongly correlated crystal. This implies that a series of steps are necessary to select a Fermi-Hubbard model that can be representative of an actual material, the final goal being of testing predictions of the model against experimental results.
From a computational perspective, calculating the physical properties of a strongly correlated material, using a realistic parameter estimation procedure from the the knowledge of a crystal structure, obeys a particular workflow that reads at a high-level:
\begin{enumerate}
    \item Pick a crystal structure representative of the material
    \item Study the orbital overlap via atomic wave function or similar and equivalent methods
    \item Derive a Fermi-Hubbard model
    \item Choose the parameters of the model (electron filling, temperature) \label{lab:workflow_input}
    \item  Pick a method or approximation scheme to study the Fermi-Hubbard model\label{lab:workflow_kernel}
    \item Investigate the properties of the Fermi-Hubbard model (phases, conductivity, etc)\label{lab:workflow_output}
\end{enumerate}
In this workflow, quantum computing can help by replacing Step~\ref{lab:workflow_kernel} with a quantum kernel (either ground state estimation or thermal sampling, as we will see later). The input and observables of this kernel are specified in Step~\ref{lab:workflow_input} and Step~\ref{lab:workflow_output} respectively. 


The generic form of the Hamiltonian of the Fermi-Hubbard model reads for a single band~\cite{Hubbard1963-qz,Gutzwiller1963-ut, Kanamori1963-yl}:
\begin{equation}
    H = - \underbrace{\sum_{\langle i,j\rangle,\sigma }}_{\langle i,j\rangle} t_{ij} (\hat{c}^\dagger _{i,\sigma} \hat{c}_{j,\sigma}+h.c.) + U \sum_{i} \hat{n}_{i,+} \hat{n}_{i,-}\label{eq:fh}
\end{equation}
where the first term is a tight-binding Hamiltonian which leads to standard band theory of itinerant electrons, while the second term represents the repulsive interaction between the electrons. Another way of interpreting this Hamiltonian is to consider that the first term represents the kinetic energy of electrons hopping between neighboring lattice sites, while the second is an onsite interaction energy. 
The operators $\hat{c}^\dagger_{i,\sigma}$ and $\hat{c}_{j,\sigma}$ are the creation and annihilation ladder operators for an electron at site i and j, respectively, with spin $\sigma\in\{-,+\}$. They satisfy the anti-commutation relationships $\{\hat{c}_{i\sigma}, \hat{c}_{j\sigma^\prime}^\dagger\} = \delta_{ij}\delta_{\sigma \sigma^\prime} \hbar$, where, for numerical simulations, units are customarily chosen such that $\hbar=1$. This choice is not restrictive, as we will explain in the applications section. The summation $\sum_{\langle i,j\rangle}$ runs over pairs of neighboring sites.  The term $\hat{n}_{i,+}\hat{n}_{i,-}$, where $\hat{n}_{i,\sigma} = \hat{c}_{i\sigma}^{\dagger}\hat{c}_{i\sigma}$, ensures that there is a repulsion between electrons. A graphical representation of the meaning of the parameters $t_{ij}$, and $U$ are shown in Fig. \ref{fig:hoppingf}. 
\begin{figure}
    \centering
    \includegraphics[scale=0.3]{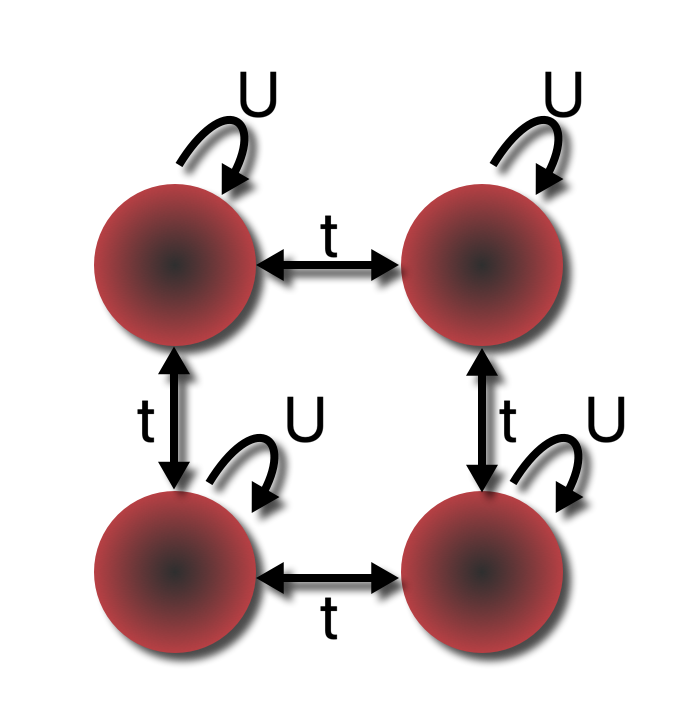 }
    \caption{A graphical representation of the Fermi-Hubbard model. The parameter $t$ is the energy associated with the hopping between sites, while the parameter $U$ represents the energy associated to the particle remaining on the same site.}
    \label{fig:hoppingf}
\end{figure}

The complex behavior of the Fermi-Hubbard model arises from the competition between localization and itinerancy: it captures the interplay between electron hopping and on-site interactions, which can give rise to interesting phenomena like metal-insulator transitions, magnetic ordering, and, in certain cases, superconductivity. The Fermi-Hubbard model is expected to display a broad set of distinct behavior, or phases, for different values of electron filling and temperature. The phases of interest are the superconductive, pseudogap, Fermi Liquid, strange metal and antiferromagnetic phases which we will describe in more detail latter. In addition, the Fermi-Hubbard Hamiltonian can be generalized to different bands, as we discuss below. The number of bands in the model refers to the energy bands that the particles can occupy as they move through the lattice. This makes the problem to study computationally even more complicated.

It is worth stressing at this point a difference between conventional superconductivity and the type of superconductivity predicted by the Fermi-Hubbard model. In both cases, it is the formation of Cooper pairs, which are pairs of electrons with opposite spin and momentum that experience an effective attractive interaction, that leads to superconductivity. For conventional superconductors, this effective attractive interaction is mediated by the lattice vibrations (phonons) and appears explicitly in the Hamiltonian.

We would like to point out that the Fermi-Hubbard Hamiltonian of Eqn. (\ref{eq:fh}) is closely related to the t-J model, which is mapped via the Schrieffer-Wolff transformation in the limit $U\gg t$ with $J=4t^2/U$  \cite{Chao1978-du,Ogata2008-qw}. Thus, many of the comments below will be valid for the t-J models as well in the limit described above.


\subsection{Methods used to derive the model Hamiltonians}\label{sec:der}
The Fermi-Hubbard Hamiltonian of Eqn.~(\ref{eq:fh}) is a generic description of a system of fermions. To make concrete predictions about physical systems, however, it is paramount that the parameters $t_{ij},U_{ij}$ are estimated from first principles, as it is the case in \textit{ab initio} methods such as DFT. For instance, the first effective Hamiltonian introduced for the study of cuprate oxides was the single band Zhang-Rice hamiltonian \cite{Zhang1988-cz}. It is commonly believed that multi-band models are necessary to describe the superconductive phase. However, certain properties such as antiferromagnetism can be understood from fewer band models. 

To see the origin of the Fermi-Hubbard Hamiltonian, let us start by considering the many-body Hamiltonian that fully describes the interactions between electrons and nuclei. This is of the form \cite{Zhu2016},
\begin{eqnarray}
    \mathcal H=\underbrace{\sum_i\frac{p_i^2}{2m_e}}_{\substack{electron\\ kinetic\\ energy}}-\underbrace{\sum_{i,a} \frac{Z_a e^2}{|r_i-R_a|}}_{\substack{electron-nuclei\\interactions}}+\underbrace{\frac{1}{2}\sum_{ij}\frac{e^2}{|r_i-r_j|}}_{\substack{electron-electron\\interaction}}+\underbrace{\sum_a \frac{P^2_a}{2M_a}}_{\substack{nuclei\\kinetic\\energy}}+\underbrace{\frac{1}{2}\sum_{ab} \frac{Z_a Z_b  e^2}{|R_a-R_b|}}_{\substack{nuclei-nuclei\\interaction}}\label{eq:continuum}
\end{eqnarray}
Solving such Hamiltonian, e.g. eigenfunctions and respective energies, would imply knowing essentially everything about the solid state.  Since this task is unreasonably ambitious, most of the results are typically obtained in some form of approximation. One of the first approximations that one encounters is the Born-Oppenheimer approximation, in which the wavefunction of the entire system is approximated as the product of wavefunctions for the nuclei and the electrons. Perturbations over the nuclei's interaction lead to phononic modes. In most cases, these interactions are neglected. In Sec.~\ref{sec:fermihubbardder} we provide some details on the derivation of the Fermi-Hubbard model from Eqn.~(\ref{eq:continuum}), but the final result is a discrete Hamiltonian, much easier to handle, of the form:
\begin{eqnarray}
    H_{FH}=-\sum_{ij} t_{ij} \hat c^\dagger_{i\sigma}\hat c_{j\sigma}+\frac{1}{2} \sum_{i,\sigma} U_i \hat n_{i\sigma} \hat n_{i,-\sigma}.\label{eq:fermihubb}
\end{eqnarray}
The model above is the Fermi-Hubbard model, and describes the hopping of electrons on the orbitals of the atoms in the particular molecular structure of interest. The coefficients $t_{ij}, U$ are given by
\begin{eqnarray}
    t_{\langle i,j\rangle}&=&-\int d\vec r \phi^*(\vec r-\vec R_i) h(\vec r)\phi(\vec r-\vec R_j) \label{eq:tijei}\\
    U_i&=&e^2 \int d\vec r d\vec r^\prime \frac{|\phi(\vec r-\vec R_i)|^2|\phi(\vec r^\prime-\vec R_i)|^2}{|\vec r-\vec r^\prime|}\label{eq:uei}
\end{eqnarray}
where $\phi(\vec r-\vec R_i)$ are the localized orbitals at atomic sites, associated with the probability $|\phi(\vec r-\vec R_i)|^2$ of finding an electron at a distance $|\vec r-\vec R_i|$ from atom $i$. The key assumption is that these are localized, and thus only $t_{ij}$ for neighboring atoms are effectively nonzero, inducing a local graph structure. Depending on the orbitals involved, and how extended they are, one can have not only nearest neighbors (N.N.) but also next-to-nearest neighbors interactions (N.N.N.). However, if the orbital wavefunctions are localized, they overlap only for nearby sites $\langle i,j\rangle $ and thus $t_{ij}$ can be represented by a local Hamiltonian, e.g. a graph representing the interactions. Also, if the system is translational invariant, then $U_i\equiv U$.  The question is then how to obtain representative numbers for $t_{ij}$ and $U$. These parameters $U,t_{ij}$ can be obtained by computing the so-called exchange integrals~(\ref{eq:tijei}) and~(\ref{eq:uei}) directly from first principles and using prior knowledge about the crystal structure. Alternatively, one can choose $t_{ij}$ as free parameters, possibly reducing it to a smaller number by symmetries, and then fit the Fermi surface of the model to the experiment. For materials that are two-dimensional, the Fermi surface can be obtained via angle-Resolved photoemission spectroscopy (ARPES).
One can fix the number of particles by looking at a subspace of the Hilbert space with fixed number of particles, $N=2Lf$ where $L$ is the number of sites and $f$ is the total fraction ($0<f<1$) of electrons, or electron filling. The interesting case is typically near half filling, $f=1/2$. One can solve for the system fixing this number, e.g. keeping all possible eigenstates $|n\rangle$ such that $\frac{\langle n| \hat N|n\rangle}{2L}=f$, where $\hat N$ is the number operator, defined as $\hat N=\sum_{i,\sigma=\pm} \hat c_{i,\sigma}^\dagger \hat c_{i,\sigma}$.  
The the last term in the Hamiltonian of Eqn.~\eqref{eq:fermihubb} is a two-body interaction, which cannot be treated with analytical methods. One of the key approximations in the study of superconductors and superfluids is, for the sake of tractability, to replace the two-body term
\begin{eqnarray}
    \hat n_{i\sigma}\hat n_{i,-\sigma}=\hat c_{i\sigma}^\dagger \hat c_{i\sigma}\hat c_{i,-\sigma}^\dagger \hat c_{i,-\sigma}  
\end{eqnarray}
with a single particle under the influence of a mean-field. As a last comment, it is often useful to work in momentum space for systems with translational symmetry e.g. periodic lattices that are found in nature, by working with operators defined in terms of the momentum $\vec k$
\begin{eqnarray}
\hat c_{\vec k\sigma}=\frac{1}{\sqrt{L}}\sum_{\vec r} e^{-i \vec k \cdot \vec r} \hat c_{\vec r\sigma}, 
\end{eqnarray}
which satisfy $\{\hat c_{ k\sigma}, \hat c_{k^\prime\sigma^\prime} ^\dagger\} = \delta_{kk^\prime}\delta_{\sigma \sigma^\prime}$, again in units $\hbar=1$.
 In momentum space, the Fermi-Hubbard Hamiltonian becomes
\begin{eqnarray}
    H_{FH}=\sum_{\vec k,\sigma} \xi_{\vec k} \hat c^\dagger_{\vec k,\sigma}\hat c_{\vec k,\sigma}+\frac{U}{L}\sum_{\vec k,\vec k^\prime,\vec q} \hat c^\dagger_{k+q,+} \hat c^\dagger_{k^\prime-q,-} \hat c_{k^\prime,-}  \hat c_{k,+} \label{eq:2dsss2}
\end{eqnarray}
\textcolor{red}{}
Using the Bogoliubov-de Gennes mean field theory \cite{Zhu2016}, the Hamiltonian above is reduced to a quadratic and numerically treatable Hamiltonian of the form:
\begin{eqnarray}
    H_{BdG}=\sum_{\vec k,\sigma} \xi_{\vec k} \hat c^\dagger_{\vec k,\sigma}\hat c_{\vec k,\sigma}+\sum_{\vec k} (\Delta_{\vec k } \hat c_{\vec k,+}^\dagger \hat c_{-\vec k,-}^\dagger  +h.c.)
\end{eqnarray}
where $\Delta_{\vec k}$ is the superconducting order parameter. The derivation involves the replacement of of two-body operators with single-body operator under the influence of mean-field and then the expectation values are then through by self-consistency equations.

In the literature, the insertion by hand of a non-zero superconducting order parameter allows to study the phases of the model, but this relies on a series of approximations. This form is typically used when trying to fit the Fermi surface from the ARPES experiments. We provide more details on how to derive the coefficients $t_{ij}$ and $U_i$ from the molecular structure in Appendix \ref{sec:methodsder}.

It is common to also consider, in addition to the hopping term, an exchange interaction between the spins of the electrons. The spin operators are defined, on each site, as 
\begin{eqnarray}
    S^*_{i}=\frac{\hbar }{2} \sigma^*_{i},\ 
\end{eqnarray}
where $\sigma^x,\sigma^y,\sigma^z$ are the Pauli operators. On a lattice, it is also common to introduce the notation for the site indices $\bm{i}=(i_x,i_y)$, which label the lattice sites on a square lattice.
An important remark, as it will be important later, is that an anti-ferromagnetic phase is also often described by the Heisenberg-like Hamiltonian on the spin degrees of freedom:
    \begin{equation}
        H_{\text{AF}} = \tilde J \sum_{\langle \bm{i}, \bm{j} \rangle} \mathbf{S}_{\bm i} \cdot \mathbf{S}_{\bm j},
    \end{equation}
    where $J$ represents the antiferromagnetic exchange coupling between neighboring spins indexed by the site vectors $\bm{i},\bm{j}$. The total Hamiltonian is then given by
\begin{eqnarray}
    \mathcal H=H_{FH}+H_{\text{AF}}.
\end{eqnarray}
Although in the following the second term will be neglected, the anti-ferromagnetic phase can arise from the second term alone.
\begin{figure}
    \centering
 
    \includegraphics[scale=0.45]{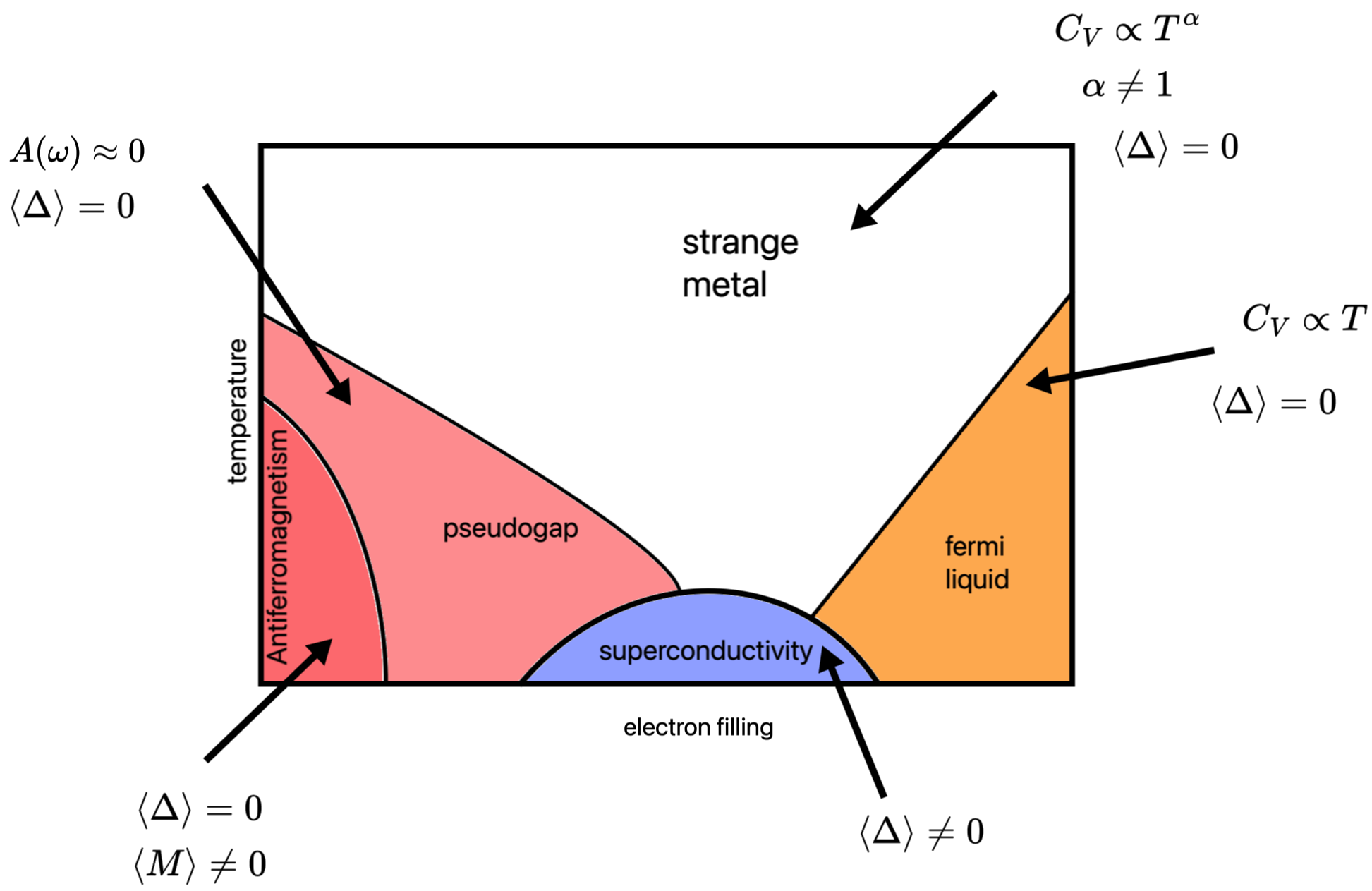}
    \caption{Pseudo-phase diagram for the Fermi-Hubbard model in two dimensions. The pseudogap phase is often measured via the spectral function as a pronounced dip in the spectral function. The other phases have a well defined combinations of the order parameters, as highlighted above.}
    \label{fig:phasediagram}
\end{figure}

\subsection{Phases of the model via observables, and their properties}\label{sec:observ}
Although, there is still a fair amount of debate \cite{Keimer2015-zz} on the exact location of the transitions lines between different phases of the model, there appears to be a consensus on the existing phases in general. A pictorial representation of such diagram is shown in Fig. \ref{fig:phasediagram}. Well-established and agreed upon local order parameters do not exist for all the phases. Another important limitation is that fixing the electron filling sometimes complicates things from numerical standpoint. This limitation, however, can be overcame through fixing the average electron density via a chemical potential $\mu$, e.g. $H\rightarrow H-\mu \hat{N}$ where $\hat{N}$ is the number operator (the sum of electron spin up and spin down). For this reason, often in the literature ones sees that the phase are sometimes plotted against a function of the energy and chemical potential instead of the electron filling (or doping).

An important aspect to clarify is that conductivity, the ultimate quantity that one wishes to understand, is not a static property, but a dynamical one. This aspect is clarified in Sec. \ref{eq:kineticdyn}. However, static observables as the ones described below provide key insights into the state (phase) of the system. Let us make some clarifying comment on how these quantities are calculated at zero vs finite temperature. Let us denote the ground state $|0\rangle$. We refer to the zero temperature expectation value of a generic observable $\hat O$ as
\begin{eqnarray}
    \langle\hat O\rangle_0=\langle 0|\hat O|0\rangle.
\end{eqnarray}
In the finite temperature setting, one simply extends the formula above to the thermal state average,
\begin{eqnarray}
    \langle \hat O\rangle_T=\text{Tr}(\rho \hat O)=\sum_n \frac{e^{-\beta E_n}}{Z}\langle n|\hat O|n\rangle.
\end{eqnarray}
where $Z=\sum_n e^{-\beta E_n}$ and $E_n$ are the energies of the $|n\rangle$ eigenstate of the target Hamiltonian, and $\rho=e^{-\beta H}/Z$.
We will often drop the explicit $T$ dependence in what follows as it is understood that either the ground state or thermal average is used.

Here we present some options given in the literature.
\begin{itemize}
\item \textbf{Antiferromagnetic Phase (AFM)}: The antiferromagnetic phase of the Fermi-Hubbard model is characterized by a staggered pattern in the spin density $\langle S_i^z \rangle$ with adjacent spins aligning antiparallel to each other \cite{Damascelli2003-bh}. The mean staggered magnetization  
\begin{eqnarray}
    M=\frac{1}{L^2}\sum_{\bm{i} }\langle (-1)^{i_x+i_y} \hat S_{i_x,i_y}^z\rangle \label{eq:afm}
\end{eqnarray} 
is an order parameter, which is non zero in this phase.  This phase shows no superconductivity ($\langle \Delta\rangle=0$, see below). This phase generally occurs at half-filling. 

\item \textbf{Fermi Liquid Phase}: The Fermi liquid phase of the Fermi-Hubbard model is characterized by well-defined quasiparticles near the Fermi surface \cite{Bruus2004-cd}, e.g. a theory of free fermions with an effective fermion mass. From a computational and non-perturbative perspective, diagonalizing the Hamiltonian is important, as $C_V(T)$ can be calculated from the derivatives of the free energy \cite{Huang2009-dr}, $F=-\kappa_BT \log Z$ with $Z=\text{Tr}(e^{-\beta H})$. Then, one can look at $\lim_{T\rightarrow 0} C_V(T)/T$ to understand in which phase the system is in \cite{Sachdev2016-xl}. If the energy states of the Hamiltonians are known, then one can calculate $C_V$ at a certain temperature $T$ from basic statistical mechanics as
\begin{eqnarray}
C_V=\kappa_B\beta^{2} \partial_\beta^2 \log \sum_{n} e^{-\beta E_n}=\frac{\partial}{\partial T} \bar E \label{eq:diff1}
\end{eqnarray}
where $E_n$ are the energy of the state $|n\rangle$ of the Fermi-Hubbard Hamiltonian, while $\beta=(\kappa_BT)^{-1}$, and 
where 
\begin{eqnarray}
\bar E=\frac{1}{Z} \sum_{n} e^{-\beta E_n} E_n= \text{Tr}(\rho \hat H).
\end{eqnarray}
Another formula for the calculation of $C_V$ is given by the thermal expectation value
\begin{eqnarray}
C_V=\frac{1}{\kappa_BT^2 Z } \sum_{n} e^{-\beta E_n} (E_n^2-\bar E^2)=\frac{1}{\kappa_BT^2  }\text{Tr}\Big((\hat H^2-\bar E^2 \hat I)\rho\Big)\label{eq:cv_var}
\end{eqnarray}
where $\hat I$ is the identity operator. From a computational perspective, the difference between Eqn. (\ref{eq:cv_var}) and Eqn. (\ref{eq:diff1}) is that Eqn. (\ref{eq:diff1}) is evaluated in terms of second order derivatives, while the second formula is a thermal expectation value. The Fermi-Liquid phase is characterized by the conductivity scaling as $\rho(T)\propto T^2$. It can be measured from $C_V$, in particular observing a linear scaling in temperature in the heat capacity, typical of metals
\begin{eqnarray}
    C_V\propto T.\label{eq:fermiliquid}
\end{eqnarray}

\item \textbf{Strange Metal Phase}:  The strange metal phase within the Fermi-Hubbard model does not possess a distinct order parameter \cite{Greene2020-hx}. It is characterized by anomalous transport properties, such as non-Fermi liquid behavior in the electrical resistivity $\rho(T)$ or violation of conventional scaling laws in the specific heat $C_V(T)$ at temperature ($T$). For Fermi liquids, the theory establishes that $\rho\propto T^2$, while in the strange metal phase one expects $\rho\propto T^\alpha$ with $\alpha$ typically in the range $[1,2)$.  To study the transport properties, in numerical simulations the Boltzmann transport equation or the (perturbative) Kubo methods are often used. However, the easiest thing to calculate in numerical experiments is
\begin{eqnarray}
    C_V\propto T^{\alpha^\prime}\label{eq:strangemetal}
\end{eqnarray}
and check that $\alpha^\prime\neq 1$.

\item \textbf{Superconducting Phase}:   The prevailing belief is that the most promising approach to achieving high-Tc superconductivity involves generating robust spin fluctuations by closely associating with antiferromagnetic-ordered phases.  Examples of this approach include the well-known cuprates and iron-based superconductors. However, as we see below, some proposals have investigated graphene-based superconductors, which is a fresh area of research. It is fair to say that there have been thousands of papers and hundreds of proposals of possible superconductive states. However, the two key proposals are the perturbations of the striped ground state, e.g. antiferromagnetic state arranged in stripes near the half filling electron filling level (with pairs of holes behaving as Cooper's pairs), or longer chains of holes at lower electron filling.  
From a computational perspective, in Fermi-Hubbard models with different number of bands is suggested to exhibit different type of superconductivity. This can be captured by combinations and observations of order parameters. For the superconductivity, typically, this is studied via the correlation function, and in particular via operators of the form
\begin{eqnarray}
    \hat \Delta_{ij}= \big(\hat{c}_{i+}\hat{c}_{j-}-\hat{c}_{i-}\hat{c}_{j+} \big)
\end{eqnarray}
adapted to the problem at hand, and the two observables 
\begin{eqnarray}
    C_1=\Delta_{ij}&=&\frac{1}{2}\langle \hat \Delta_{ij}+\hat \Delta_{ij}^\dagger\rangle \label{eq:sc1}\\
    C_2=P_{ab,ij}&=&\langle\hat  \Delta_{ab}^\dagger \hat \Delta_{ij}\rangle \label{eq:sc2}
\end{eqnarray}
to understand the scaling of the correlation, for instance power laws as a function of the distance $d=|i-j|$. 
In two or higher dimensions, this can be replaced by the Manhattan distance on the lattice. In particular, the numerical calculation can be simplified by looking at $i=j$ in $C_1$ and $a=b$, $i=j$ in $C_2$. In the literature, it is often observed that superconductivity is due to spin fluctuations near an antiferromagnetic order, which is measured by $C_1$. Power law correlations, symptoms of a critical phase, are measured by $C_2$ \cite{Maier2000-uc}. The observable $C_1$ and $C_2$ can be used as an indicator of d-wave superconductivity \cite{schrieffer} in the 2D Fermi-Hubbard model. In fact, in d-wave superconductivity the order parameter $\Delta_{ij}$ has a specific modulation in the angle (See Appendix).
\item \textbf{Pseudogap}: The pseudogap phase in the Fermi-Hubbard model lacks a universally agreed-upon order parameter, and it can be measured experimentally \cite{Timusk1999} via ARPES in the spectral function. This phase is characterized by a substantial suppression or depletion of the electronic density of states (DOS $D_E$) near the Fermi level ($D_E(0)$) without displaying any long-range order \cite{Keimer2015-zz}. The spectral function, the Green function and the density of states are related. 
From an operational perspective, from the Lehmann representation one can obtain the following expression of the spectral density in the frequency domain \cite{Bruus2004-cd} (pp 129, eq 8:50) reduced to a finite system:
\begin{eqnarray}
A(i,\sigma,\omega)=\frac{4\pi}{Z(1+e^{-\eta \omega})} \sum_{n} e^{-\beta E_n }\sum_{n^\prime} \langle n|c_{i,\sigma}|n^\prime\rangle\langle n^\prime|c_{i\sigma}^\dagger|n\rangle \delta(E_n-E_{n^\prime}+\omega)\label{eq:pseudogap}
\end{eqnarray}
where $|n\rangle $ are the eigenstates of the Hamiltonian. This quantity has to be scanned over the frequency $\omega$ to observe a suppression $A(\omega)\approx 0$. Of all the phases above, the pseudo-gap is a dynamical phase as $\omega$ arises from the Fourier transform in time. Thus, this phase requires an analysis of the dynamics of the system.

\end{itemize}

The reduction from a five bands to a single bands goes through various steps of integrations of degrees of freedom in the original crystal to a single atomic structure on a lattice sub-lattice. For instance, for La$_2$CuO$_4$, the reduction from the full crystal to a single layer and single atom and band (where La can be replaced for instance with Sr).
\section{Computational applications}\label{sec:application}
Below, we provide various concrete examples taken from the literature of relevant problems of interest for scientific discoveries. In all the Fermi-Hubbard models below, classical algorithms have a variety of shortcomings associated with the numerical evaluation of the ground state, as described in Sec. \ref{sec:FH_classical_methods}. It is important to stress that it is hard to image a self-contained algorithm which fully replaces the human expertise. In many of the methods, it is necessary to look at a variety of order parameters to suggest a phase diagram and state of the system, as in Fig. \ref{fig:phasediagram}. The range of the parameters is typically $f=[0,1]$ for the electron filling, while the temperature is kept between $T\in[0,300]$ Kelvin, this range can be restricted further to observe the strange metal phase. The order parameters that one needs to look at in numerical simulations are of the form of Eqn. (\ref{eq:pseudogap}) for the pseudogap, and Eqn. (\ref{eq:afm}) for the antiferromagnetic phase, Eqn. (\ref{eq:fermiliquid}) or  (\ref{eq:strangemetal}) for the Fermi liquid or strange metal phase, Eqn. (\ref{eq:sc1}),(\ref{eq:sc2}) for the superconducting phase.

Let us now describe the units of the parameters of the Fermi-Hubbard model in adimensional units.
The FH is the sum of two terms
\begin{eqnarray}
    H_{FH}=\bar t\times H_{Kin}+\bar U\times H_{int},
\end{eqnarray}
where $\bar t$ is a scale, satisfying $\bar t\geq\text{max}_{ij} t_{ij}$ and $\bar U\geq\text{max}_i U_i$. The onset of superconductivity in cuprates in the temperature range of $100-200$ Kelvin, which corresponds to $8-16$ milli electron Volts ($\sim 10\ $meV). The pseudo-gap phase is in the order of also tens of meV in temperature. The range of interesting physics for the parameter $\chi=U/t$ is in the order of also $5-20$. For instance, a common choice for the $2d$ Fermi-Hubbard model is $t=1$ and $\chi=8$. When computing the phase diagram, there is no time evolution, and it is not restrictive to also choose $\hbar=1$ and impose $\{c^\dagger_{i\sigma},c_{j\sigma^\prime}\}= \delta_{ij}\delta_{\sigma \sigma^\prime}$. 

In the following, one can set $\bar t=1$, set $\chi\in[0,20]$ in adimensional units, and the temperature in adimensional units to range in the $[0,200]$ range, to make sure that it contains the strange metal phase. 
Using these adimensional units, we can introduce $\kappa  T$ measured in meV, $\bar t_{max}=1$ and $\chi=[0,2]$. The precision on the phase diagram variables varies. Typical choices are $df_c=0.1$, $d\chi_c=0.5$, $d T_c=10$ with a refinement close to the phase diagram boundaries, with the underline $c$ notation means coarser.

Thus, one way to think of the phase diagram in terms of discretization of the parameter space is as follows, in terms of coarser and finer phase diagrams. At a coarser level a homogeneous discretization for $f$, $T$ respectively  contained in the box $[0,1]\times [0,50]$. In this case, the half filling line is $f=1/2$. Near the phase diagrams lines, it is common to pick a finer value of $df_f=df_c/10$ and $d T_f=d T_c/100$, with the underline $f$ notation means finer in the vicinity of phase lines. We now present a list of concrete computational workflows and application by increasing order of complexity.


\begin{tcolorbox}[breakable,width=\columnwidth,
                  colback=white]

\subsection*{Application 1: Zero-Temperature Superconductivity}
{The quantity $C_1$ cannot be measured if N is fixed but the measurement of $\Delta^2_0$ is still possible. }
The main goal of this application is to find signatures of superconductivity of a Fermi-Hubbard model at zero temperature. The core quantum capabilities required for this task is ground-state preparation. The zero temperature superconductivity is the simplest of all the tasks presented in this document, and is important to infer the exact critical density for the antiferromagnetic to superconductive phase.

\subsubsection*{Specific background of the application}
There are two different techniques for specifying the electron filling $f$. The first approach is to restrict computations to states that are eigenvectors of the electron number operator $\hat{N}=\sum_{i,\sigma} \hat{c}^\dagger_{i\sigma}\hat{c}_{i,\sigma}$ with value $\langle \hat{N}\rangle$ satisfying $\langle N\rangle =2fL$. The second approach consists of modifying the original Hamiltonian by $H\rightarrow H-\mu \hat{N}$, and varying the chemical potential $\mu$ until the average electron filling of the ground state has the desired property $\langle N\rangle =2fL$ on average. $L$ is the number of lattice sites.

\subsubsection*{Objectives}
For simplicity, we consider the superconductivity parameter~\eqref{eq:sc1} at a fixed location $i=j=0$ in the center of the lattice, i.e., $\hat{\Delta}_{0}=\frac{1}{2} \big(\hat{c}_{0+}\hat{c}_{0-}-\hat{c}_{0-}\hat{c}_{0+} +\hat{c}^\dagger_{0-}\hat{c}^\dagger_{0+}-\hat{c}^\dagger_{0+}\hat{c}^\dagger_{0-}\big)$.

The objective is to compute for a given ground state $|\phi\rangle$, the mean and fluctuation of the superconductivity parameter, i.e., $C_1 = \langle \phi|\hat{\Delta}_{0}|\phi\rangle$ (This quantity is only defined if filling is set through chemical potential) and $C_2 = \langle \phi|\hat{\Delta}^2_{0}|\phi\rangle$.
The precision required on the dimensionless quantities $C_1$ and $C_2$ is $10^{-4}$.

This computation is realized for different values of the electron filling $f$ and potential to kinetic energy ratio $\chi = U/t$. The minimal target size of the lattice is $L=20\times 20$.

\subsubsection*{End-to-end computational workflow}
 The inputs to the problem are the coupling parameters of the Fermi-Hubbard model and the required precision on the observables. 
\begin{verbatim}
    for chi : 0.5 to 20, step dchi=0.5
      for f : .1 to 1, step df=0.1
      (alternatively, loop over mu, mu=0... 100, dmu=1)
          # Set Hamiltonian  (alternatively, with chemical potential)
          # Set Hilbert space with fixed number of electrons filling 
          (skip this step with chemical potential)
          # Find ground state
             - Evaluate C1 or/and C2 parameter up to prescribed precision
                (with chemical potential, report average electron filling)
             Is it nonzero?
              yes -> potential superconducting
              no -> no superconductivity
      end
    end
\end{verbatim}
As described earlier, the for loop in $f$ can be replaced, alternatively, with a loop on the chemical potential $\mu$.


\subsubsection*{Concrete problem instantiations}
This workflow is applicable to all Hamiltonians described in Section~\ref{sec:problem_instances}.

\end{tcolorbox}

\begin{tcolorbox}[breakable,width=\columnwidth,
                  colback=white]

\subsection*{Application 2: High-Temperature Superconductivity}
The main goal of this application is to find signatures of superconductivity of a Fermi-Hubbard model at non-zero temperature. The core quantum capabilities required for this task is thermal-state preparation. 

\subsubsection*{Specific background of the application}
A thermal state, at the temperature $T$, is described by the density matrix $\rho =\frac{e^{- H/\kappa_B T}}{Z}$, where $\text{Tr}(\rho)=1$. There are again two different ways of specifying the electron filling $f$. The first is to construct the density matrix only states that are eigenvectors with eigenvalue $f$ of the electron number operator $\hat{N}=\sum_{i,\sigma} \hat{c}^\dagger_{i\sigma}\hat{c}_{i,\sigma}$. The second approach consists in modifying the original Hamiltonian by $H\rightarrow H-\mu \hat{N}$, and vary the chemical potential $\mu$ until the average electron filling of the ground state has the desired property $\text{Tr}(\rho \hat{N}) = 2fL$ on average.
\subsubsection*{Objectives}
The objective is to compute for a given thermal state $\rho$, the mean and fluctuation of the superconductivity parameter, i.e., $C_1 = \text{Tr}(\rho \hat{\Delta}_{0})$ (This quantity is only defined when filling fraction is adjusted through $\mu$) and $C_2 =  \text{Tr}(\rho \hat{\Delta}^2_{0})$. 

The precision required on the dimensionless quantities $C_1$ and $C_2$ is $10^{-4}$. The minimal target size of the lattice is $L=20\times 20$. This computation is realized for different values of the electron filling $f$, potential to kinetic energy ratio $\chi = U/t$ and temperature $T$.

\subsubsection*{End-to-end computational workflow}
 The inputs to the problem are the coupling parameters of the Fermi-Hubbard model and the required precision on the observables. 
\begin{verbatim}
   for T : 1 to 200, step dT = 10
    for chi : 0.5 to 20, step dchi = 0.5
      for f : .1 to 1, step df = 0.1
      (alternatively, loop over mu, mu=0... 100, dmu=1)
          # Set Hamiltonian  (alternatively, with chemical potential)
          # Set Hilbert space with fixed number of electrons filling 
          (skip this step with chemical potential)
          # Find thermal state
             - Evaluate C1 or/and C2 parameter up to prescribed precision
                (with chemical potential, report average electron filling)
             Is it nonzero?
              yes -> potential superconducting phase
              no -> return (see application 3)
      end
    end
  end
\end{verbatim}
In typical applications, if a critical line is observed in a particular interval, precision in $df$ is increased and centered around the critical line by a factor or ten or more, e.g. $df=0.01$.

Note that the loop on the in electron filling can be replaced by a loop on the chemical potential $\mu$ as described earlier.


\subsubsection*{Concrete problem instantiations}
This workflow is applicable to all Hamiltonians described in Section~\ref{sec:problem_instances}.

\end{tcolorbox}

\begin{tcolorbox}[breakable,width=\columnwidth,
                  colback=white]

\subsection*{Application 3: Phase-Diagram of Fermi-Hubbard Models}

The main goal of this application is to identify the full phase-diagram (Fermi-liquid, strange metal, AFM, superconductor) of the Fermi-Hubbard model. The core quantum capabilities required for this task is thermal-state preparation. 

\subsubsection*{Specific background of the application}
A thermal state, at the temperature $T$, is described by the density matrix $\rho =\frac{e^{- H/\kappa_B T}}{Z}$, where $\text{Tr}(\rho)=1$.

The two different ways of specifying the electron filling $f$ in the range $[0,1]$ are: 1) Constructing the density matrix only states that are eigenvectors with eigenvalue $f$ of the electron number operator $\hat{N}=\sum_{i,\sigma} \hat{c}^\dagger_{i\sigma}\hat{c}_{i,\sigma}$; 2) modifying the original Hamiltonian by $H\rightarrow H-\mu \hat{N}$, and vary the chemical potential $\mu$ until the average electron filling of the ground state has the desired property $\text{Tr}(\rho \hat{N}) = 2fL$ on average.

\subsubsection*{Objectives}
The objective is to compute for a given thermal state $\rho$ the superconducting order parameters $C_1$ from Eqn.~\eqref{eq:sc1} and $C_2$ from Eqn.~\eqref{eq:sc2} at a given site $i=j$, the AFM order parameter $M$ from Eqn.~\eqref{eq:afm} and $C_v(T)$ from Eqn.~\eqref{eq:cv_var}. 
This is done for different electron filling $f$, for different values of $\chi=U/t$ and temperature $T$. The precision on $C_1$ and $C_2$ is $10^{-5}$, while on $M$ is $10^{-2}$. The precision on the scaling exponent of $C_v(T)$ is $10^{-2}$, (or alternatively choose a relative precision on $C_v(T)$ of $10^{-3}$).

\subsubsection*{End-to-end computational workflow}
 The inputs to the problem are the coupling parameters of the Fermi-Hubbard model and the required precision on the observables. The minimal target size of the lattice is $L=20\times 20$.

A pseudo-code describing this procedure is provided below. 
\begin{verbatim}
   for T : 1 to 200, step dT = 10
    for chi : 0.5 to 20, step dchi = 0.5
      for f : .1 to 1, step df = 0.1
      (alternatively, loop over mu, mu=0... 100, dmu=1)
          # Set Hamiltonian  (alternatively, with chemical potential)
          # Set Hilbert space with fixed number of electrons filling 
          (skip this step with chemical potential)
          # Find thermal state
             - Evaluate C1 or/and C2 parameter up to prescribed precision
                (with chemical potential, report average electron filling)
             Is it nonzero?
              yes -> potential superconducting phase
              no -> continue
             - Evaluate AFM order paramater
              # Is it AFM?
              yes -> return
              no -> continue
             - Evaluate Cv
              Does Cv scale as T?
              -> Fermi liquid?
              Does it scale as T^alpha, alpha>1?
              -> Strange metal
             - (Optional: calculate the pseudogap parameter A)
      end
    end
  end
\end{verbatim}
In typical applications, if a critical line is observed in a particular interval, precision in $df$ is increased and centered around the critical line by a factor or ten or more, e.g. $df=0.01$.

Note that the loop on the in electron filling can be replaced by a loop on the chemical potential $\mu$ as described earlier.

\subsubsection*{Concrete problem instantiations}
This workflow is applicable to all Hamiltonians described in Section~\ref{sec:problem_instances}.

\end{tcolorbox}











    
\section{Models instantiations}\label{sec:problem_instances}
In this section, we discuss some of the models that are Fermi-Hubbard based. On one end, we introduce models that are based on highly novel 2-D materials like Graphene. One the other end, iron based models known to describe superconductivity in many functional superconduting materials are also given. These models instantiations can be used as Hamiltonian of any specific application mentioned in the last section. 
\subsection{Two dimensional lattice models}
The specific instance of interest is a superconductor model at low-electron filling  in 2-dimensions (lattice crystal) that was recently studied \cite{Zheng2017-ou} but has been known for years \cite{Arovas2022-mt} for the model of Eqn. (\ref{eq:fh}) with $U=8t$ and at $f=1/8$. A stripe phase was observed with approximate methods described in Sec. \ref{sec:FH_classical_methods}
\begin{eqnarray}
    H=-t \sum_{\langle ij\rangle,\sigma}(\hat c_{i\sigma}^\dagger \hat c_{j\sigma}+\hat c_{j\sigma}^\dagger \hat c_{i\sigma})+U\sum_i \hat n_{i+} \hat n_{i-}
\end{eqnarray}
for $U=8t$ and at $1/8$ electron filling. Thus, for this particular instantiation, the parameter scan over $f$ and $\chi$ is omitted in the application boxes. The Hamiltonian of interest is \cite{Keimer2015-zz} for a 2-dimensional lattice
\begin{equation} \label{eq:single_band_FH_model}
    \hat H=-\sum_{\langle i,j\rangle} \sum_{\sigma=\pm} t_{ij} \hat c^\dagger_{i\sigma} \hat c_{j,\sigma}+U \sum_{i} \hat c_{i+}^\dagger \hat c_{i-}^\dagger \hat c_{i-}\hat c_{i+}
\end{equation} 
with $t_{ij}=t^*_{ij}=t \delta_{|\vec i-\vec j|,1}$. This is the generalization of Fermi-Hubbard model given in Eqn. ~\eqref{eq:fh}
\subsubsection*{Hamiltonian in momentum space}
We start the Fermi-Hubbard Hamiltonian in momentum-space given in Eqn. (\ref{eq:2dsss2}) and assume that interactions happens for zero center of mass momentum i.e. $\vec{k}+\vec{k}'=0$ or $\vec{k}=-\vec{k}'$. We obtain the Hamiltonian
\begin{eqnarray}
    H=\sum_{\vec k,\sigma} \xi_{\vec k} \hat c^\dagger_{\vec k,\sigma}\hat c_{\vec k,\sigma}+\frac{U}{L}\sum_{\vec k,\vec q} \hat c^\dagger_{q,+} \hat c^\dagger_{-q,-} \hat c_{-k,-}  \hat c_{k,+} \label{eq:mf2}
\end{eqnarray}

The above Hamiltonian describes zero-center of momentum superconductivity and as mentioned previously, the treatments to solve the above Hamiltonian are limited. We can also change the second term in Eqn. (\ref{eq:mf2}) to include the symmetry of the interaction in momentum space as under 
\begin{eqnarray}
    H_{FH}=\sum_{\vec k,\sigma} \xi_{\vec k} \hat c^\dagger_{\vec k,\sigma}\hat c_{\vec k,\sigma}+\frac{U}{L}\sum_{\vec k,\vec q} \gamma_{\vec{k}} \gamma_{\vec{q}}\hat c^\dagger_{q,+} \hat c^\dagger_{-q,-} \hat c_{-k,-}  \hat c_{k,+} \label{eq:hdw}
\end{eqnarray}
When $\gamma_{\vec {k}}\sim cos(k_x)-cos(k_y)$, we have high Tc superconducting model Hamiltonian known to represent d-wave superconductivity. A mean-field Hamiltonian derived from Eqn.~\eqref{eq:hdw} is studied in \cite{Norman1995,Zhu2006}. A similar phase diagram can be studied also for other models, in particular with different parameterizations and dimensionalities. For example, a mean-field like model is given by the Hamiltonian on a 2-dimensional lattice written in Fourier space \cite{Norman1995,Zhu2006}. With the lattice constant $a$, the discrete values of momentum are $\vec k(i,j)=(\frac{2\pi}{a} i,\frac{2\pi}{a} j)$ where $i,j$ are integers. \footnote{In the Bogoliubov-de Gennes approximation, the linear size of the lattice can be large e.g. 100 site chain or multiple thereof. This is due to the fact interactions part has been simplified in the Bogoliubov-de Gennes approximation. However, when the quartic interactions are present, at most the system's size is between 16-20 sites in total. } Then, the sum $\sum_{\vec k}=\sum_{i,j=0}^{L-1}$, and
\begin{eqnarray}
    H=\sum_{\vec k,\sigma} \xi_{\vec k} \hat c^\dagger_{\vec k,\sigma}\hat c_{\vec k,\sigma}+\sum_{\vec k} (\Delta_{\vec k } \hat c_{\vec k,+}^\dagger \hat c_{\vec k,-}^\dagger  +h.c.) \label{eq:2dsss}
\end{eqnarray}
with 
\begin{eqnarray}
    \xi_k&=&-2t_1 (\cos k_x+\cos k_y)-4 t_2 \cos k_x\cos k_y -2 t_3 (\cos 2k_x+\cos 2k_y)\nonumber \\
    &+&-4 t_4(\cos 2k_x \cos k_y+\cos k_x \cos 2k_y)-4 t_5 \cos 2k_x \cos 2k_y-\mu
\end{eqnarray}
with $t_1=1$, $t_2=-0.2749$, $t_3=0.0872$, $t_4=0.0938$, $t_5=-0.0857$, $\mu=-0.8772$, and 
\begin{eqnarray}
    \Delta_k=\frac{\Delta_0}{2}(\cos k_x-\cos k_y)
\end{eqnarray}
with $\Delta_0=0.1$. The parameters were obtained in \cite{Zhu2006} fitting ARPES data for Bi-2212. The asymmetry between the x and y direction is due to the fact that the square lattice Hamiltonian is inherited from an original orthorombic symmetry \cite{Norman1995}, and is in real space such asymmetry is shown Fig. \ref{fig:asymsquare}.
\begin{figure}
    \centering
    \includegraphics[scale=0.2]{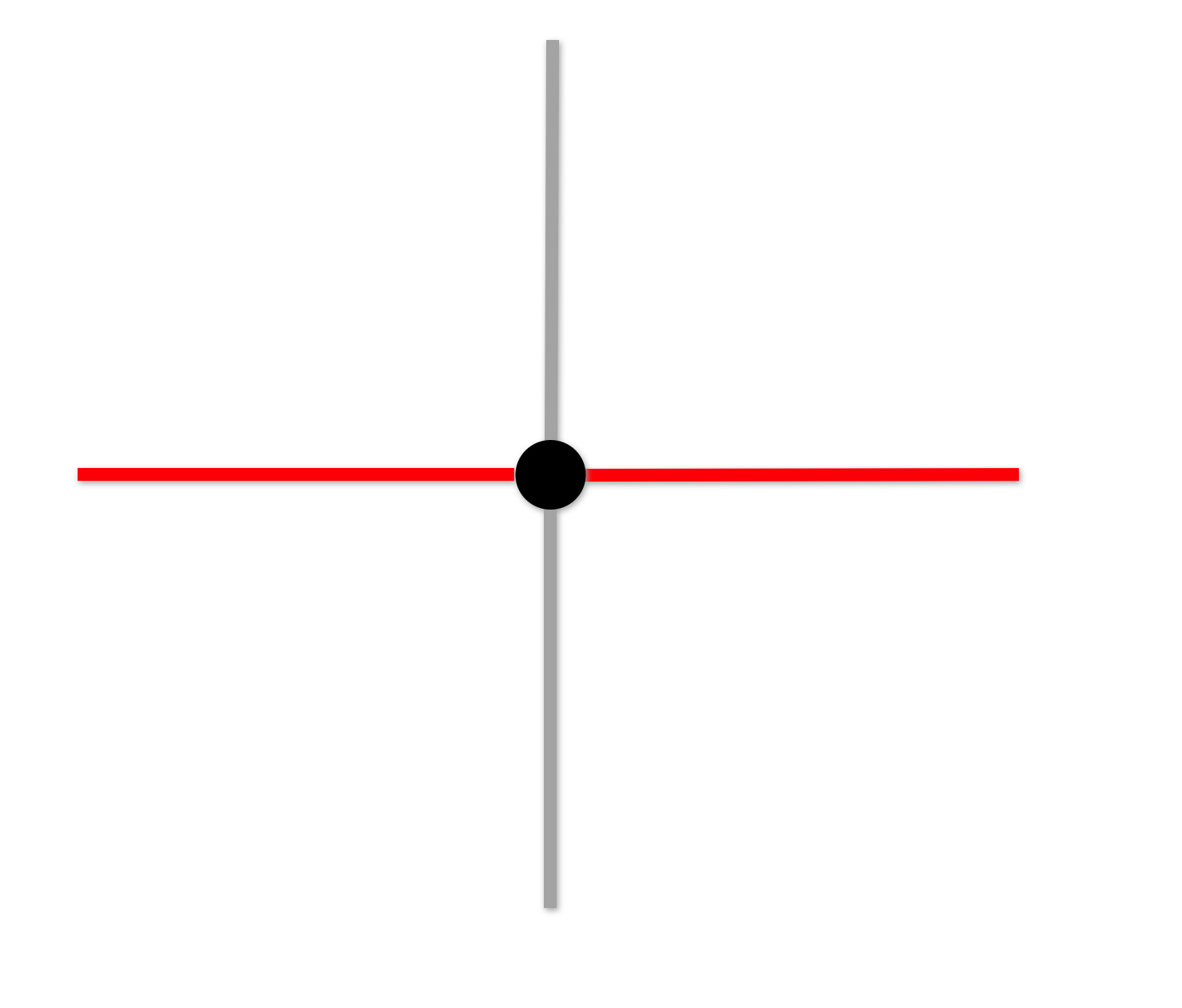}
    \caption{Asymmetry in the couplings along the x- and y- direction of Eqn. (\ref{eq:2dsss}).}
    \label{fig:asymsquare}
\end{figure}
The quadratic Hamiltonian above can be converted to Fermi-Hubbard model by keeping the kinetic term identical term, and adding an interaction term:
This Hamiltonian has to be scanned in the parameter $U$.

\subsubsection{Graphene-based examples}
We also report the graphene-based model in 2d \cite{Kang2019-et} with a non-standard 2d lattice.
\begin{equation}
H=H_{TB}+U \sum_i \hat n_{i+} \hat n_{i-}
\end{equation}
with $U\approx 10$ eV, while
\begin{eqnarray}
H_{TB}&=&-t_1 \sum_{\langle i,j\rangle, \sigma} (\hat c_{i\sigma}^\dagger \hat c_{j\sigma}+h.c.)\\
&&-t_2\sum_{\langle i,j\rangle, \sigma} (\hat c_{i\sigma}^\dagger \hat c_{j\sigma}+h.c.)
\end{eqnarray}
where the first and second sum is intra-square NN and inter-square NN hopping, with $t_1\approx 2.5$ eV while $t_2\approx 2.9$ eV, and $U=10$ eV. The lattice is shown in Fig. \ref{fig:graphene}, with $t_1$ for the black bonds and $t_2$ for the red bonds.

\subsection{Extended-Hubbard Model}
The model given in Eqn.~\eqref{eq:fh} is the minimal model for superconductivity that describes electron correlations through on-site interaction parameter U and can be extended to include further interactions. This simple model already can explain the behaviors of a Mott insulator, stripe order, strange metal, and to some extent d-wave superconductivity \cite{ExtendedFH2D}. For Copper-based superconductors, we can also consider nonlocal Coulomb interaction $V$ between neighboring sites. Hubbard model with Coulomb interactions between neighboring sites is often names as extended Hubbard model. This model incorporates additional physics not present in the Hubbard model. A repulsive $V$ can induce a charge density wave (CDW), and an attractive V was argued to favor an exotic form of superconductivity called spin-triplet superconductivity \cite{Cuprate1D}. 1D cuprate chains have revealed the presence of attractive Coulomb interaction $V<0$ between nearest-neighbor (NN) electrons. Thus, extended Hubbard model is believed to give more faithful description of Copper based superconductors and can lead to potentially new superconducting phases not captured by simple Hubbard model. The Hamiltonian for extended-Hubbard model is given as \cite{ExtendedFH2D}
\begin{eqnarray}\label{Eq:FH1}
\hat{H}_{EFH}=-t_h\sum_{\langle i,j \rangle,\sigma}\bigg(\hat{c}_{i,\sigma}\dg \hat{c}_{j,\sigma}+H.c.\bigg)+U\sum_{i} \hat{n}_{i,\uparrow} \hat{n}_{i,\downarrow}+V\sum_{\langle i,j \rangle,\sigma,\sigma'}\hat{n}_{i,\sigma}\hat{n}_{j,\sigma'}
\end{eqnarray}
As before, $t_h$ is the hoping (hoping integral) amplitude. $U$ is the onsite Coulumb interaction and $V$ describes the additional interactions between the nearest neighbors (NN).
\subsection{Two band superconductivity via transverse Ising mapping}
Although for the following example a set of parameters was not found which can be used to model a real material, we would like to point out a mapping in the strong coupling limit of a two band Fermi-Hubbard model to a transverse quantum Ising model \cite{Wang2021-og}. 
The original Hamiltonian is a two-band model of the form
\begin{eqnarray}
    H&=&H_2+H_4\\
    H_2&=&\sum_{ij,\sigma}\Big((t_{ij}^c -\mu^c \delta_{ij})\hat c_{i\sigma}^\dagger \hat c_{j\sigma}+(t_{ij}^d-\mu^d\delta_{ij})\hat d_{i\sigma}^\dagger \hat d_{i\sigma}\Big)\\
H_4&=&\sum_{i,\sigma,\sigma^\prime}\Big(U_1 \hat c_{i\sigma}^\dagger \hat c_{i\sigma} \hat d_{i\sigma^\prime}^\dagger\hat d_{i\sigma^\prime} +\frac{U_2}{2} \hat c_{i\sigma}^\dagger \hat c_{i\sigma^\prime}^\dagger \hat d_{i\sigma}\hat d_{i\sigma^\prime}+h.c.\Big)
\end{eqnarray}
where the operators $\hat c_{i\sigma}$ and $\hat d_{i\sigma}$ are ladder operators between electrons in bands $c$ and $d$ of a square lattice $i$ with spin $\sigma$. The parameters $U_1$ represents inter-band repulsion energy,  while $U_2$ is a pair hopping energy. In the strong coupling limit and for a particular choice of parameters $U_1=4U(1-\delta_{\sigma\sigma^\prime})$, and $U_3=4U$ this model is interesting because it can be studied via QMC \textit{without} sign problem.  Also, the model can be mapped to a so called $J_1$-$J_2$ quantum Ising Hamiltonian of the form with a homogeneous transverse field if the $\mu$'s are homogeneous (after a rotation)
\begin{eqnarray}
    H=\sum_{ij} J_{ij} \sigma_i^z \sigma_j^z-h\sum_{i}  \sigma_i^x \label{eq:dwaveham}
\end{eqnarray}
where $h=\frac{(\mu^c-\mu)^2}{8 U}$ and $J_{1}=\frac{t_1^c t_{1}^d}{3 U}$ ($J_{i,j}$ for $i$, $j$ in nearest neighbors) and $J_{2}=\frac{t_2^c t_{2}^d}{3 U}$ ($J_{i,j}$ for $i$, $j$ in neat-nearest neighbors). $t_{1}$ represents uniform nearest-neighbor couplings $t_{ij}$ while $t_{2}$ represent next-nearest-neighbor couplings and 2D square lattice is assumed. QMC investigations has shown that this model has a paramagnetic to Neél transition. From a computational methodological perspective, the interesting properties of this model is the phase space in the plane $a=h/J_2$ and $b=J_1/J_2$ at zero temperature, and in three dimensions including temperature. The input is thus a Hamiltonian without well defined parameters but at zero temperature. One can either fix the electron filling or scan it.
Contrarily to the previous examples, this is a case that can be studied efficiently using Quantum Monte Carlo \cite{Gubernatis2016} without running into the sign problem. Although Monte Carlo methods can be used in this case, in principle \emph{the bottleneck of MC is the generation of random number and samples}. Interestingly, the problem of Eqn. (\ref{eq:dwaveham}) can be studied using the current family of Ising annealers provided by D-Wave that can generate samples from the Hamiltonian of Eqn. (\ref{eq:dwaveham}).
\begin{figure}[!htb]
    \centering
    \includegraphics[scale=0.4]{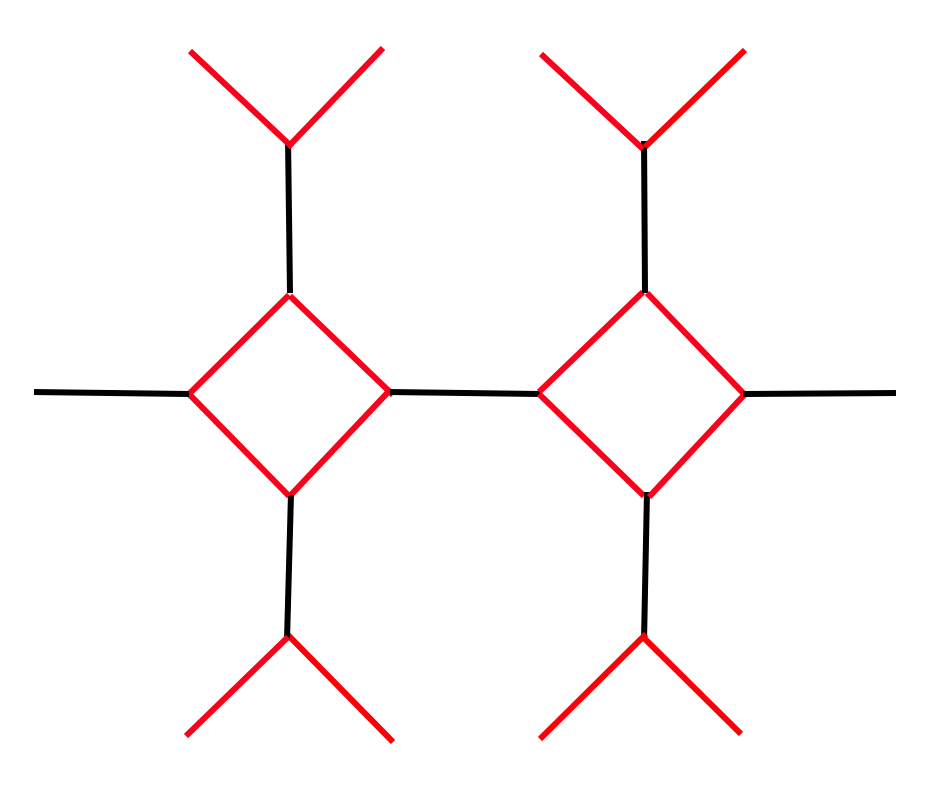}
    \caption{The lattice considered in graphene-based superconductors of \cite{Kang2019-et}.}
    \label{fig:graphene}
\end{figure}

\subsection{Iron-based models}
Aside from cuprates, there is also an interest in iron-based superconductors offering critical temperatures between $30-60K$. The Hamiltonians presented below are obtained via the Slater-Koster approximation, with the extra parameters fitting experimentally observed Fermi surfaces \cite{Dagotto1994-tz}. In this section, we consider the models of superconductivity for which multi-orbital methods are needed. This is the minimal model representing iron-based superconductivity, although more realistic models should involve five 3d orbitals of the Fe. A detailed analysis of the many possible hopping channels for electrons when only the two orbitals $d_{xz}$ and $d_{yz}$ are involved in the process, led to the construction of a tight-binding model Hamiltonian for this case \cite{Dagotto1994-tz}.

In this instance (referred as two-orbital model), the Hamiltonian involves only the $3d_{xz}$ and $3d_{yz}$ orbitals is given by the two-dimensional Hamiltonian shown in Eqn.~\eqref{eq:dxy2}. The operators $\hat d^{\dagger}_{\alpha,{\boldsymbol i} ,\sigma}$ is a fermionic ladder operator which creates an electron at site $i$ of the lattice, at the orbital $\alpha$, with spin (projection on the z-axis) $\sigma$. The hopping is  both between nearest-neighbor (NN) atoms, and along the plaquette diagonals or next nearest-neighbour (NNN) sites \cite{Dagotto2011-pu}. 
\begin{align}\label{eq:dxy2}
H^{xz,yz} &=-t_1\sum_{i,\sigma} \bigg(\hat d_{xz,i,\sigma}^\dagger \hat d_{xz,i+\hat y,\sigma} + \hat d_{yz,i,\sigma}^\dagger \hat d_{yz,i+\hat x,\sigma }+h.c.\bigg)\nonumber \\
&-t_2\sum_{i,\sigma} \bigg( \hat d_{xz,i,\sigma}^\dagger \hat d_{xz,i+\hat{x},\sigma}+\hat d_{yz,i,\sigma}^\dagger \hat d_{yz,i+\hat{y},\sigma}+h.c. \bigg)\nonumber \\
&-t_3\sum_{i,\mu,\nu,\sigma}\bigg(\hat d_{xz,i,\sigma}^\dagger \hat d_{xz,i+\mu+\nu,\sigma}+\hat d_{yz,i,\sigma}^\dagger \hat d_{yz,i+\mu+\nu,\sigma}+H.c. \bigg)\nonumber \\
&+t_4\sum_{i,\sigma}\bigg(\hat d_{xz,i,\sigma}^\dagger \hat d_{yz,i+\hat{x}+\hat{y},\sigma}+\hat d_{yz,i,\sigma}^\dagger \hat d_{xz,i+\hat{x}+\hat{y},\sigma}+H.c\bigg)\nonumber\\
&-t_4\sum_{i,\sigma}\bigg(\hat d_{xz,i,\sigma}^\dagger \hat d_{yz,i+\hat{x}-\hat{y},\sigma} + \hat d_{yz,i,\sigma}^\dagger \hat d_{xz,i+\hat{x}-\hat{y},\sigma} +H.c.\bigg)-\mu \sum_i \bigg( \hat{n}_{xz,i} + \hat{n}_{yz,i} \bigg)
\end{align}
Where $\hat{n}_{xz,i}=\hat{n}_{xz,i,\uparrow}+\hat{n}_{xz,i,\downarrow}=\hat{d}_{xz,i,\uparrow}^\dagger \hat{d}_{xz,i,\uparrow}+\hat{d}_{xz,i,\downarrow}^\dagger \hat{d}_{xz,i,\downarrow}$ is the occupancy of $d_{xz}$ orbital at site i including both spin-projections. The similar expression for $\hat{n}_{yz,i}$ can be written. The parameters fit using experimental data are $t_1=-1.0$, $t_2=1.3$, $t_3=t_4=-0.85$ in electronvolt units \cite{Dagotto2011-pu}. The interaction part of the two-band model is given by 
\begin{align}\label{eq:intdxy2}
H_{int}^{2} &=U\sum_{i}\bigg(\hat{n}_{xz,i,\uparrow}\hat{n}_{xz,i,\downarrow}+\hat{n}_{yz,i,\uparrow}\hat{n}_{yz,i,\downarrow}\bigg)+\bigg( U'-J/2\bigg)\sum_{i}\hat{n}_{xz,i}\hat{n}_{yz,i}\nonumber\\
& -2J\sum_{i}{\bf S}_{xz,i}.{\bf S}_{yz,i}+J\sum_{i}\bigg( \hat{d}_{xz,i,\uparrow}^\dagger \hat{d}_{xz,i,\downarrow}^\dagger \hat{d}_{yz,i,\downarrow}\hat{d}_{yz,i,\uparrow}+\hat{d}_{yz,i,\uparrow}^{\dagger}\hat{d}_{yz,i,\downarrow}^{\dagger} \hat{d}_{xz,i,\downarrow}\hat{d}_{xz,i,\uparrow}\bigg)
\end{align}
Where $U'=U-2J$, $J$ being Hund's coupling. ${\bf S}_{xz,i}$ is the spin of the electron occupying $d_{xz}$ orbital at site i. The two-orbital model in the sum of single particle hopping of Eqn.~\eqref{eq:dxy2} and the interaction part given in Eqn.~\eqref{eq:intdxy2}. The interaction parameters can be set as $J/U=1/8$ and $U$ can take values $0, 2.8, 4$ eV for three different interaction strength runs. We consider half-filling for our two-band model i.e. there is one electron in each $d_{xz}$ and $d_{yz}$ orbitals at individual Fe atom arranged in 2D lattice.

\subsubsection*{List of other candidate instances}
The two orbital model agrees with many aspects of experiments. However, three orbital model is more faithful model for iron-based superconductors. For three orbital model, we need to add single-particle processes corresponding to $d_{xy}$ in addition to single-particle processes given in Eqn. \eqref{eq:dxy2}
\begin{align}\label{eq:dxy31}
    H^{xy} &=t_5\sum_{i,\mu,\sigma} \bigg(\hat{d}_{xy,i,\sigma}^\dagger \hat{d}_{xy,i+\mu,\sigma}+h.c. \bigg)-t_6 \sum_{i,\mu,\nu,\sigma} \bigg( \hat{d}_{xy,i,\sigma}^{\dagger}\hat{d}_{xy,i+\mu+\nu,\sigma}+h.c. \bigg) \nonumber \\
    & +\Delta_{xy}\sum_i \hat{n}_{xy,i}-\mu \sum_i \hat{n}_{xy,i}
\end{align}
Where $\Delta_{xy}$ is the energy difference between $3d_{xy}$ and degenerate $3dxz/3dyz$ orbitals. Adding $d_{xy}$ orbital leads to additional processes between $d_{xy}$ and $d_{xz}/d_{yz}$ orbitals given as under
 \begin{align}\label{eq:dxy32}   
    & H^{xz,yz,xy}=-t_7\sum_{i,\sigma} (-1)^{|i|}\bigg(\hat{d}_{xz,i,\sigma}^{\dagger}\hat{d}_{xy,i+\hat{x},\sigma}+ \hat{d}_{xy,i,\sigma}^{\dagger}\hat{d}_{xz,i+\hat{x},\sigma} +\hat{d}_{yz,i,\sigma}^{\dagger} \hat{d}_{xy,i+\hat{y},\sigma}+\hat{d}_{xy,i,\sigma}^\dagger \hat{d}_{yz,i+\hat{y},\sigma} +H.c.\bigg)\nonumber\\
    &-t_8\sum_{i,\sigma}(-1)^{|i|}\bigg(\hat{d}_{xy,i,\sigma}^{\dagger} \hat{d}_{xy,i+\hat{x}+\hat{y},\sigma} -\hat{d}_{xy,i,\sigma}^{\dagger} \hat{d}_{xz,i+\hat{x}+\hat{y},\sigma}+\hat{d}_{xz,i,\sigma}^\dagger \hat{d}_{xy,i+\hat x -\hat y,\sigma}-\hat{d}_{xy,i,\sigma}^\dagger \hat{d}_{xz,i+\hat x -\hat y,\sigma}+H.c.\bigg)\nonumber\\
    &-t_8\sum_{i,\sigma}(-1)^{|i|} \bigg(\hat{d}_{yz,i,\sigma}^{\dagger}\hat{d}_{xy,i+\hat x+ \hat y}-\hat{d}_{xy,i,\sigma}^{\dagger}\hat{d}_{yz,i+\hat x +\hat y,\sigma}-\hat{d}_{yz,i,\sigma}^{\dagger} \hat{d}_{xy,i+\hat x -\hat y,\sigma}+\hat{d}_{xy,i,\sigma}^{\dagger}\hat{d}_{yz,i+\hat x - \hat y,\sigma}+H.c.\bigg)
\end{align}
\begin{align}\label{eq:intdxy3}
H_{int}^{3}&=U\sum_{i}\bigg(\hat{n}_{xz,i,\uparrow}\hat{n}_{xz,i,\downarrow} +\hat{n}_{yz,i,\uparrow}\hat{n}_{yz,i,\downarrow}+ \hat{n}_{xy,i,\uparrow}\hat{n}_{xy,i,\downarrow}\bigg)\nonumber\\
& +\bigg( U'-J/2\bigg)\sum_{i}\bigg(\hat{n}_{xz,i}\hat{n}_{yz,i}+\hat{n}_{xz,i}\hat{n}_{xy,i}+\hat{n}_{yz,i}\hat{n}_{xy,i}\bigg)\nonumber\\
& +J\sum_{i}\bigg(\hat{d}_{xz,i,\uparrow}^\dagger \hat{d}_{xz,i,\downarrow}^\dagger \hat{d}_{yz,i,\downarrow}\hat{d}_{yz,i,\uparrow}+\hat{d}_{xz,i,\uparrow}^\dagger \hat{d}_{xz,i,\downarrow}^\dagger \hat{d}_{xy,i,\downarrow}\hat{d}_{xy,i,\uparrow}\nonumber+\hat{d}_{yz,i,\uparrow}^\dagger \hat{d}_{yz,i,\downarrow}^\dagger \hat{d}_{xy,i,\downarrow}\hat{d}_{xy,i,\uparrow}+H.c.\bigg)\\
& -2J\sum_{i}\bigg({\bf S}_{xz,i}.{\bf S}_{yz,i}+{\bf S}_{xz,i}.{\bf S}_{xy,i}+{\bf S}_{yz,i}.{\bf S}_{xy,i}\bigg)
\end{align}
A three orbital generalization of the two orbital model is presented when Eqn.~\eqref{eq:dxy2}, ~\eqref{eq:dxy31}, ~\eqref{eq:dxy32} and ~\eqref{eq:intdxy3} are added together ~\cite{Dagotto2011-pu}. With $t_1=0.02$, $t_2=0.06$, $t_3=0.03$, $t_4=-0.01$, $t_5=0.2$, $t_6=0.3$, $t_7=-0.2$, $t_8=0.1$ and $\Delta_{xy}=0.4$, with all the parameters in eV. In this case, the orbitals $d_{xz},d_{yz}$ and $d_{xy}$ are involved in the hopping of the electrons. For three-orbital model, the ratio $J/U$ is taken between $0-0.35$ and $U$ can be sweeped between $0-2.5$ electronvolt. The filling can be set such that there are 4 electron in 3 orbitals at each Fe atom.

In the case with five iron orbitals, $3d_{xz}$, $3d_{yz}$, $3d_{x^2-y^2}$, $3d_{xy}$, and $3d_{3z^2-r^2}$ a Hamiltonian of the type $t-J_1-J_2$ has been studied in \cite{Yu2013-re}, and obtained via an expansion of the five-orbital Fermi-Hubbard model with respect to the Mott transition point \cite{Goswami2010-vh}.

\section{Requirements summary}
This section summarizes the application requirements for all the applications described in this chapter.
\subsection*{Application 1: Zero-Temperature Superconductivity}
\begin{small}
\begin{tabular}{ |l l l| }
    \hline
        & & \\
     & Total time limit & 1 month \\ [1ex]
    \textbf{Workload:} & Number of subroutine calls required & $40 \times 10$ \\ [1ex]
     & Maximum subroutine time limit & NA \\ [1ex]
     & & \\[1ex]
     & Model type & Fermionic Hamiltonian \\[1ex]
     & Size & minimum $20 \times 20$ lattice sites $\times$ $1$ orbital\\
     &&target $100 \times 100$ lattice sites $\times$ $3$ orbitals\\[1ex]
     \textbf{Problem specifications:} &  Interaction Structure & Sparse regular (2-D NNN square latice) \\[1ex]
     & Computational target &Superconducting order parameters\\[1ex]
     & Accuracy requirement & minimum $10^{-3}$, target $10^{-5}$\\[1ex]
     & & \\
     \hline
\end{tabular}
\end{small}
\subsection*{Application 2: High-temperature Superconductivity}

\begin{small}
\begin{tabular}{ |l l l| }
    \hline
        & & \\
     & Total time limit & 1.5 month \\ [1ex]
    \textbf{Workload:} & Number of subroutine calls required & $40 \times 10 \times 20$  \\ [1ex]
     & Maximum subroutine time limit & NA \\ [1ex]
     & & \\[1ex]
       & Model type & Fermionic Hamiltonian \\[1ex]
     & Size & minimum $20 \times 20$ lattice sites $\times$ $1$ orbital\\
     &&target $100 \times 100$ lattice sites $\times$ $3$ orbitals\\[1ex]
     \textbf{Problem specifications:} &  Interaction Structure & Sparse regular (2-D NNN square lattice) \\[1ex]
     & Computational target &Superconducting order parameters\\[1ex]
     & Accuracy requirement  & minimum $10^{-3}$, target $10^{-5}$\\[1ex]
     & & \\
     \hline
\end{tabular}
\end{small}
\subsection*{Application 3: Phase-Diagram of Fermi-Hubbard Models}
\begin{small}
\begin{tabular}{ |l l l| }
    \hline
        & & \\
     & Total time limit & 2 month \\ [1ex]
    \textbf{Workload:} & Number of subroutine calls required & $40 \times 10 \times 20$  \\ [1ex]
     & Maximum subroutine time limit & NA \\ [1ex]
     & & \\[1ex]
       & Model type & Fermionic Hamiltonian \\[1ex]
     & Size & minimum $20 \times 20$ lattice sites $\times$ $1$ orbital\\
     &&target $100 \times 100$ lattice sites $\times$ $3$ orbitals\\[1ex]
     \textbf{Problem specifications:} &  Interaction Structure & Sparse regular (2-D NNN square lattice) \\[1ex]
     & Computational target &Superconducting order parameters\\[1ex]
     &                      &Specific heat scaling coefficient\\[1ex]
     &                      &Mean staggered magnetization\\[1ex]
     & Accuracy requirement & minimum $10^{-3}$, target $10^{-5}$\\[1ex]
     & & \\
     \hline
\end{tabular}
\end{small}
\newpage 
\subsection{Supplementary material for Fermi-Hubbard models} \label{sec:FH_suppmat}
\subsubsection{More on the underlying Fermi-Hubbard}\label{sec:fermihubbardder}
Let us mention some details on the derivation of the Fermi-Hubbard model of the main text. First of all, let us stress that the shortcomings of the tight-binding Hamiltonian is that this is only an approximation to the full problem, as this section shows. At the end of several of these approximations, one is \textit{usually} left with a final many-body problem of the form ignoring terms like L-S coupling etc for low atomic number systems.
\begin{eqnarray}
    \Big( -\frac{\hbar^2}{2m_e} \nabla_i^2-\sum_{i,a} \frac{Z_a e^2}{|\vec r_i-\vec R_a|}+\frac{1}{2}\sum_{ij} \frac{e^2}{|\vec r_i-\vec r_j|}\Big)\Psi_\alpha(\{\vec r_i\};\{\vec R_a\})=E_\alpha \Psi_\alpha(\{\vec r_i\};\{\vec R_a\})
\end{eqnarray}
where the coordinate derivative $-i\hbar \vec \nabla_{r_i}$ is the momentum of i-th electron, and $\Psi_\alpha(\{\vec r_i\};\{\vec R_a\})$ is wave-function for the many-electron problem and is antisymmetric when two electron are interexchnaged. A common approach to solve the above mentioned Hamiltonian, is to use methods the Kohn-Sham density functional method, which is generally known to work well for static problems, but not for dynamic ones except in the adiabatic regime (Time Dependent DFT). It is very hard to study the continuous Hamiltonian if not using renormalization group method. This method is typically applied to strongly correlated systems to study the low-energy behavior of the Hamiltonian, directly at the level of the second quantized Hamiltonian of Eqn. (\ref{eq:continuum2}). Analytical techniques encompass both perturbative and functional techniques and require a parametrization of the Fermi surface \cite{Metzner2012-gl} and nonetheless the integration of differential equations for the flow \cite{Bulla2008-ns,Kugler2020-uc}.

\noindent To derive the Fermi-Hubbard model, one has to write the wavefunction in a second quantization scheme:
\begin{equation}
    \mathcal H=\int dx\ \hat \psi^\dagger({\mathbf x}) \sum_i h_i({\mathbf x}) \hat \psi({\mathbf x})+\frac{1}{2}\int d{\mathbf x}d{\mathbf x}'\ \hat \psi^\dagger({\mathbf x})\hat \psi^\dagger({\mathbf x}') v({\mathbf x},{\mathbf x}') \hat \psi({\mathbf x}')\hat \psi({\mathbf x})\label{eq:continuum2}
\end{equation}
where ${\mathbf x}=({\vec r},s)$ representing the coordinate and spin of the electron
\begin{eqnarray}
    h_i({\vec r_i})&=&-\frac{\hbar^2}{2m_e} \nabla_i^2-\sum_a \frac{Z_a e^2}{|\vec r_i-\vec R_a|}\\
    v({\vec r}_i,{\vec r}_j)&=&\sum_{ij} \frac{e^2}{|{\vec r}_i-{\vec r}_j|}
\end{eqnarray}
\noindent For a solid, the wavefunctions are written in terms of localized orbitals, 
\begin{eqnarray}
    \hat{\psi}_\sigma^\dagger=\sum_{a} \phi(\vec r-\vec R_a) \hat{c}^{\dagger}_{a\sigma}
\end{eqnarray}
Where $\hat c_{\vec k,\sigma}$, $\hat c_{\vec k,\sigma}^\dagger$ are the electron destruction and creation operators. $\phi(\vec r -\vec R_a)$ are the localized orbitals for the atom at site $a$. Inserting this expression into the second quantized Hamiltonian, we get the Fermi Hubbard Hamiltonian of the main text, Eqn. (\ref{eq:fermihubb}) with quantities as
\begin{eqnarray}
    t_{i,j}=-\int d{\vec r} \phi(\vec r-\vec R_i)h({\vec r})\phi(\vec r - {\vec R}_j)\\
    U=e^2\int d\vec r d \vec r' \frac{|\phi(\vec r-\vec R_i)|^2 | \phi(\vec r'-\vec R_i)|^2}{|\vec r-\vec r'|}
\end{eqnarray}

\noindent We can also repeat the above mentioned calculations for electron gas (jellium model), by defining the field operator as
\begin{eqnarray}
    \psi_{\sigma}^{\dagger}(\vec r)=\frac{1}{\sqrt{L}} \sum_{\vec k} e^{i \vec k\cdot \vec r} \hat c_{\vec k,\sigma}
\end{eqnarray}
\subsubsection{Methods to derive the parameters of the Fermi-Hubbard models}\label{sec:methodsder}
In the main text we have mentioned fitting the Fermi surface via ARPES as a method to estimate the parameters. Another way of deriving the Fermi-Hubbard parameters is by performing the integrals knowing the lattice structure. There are various methods to do this. We briefly mention the Wannier functions and the Slater-Koster methods.

Wannier functions play an important role in the derivation of the hopping parameters for the Fermi-Hubbard model \cite{Marzari2012-lt}. The Bloch theorem in solid state physics establishes that in a periodic potential, the wave function describing an electron can be written as a product of a periodic function and a plane wave \cite{Kittel2004,Ashcroft76}.  Their computational complexity can vary depending on the size of the basis set and the system under consideration. Often linear or slightly superlinear scaling with the number of basis functions or atoms (N). Wannier functions are derived from Bloch states through a transformation, are localized electron wave functions (over one or more lattice sites to mimic hybridization).

The Wannier functions are used  of an isolated band is defined as
\begin{equation}
    w_n(\vec r-\vec R)=\frac{V}{(2\pi)^d} \int _{BZ} \int d^dk\ e^{-i \vec k\cdot \vec R}\psi_{n,k}(\vec r)
\end{equation}
where $\psi_{n,k}(\vec r)$ are the localized Bloch functions, and BZ stands for Brillouin zone. Although it is not immediately obvious, it is important for what follows that Wannier functions are localized. For instance, see Fig. \ref{fig:cup} for the cuprates. The overlap of the orbital $3d$ and $2p$ for copper and oxygen leads to a non-zero hoping $t_{ij}$ between these two atoms.
Let us briefly list the existing numerical methods that allow to construct the Wannier function.
A method related to Wannier wavefunctions described above is the Maximally Localized Wannier Functions, whose aim is to maximize the localization of Wannier functions by minimizing the spread of the functions in real space \cite{Marzari2012-lt}. The computational complexity can vary based on the algorithm used for optimization, based on  the minimisation of the Marzari–Vanderbilt quadratic spread functional, often involving iterative optimization schemes. The computational complexity depends on the grid size and the accuracy required. Typically cubic scaling with the number of grid points or spatial resolution ($N_{G}^3$) \cite{Vitale2020-qn}. In scenarios where the Wannier functions are fewer than the Bloch bands there are entangled bands due to intricate mixing of electronic states \cite{Souza2001-ki}; in these case, some methods that work specifically for this case.

The Projector Augmented Wave (PAW) Method
is an approach used in ab initio methods like density functional theory (DFT) to construct Wannier functions \cite{Blochl1994-fk}. It involves the projection of Bloch functions onto atomic-like functions, which helps in obtaining localized Wannier functions. From a complexity standpoint, the scaling is often quadratic or slightly higher than linear, with respect to the number of atoms or basis functions.


There exist several software packages  used to perform the calculations related to Wannier functions, facilitating the construction and analysis of these fundamental electronic states in solid-state materials. Wannier90 \cite{wannier90} is still the primary software designed explicitly for computing Wannier functions. Its functionality extends to interfacing with various electronic structure codes like Quantum ESPRESSO \cite{Giannozzi2009-ea}, VASP (Vienna Ab initio Simulation Package) \cite{Hafner2008-nd}, ABINIT \cite{Romero2020-dl}, CRYSTAL \cite{Bush2011-nn}, SIESTA \cite{Soler2002-qx}, and OpenMolcas \cite{Aquilante2020-cg}.  Among these first-principles codes, Quantum ESPRESSO is widely used and is known for its capabilities in electronic structure calculations.

The Slater-Koster approximation (or tight-binding), is a method used in solid-state physics to describe the electronic structure and properties of crystalline materials \cite{Papaconstantopoulos2003-ly}. It is a simplified yet effective way to calculate the electronic band structure and electronic properties of a crystal based on the interactions between neighboring atoms. In the Slater-Koster approximation, the electronic wavefunctions are expressed as linear combinations of atomic orbitals (LCAO) localized on individual atoms. To overcome the computational constraints, they proposed treating these integrals as adjustable parameters, to be determined based on results obtained from other, more efficient calculations. Similarly to the Wannier function method described above, the Slater-Koster approximation assumes that the interactions between atoms are predominantly due to the overlap of atomic orbitals and neglects other effects such as long-range electrostatic interactions. It provides a computationally efficient approach to calculate the electronic band structure and other electronic properties of materials.  The method is used in one of the applications described below as a method to derive the hopping terms of the Hamiltonian of a Fermi-Hubbard model.
The general form of the Hamiltonian matrix elements in the Slater-Koster approximation can be written as:
\begin{equation} t_{ij}(\mathbf{k}) = \sum_{\mathbf{R}} e^{i\mathbf{k} \cdot \mathbf{R}} \sum_{\mu,\nu} \left[ V_{\mu\nu}(\mathbf{R}) S_{\mu\nu}(\mathbf{R}) + W_{\mu\nu}(\mathbf{R}) \right] \end{equation}
In the equation above, $t_{ij}(\mathbf{k})$ represents the matrix element between orbitals of the sites $i$ and $j$, at wavevector $\mathbf{k}$. The sum over $\mathbf{R}$ runs over the lattice vectors of the crystal, and $e^{i\mathbf{k} \cdot \mathbf{R}}$ accounts for the phase factors associated with the periodicity of the crystal lattice. The terms $V_{\mu\nu}(\mathbf{R})$ and $W_{\mu\nu}(\mathbf{R})$ represent the overlap and exchange integrals, respectively. They depend on the specific form of the atomic orbitals $\mu$ and $\nu$ at atomic sites separated by the lattice vector $\mathbf{R}$. The overlap integral $S_{\mu\nu}(\mathbf{R})$ quantifies the overlap between the orbitals.

The empirical parameters $V_{\mu\nu}(\mathbf{R})$ and $W_{\mu\nu}(\mathbf{R})$ in the Slater-Koster approximation are obtained through fitting to experimental data or other theoretical calculations. These parameters capture the essence of the orbital interactions and depend on the specific elements and crystal structure under consideration.

\subsubsection{Numerical methods used to study the Fermi-Hubbard model} \label{sec:FH_classical_methods}
The study of Fermi-Hubbard Hamiltonian is, as often for the case of quantum many-body systems, plagued by the exponential growth of the Hilbert space with the size of the system. While mean-field techniques mentioned below can be employed, they do not account for strong correlations between electrons. Analytical methods are available in certain parameter regimes (see below). Analytical solutions like Bethe anstaz can only be used in 1D for a single-band systems \cite{Lieb1968-vn,Essler1991-ct,Franchini2017-jo,Gaudin2014-xz}, where physics like the Mott transition is absent. It is usually necessary to use computational methods to study the behavior of the model \cite{Bonitz2006-jn}. 

\begin{itemize}
    \item \textbf{Exact Diagonalization:} This method involves numerically diagonalizing the Hamiltonian matrix for small system sizes. For a system with L sites (or orbitals) and Hilbert-space size of $N_H$, the computational effort scales as $O(N_H^3)$ for a dense-matrix of dimentions $N_H\times N_H$. Moreover, as the number of sites L is increased, $N_H$ increases exponentially and performing exact diagonaliztion is not feasible due to limited memory and computation resources. To make the diagonalization more efficient, other techniques like Lancoz algorithm is used. The Lanczos algorithm starts with an initial vector $\mathbf{v}_1$, which is chosen randomly or with some prior knowledge about the system being studied. The Lanczos algorithm starts with an initial vector $\mathbf{v}_1$, which is chosen randomly or with some prior knowledge about the system being studied. The first Krylov vector is then defined as $\mathbf{w}_1 = \mathbf{H}\mathbf{v}_1$, where $\mathbf{H}$ is the matrix to be diagonalized. The Lanczos process then proceeds recursively as follows:
a) Normalize the Krylov vector: $\mathbf{v}_{k+1} = \frac{\mathbf{w}_k}{|\mathbf{w}_k|}$. b) Apply the matrix to the Krylov vector: $\mathbf{w}_{k+1} = \mathbf{H}\mathbf{v}_{k+1}$. c) Orthogonalize the new Krylov vector with respect to the previous ones: $\mathbf{w}_{k+1} = \mathbf{w}_{k+1} - \sum_{i=1}^k \langle \mathbf{w}_{k+1} | \mathbf{v}_i \rangle \mathbf{v}_i$. At each iteration, the resulting Krylov vectors form an orthonormal basis set of the Krylov subspace, which is a small subspace of the full vector space spanned by the original matrix. The Lanczos process can be truncated after a certain number of iterations or when the Krylov vectors stop improving the approximation to the eigenvalues. Once the Krylov subspace has been constructed, the matrix becomes tridiagonalized in this subspace and can be easily diagonalized.

\item \textbf{Monte Carlo methods}. There are many Quantum Monte Carlo  (QMC) methods, of which a few we list below. From a computational perspective, the issue with QMC is the so called fermionic sign problem which makes the convergence of the method very slow. \textit{Variational Monte Quantum  Carlo}. The Variational Quantum Monte Carlo (VQMC) algorithm involves the use of a trial wave function (WF) and the estimation of the expectation value of the energy of a quantum many body system. The algorithm involves the use of a trial WF $ |\Psi_T(\boldsymbol{\alpha})\rangle$, the evaluation of the expectation Value of Energy via $\langle E \rangle = \frac{\langle \Psi_T | H | \Psi_T \rangle}{\langle \Psi_T | \Psi_T \rangle}$ and then Monte Carlo sampling $\langle O \rangle \approx \frac{1}{N} \sum_{i=1}^{N} O(x_i)$. The key idea behind VQMC is to optimize the parameters  of the trial wave function to minimize 
$\langle E\rangle $ and consequently obtain an approximation to the ground state energy of the quantum system. This method can be used, for instance, for ground state preparation. The scaling behavior can range from linear to polynomial, depending on the complexity and quality of the trial wave function and the system being studied.  \textit{Quantum Monte Carlo (QMC) Methods}.
For fermionic models, there are specific methods that take advantage of constraints due to properties of fermions in the partition function. We mention in particular the work \cite{Zhang1997-ya} which is commonly used for strongly correlated fermions. 
\textit{Continuous-time Quantum Monte Carlo (CTQMC)}: This is an efficient method for studying equilibrium properties at finite temperatures \cite{Rubtsov2005-bb} when the sign problem is absent, which is true for particular choices of parameters.
\textit{Auxiliary Field Quantum Monte Carlo (AFQMC)} is also particularly useful for ground state properties of the Fermi-Hubbard model.
AFQMC is used to simulate systems with fermionic degrees of freedom by introducing auxiliary fields to represent the interaction terms (via the so called Hubbard-Stratonovich transformation). It involves decomposing the interactions into auxiliary fields and transforming the fermionic problem into a simpler bosonic problem \cite{Blankenbecler1981-xh}.
 \textit{Determinantal Quantum Monte Carlo} (DMC) is a quantum Monte Carlo method specifically designed for systems of interacting fermions \cite{Chakraborty2007-oq}. DMC represents the many-body wave function in terms of Slater determinants, which describe the configuration of occupied single-particle orbitals for fermions. These determinants capture the antisymmetric nature of fermionic wave functions.  It involves an imaginary-time propagation of the system starting from a trial wave function. The system's ground state is approached by simulating the time evolution of the wave function in imaginary time using the Schrödinger equation. DMC employs importance sampling, another Monte Carlo technique, to sample configurations of particles according to their contributions to the total energy, focusing on more relevant configurations.

\item \textbf{Analytical techniques and Mean-Field Theories}. Mean-field theory has been very useful in analyzing the behavior of the Fermi-Hubbard model. The \textit{ Hartree-Fock Approximation} (HF) simplifies the many-body problem by assuming a mean-field description of interactions of the form $\hat n_{i} \hat n_{j}$ which are composed of four ladder operators, while keeping the rotational spin symmetry preserved turning it effectively into a simpler (quadratic) model. Expressing interactions as a mean-field enables one to write many-body wavefuntion in the terms of determinant (usually called Slater determinant) of sigle-particle wave functions. HF is also often utilized as a starting point to describe the electronic band structure \cite{noauthor_1999-ra}. This method, is often used to infer the location of the transitions between paramagnetic, antiferromagnetic and ferromagnetic phases \cite{Penn1966-wb}. This method can be extended to study also dynamical mean-fields to study the Mott transition \cite{Vollhardt2017-el,Pavarini2011-fv}. Also, another method is the Cellular dynamical mean-field theory (CDMFT) \cite{Kotliar2001-uz} which is obtained by including in the cluster Hamiltonian an external bath of uncorrelated electrons that mimics the rest of the lattice. Regardless of mean-field nature of the approached a lot of additional computer calculations are necessary e.g. to solve self-consistent mean-field equations etc. The \textit{Gutzwiller Approximation} introduces a variational wavefunction by projecting from a product state. The general idea behind the Gutzwiller and derivative methods is that electronic correlations impose constraints on reachable Hilbert space. this is particularly true in the Fermi-Hubbard model at large values of $U$, with the double occupations on the lattice sites suppressed. This method can be used to study the Fermi-liquid regime, and there are methods to extend it at finite U \cite{Kotliar1986-ky}. \textit{Peculiar limits}. The behavior of the Fermi-Hubbard model can analyzed analytical in certain limits, for instance when $U=0$ we have a Fermi gas, and for $t/U$ small \cite{Gebhard1997-td}.

\item \textbf{Density Matrix Renormalization Group (DMRG)} Originally developed for 1D systems for capturing the ground state properties, low-energy excitations, and correlation functions of one-dimensional strongly correlated quantum systems. DMRG is quite usefull for studying ground states and low-energy excitations of the Fermi-Hubbard model in 1D \cite{White1992-ir}. In DMRG, tensor networks play a crucial role in representing quantum states of many-body systems. The tensor network representation in DMRG, often in the form of Matrix Product States (MPS), provides an efficient way to encode and manipulate quantum states, particularly in one-dimensional systems. Leveraging on the Matrix Product States (MPS) representation, DMRG can encode intricate entanglement structures and correlations between particles within these 1D arrangements while keeping Hilbert space small. The model is hardly usable beyond 2D, but in two-dimensions it can be used in a variety of contexts \cite{White1998-mq}. DMRG's computational cost typically scales polynomially or sub-exponentially with the system size for one-dimensional systems. Specifically, it often exhibits a scaling of $O(m^3 \cdot L)$ or $O(m^3 \cdot L^2)$, where $m$ is the number of states kept in the matrix product state (MPS) approximation and $L$ is the number of lattice sites. Another way of seeing this is that it scales linearly with the number of sites and in the number of tensorial legs. Thus, the method is efficient in 1D (2 legs) and in 2D the method can be reduced, using an approximation, to a 1D model (2 legs), while in principle one should use a 4 legged tensor representation. 
\end{itemize}

As an example of how some of these methods are employed, let us consider the Fermi liquid theory. Parameters like the quasiparticle weight $Z$ or the effective mass of the particles near the Fermi surface $m^*$ serve as indicators of this phase, which follows Landau's theory of Fermi liquids. In numerical experiments and analytical methods, there are various methods that can be used to estimate these order parameters. For instance, in Dynamical Mean-Field Theory (DMFT) and in the Numerical Renormalization Group (NRG), the self-energy is evaluated directly $\Sigma(\mathbf{k}, \omega)$, which is the self-energy, describing the energy shift and damping of a quasiparticle excitation at momentum $\mathbf{k}$ and frequency $\omega$.
Quantum Monte Carlo (QMC) methods directly compute spectral functions, revealing information on the residue at the Fermi level, and therefore $Z$ can be evaluated. 
In Density Matrix Renormalization Group (DMRG) calculations, especially in finite-size systems, provide spectral information near the Fermi surface, giving indications of $Z$ and the effective mass renormalization. For the case of Exact Diagonalization (ED) techniques, applicable to smaller systems, the green function can be calculated via matrix inversion. 
\subsection{D-Wave superconductivity}
D-Wave superconductivity is a modulation of the order parameter $\langle \Delta_{ij}\rangle$. To express the order parameter $\Delta_{ij}$ in terms of momentum space variables $k$, we can first Fourier transform the lattice site indices $i$ and $j$. We assume a simple square lattice for clarity. In the case of square lattice, we have $i = (i_x, i_y)$ and $j = (j_x, j_y)$. The lattice constant is denoted by $a$. Then, the Fourier transform of the order parameter $\Delta_{ij}$ is given by:

$$
\langle \Delta(\mathbf{k})\rangle = \sum_{i, j} e^{-i\mathbf{k}\cdot(\mathbf{R}_i - \mathbf{R}_j)} \langle \Delta_{ij}\rangle
$$
where $\mathbf{R}_i$ and $\mathbf{R}_j$ are the position vectors corresponding respectively to lattice sites $i$ and $j$, respectively, and $\mathbf{k}$ is the momentum vector.

In momentum space, the lattice vectors are given by

$$
\mathbf{R}_i = i_x \mathbf{a}_x + i_y \mathbf{a}_y
$$

where $\mathbf{a}_x$ and $\mathbf{a}_y$ are the primitive lattice vectors.

Now we can substitute $i$ and $j$ into the Fourier transform expression:

$$\Delta(\mathbf{k}) = \sum_{i_x, i_y, j_x, j_y} e^{-i\mathbf{k}\cdot((i_x - j_x)\mathbf{a}_x + (i_y - j_y)\mathbf{a}_y)} \Delta_{i_x, i_y, j_x, j_y}$$

We can express $i$ and $j$ in terms of momentum space variables. The momentum components $k_x$ and $k_y$ are related to the lattice vectors by $k_x = \frac{2\pi}{a} n_x$ and $k_y = \frac{2\pi}{a} n_y$, where $n_x$ and $n_y$ are integers. Therefore, we can rewrite the sum as integrals over the Brillouin zone:

$$
\Delta(\mathbf{k}) = \sum_{n_x, n_y} e^{-i\mathbf{k}\cdot((n_x - n_x')\mathbf{a}_x + (n_y - n_y')\mathbf{a}_y)} \Delta_{n_x, n_y, n_x', n_y'}
$$

Now, to get the expression for $\Delta(\mathbf{k})$, we need to evaluate the sum over $n_x$ and $n_y$, which will involve the lattice geometry and the specific form of $\Delta_{ij}$. The result will be a function of $\mathbf{k}$, giving us $\Delta(\mathbf{k})$, the Fourier transform of the order parameter.

D-Wave superconductivity is observed if $\langle \Delta(\mathbf{k})\rangle $ is modulated, e.g.
\begin{eqnarray}
    \langle \Delta(\mathbf{k})\rangle=\Delta_0 (\cos(k_x a)-\sin(k_y a)),
\end{eqnarray}
where $a$ is the lattice spacing, for a certain value $\Delta_0$ which sets the energy scale.

\subsection{Statics vs Dynamics in superconductivity}
\label{eq:kineticdyn}

It is important to stress that conductivity is the dynamical response of a system and not a static quantity. The phase diagram of the Fermi-Hubbard model provides indications of the physical states in equilibrium, via the correlations of the material. However, many quantities of interest in the context of conductors are dynamical quantities, such as the conductivity (which is usually a tensor), which we discuss briefly below in the context of FH for simplicity. In the Fermi-Hubbard model, the current operator (sometimes called local-current density) $J_{\alpha}$ along the $\alpha$-direction on a lattice can be expressed in terms of creation and annihilation operators for electrons hopping between neighboring lattice sites after series of steps that model the effect of electrical field. In general, an electromagnetic field is introduced to the Hamiltonian in Eqn.~\eqref{eq:fh} via a time-dependent vector potential $\mathbf{A}(\mathbf{r},t)$. In the Coulumb gauge, which dictates $\mathbf{\nabla}.\mathbf{A}=0$, the electric-field $\mathbf{E}$ is given by $-\partial\mathbf{A}(\mathbf{r},t)/\partial t=\mathbf{E}(\mathbf{r},t)$. For our 2D system in Eqn.~\eqref{eq:fh}, the electric-field at site $j$ is simply given by $-\partial\mathbf{A}(\mathbf{r}_j,t)/\partial t=\mathbf{E}(\mathbf{r}_j,t)$. The addition of the vector potential change the hopping amplitudes $t_{i,j}\to \tilde{t}_{i,j}$ where $\tilde{t}_{i,j}$ are the new amplitudes between sites $i$ and $j$. To obtain the specific form of $\tilde{t}_{i,j}$, some simplifications are required. The current operator depends upon the direction of vector potential and can be obtained via
\begin{eqnarray}
J_{\alpha}(\mathbf{r}_j)= -\frac{\delta H}{\delta A_{\alpha}(\mathbf{r}_j)}\implies J_{\alpha}(\mathbf{r})\propto(\tilde{t}_{i,j}\hat{c}_{i,\sigma}^\dagger \hat{c}_{j,\sigma}-H.c.\big)
\end{eqnarray}

The operator above measures the flow of charge in the FH model. Above $\tilde{t}_{ij}$ are the hopping amplitude between sites $i$ and $j$ under the influence of the field. The first thing to note is that the current is proportional to the (derivative of) kinetic term of the FH model, since the introduction of the vector potential only influences the single-particle processes. The conductivity can be written in terms of the current operator via linear response \cite{Bruus2004-cd}.
 The Kubo formula, derived from perturbation theory, provides a framework for understanding electrical conductivity in materials operationally. It relates the dynamics of electrons under the influence of an external electric field.  The Kubo formula expresses the electrical-conductivity tensor $\sigma_{\alpha\beta}$ (conductivity along the direction $\alpha$ and $\beta$) as:
 
$$\sigma_{\alpha\beta}(\mathbf{r}) = \lim_{\omega \to 0} \frac{i}{\omega} \left( \frac{e^2}{L} \right) \int_{0}^{\infty} dt \, e^{i \omega t} \langle [J_{\alpha}(t) ,J_{\beta}(0)] \rangle$$

We see that the expression above is the Fourier transform of the quantity
$ \Pi_{\alpha\beta}(t)=\langle [J_{\alpha}(t), J_{\beta}(0)]\rangle$
Above, $e$ is the elementary charge, $L$ is the number of sites in the system and $\langle [J_{\alpha}(t) ,J_{\beta}(0) ]\rangle$ is the current-current correlation function. We see that conductivity in principle requires the knowledge of the time-evolved current operator, $U^\dagger(t) J_\alpha(0) U(t)$. The limit $\omega\rightarrow 0$ implies that one focuses on long times (or rather equilibrium), but are derived from a dynamical quantity.
By substituting the expression for the current operator into the Kubo formula and evaluating the relevant correlation functions, one can in principle calculate the conductivity of the Fermi-Hubbard model system. For Fermi-Hubbard, the expression above provides ultimate insights into  the interplay between electron hopping, on-site interactions. However, since quantity depends on the expectation values, the state of the system is a key indicator of how the system is behaving. 

\section{Quantum implementation}
A quantum implementation and resource estimation for the zero-temperature superconductivity for the single band Fermi-Hubbard model from Eqn.~\eqref{eq:single_band_FH_model} and the two and three bands models from Eqs.~\eqref{eq:dxy2}, ~\eqref{eq:dxy31} and ~\eqref{eq:dxy32} can be found in \cite{qc-applications-notebooks}.

\newpage
\end{chapter}

\begin{chapter}{Computational catalysis for artificial photosynthesis} \label{ch:q_chem}

This chapter discusses the identification of optimal catalysts in artificial photosynthesis, focusing on enhancing the efficiency of water oxidation and carbon dioxide reduction reactions. The preliminary phase of catalyst selection requires determining the reaction pathway and leveraging computational capabilities to estimate the ground state energies, thereby calculating the reaction barrier. Present-day classical computers fall short of accurately solving the Hamiltonians that encompass complex static and dynamic electronic correlations within molecular systems with trade-offs between accuracy and numerical efficiency. In contrast, a quantum computer's distinct advantage lies in its ability to handle huge Hilbert space, potentially enabling precise estimations of ground state energy and beyond efficiently without sacrificing accuracy. Subsequently, quantum molecular dynamics are employed on a subset of catalysts to understand the reaction's dynamics and further optimize the catalysts. This includes propagating classical Newton's equations for nuclei with ground state energies and the forces (the gradient of energy with respect to nuclear coordinates) calculated ``on-the-fly" using a quantum computer. Moreover, the nuclear quantum effects in promising catalysts are examined through nuclear quantum dynamics simulations, where nuclear wavepackets are evolved using the time-dependent Schro\"odinger equation.\\

\noindent
\textbf{Hamiltonian Type}:  Fermionic Hamiltonian; Nuclear Hamiltonian.\\
\textbf{Quantum Computational Kernel}: Ground State Preparation, Molecular Dynamics, and Hamiltonian Simulation.

\clearpage

\section{Application area overview: computational catalysis in quantum chemistry}

\subsection{High-level description of the application area}

In this application area, we discuss an intriguing field where quantum chemistry and computational catalysis combine to address ever-growing energy challenges. Quantum chemistry is essential for understanding how natural processes, such as photosynthesis, function in plants ~\cite{hall1999photosynthesis}, where sunlight is converted into energy. An in-depth understanding of photosynthesis requires a detailed examination of underlying chemical reactions, exploring how atoms and molecules interact at atomic and subatomic levels (electrons and protons). The gained insight is not only pivotal for photosynthesis but has broader applications in chemistry~\cite{ap1}. By applying these principles, we can revolutionize our approach to energy, climate, and material challenges in a sustainable and efficient manner~\cite{atwater2023artificial}. 

Our aim to unravel the intricate mechanisms of photosynthesis directs us to computational quantum chemistry that predicts the electronic structures, response properties, chemical bonding, and nuclear dynamics by solving the Schr\"odinger equations from the first principles. It facilitates a fundamental understanding of chemical reactions, explaining reaction mechanisms, rates, and environmental influences without conducting costly and time-consuming experiments. Crucially, it helps predict reactions and identify superior catalysts~\cite{Chen:2021cr}, which are vital for enhancing processes like photosynthesis.

In this chapter, we specifically focus on how quantum chemistry can address the ever-growing energy demand and climate change challenges by understanding and developing the \emph{artificial photosynthesis} process~\cite{yang2023, atwater2023artificial}. Artificial photosynthesis aims to replicate the natural process of converting sunlight, water, and carbon dioxide into energy-rich fuels, revolutionizing how we generate and use energy~\cite{reyes2022molecular}. By harnessing solar energy to produce clean fuels, artificial photosynthesis represents a transformative approach to generating energy, significantly reducing greenhouse gas emissions and advancing towards a carbon-neutral energy cycle~\cite{atwater2023artificial, Kim:2015uj}. 

However, the efficiency of artificial photosynthesis systems needs improvement, primarily due to challenges in controlling the rate of involved chemical reactions due to the increased energy barrier for converting reactants to products. In this context, catalysts play a crucial role in accelerating reactions by reducing the energy barrier without being consumed in the process~\cite{costentin2013catalysis}. These catalysts are essential for facilitating and directing the specific reactions needed to efficiently convert solar energy into chemical energy, which can be easily stored and transported. The effectiveness of the process relies on the ability to design highly efficient but selective and stable catalysts. For instance, water (\ce{H2O}) oxidation and carbon dioxide (\ce{CO2}) reduction reactions are essential and often limiting steps in artificial photosynthesis~\cite{ap1,ap2}. The ideal catalyst must selectively steer the reaction towards desirable products rather than other less valuable by-products.

To improve the efficiency of artificial photosynthesis, the fundamental focus is on designing catalysts for both reactions, i.e., water oxidation~\cite{Ye:2019wj, Karkas:2014wv} and \ce{CO2} reduction~\cite{Fang:2023tf, Li:2020wr, C3CS60405E}. The scientific questions we ask here are complex and critical: How can we screen the highly selective and efficient catalysts that drive the necessary reactions? Can we accurately estimate the energies of involved molecules to identify favorable reaction pathways? How do we ensure the system favors producing desired products over unwanted ones? Answering these questions is vital as they affect artificial photosynthesis's feasibility, efficiency, selectivity, and environmental impact, enabling the transition from theory to a practical, scalable technology.

\subsection{Utility estimation}

\subsubsection{Overview of the value of the application area}

The development of artificial photosynthesis has immense utility value for both the scientific community and humanity. It enables us to better understand molecular interactions and solar-to-chemical energy conversion processes, leading to breakthroughs in catalyst design and the optimization of artificial photosynthesis processes~\cite{ap1,ap2}. These advancements can unlock new ways to efficiently harness and store solar energy, potentially tapping into a multi-billion-dollar renewable energy market. Artificial photosynthesis has even more profound implications for humanity. It could offer a promising solution to climate change and fossil fuel shortage with its role in carbon capture and utilization, which could significantly impact the growing carbon credit market. Developing sustainable and eco-friendly alternatives to traditional energy sources can reduce our dependence on fossil fuels, significantly decrease greenhouse gas emissions, and combat global warming. In addition, the computational workflow used in designing catalysts for photosynthesis can be readily applied to other areas of catalyst development, such as fuel cells~\cite{list6a}, batter chemistry, ammonia catalyst~\cite{mechno2}, automotive catalytic converters, pharmaceuticals, \emph{etc.} (Note: the global Catalyst market size was estimated at about \$39 billion in 2021~\cite{linkedinCatalystMarket}). 
 
Artificial photosynthesis holds great promise for transforming our energy generation and usage approaches. If successful, we're looking at a future where energy could be clean and abundant, easily accessible to everyone. This could be a game-changer in global energy dynamics, significantly reducing energy deprivation and paving the way for impartial energy access worldwide. Such a shift would profoundly impact the health of our planet and improve the quality of life for people everywhere. But pursuing these questions goes beyond just expanding our knowledge. It's about contributing to a more sustainable future filled with hope for humanity. These scientific endeavors stretch far beyond labs and academic efforts, which are crucial to our environmental 

\subsubsection{Concrete utility estimation}

Problems described in this chapter have a computational bottleneck beyond the reach of current computing methods. We expect quantum computers to provide utility for these quantum chemistry problems due to their scalable nature and exponentially larger expression capability. In the absence of those today, we can provide a loose lower bound on the cost of computations with existing approximate/heuristic methods on high-performance computing platforms. For the examples of the concrete applications presented in the following sections ({\it Application 1}), the hybrid DFT, CASSCF, or other classical many-body methods using the CC-PVDZ basis set takes at least 20 hours per state for a feasible size of an active space due to the geometry optimization. The inclusion of other effects (such as solvents, dispersion, periodic boundary, \emph{etc}) will further increase the computational cost. {\it Application 2} focused on the quantum molecular dynamics (QMD) (explained in the subsequent sections below) is dominating the computational volume considered here. QMD with $10^3$ steps and 100 initial conditions for 10 most promising screened catalyst candidates would require $20 \times 10^3 \times 100 \times 10 = 20,000,000$ core-hours. Using a typical cloud computing cost available from commercial vendors of \$0.03 per core-hour, we obtain that the computations under Application 2 for one of the reactions involved in artificial photosynthesis would cost on the order of \$600,000 in terms of the classical compute-time using existing approximate/heuristic methods. {\it Application 3} is to further fine-tune the catalyst design by examining the nuclear quantum effects. On classical computers, nuclear quantum dynamics of a system with 10 nuclear Degrees of Freedom (DOFs) requires $\sim 256^{10}$ grids.

\section{Problem and computational workflows}

\subsection*{Detailed background of the application area: artificial photosynthesis}

In nature, plants, algae, and some bacteria undergo this incredible transformation into energy-rich carbohydrates and oxygen through photosynthesis~\cite{hall1999photosynthesis}. This process is fundamental to life on Earth, serving as the primary energy source for nearly all organisms. An essential step in photosynthesis is the charge separation through the absorption of photons, creating electron-hole pairs. This step is crucial as it generates energy and reduces the energy necessary for driving the reaction to synthesize carbohydrates. Drawing inspiration from this natural blueprint (photosynthesis), artificial photosynthesis aims to mimic and refine the process~\cite{meyer1989chemical,benniston2008artificial,barber2013natural}. It focuses on generating renewable energy and capturing \ce{CO2}, which is vital in addressing global energy and environmental challenges.

Artificial photosynthesis is a multi-step process that begins with the absorption of sunlight, followed by charge separation and the oxidation of water (\ce{H2O}), producing oxygen (O$_2$), protons (H$^+$), and electrons (e$^-$). These electrons and protons are utilized in the subsequent step for the \ce{CO2} reduction to facilitate the production of fuels. This step involves reducing \ce{CO2} into either CO, hydrocarbons like methane (CH$_4$), or other carbohydrates such as glucose (C$_6$H$_{12}$O$_6$)~\cite{wasielewski1992photoinduced}. Water oxidation not only supplies the initial electron and proton necessary to drive the \ce{CO2} reduction but also provides additional ones, facilitating the generation of various products during the reduction reaction. Alternatively, the hydrogen evolution reaction (HER) presents a pathway where the electrons and protons from water oxidation could be used to generate hydrogen gas (H$_2$). However, due to the challenges associated with storing and transporting H$_2$ gas, artificial photosynthesis processes are designed to avoid HER. \textit{Thus, water oxidation and subsequent \ce{CO2} reduction reaction are core of artificial photosynthesis}~\cite{Ye:2019wj, Karkas:2014wv, Fang:2023tf, Li:2020wr, C3CS60405E}. \\

\textbf{Water oxidation reaction:}
\begin{align*}
    2\ce{H2O} \xrightarrow{\text{Catalyst}} \ce{O2} + 4H^+ + 4e^-
\end{align*}

\textbf{Example \ce{CO2} reduction reactions:}
\begin{align*}
 \text{Reduction I:} && \ce{CO2} + 2H^+ + 2e^- \xrightarrow{\text{Catalyst}} \ce{CO} + \ce{H2O}  \\
 \text{Reduction II:} && \ce{CO2} + 6H^+ + 6e^- \xrightarrow{\text{Catalyst}} \ce{CH3OH} + \ce{H2O}  \\
 \text{Reduction III:} && \ce{CO2} + 8H^+ + 8e^- \xrightarrow{\text{Catalyst}} \ce{CH4} + 2\ce{H2O} 
\end{align*}

\noindent All of these resulting chemicals from \ce{CO2} reduction have a higher energetic content and can be used as fuel. These solar fuels circumvent the above-mentioned conventional challenges associated with hydrogen storage and transportation. Instead, \ce{CO2} reduction focuses on creating fuels that can be more easily integrated into the existing energy infrastructure, offering a more practical and immediate application. 

Although artificial photosynthesis offers a promising solution for global energy and environmental needs, its efficiency and scalability face significant challenges. Artificial photosynthesis requires the materials to a) efficiently absorb a broad spectrum of sunlight and b) effectively separate and transfer photo-excited electron holes to active sites for chemical reactions, which are often limited by the materials used. However, from a chemistry perspective, the significant challenges lie in identifying a catalyst that can effectively drive the water oxidation reaction by lowering the reaction barriers and selecting a catalyst tailored for the \ce{CO2} reduction reaction. The catalyst for water oxidation must be highly active and stable over time to increase reaction kinetics, thus producing electrons and protons for \ce{CO2} reduction. Meanwhile, the catalyst for the reduction reaction should exhibit selectivity and efficiency in converting \ce{CO2} into valuable fuels with maximal yield. Therefore, an ideal catalyst needs to address efficiency, selectivity, and durability. Furthermore, these catalysts should be cost-effective and made from abundant materials to ensure the scalability and economic viability of the technology.
\textit{The initial step involves carefully selecting potential catalyst candidates, focusing on those capable of efficiently driving the preferred reaction. This requires the characterization of energetically favorable reaction pathways and a calculation of activation energies. 
This assessment should focus on the catalyst's optimality, potential by-products, selectivity, and stability during the reaction.} 

Experiments aimed at understanding the reaction mechanisms of artificial photosynthesis are often time-consuming and expensive~\cite{yang2023landscape,atwater2023artificial}. Moreover, capturing pathways opted during the reaction through in-situ~\cite{lyu2023time} methods is challenging, given the timescale over which these reactions occur. Therefore, computational modeling offers a more comprehensive approach to gain in-depth insight into the reaction mechanisms. By using computational methods, such as ground state energy estimation, QMD, or advanced nuclear quantum dynamics methods, we can unravel the energetically favorable reaction pathways to select promising catalysts for water oxidation and \ce{CO2} reduction. Additionally, computational modeling allows exploring a vast range of catalyst and reaction conditions, which might be impractical or unfeasible to test experimentally, leading to the discovery of efficient, sustainable, cost-effective catalysts with earth-abundant elements. Hence, this accelerates the development of energy solutions, bringing us closer to the practical implementation of sustainable energy production.

\subsection*{Challenges of simulating chemical reactions on classical computers} 
Understanding the physical process of any chemical reactions generally requires the solutions to the Time-dependent Schr\"odinger equation (TDSE) of both electrons and nuclei, 
\begin{equation}
    i\hbar \frac{\partial \Theta(\br,\bR)}{\partial t} = \hat{H}(\br,\bR)\Theta(\br,\bR).
    \label{eq:tdse_enuc}
\end{equation}
Where $\Theta(\br,\bR)$ is the total wavefunction of electrons and nuclei. $\boldsymbol{r}$ and $\boldsymbol{R}$ are the coordinates of electrons and nuclei, respectively. The molecular Hamiltonian is given as
\begin{align}
    \hH(\br,\bR)=&\hT_n + \hT_e + \hV_{ee}(\br)+\hV_{en}(\br,\bR)+\hV_{nn}(\bR)
    \nonumber\\
    =& \hT_n + \hH_{e}(\br;\bR).
\end{align}
$\hT_n=-\sum_A\frac{1}{2m_A}\nabla^2_{\bR_A}$ is the nuclear kinetic energy operator and $m_A$ is the mass of nuclei $a$. $\hat{T}_e$ is the electronic kinetic operator. The electronic Hamiltonian $\hH_e(\br;\bR)$ contains the electronic kinetic energy operator as well as all the Coulomb operators involving electrons and nuclei.
Since there is a large difference between nuclear and electronic mass, the Born-Oppenheimer approximation, which assumes that electrons adjust instantaneously to the slower motion of the nuclei so that a single adiabatic potential energy surface (PES) governs the motion of the latter, is used to factorize the total wavefunction as $\Theta(\br,\bR)\approx \Psi(\br;\bR)\chi(\bR)$. Note that the Born-Oppenheimer approximation is a single decomposition approximation to the general Born-Huang representation. Such an approximation usually breaks down when two PESs are close, and the nonadiabatic effect becomes essential in these regimes~\cite{Rachel:2018cr}. Consequently, the Schr\"odinger equation can be divided into two equations for electrons and nuclei, respectively,
\begin{align}
    i\hbar \frac{\partial \Psi(\br; \boldsymbol{R})}{\partial t} &= \hH_e(\boldsymbol{r}; \bR)\Psi(\br;\bR), \text{ or }  \hH_e(\boldsymbol{r}; \bR)\Psi(\br; \bR)=E(\bR)\Psi(\br;\bR), \label{eq:electronse}
    \\
     i\hbar \frac{\partial \chi(\bR)}{\partial t} &= [\hT_n + E(\bR)]\chi(\br,\bR),\label{eq:nucdyn}
\end{align}
Where $E(\bR)$ is the ground state of $\hH_e(\boldsymbol{r}; \bR)$ for a given nuclear configuration $\boldsymbol{R}$, determining the PES for chemical reactions along certain reaction coordinates. Hence, in principle, two steps are required to simulate chemical reactions: (1) compute the energy $E(\bR)$ from the electronic Hamiltonian $\hH_e(\br;\bR)$ for given nuclear coordinates (Eq.~\ref{eq:electronse}), (2) solve the {\it nuclear quantum dynamics} governed by Eq.~\ref{eq:nucdyn} or {\it quantum molecular dynamics} which treats nuclei classically by taking the classical limit of Eq.~\ref{eq:nucdyn} with quantum forces computed from ab initio PES $E(\bR)$. Unfortunately, neither of the two steps is trivial on classical computers due to the infeasible scaling of solving quantum many-body problems governed by Eqs.~\ref{eq:electronse} and ~\ref{eq:nucdyn}. Consequently, Density functional theory (DFT) or semi-empirical methods are widely used to solve Eq.~\ref{eq:electronse}. However, DFT or semi-empirical methods underestimate the electron correlations, particularly the dynamic correlations usually involved in chemical reactions, leading to inaccurate potential energy surface $E(\bR)$. Hence, the challenging nature of accurately simulating quantum systems led Richard Feynman to propose the development of quantum computers~\cite{feynman:1982qc}. Today, quantum chemistry is recognized as one of the foremost applications of quantum computers~\cite{cao2019quantum, Bauer:2020ud, McArdle:2020we}.

\subsubsection*{Current state-of-art in computational catalysis}
Computational catalysis studies employ electronic structure methods to optimize stationary structures on Born-Oppenheimer potential energy surfaces (PES). This optimization identifies stable reactants, products, and intermediates essential to chemical processes. Optimizing first-order saddle points on these surfaces reveals transition state structures. Due to the computational challenges inherent in these studies, density functional theory (DFT) is widely used in the community. Significant efforts have been directed towards enhancing the accuracy and efficiency of DFT for catalyst screening. These improvements include the development of advanced solvent models, functionals beyond generalized gradient approximation (GGA), linear scaling methods, machine-learning potentials, and empirical models~\cite{Manos:2021cr}. Additionally, first-principles-based kinetic methods have been developed to address kinetic effects in catalysis, such as Mean-Field Microkinetic Modeling, kinetic Monte Carlo, and quantum molecular dynamics. Further details are available in several reviews~\cite{Manos:2021cr, Pineda:2022vg}. Despite these advancements, the electronic structure remains a critical component of many models, and accurately computing this structure continues to pose significant challenges in classical computational catalysis.

\subsection{Application 1: Energetics of reaction pathways for down selection of catalysts for artificial photosynthesis}
\subsubsection{Specific background of the application}

In artificial photosynthesis, optimizing the reaction mechanism is intrinsically linked to the design or identification of catalysts that can efficiently guide/accelerate reactions to yield desired products. For example, identifying catalysts for water oxidation and subsequent \ce{CO2} reduction is vital.  The down-selection of ideal catalysts for oxidation and reduction reactions depends on characterizing the energetically favorable reaction pathways with higher reaction rates for the desired product. For example, producing oxygen, electrons, and protons from water oxidation and useful fuels from \ce{CO2} reduction. This involves accurately describing transition states and calculating activation energies. Transition states, as critical intermediates, are essential in mapping reaction pathways and understanding the energetics. They are the transformation step from reactants to products. In addition, activation energy, indicative of the energy barriers for reactions, is essential in assessing the feasibility and kinetics of a reaction and the efficiency of a catalyst.

Given the challenges and limitations of experimental methods in capturing these detailed aspects of catalyst, computational modeling becomes an essential tool. It allows for predicting catalyst behavior, describing transition state, and calculating activation energies. Through computational modeling, we can screen various catalyst candidates and identify the most effective one for driving the desired reactions. This approach is essential for both water oxidation and \ce{CO2} reduction, where the choice of catalyst has a profound impact on the efficiency and practicality of the artificial photosynthesis process.

\subsubsection{Objectives}
In this section, we propose a workflow to examine the potential of quantum computers to enhance our understanding of reaction mechanisms involving catalysts~\cite{reiher2017elucidating}. By quantum computation of transition states and corresponding activation energies, we aim to significantly improve the accuracy and efficiency of catalyst design and optimization processes. Based on the challenges outlined above, the primary objective is to \emph{down-select a pool of catalysts} for water oxidation and \ce{CO2} reduction reaction. The catalysts for — water oxidation and \ce{CO2} reduction — could either be the same or different. This objective entails identifying catalysts that can improve the efficiency of the reaction and are selective toward desired outcomes, such as water oxidation to produce electrons and protons or \ce{CO2} reduction to produce CO, hydrocarbons, carbohydrates, etc. 

The ultimate goal is precisely estimating the ground state electronic structure for transition states to calculate the energy barrier for the reaction pathway and select superior catalysts. Such calculations are usually non-trivial on classical computers because a) many catalysts involve transition metals with strong electronic correlations, and b) precise treatment of bond formation or breaking during the reactions. For identifying catalysts that enhance the efficiency of artificial photosynthesis, it is crucial to understand specific aspects of the reaction mechanisms for each catalyst, and reaction type (water oxidation or \ce{CO2} reduction) is essential.

\begin{itemize}
    \item \textbf{Reaction Pathway}: Identify the sequence of all intermediates or transition states (TS) characterizing the reaction pathways. Starting with a catalyst [K], we need to select the pathway.\\
    \begin{align*}
     [K] + \text{R} \rightarrow [\text{TS$_1$}] \rightarrow [\text{TS$_2$}] \rightarrow  \ldots \rightarrow [\text{TS$_n$}] \rightarrow [K] + \text{P}. \\
    \end{align*}
     This involves determining each intermediate stage, from the initial interaction of the catalyst with reactants ([K] + R) through various transition states (TS$_1$, TS$_2$, ..., TS$_n$) to the final product ([K] + P).
     
     \item \textbf{Activation Energy }: For the established reaction pathway, \textit{calculate the activation energy ($E_a$)} that is crucial for determining rate constants. The Arrhenius equation defines the rate constants (k): 
     \( k \propto \exp(-\text{E$_a$}/\textit{\textbf{R}}T) \), where \textit{\textbf{R}} is the gas constant and T is the temperature. We can predict the rate at which the reaction will proceed by estimating the activation energy of each reaction pathway and determining the rate-limiting step. However, these estimations are made under the approximation that nuclear dynamics can be neglected. This is an important and initial step for catalyst down-selection.
\end{itemize}

These steps are critical for designing efficient and selective catalysts for desired products. Understanding these processes at a molecular level allows for precise modifications to the catalysts, enhancing their overall behavior and efficiency in artificial photosynthesis.  

\subsubsection{End-to-end computational workflow}
\begin{figure}[htbp]
\centering
\includegraphics[width=0.95\linewidth]{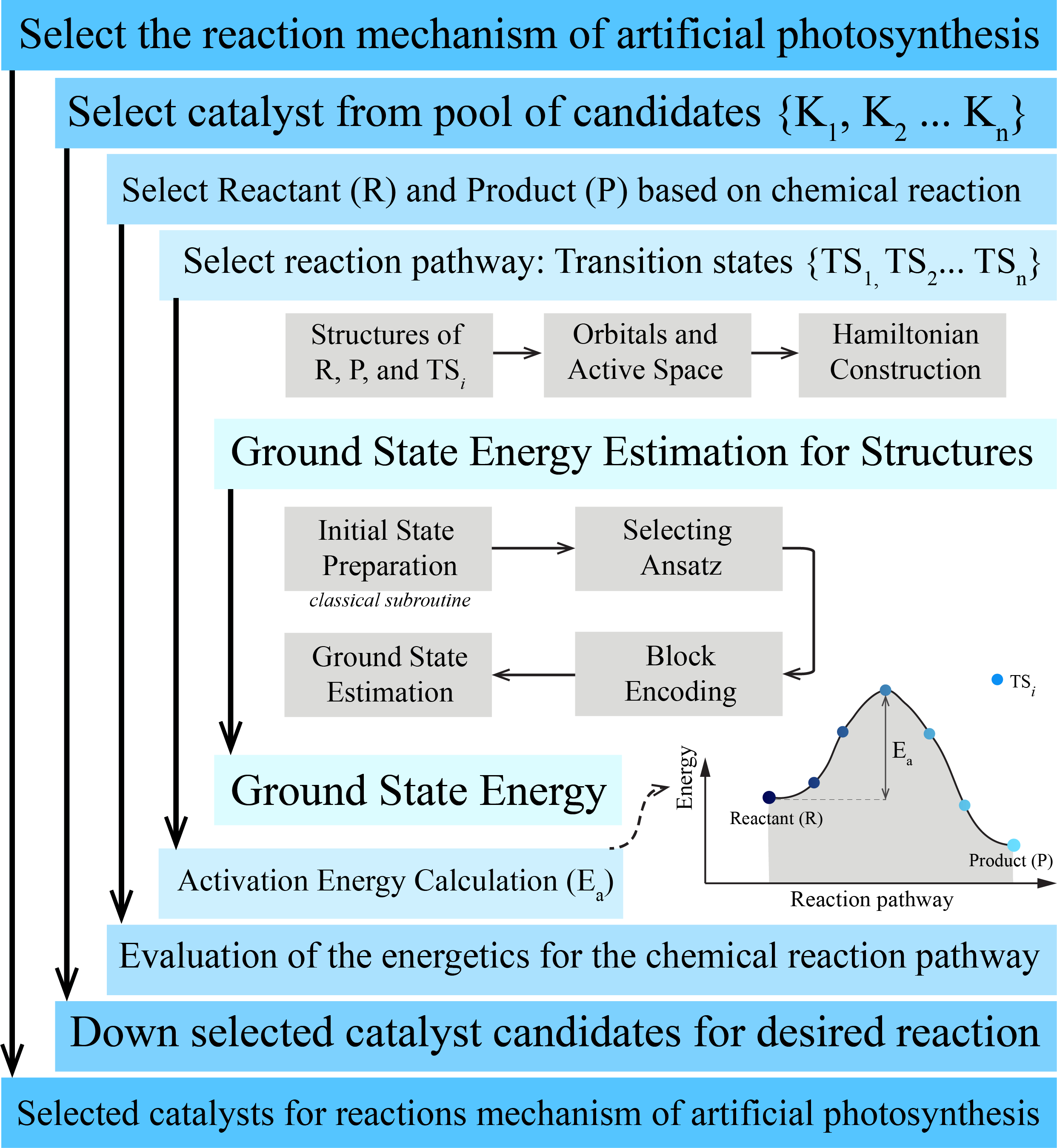}
\caption{\label{flowchart} Flowchart of selecting catalysts for different reaction mechanisms of artificial photosynthesis using computational modeling.}
\end{figure}
\textbf{Computational Workflow:}  Figure \ref{flowchart} shows the computational workflow to down-select the most suitable catalyst for water oxidation and \ce{CO2} reduction reactions from a pool of candidates by leveraging computational modeling. It begins by selecting the reaction mechanisms of artificial photosynthesis and evaluating a comprehensive pool of catalyst candidates (K$_1$ through K$_n$), followed by selecting the reactants (R) and the products (P) based on the chemical reaction (for example, \ce{H2O} is the reactant and electrons and protons are the desired product in the water oxidation reaction). Next, potential reaction pathways are explored by identifying possible transition states (TS$_1$ through TS$_n$) with the nudged elastic band (NEB) or other counterparts~\cite{Mills:1005sc}, which are intermediate configurations leading from reactants to products. This exploration is facilitated using approximate quantum chemistry methods, such as DFT and CC methods, on classical computers. 
Such exploration of calculating energies for transition states is replaced by the quantum algorithms in the quantum computing workflow depicted in Figure~\ref{flowchart}.

The workflow considers the molecular structures of the reactant (R), product (P), and each transition state (TS$_i$). Subsequently, a subset of these orbitals (active space) is selected for high-level electronic structure calculations, which is crucial for defining the electronic Hamiltonian's energy spectrum and balancing computational accuracy and efficiency.  While molecular Hamiltonians are theoretically exact, they still depend on the wavefunction Ansatz to strike a balance between high accuracy and computational efficiency. The selected Hamiltonian within a certain active space is then used to estimate the ground state energy with the corresponding wavefunction Ansatz.  Next, the wavefunction Ansatz is mapped onto a computational circuit, which can be executed on either classical computers for less demanding calculations or quantum computers for more complex problems. This encoding allows for estimating the ground state energies via either quantum phase estimation or other quantum algorithms~\cite{kitaev1995}. The workflow then computes the energy for the reactant, product, and transition states, creating a potential energy surface along the reaction pathway (shown as inset in Figure~\ref{flowchart}). The workflow uses these energy values to estimate the activation energy (E$_a$) and reaction barrier (reaction rates). These insights allow the workflow to characterize favorable pathways with the lowest barrier (highest reaction rate) for the selected catalyst, facilitating the identification of the most efficient catalyst based on the reaction rates through a down-selection process. This preferred catalyst is expected to improve the effectiveness and feasibility of artificial photosynthesis. As depicted in Figure \ref{flowchart}, this workflow represents the integration of chemistry, quantum chemistry, and computational modeling to down-select effective catalysts for the different reaction mechanisms in artificial photosynthesis.

\textbf{Hard Computational Module:} The challenging aspect of the computational workflow is estimating the ground state energy. This involves complex calculations and optimization related to electronic correlations. The primary challenge in these calculations lies in accounting for all essential electronic configuration interactions (CIs) of a molecule in an exponentially large Hilbert (or CI) space, which becomes increasingly demanding with large numbers of electrons. This requires significant computational power, especially for catalysts with transition metals (that have strong electron correlations and spin polarization), where precision in calculating energy barriers is essential to determine the feasibility and rate of chemical reactions. Classical computational resources often fall short in processing these complex tasks due to their limitations in handling the huge CI space of large systems (curse of dimension). Quantum computing emerges as a solution to these challenges with its inherent exponentially large representation capability to encode quantum many-body wavefunction~\cite{feynman:1982qc}(or high-dimensional data), offering a much more efficient approach to these complex problems. They can handle the intricacies of quantum mechanics more naturally, potentially allowing for more accurate and economical solutions to problems. Integrating quantum computing into such workflows will significantly accelerate the research in quantum chemistry~\cite{von2021quantum}. It enables the exploration of more complex molecular systems and reaction mechanisms that were previously inaccessible by classical computing, leading to more advanced and precise catalyst design.

\noindent The whole workflow shown in Figure~\ref{flowchart} can be summarized as follows:
\begin{enumerate}

    \item \textbf{Select Reaction Mechanism}: Begin by choosing the reaction mechanism involved in artificial photosynthesis, water oxidation reaction, or \ce{CO2} reduction reactions.
        
    \item \textbf{Select Catalyst from Pool of Candidates}: Accordingly choose the potential catalyst candidates,  $K_1, K_2, \ldots, K_n$.
    
    \item \textbf{Select Reactant (R) and Product (P)}: For the water oxidation reaction, \ce{H2O} is the reactant, and O$_2$ (along with electrons and protons) is the product. For the \ce{CO2} reduction, \ce{CO2} is the reactant, and desired products are CO or other fuels.
    
    \item \textbf{Select Reaction Pathway}: Identify possible transition states, $TS_1, TS_2, \ldots, TS_n$, which are intermediate states between reactants and the products.
    \begin{enumerate}
    \item \textbf{Structures of R, P, and $TS_i$}: Determine the molecular structures of the reactant, product, and each transition state.

    \item \textbf{Orbital and Active Space}: Determine the relevant orbital and select active space to define the scope of the Hamiltonian.
    
    \item \textbf{Hamiltonian Construction}: Construct the electronic Hamiltonian for these molecules.

    \begin{enumerate}
    \item \textbf{Quantum Ground State Estimation}: Employ quantum algorithms to approximate the ground state energies and eigenfunctions.
    
    \begin{enumerate}

    \item \textbf{Initial State Preparation}: Prepare initial states or wavefunctions for each molecular structure.
        
    \item \textbf{Selecting Ansatz}: Selecting the initial quantum state to efficiently navigate the solution space of a problem on the computer.
        
    \item \textbf{Block Encoding}: Encode the state Ansatz (including the initial states and unitaries for uncovering the correlations) of the molecules onto a computer.
    
    \item \textbf{Ground State Measurement}: Perform the ground state energy measurements on computers,  for example, on quantum computers, by measuring each individual Pauli string in the qubit Hamiltonian. 
    
    \end{enumerate}
     \end{enumerate}
    \item \textbf{Ground State Energy for R, P, and $TS_i$}: Compute the energy levels for the reactant, product, and each transition state.
    
    \end{enumerate}
    
    \item \textbf{Activation Energy Calculation ($E_a$)}: Calculate the activation energy required to convert reactants into products through the transition states.
    
    \item \textbf{Energetic Pathway Evaluation}: Based on the calculated energies, identify an energetically favorable reaction pathway for the selected catalyst.
    
    \item \textbf{Catalyst Selection for Reactions}: From the initial pool of candidates, select the ideal catalyst that best facilitates the reaction.

    \item \textbf{Selected Catalysts for Artificial Photosynthesis}: List all down-selected catalysts for water oxidation reaction and \ce{CO2} reduction reactions.
\end{enumerate}

\subsubsection{Why classical methods are not sufficient to perform the hard computational module}

The hardness of classical computational methods in performing demanding tasks in quantum chemistry, particularly for down-selecting catalysts, can be attributed to system size, electronic structure complexity, and the inherent computational cost of these methods. Classical computational methods often struggle with large molecular systems or those with complex electronic structures. As the size of the system increases, the number of possible electronic CIs escalates, or an extended basis is needed to represent the wavefunctions, making the calculations more challenging.

The computational cost of DFT, a widely used method in quantum chemistry, scales cubicly with the number of system electrons O(N$^3$) with density fitting technique~\cite{birkenheuer2005model}. Though DFT is efficient on classical computers, it has difficulty in accurately handling localized charges and strong electron correlation (a common scenario in catalysts) due to the lack of exact functional~\cite{Becke:2014jcp}, leading to less reliable results. The coupled-cluster (CC)~\cite{Bartlett:2007wj} method, which represents the wavefunction as an exponential of cluster operators and can be systemically improved by truncating at different orders ($M$), is accurate for many systems, its steep computational scaling ($O(N^{2(M+1)})$) makes it impractical for large systems. Moreover, CC is not particularly effective for strong static correlations~\cite{Bartlett:2007wj}. For systems with strong electron interactions, the Density Matrix Renormalization Group (DMRG)~\cite{Chan:2011vs} method offers significant advantages. DMRG approximates the wavefunction as a low-rank tensor product, enhancing its ability to capture strong electron correlations with greater accuracy. Still, it is subject to the area law~\cite{Hastings_2007}, is less practical for higher-dimensional systems, and is limited to small systems due to its computational complexity. Nevertheless, as the dimensions increase, the number of possible CIs grows exponentially, making classical calculations increasingly difficult and resource-heavy.

Given these limitations, classical computational methods often fall short in either computational efficiency or accuracy in capturing the challenging electron correlation problems in quantum chemistry, particularly for large or complex molecular systems. This is where quantum computing, with its different approach to handling calculations involving quantum states, offers a promising alternative capable of overcoming these limitations~\cite{chemrev.8b00803, lee_evaluating_2023, Bauer:2020ud, McArdle:2020we}.

\begin{figure}[htb]
    \centering
    \includegraphics[width=0.98\textwidth]{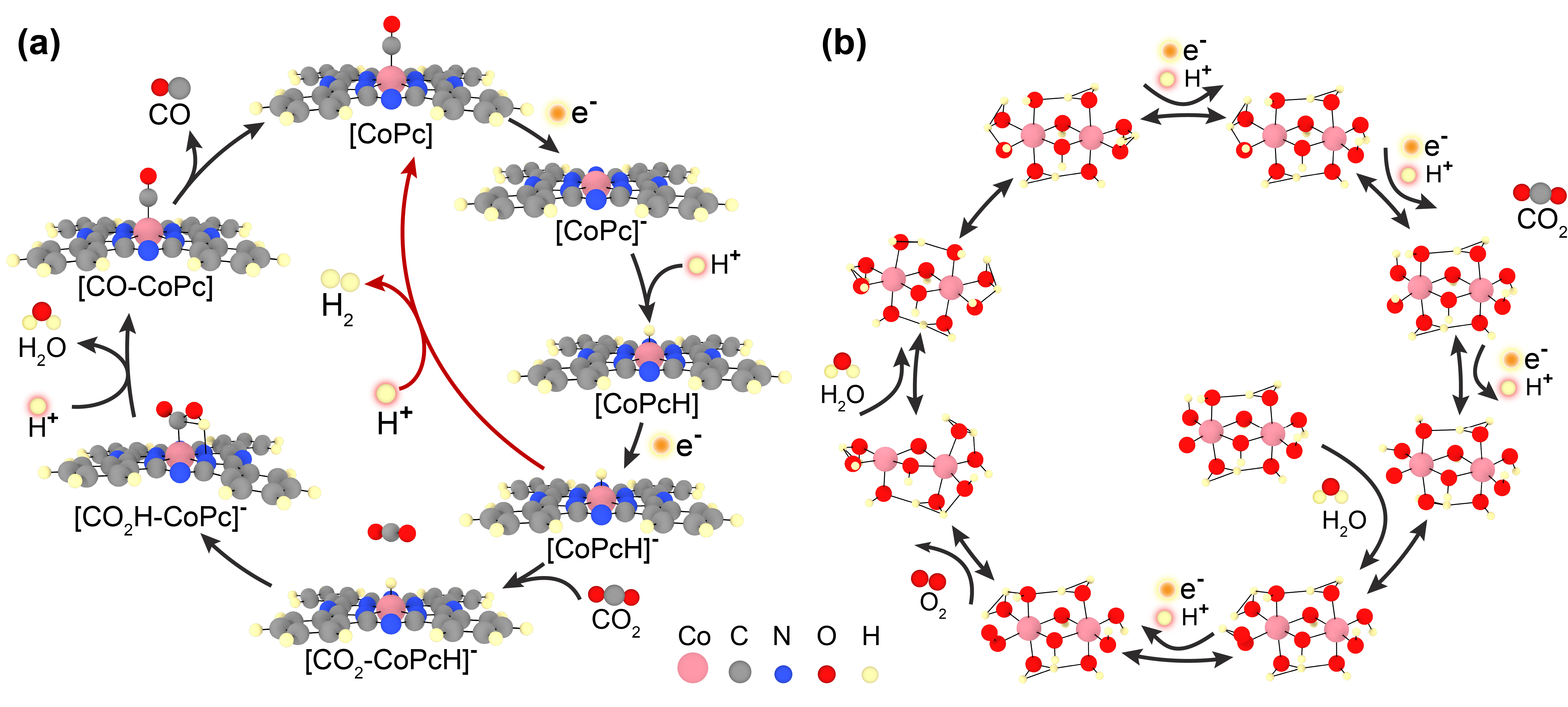}
    \caption{(a) One proposed \ce{CO2} reduction mechanisms with \ce{CoPc} catalyst~\cite{Liu:2019nc}.
    (b) One possible water oxidation reaction pathway with the cobalt dimer catalyst~\cite{jp511805x}.  
    }
    \label{fig:co_water}
\end{figure}

\subsubsection{Concrete problem instantiations}
Here, we use transition metal complexes as the example catalyst to demonstrate the application of quantum computation to computational catalysis.

{\bf \ce{CO2} Reduction: } Cobalt phthalocyanine (CoPc, molecular formula is \ce{CoC32N8H16}) molecule is one of the high-performance \ce{CO2} reduction catalysts~\cite{Zhang:2017tn}. The CoPc can also be used to make a metal-organic framework (MOF) for both high selectivity and \ce{CO2} capture capability~\cite{Li:2020wr, Fang:2023tf}. Nevertheless, we use the basic unit, CoPc, as the sample to demonstrate the reaction pathway of \ce{CO2} reduction. One of the proposed mechanisms of \ce{CO2} reduction with the CoPc catalyst is shown in Figure~\ref{fig:co_water}(a).

{\bf Water Oxidation with metal oxide catalyst:}
The other sample application we consider is water oxidation to molecular oxygen mediated by model cobalt oxide. Cobalt oxide has recently gained wide popularity for its remarkable catalytic activity toward water splitting~\cite{Kanan2008sci}. The structure of this molecular catalyst was also found to consist of \ce{Co4O4} cubane-shaped metal oxide units. Without losing generality, we consider a model cobalt oxide dimer \ce{Co2O9H12}~\cite{jp511805x}, which is a minimal unit of the cubane complex and serves as a typical model system for studying the paramagnetic intermediate in Cobalt-Ctalyzed water oxidization, as shown in Figure~\ref{fig:co_water}(b). This cobalt dimer compound is characterized by two aqua ligands and two hydroxyl ligands attached to each cobalt center and Two $\mu$-hydroxo groups linking the two cobalt centers. The dimer compound can have many possible isomers with different positioning of these two aqua and hydroxyl ligands around the cobalt atoms. 

It should be noted that both the water oxidation reactions pathways on the cobalt dimer and \ce{CO2} reduction of CoPc catalyst are characterized by the proton-coupled charge transfer (PCET) as shown by experimental evidence~\cite{Dau2010}, similar to PCET reactions in the oxygen-evolving complex of photosystem II in the nature. The PCET encompasses nuclear quantum effects such as quantum tunneling and quantum interference, which will be discussed in {\it Application 3}.

{\bf Molecular structure and classical hardness.}
The molecular structures of the \ce{CO2} reduction and water oxidation along one reaction pathway are shown in Figure~\ref{fig:co_water}(a) and (b), respectively. The number of atoms, total electrons, valence electrons, and the number of molecular orbitals with different basis sets are shown in Table~\ref{tab1}. From the molecular structures, we are readily able to obtain electronic energies using quantum chemistry methods. However, the electronic structure of such transition metal compounds remains poorly understood due to the complexity of the multiple transition metal ions with their multiple charge states and complicated spin-couplings. 
The presence of a larger number of electrons/orbitals and strong correlation in $d$ orbital electrons makes it hard to compute the energies with DMRG methods on classical computers if all the electrons and orbitals are considered. Even after removing the core electrons, there are still 92 valence electrons in 120 spatial orbitals for the water oxidization reaction. Hence, classical algorithms usually prune the number of orbitals/electrons further to construct small active valence electrons in the electronic structure calculations. However, even with this active space of 48 electrons in 32 orbitals (64 spin orbitals),i.e., (48e, 32o) as shown in Table~\ref{tab1}, the dimension of the Full configuration interaction (FCI) is on the order of $10^{14}$ (i.e., $\binom{32}{24}\times\binom{32}{24}\sim 10^{14})$ on classical computers. Even the efficient DMRG method has a complexity of $O(ND^3)$ where $N$ and $D$ are the number of orbitals and bond dimensions, respectively. 
In the typical simulation of transition metal with a bond dimension of 5000, DMRG simulation requires a dimension of $64\times (5000)^3\sim 10^{13}$, which requires more than 80 TB RAM memory.

\begin{table}[h]
  \centering
  \caption{Molecular information of \ce{CO2} reduction and water oxidation.}
  \label{tab1}
  \begin{tabular}{c|c|c|c}
   \\
    \hline\hline
    {\bf System} & {\bf Atoms} & {\bf Total Electrons}
    & {\bf Active space}\\
    \hline
    \ce{CO2}-CoPc                 & 60 & 269
    & (45e, 35o) \\
    \hline
    \ce{H2O}-\ce{Co2}O$_9$H$_{12}$& 26 & 148 
    & (48e, 32o) \\
    \hline
  \end{tabular}
\end{table}

{\bf Hamiltonian generation and encoding in quantum algorithms for artificial photosynthesis.} A good starting point for solving chemistry problems is the mean-field Hartree-Fock (HF) method. This method approximates the many-electron problem by effective one-body problems where each electron evolves in the mean field of the others. The HF method is very efficient on classical computers ($O(N^3)$ scaling if density fitting for the electron repulsion integrals is used). It includes the exact exchange, but the electron correlations are missing. Therefore, the HF method is usually a good starting point for post-HF methods to include the missing correlations. With the HF method, we can rewrite the electronic Hamiltonian in the basis set of HF orbitals (namely, molecular orbitals (MOs)) in the second-quantization form (termed as molecular Hamiltonian). Since not all the electrons/orbitals contribute to the correlation energy, we usually only need to select a set of electrons/orbitals (namely active space) in the simulation. In the selection of active space, the key principle is that all strongly correlated orbitals must be identified and selected to active space, which can be accomplished by the so-called mutual information that measures the entanglement among orbitals~\cite{Rissler2006, tkachenko2021correlation, clusterVQE}. With the selection of active space, we can effectively reduce the size of the molecular Hamiltonian and the subsequent qubit Hamiltonian described below.

However, the Fermionic operator in the molecular Hamiltonian cannot be directly simulated on quantum computers that work on qubits. The molecular Hamiltonian has to be mapped into its qubit counterpart by using either Jordan-Wigner (JW)~\cite{JW:1928}, Bravyi-Kitaev (BK)~\cite{BRAVYI2002210}, or parity~\cite{parity2012} encoding. Hence, different encoding techniques lead to various numbers of qubits requirement and sparsity in the resulting qubit Hamiltonian~\cite{chemrev.8b00803} and consequently affect the gate complexity in the QPE. Nevertheless, the resulting qubit Hamiltonian can contain $O(N^4)$ Pauli strings. Although the ultimate goal is to include all valence electrons and beyond (which requires hundreds of spin orbitals) in quantum simulations, we selected a restricted active space corresponding to orbitals near the Fermi energy surface, as shown in Table~\ref{tab1}. This is due to the computational expense of generating two-body integrals and the associated mapping into qubit Hamiltonians. In fact, the large number of two-body integrals and Pauli strings in quantum chemistry problems stimulates the development of pre-processing techniques~\cite{Rocca2024} and measurement reduction techniques~\cite{Ching:21lc, Huang:2020wo, Xavier:2020prx}. For the specific cases of water oxidation and CO2 reduction, we chose a small active space of (48e, 32o) and (45e, 35o), leading to 64 and 70 spin orbitals (qubits), respectively. But our target is to simulate a large active space ($\sim$ 650 spin-orbitals for the CoPc molecule) in order to treat either all electrons in a larger basis set (such as CC-PVQZ) or larger molecules. It should be noted that selecting the active space is a nontrivial task. In this context, the active space is determined by imposing an energy window that truncates according to the energy difference between the molecular orbital energies and the Fermi energy. This method represents one of the simplest approaches to select the active space. To enhance the quality of the active space, techniques inspired by quantum information theory~\cite{clusterVQE, Ding:2023wl} or based on natural orbitals can be utilized~\cite{Stein:2016ve}.
The resulting qubit Hamiltonians have 1,542,149  and 930,650 Pauli strings, respectively. 

{\bf Initial state preparation.} The choice of initial states in quantum computation plays a pivotal role in achieving accurate results, particularly in capturing quantum systems' true ground state energies. In many cases, a single-determinant Hartree-Fock (HF) state is a good initial approximation, as it provides a reasonable starting point with a relatively large overlap with the actual ground state. However, in complex reactive systems like photosynthesis, which is characterized by bond breaking and formation, the static correlation effects become significant, and the single-determinant HF state usually has poorer overlap with the grun ground states. In such instances, the wavefunction has a multireference nature, i.e., more than one electronic configuration contributes significantly to the ground state energies. Consequently, the low-complexity multireference can be employed, such as the linear combination of the HF state and selected configuration interaction singles (CIS) states~\cite{tkachenko_quantum_2022}. Such multireference states can be easily prepared on quantum computers with rotation and entanglement gates~\cite{tubman_postponing_2018}, and it was found that such low-complexity multireference state can usually provide sufficient overlap with the true ground states, boosting the success probability of the phase estimation for the complex reactive systems~\cite{tubman_postponing_2018}.

\subsubsection{List of candidate systems where a similar process is relevant?}

A similar workflow can be used for cases where ground-state energy estimation is required. For example, 

\begin{itemize}
    \item Oxygen Evolution Reaction on metal oxides, such as Cobalt Oxide~\cite{koza2012electrodeposition} and titanium Oxide~\cite{pei2015defective}. 
    \item Hydrogen Evolution Reaction or protonation for H$_2$ fuel generation and storage~\cite{roubelakis2012proton}.
    \item In materials science: Defect migration within the crystal~\cite{johansen2015aluminum} and structural phase transitions~\cite{parida2020vertically,mishra2017atomistic}. 
\end{itemize}


\subsection{Application 2: Reaction dynamics through quantum molecular dynamics simulations}

\subsubsection{Specific background of the application}
The quantum computation of reaction energies in \textit{Application 1} leverages pre-determined reaction pathways in reduced dimensions. Among the various strategies for identifying these pathways, the nudged elastic band (NEB)~\cite{Mills:1005sc} method, supported by chemical intuition, stands out. Although proficient, the NEB method recognizes that it may not always determine the lowest barrier path. Such discrepancies arise because the method's success relies on the chemist's insight and experience, which guides the selection process but does not guarantee the most energetically favorable pathway. Furthermore, despite its effectiveness, the NEB method incorporates approximations to determine the reaction pathways, which may not fully capture the complexity of the quantum mechanical landscape~\cite{Bernhard:2003ep}. 

Besides, \textit{Application 1} primarily uses a static method to identify reaction barriers, but it doesn't provide detailed insights into reaction dynamics~\cite{Bernhard:2003ep,hynes1985chemical,collins2002molecular}, such as kinetic effects and efficiencies of chemical reactions. These drawbacks, inherent in traditional static quantum chemistry methods for transition states, can be mitigated by the adiabatic Quantum Molecular Dynamics (QMD)~\cite{andp.19273892002, PhysRevLett.55.2471, RevModPhys.64.1045,marx2000ab,tuckerman2002ab} or nonadiabatic molecular dynamics (NAMD)~\cite{Nelson:2018cr, Rachel:2018cr}. QMD propagates the nuclei classically according to Newton's equations using forces from ab initio calculations, i.e., classical molecular dynamics using quantum forces. Using the QMD allows the system to evolve from the reactant state to the product state via the ``on-the-fly" update of the potential energy surfaces, involving extensive sampling of initial conditions and/or long simulation times~\cite{RevModPhys.64.1045}. This approach enables comprehensive analysis of reaction dynamics, capturing intricate details of electron interactions and molecular kinetics. Consequently, QMD's ability to accurately simulate the evolution of transition states is fundamental for elucidating the chemical reaction mechanisms of complex structures. Furthermore, QMD also helps identify intermediates that impact product formation, allowing for optimization of reaction conditions to improve selectivity and yield in chemical processes.

Thus, QMD is a pivotal method in assessing catalysts for water oxidation and \ce{CO2} reduction in artificial photosynthesis. For water oxidation, QMD can identify the most efficient pathways for breaking water molecules into oxygen and hydrogen, highlighting catalysts with higher reaction rates, efficiency, and speed. Similarly, for \ce{CO2} reduction, QMD can reveal how catalysts can effectively convert \ce{CO2} into useful carbon-based fuels or chemicals, pinpointing reactions with lower energy costs and higher selectivity. This detailed insight is important for designing effective, sustainable, and cost-effective catalysts, accelerating the development of artificial photosynthesis. Moreover, QMD allows the investigation of the role of temperature, pressure, and external stimuli on the reaction mechanism. This also provides the opportunity to comprehensively understand how these factors influence the reaction rate and mechanism. Additionally, QMD can mimic realistic chemical environments, extending the scope of the investigation to conditions that would be challenging to achieve in \textit{Application 1} and experiments. \textit{Application 2} provides the ability to explore a wide range of conditions, including
\begin{itemize}
    \item \textbf{Effect of External Fields}: Dynamics electric or magnetic fields to enhance catalytic efficiency and selectivity~\cite{list1a,list1b,list1c}.
    \item \textbf{Dynamic Effect on Reactivities}: Non-adiabatic effects, which coupled electronic and nuclear motions, are crucial in a wide range of chemical processes~\cite{list2a,list2b}.
    \item \textbf{Kinetic and Thermodynamic Properties}: Extreme temperatures and pressures that are often encountered in industrial settings influence reaction rates and equilibrium~\cite{mechno1,mechno2,mechno3}.
    \item \textbf{Surface and Interface Phenomena}: The role of surfaces and interfaces in catalyzing reactions, including adsorption and desorption processes and surface-induced catalytic activity~\cite{list5a,list5b,list5c}.
    \item \textbf{Material Degradation and Stability}: Understanding how materials degrade over time under various conditions and strategies for improving the longevity and durability of catalysts~\cite{list6a,list6b}.
\end{itemize}

Though we only discussed the adiabatic QMD method in the following, the nonadiabatic molecular dynamics (NAMD) can also be applied to study the kinetics of catalysis with additional computation of excited states, nonadiabatic couplings, and propagation of electronic wavefunction~\cite{Rachel:2018cr, Nelson:2018cr}, which will be discussed in later applications. In particular, NAMD plays an essential role in modeling exciton migration and separation into charges (electrons and holes)~\cite{Nelson:2018cr} before they are utilized in catalysis.

\subsubsection{Objectives}
\label{app2-objective}
As discussed, dynamics simulations are crucial for revealing the complexity of chemical reactions, including both kinetics and dynamical effects. Unlike \textit{Application 1}, which is limited to predefined reaction pathways, QMD simulations explore the most effective pathways by analyzing efficiency and reaction rates with various catalysts. Therefore, QMD improves the methodology for selecting appropriate catalysts for artificial photosynthesis. The catalyst selection process within the QMD framework involves determining the number of reacted trajectories ($N_{\text{reacted}}$) out of the total trajectories ($N_{\text{all}}$). The ratio of these trajectories determines the quantum yield ($Q$) or reaction efficiency, while reacted trajectories are examined to determine the reaction rate ($k$) similar to that of in \textit{Application 1}. The reaction rate is defined by the inverse of the average time ($\langle t \rangle$) required for a reactant (R) to convert into a product (P) under specified reaction conditions. In this context, a ``trajectory'' refers to the path molecules follow on their PES over time. Each trajectory outlines the evolution of the molecule, beginning with a specific set of initial conditions, including the positions and velocities of all particles. A trajectory is categorized as a reacted trajectory if it transforms the reactant into the desired product.

For a down-selected candidate catalyst from \textit{Application 1} and specific reaction mechanism, such as water oxidation or \ce{CO2} reduction, the reaction efficiency or quantum yield ($Q$) can be estimated using the formula $Q =\frac{N_{\text{reacted}}}{N_{\text{all}}}$. This ratio measures the probability that reactants convert to products under given conditions. 
By calculating the average time for reacted trajectories, $\langle t \rangle$, the reaction rate can be defined as $k\sim\frac{1}{{\langle t \rangle}}$. This approach describes a reaction rate in kinetic terms, offering a quantifiable measure of the efficiency and effectiveness of a catalyst in facilitating a particular chemical reaction. Thus, QMD enables the design and optimization of catalysts for artificial photosynthesis under diverse conditions. Accurate QMD simulations require selecting appropriate quantum methods to estimate Potential Energy Surfaces (PES) and forces alongside a classical approach for modeling nuclear motion. Determining the ideal simulation parameters, such as the time step and total simulation time, is also crucial. Additionally, it is essential to select appropriate initial conditions for reaction trajectories. This involves sampling from a distribution of initial velocities and positions representing a wide range of possible reactant states.

\subsubsection{End-to-end computational workflow}
\begin{figure}[htbp]
\centering
\includegraphics[width=0.95\linewidth]{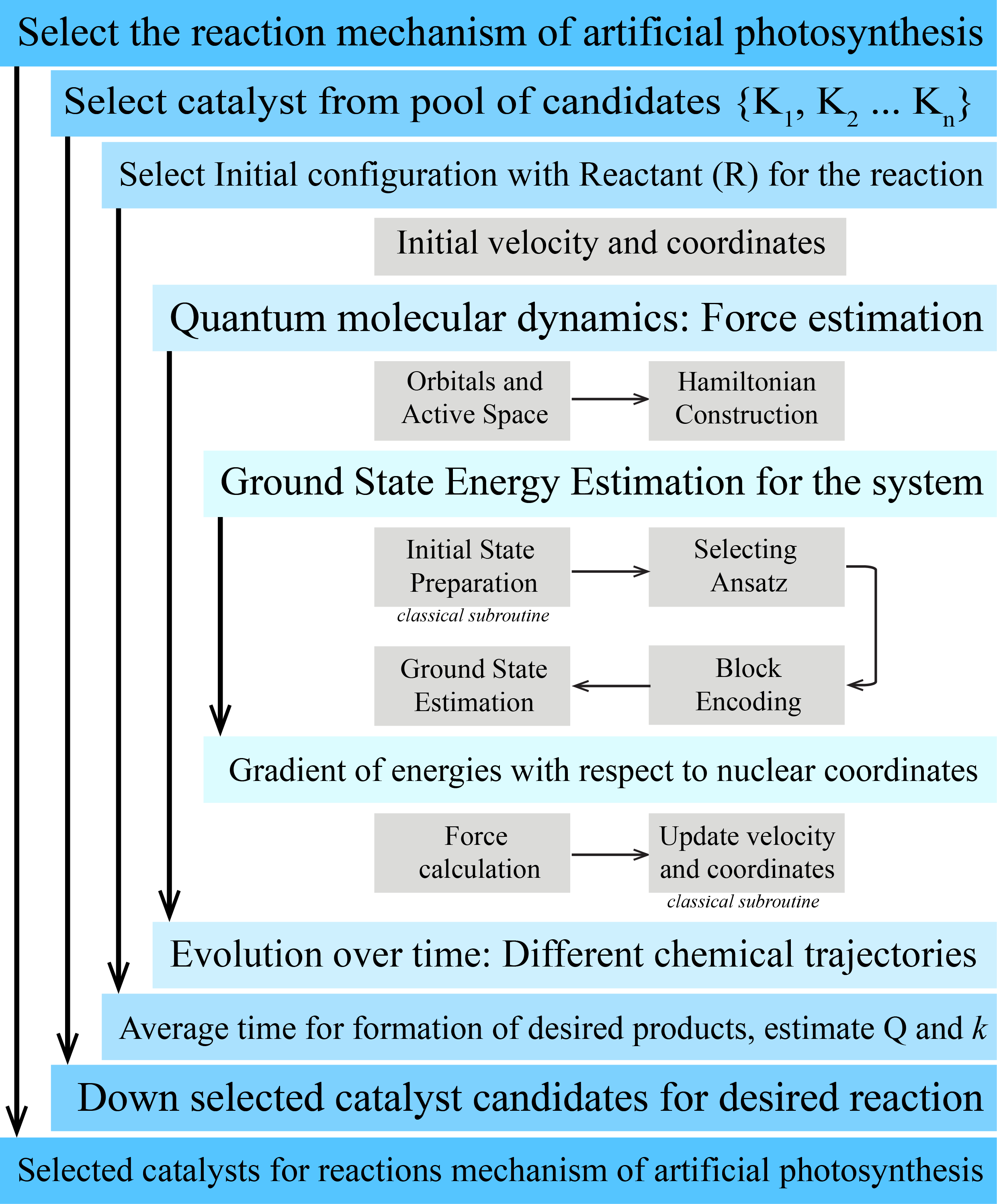}
\caption{\label{flowqmd} Flowchart of selecting catalysts for different reaction mechanisms of artificial photosynthesis using QMD.}
\end{figure}
\textbf{Computational Workflow:}  Figure \ref{flowqmd} shows a process for performing QMD simulations to identify the most effective catalysts for water oxidation and various \ce{CO2} reduction reactions. Analogous to \textit{Application 1}, this workflow involves choosing a reaction for artificial photosynthesis (either oxidation or reduction) and a set of down-selected catalyst candidates from \textit{Application 1} (ranging from K$_1$ to K$_n$). Subsequently, it involves selecting the initial configuration for reactant (R) and catalyst (K). Unlike the previous workflow, this method does not require a predefined product (P) or reaction pathway. After assigning the initial velocity and coordinates for the molecular system, the next steps, akin to \textit{Application 1}, involve selecting the appropriate valence orbitals within the system and choosing the active space for simulation. Subsequently, the electronic Hamiltonian is determined to achieve a balance between high accuracy and computational efficiency. 
The Hamiltonian and wavefunction for initial configuration are mapped onto a computational circuit. Further, the workflow maps the state of the initial configuration  onto a computer to estimate the ground state energies. Next, the quantum circuit that represents the converged ground state Wavefunction is used to measure the forces (gradient of energies with respect to atomic coordinates). Using the computed force, the coordinate and velocity are updated, and the system is then propagated along the PES by Newton’s equation of Motion~\cite{Curchod2018CR}. 

Next, for estimating the quantum yield ($Q$), the workflow determines the ``reacted'' trajectories that result in the creation of desired products. For example, for a water oxidation reaction, electrons and protons are the desired product, while for a \ce{CO2} reduction reaction, CO or other fuels are the desired product. The inverse of the average time for the `reacted'' trajectories is used to calculate the reaction rate ($k$). In the next step, following the comparison of reaction rates for various catalysts (K$_1$ through K$_n$), the workflow selects the most effective catalyst based on its efficiency and selectivity in driving the reactions (higher quantum yield and faster rates). This selected catalyst is anticipated to enhance the efficiency and viability of artificial photosynthesis. As illustrated in Figure \ref{flowqmd}, this procedure includes chemistry, quantum chemistry, and both quantum and classical modeling. This workflow helps advance artificial photosynthesis by identifying catalysts for different reaction mechanisms.

\textbf{Hard Computational Module:} The initial ``hard computational module'' of QMD simulation is the computationally demanding task of estimating the ground state energy (similar to \textit{Application 1}). Quantum computing is a pivotal solution to these challenges, offering a pathway to solve problems involving electron correlations with high accuracy and efficiency. The complexity of this task is particularly pronounced in QMD simulations, which require dynamic updates to the system's state. Each update of coordinates necessitates a subsequent estimation of the ground state energy to reflect the system's current state. This emphasizes both the computational intensity and the necessity for precise capture of quantum mechanical behaviors of interacting electrons.

The other challenge in performing large-scale QMD simulations is accurately estimating the forces during the simulation. The difficulty in computing forces lies in calculating the gradients of the ground state wavefunction with respect to nuclear coordinates. The numerical gradients, which require $6N$ computational of ground states with a finite difference (where $N$ is a number of atoms), make it almost impossible to propagate the QMD. Analytical gradient, while much more efficient than its numerical counterpart, is hard to develop due to the nested chain rule of derivatives of intermediates parameters.
The derivatives are further complicated by the exponentially large wavefunctions in some cases, which impede efficient computation on classical computers. Thus, quantum computing holds the potential to revolutionize QMD simulations, enabling significant advancements in quantum chemistry.

\noindent The whole workflow shown in Figure~\ref{flowqmd} can be summarized as follows:
\begin{enumerate}
    \item \textbf{Select Reaction Mechanism}: Begin by choosing the reaction mechanism involved in artificial photosynthesis, water oxidation reaction, or \ce{CO2} reduction reactions.
        
    \item \textbf{Select Catalyst from Pool of Candidates}: Accordingly choose the potential catalysts,  $K_1, K_2, \ldots, K_n$.
    
    \item \textbf{Select Initial configuration}: Select initial configuration with reactant and catalyst. For the water oxidation reaction, \ce{H2O} is the reactant, and for the \ce{CO2} reduction, \ce{CO2} is the reactant. 

    \begin{enumerate}
    \item \textbf{Initial velocity and coordinates}: Determine the initial condition for molecular structures of the reactant and catalyst.
    \end{enumerate}

    \item \textbf{Quantum Molecular Dynamics}: Calculate forces using ground state energies and update the velocity and coordinates for the system. 
        
    \begin{enumerate}
    \item \textbf{Orbital and Active Space}: Determine the relevant orbital and select active space to define the scope of the Hamiltonian.
    
    \item \textbf{Hamiltonian Construction}: Construct the electronic Hamiltonian for the molecular system.
    \begin{enumerate}
    \item \textbf{Ground State Energy Estimation}: Employ quantum algorithms to approximate the ground state energies and eigenfunctions.
    
    \begin{enumerate}
    \item \textbf{Initial State Preparation}: Prepare initial states or wavefunctions for each molecular structure.
        
    \item \textbf{Selecting Ansatz}: Selecting the initial quantum state to efficiently navigate the solution space of a problem on the computer.
        
    \item \textbf{Block Encoding}: Encode the state Ansatz (including the initial states and unitaries for uncovering the correlations) of the molecules onto a computer.
    
    \item \textbf{Ground State Energy}: Perform the ground state energy estimation on computers,  for example, on quantum computers, by measuring each individual Pauli string in the qubit Hamiltonian. 
    \end{enumerate}
    \end{enumerate}

    \item \textbf{Force Calculation}: Compute forces by taking the gradient of energies with respect to nuclear coordinate using ground state energy computed in the previous step. 

    \item \textbf{Update the Velocity and Coordinate}:  The system with catalyst is then propagated along the potential energy surface (PES) by Newton’s equation of Motion. 
    \end{enumerate}
    
    \item \textbf{Evolution Over Time}: Identify reaction trajectories that result in the formation of diverse products.
    
    \item \textbf{Reaction rate Calculation ($k$)}: Estimation of the average time for the formation of the desired products and reacted trajectories, facilitating the computation of reaction rate ($k$) and quantum yield ($Q$).
        
    \item \textbf{Catalyst Selection for Reactions}: From the initial pool of candidates, select the ideal catalyst that best facilitates the reaction.

    \item \textbf{Selected Catalysts for Artificial Photosynthesis}: List all promising catalysts for reactions, both water oxidation reaction and \ce{CO2} reduction reactions.
\end{enumerate}

\subsubsection{Why classical methods are not sufficient to perform the hard computational module}
Ab initio QMD method treats the electrons and nuclei quantum mechanically and classically, respectively, within the Born-Oppenheimer approximation. Within the Ab initio QMD, nuclei then propagated along the potential energy surface (PES) by propagating Newton's equation of Motions~\cite{Curchod2018CR}
\begin{align}
    M_I \dot{\boldsymbol{R}} &= \boldsymbol{P}_I, \quad 
    \dot{\boldsymbol{P}}_I = \boldsymbol{F}_I = -\nabla_{\boldsymbol{R}_I}\left[E_g(\boldsymbol{R})+V_{NN}(\boldsymbol{R}) \right].
\end{align}
where $\boldsymbol{R}_I$, $\boldsymbol{P}_I$, $M_I$, and $\boldsymbol{F}_I$ are positions, momenta, masses, and forces of the $I$-th atom. $E_g(\boldsymbol{R})$ and $V_{NN}(\boldsymbol{R})$ are the ground state electronic and nuclear repulsive energies, respectively. Therefore, in the propagation of classical trajectories, energy and forces (negative gradient of energies with respect to nuclear coordinates) are computed ``on-the-fly" at each MD step through the solution of the time-independent Schro\"odinger equation for the electrons. Since integrating Newton's equation is cheap on classical computers (compared to the quantum mechanical solution to the \"odinger equation), the fundamental complexity of ab initio QMD simulation is twofold: 1) the intrinsic complexity of computing energy (which has been discussed in application 1) and forces and 2) the larger number of time steps to accurately simulate the dynamics of reactions over long time scale ($>$ ps or even ns).

{\bf Larger number of time steps.}
Take a typical reaction with a picosecond time scale as an example; a QMD simulation with a 1~fs time step requires at least $10^3$ calculations of energies and forces. Even if each energy/force calculation only takes 1 CPU second, a QMD simulation in a nanosecond time scale needs ~280 CPU hours. Consequently, the majority of current QMD simulations can only handle small systems and semiempirical quantum mechanical (SQM) methods (that simplify the first-principles calculations parameterizing one-body and Coulomb integrals)~\cite{Nelson:2018cr} are usually used. 

{\bf Complexity of quantum forces.} Since we have discussed the complexity of energy calculations, here we focus on the discussion of force calculations. In general, the energy $E$ explicitly depends on $\boldsymbol{R}$ and $\theta$ (parameters of the Ansatz). 
Gradients of ground state energy with respect to nuclear coordinates are given by
\begin{equation}\label{eq:force}
\nabla_{I\alpha} E=\bra{\Phi}\nabla_{I\alpha}\hat{H} \ket{\Phi}+\left[\bra{\Phi}\hat{H} \ket{\nabla_{I\alpha}\Phi}+h.c.\right]=\frac{\partial E}{\partial \hat{H}}\frac{d\hat{H}}{d\boldsymbol{R}}+
\frac{\partial E}{\partial \theta}\frac{d\theta}{d\boldsymbol{R}}.
\end{equation}
The last two terms are Pulay forces, which do not vanish when $\Phi(\theta)$ varies with respect to nuclear coordinates (because the wavefunction or the parameterized Ansatz are dependent on $\boldsymbol{R}$). Hence, the complexity of computing forces on classical computers mainly are two-fold: 1) $\nabla_{I\alpha}\hat{H}$ requires the gradients of one-body and two-body integrals; 2) $\ket{\nabla_{I\alpha}\Phi}$ accounts the ``orbital relaxation" effect on gradients, its analytical form can be derived from the widely used ``Lagrangian formalism" or ``Z-vector" techniques~\cite{Helgaker:1982tb, Handy:1984tq}).

\subsubsection{Concrete problem instantiations}
Though substantial advancements have been achieved in the development of quantum algorithms for solving electronic structure problems, the majority of this work has been predominantly focused on the static ground state properties as shown in \emph{Application 1}, with only a limited number of studies addressing the challenge of computing forces using quantum computers. In our application, we specifically explore the use of transition metal complexes as catalysts (illustrated in Figure~\ref{fig:co_water}) to showcase the potential of quantum computing in the ab initio QMD simulations of catalysts. 

{\bf Measuring forces on quantum computers.}
$\frac{d\hat{H}}{d\boldsymbol{R}}$ in Eq.~\ref{eq:force} contains the derivatives of one-electron and two-electron integrals, which can be efficiently computed on classical computers with a computational complexity of $O(N^4)$. $\frac{\partial E}{\partial \hat{H}}$ involves the ``unrelaxed" one- and two-body reduced density matrix (1-RDM and 2-RDM), which can be readily measured on QC once the quantum circuits for the ground state wavefunction are converged,
\[
\rho^1_{pq}\equiv \frac{\partial E}{\partial h_{pq}}
=\bra{\Phi(\theta)}a^\dag_q a_p\ket{\Phi(\theta)}
\]
\[
\rho^2_{pqrs}\equiv \frac{\partial E}{\partial h_{pqrs}}
=\bra{\Phi(\theta)}a^\dag_q a^\dag_p a_r a_s\ket{\Phi(\theta)}.
\]
Though the ``orbital relaxation'' part $\left(\frac{\partial E}{\partial \theta}\frac{d\theta}{d\boldsymbol{R}}\right)$ is nontrivial, the analytical formalism can be derived via the Lagrangian formalism~\cite{Henrik:1990jcp, Parrish:2021arxiv}, ending up with a formula that is functional of 1- and 2-RDMs. As a result, the number of terms that need to be measured on quantum computers also scales as $O(N^4)$. Since the force measurements adapt the same quantum circuits as the ground state energy estimation, there is no additional phase estimation, introducing only a very low computational overhead compared to energy estimation~\cite{o2022efficient,o2019calculating}.

{\bf Sampling of initial conditions.}
In Ab Initio QMD simulation of chemical reactions, the initial conditions of reactants are sampled to reflect thermal fluctuations. Then, the QMD simulates how these conditions evolve over time, specifically focusing on the trajectories that lead to product formation. The quantum yield, representing the efficiency of the reaction, is calculated by the proportion of successful transitions to product states among all trajectories. Achieving a statistically converged quantum yield requires sampling a sufficient number of trajectories to accurately represent the thermodynamic ensemble of initial states. The exact number needed depends on the reaction complexity and the variability introduced by thermal fluctuations. Typically, hundreds of trajectories ($\sim$ 300, for example~\cite{Nelson:2018cr}) may be necessary to ensure that the calculated quantum yield is robust and reflective of the true reaction dynamics under study.

It should be noted that MD trajectories can be trivially paralleled over a larger number of processors as each trajectory is independent. In addition, the expensive measurement of energies and forces at each MD step can be bypassed by the recently proposed quantum-classical Liouvillian dynamics methods. Based on the Koopman–von Neumann formulation of classical mechanics, the Liouville equation of motion for nuclei can be implemented as unitary dynamics on quantum computers, and the resulting framework is able to propagate electronic and nuclear DOFs with quantum and classical dynamics, respectively~\cite{Simon:2024prxq}.

\subsubsection{List of candidate systems where a similar process is relevant?}

A similar workflow can be used for cases where the dynamical evolution of the system is required. For example, 

\begin{itemize}
    \item Designing catalysts for other chemical reactions~\cite{cat1,cat2,cat3}. 
    \item Mechanochemistry: Initiation of reactions under mechanical influence~\cite{mechno1,mechno2,mechno3}.
    \item In materials science: Ion diffusion for battery applications~\cite{bat1,bat2} and exfoliation of layered materials~\cite{exfo1,exfo2,exfo3}. 
    \item Nuclear fusion reaction. In fusion reactions, the nuclei need to overcome the Coulomb barrier to trigger the fusion reaction. The reaction is very similar to the chemical reactions, though the energy, time, and length scales are different~\cite{Back:2014tb}.
\end{itemize}

\subsection{Application 3. Fine-tuning Reaction mechanism through nuclear quantum dynamics Simulations} 

\subsubsection{Specific background of the application}
The ground state energy estimation (GSEE) in \textit{Application 1} helped to down-select catalysts for artificial photosynthesis. \textit{Application 2} examined the kinetic effect to understand the selectivity and stability of down-selected catalysts and potential by-products, enabling the study of external constraints on the reaction mechanisms. However, both approaches neglect the intricate nuclear quantum behavior and non-adiabatic electron-nuclei couplings, which could be crucial for reactions involving light elements (like Hydrogen) and multiple adiabatic states. 
Increasing evidence of significant quantum nuclear effects in chemical reactions highlights the need to account for quantum tunneling, especially in reactions involving light elements such as protons~\cite{Schiffer:2015jacs, Weinberg:2012cr}.

One such mechanism is the Proton-Coupled Electron Transfer (PCET)~\cite{Schiffer:2015jacs, Weinberg:2012cr, Dau2010}, an electron transfer coupled with proton transfer that occurs at lower overpotentials~\cite{matheu2015,liao2015}, is a typical example of reactions involving nuclear quantum effects. It is a critical mechanism in many biological and chemical processes, such as artificial photosynthesis~\cite{Mora:2018tq}, hydrogen oxidation and production~\cite{Horvath:2012vn}, nitrogen reduction~\cite{Garrido-Barros:2022us}, organic solar cells~\cite{Nocera:2022up}, and beyond. 
Additionally, while PCET is particularly significant for lighter elements, the zero-point energy (ZPE)~\cite{einstein1913einige} correction should ideally be applied for all elements. ZPE originates from the quantum fluctuations dictated by the Heisenberg uncertainty principle~\cite{heisenberg1927}; it is the residual kinetic energy of a system near 0~K~\cite{einstein1913einige}.

In QMD, efforts have been made to incorporate quantum effects by introducing quantum thermal baths (QTB)~\cite{Dammak:2009tb}. A stochastic force mimicking quantum fluctuations is introduced to allow nuclei to sample phase space like nuclear quantum simulations while accounting for ZPE to ensure accurate thermal distributions. The use of QTB has improved the alignment between classical MD and experimental data, effectively explaining the kinetic isotope effect~\cite{bigeleisen1949relative,melander1980reaction,simon1966isotope}, highlighting the significance of quantum effects. However, QTB does not fully account for all quantum phenomena, such as proton tunneling or temperature-dependent phase transitions in certain quantum materials. Hence, quantum nuclear effects are vital for accurately predicting properties like reaction rates, phase transitions, and material stability at low temperatures or in systems involving light elements.

Therefore, we advocate nuclear quantum dynamics (NQD) workflow to examine down-selected catalysts for artificial photosynthesis. As shown in Eq.~\ref{eq:tdse_enuc}, the governing equation for NQD is the time-dependent Schro\"odinger equation (TDSE) of nuclei. In previous applications, BO approximation was used to disentangle the electronic and nuclear DOFs, resulting in separate TDSEs for electrons, Eq.~\ref{eq:electronse} and nuclei (Eq.~\ref{eq:nucdyn}), respectively. However, the BO approximation breaks down when two PESs become closer (or even cross), and a strong ``non-adiabatic" coupling between electron and nuclei appears. Therefore, methods beyond the BO approximation are usually required for studying chemical reactions with strong non-adiabatic effects. The exact Born-Huang (BH) representation of molecular wavefunction is generally required when the nonadiabatic effect (and two PESs are involved) becomes essential,
\begin{equation}\label{eq:bornhuang}
    \Psi(\br,\bR,t)=\sum^\infty_I \chi_I(\bR,t)\Phi_I(\br;\bR).
\end{equation}
Here, $\Phi_I$ signifies the electronic state, commonly depicted in an adiabatic representation, and $\chi_I(\bR,t)$ is the time-variant nuclear wavefunction linked to the $I^{th}$ adiabatic state. This adiabatic representation suggests that the electronic foundation is composed of eigenfunctions of the steady-state electronic Schro\"odinger equation for a specific nuclear arrangement, defined as Eq.~\ref{eq:tdse_enuc}.
Though the Born-Huang expansion is theoretically precise only when an unlimited number of electronic states are encompassed, a good approximation can be achieved with a handful of electronic states~\cite{Curchod2018CR}.

Within the Born-Huang representation, the TDSE for nuclear wavefunction can be readily derived by inserting Eq.~\ref{eq:bornhuang} into TDSE (Eq.~\ref{eq:tdse_enuc}) and integrating out the electronic DOFs, leading to a set of coupled time-dependent equations,
\begin{align}\label{eq:ctqd}
   &i \frac{\partial \chi_I(\bR,t)}{\partial t}
   = \sum_J \left[\hat{H}_I\delta_{IJ} - C_{IJ} \right]\chi_J(\bR,t).
\end{align}
Where $\hat{H}_I \equiv \hT_n + E_I(\bR)$ is the nuclear Hamiltonian on a certain potential energy surface (defined by $V_I(\bR)$) with $3N$ dimensions.
$C_{IJ}$ describes coupling with other electronic states through
\begin{align}\label{eq:qnad}
C_{IJ}= &\sum_{\bR_{x}} \frac{1}{M_x}
   \left[\bra{\Phi_I} \nabla_{\bR_x}\ket{\Phi_J}\nabla_{\bR_x}
   +\frac{1}{2}\bra{\Phi_I} \nabla^2_{\bR_x}\ket{\Phi_J}
   \right],
\end{align}
where $d(\bR)_{IJ}$ and $D(\bR)_{IJ}$ are the first- and second-order nonadiabatic derivative couplings (NACs) that characterize couplings among the electronic states via nuclear motion, which are critical in nonadiabatic dynamics. 
Note that Eq.~\ref{eq:ctqd} is different from Eq.~\ref{eq:nucdyn} due to the rise of nonadiabatic couplings (NACs) beyond the BO approximation. In the absence of NACs, Equation~\ref{eq:ctqd} reduces to the adiabatic quantum dynamics on single potential energy surface $E_I(\boldsymbol{R})$. It further leads to QMD after taking a classical limit for the nuclear DOF. 

In contrast to the classical treatment of nuclei, the NQD, governed by Eq.~\ref{eq:ctqd}, captures significant quantum effects (even due to subtle changes in the reaction barrier). Incorporating NQD will enhance our understanding of reaction mechanisms, improve predictions of catalyst behavior and reaction outcomes, and identify efficient pathways and favorable by-products. However, solving the coupled electron-nuclear dynamics (Eq.~\ref{eq:ctqd}) on a classical computer is prohibitive due to the exponential growth of the Hilbert space. \textit{Application 3} provides a framework for leveraging quantum computation to address these challenges. The workflow is implemented with several approximations, keeping the complexity of the problem manageable while still crucially incorporating the effects of nuclear dynamics to accurately predict the reaction mechanism.

\subsubsection{Objectives}
Similar to \textit{Application 2}, where QMD simulations (classical dynamics using quantum forces) were performed to identify effective reaction pathways, NQD also involves exploring reaction mechanisms to determine the \textbf{quantum yield ($Q$) and reaction rate ($k$)} (for NQD it is defined in Sec~\ref{ap3:workflow} and for QMD it is defined in Sec.~\ref{app2-objective}) by treating nuclear dynamics quantum mechanically. Here, the {\it quantum} trajectory would be monitored to calculate probability flux  for a reactant (R) to convert into the desired product (P) via the evolution of reaction probability.  Moreover, given the significantly increased computational cost of NQD simulations, we will only focus on a subsect of promising catalysts identified in \textit{Application 1 and 2}. NQD simulations will allow us to investigate the nuclear quantum effect, such as quantum interference and quantum tunneling~\cite{Ryabinkin:2017uz, Mills:1005sc, Ceriotti:2016wt}, on reaction rates and catalyst selection. 

The perusal of the literature indicates that various strategies have been proposed for approximate classical NQD simulations. Popular methods, such as multiconfiguration time-dependent Hartree (MCTDH)~\cite{Beck:2000wm} and its Gaussian-based approximate counterpart~\cite{Burghardt:2008ut}, are numerically exact. However, the complexity of these methods increases rapidly with truncation order.
In this context, quantum computers may offer significant advantages for NQD that involve multiple potential energy surfaces and the evolution of nuclear degrees of freedom (DOF) by efficiently representing the nuclear wavefunction with a quantum circuit. The objective of this application is to provide workflows for systems where nuclear dynamics is important for an accurate prediction of the reaction mechanism. It is important to examine new reaction pathways in this framework. In NQD simulations, reaction pathways are characterized by the evolution of nuclear wavefunctions (wavepackets) along pre-computed PES. This differs from \textit{Application 1}, where the system follows a pre-defined classical path; here, nuclei evolve based on the TDSE. 
Since NQD simulations are computationally expensive, several approximations are often employed to reduce the computational cost for nuclear dynamics. First, the second quantization of the nuclear Hamiltonian relies on the choices of nuclear basis set function, which is always subject to certain truncation on the basis set. It is similar to the truncation in the Gaussian or Plane wave basis set used in electronic structure calculations. Second, all-atom quantum simulation requires a huge Hilbert space. Practical simulation of nuclear quantum dynamics usually reduces the dimensions by freezing certain nuclear coordinates and restricting the evolution of nuclear wavepacket regions where the desired reaction is expected. It is based on chemical intuition or simulations performed in \textit{Applications 1 and 2}. Such approximation is  analogous to the active space selection in the electronic structure problems. 

\subsubsection{End-to-end computational workflow}
\label{ap3:workflow}

\begin{figure}[htbp]
  \centering
  \includegraphics[width=0.80\linewidth]{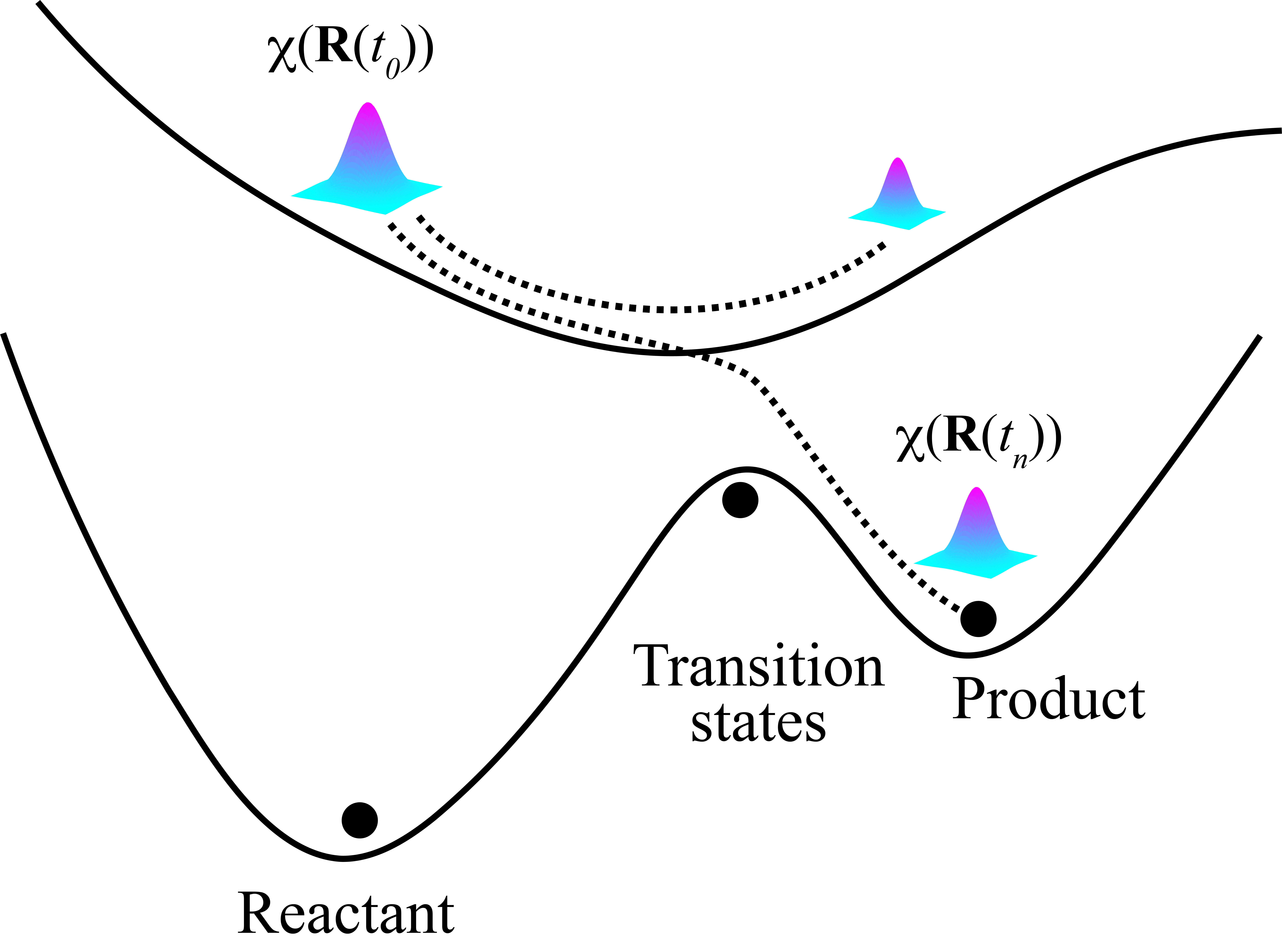}
  \caption{\label{fig:flownqd} 
  Schematic diagram illustrating the wavepacket dynamics on a multidimensional potential energy surface. Here, the reactant wavefunction at the initial time (t = 0) is represented by $\chi(\boldsymbol{R} (t_0))$, which evolves on the potential energy surfaces of low-lying (ground state and a few lowest excited states). The number of low-lying states is determined by the density of states of the molecules and photon excitation energy. For instance, if the photon excitation energy in photosynthesis is 2 eV, it is generally necessary to include excited states with energies up to around 2.2 eV—slightly higher than the excitation energy—to account for thermal fluctuations and possible initial transitions to higher-energy states.
  }
\end{figure}

\textbf{Computational Workflow.}
The computational workflow for this application is closely related to that of \textit{Application 2}. Still, with a key distinction: instead of classical nuclear dynamics, we simulate the wavepacket dynamics of nuclei on a pre-computed PES using the TDSE. 
The first approximation considered in this application is the decoupling of electron and nuclear motion (BOA), or the evolution of nuclei on pre-computed PES. 
Figure~\ref{fig:flownqd} outlines the schematic diagram of the NQD simulation process for examining the reaction pathway and products for artificial photosynthesis and any chemical reactions at large. This workflow begins by selecting a reaction involved in artificial photosynthesis and a potential catalyst, down-selected from \textit{Application 1 and 2} (K$_1$ to K$_n$). The workflow is tasked with choosing the reactant (R) and catalyst (K). Next, the multidimensional PES, due to bond stretching, bending, torsion, etc., are constructed along several possible transition states. This involves calculating ground-state and a few low-lying excited states PES. The density of states of the molecules and photon excitation energy determines the number of low-lying states. Usually, the highest state involved in the simulations should have energy larger than the photon excitation energy in order to account for the thermal fluctuation and possible initial transitions to higher-energy states~\cite{Nelson:2018cr}. For example, if the photon excitation energy is 2 eV, the excited states with energies up to around 2.2 eV are included. Here, the energy estimation subroutine mentioned in \textit{Application 1} is used to calculate PES for low-lying states. However, to keep the computational effort reasonable, the PES is generated on a sparse grid (for example, 32 per dimension~\cite{Wu:2020vc}) while ensuring that the PES remains smooth. 

A nuclear wavepacket represents the quantum state of nuclei (both position and momentum). Quantum circuits encode an initial wavepacket to represent the system's initial condition at the reactant states. For this, the nuclear coordinates of molecules are mapped using Discrete Variable Representation (DVR), a numerical method representing operators and wavefunctions on a discrete grid. This initial wavepacket is a linear combination of the DVR basis functions (each grid point). Different initial conditions are prepared to mimic the experimental conditions. The nuclear kinetic operators ($\hat{T}_n=-\frac{1}{2m}\nabla^2_{\boldsymbol{R}}$) and the multidimensional PES $\{E_I(\boldsymbol{R})\}$ are then mapped into the qubit Hamiltonian (more details are discussed in Sec.~\ref{sec:nqdtheory}), which will later be used to propagate the nuclear wavepacket on quantum devices. During the TDSE simulation, the nuclear wavepacket may undergo nonadiabatic transitions between adiabatic (ground and excited) states as it evolves.

Unlike \textit{Application 2} where \textbf{quantum yield} is defined as $Q =\frac{N_{\text{reacted}}}{N_{\text{all}}}$, measuring the probability that reactants convert to products. In \textit{Application 3}, the \textbf{quantum yield} is determined by the maximum value of the product state probability density, $\rho(\boldsymbol{R}, t) =|\chi(\boldsymbol{R}, t)|^2$, representing the likelihood of the system transitioning to the product state over time. This highlights the probability of the system being in the product state, $Q =  \int_{\text{product}} \left| \rho(\boldsymbol{R}, t) \right| \, d\boldsymbol{R}$. 
This probability flow provides insights into the minimum energy reaction pathways. It indicates whether the system remains in the initial state, transitions to an intermediate state, or evolves to the desired product state. For instance, in artificial photosynthesis, electrons and protons are the desired products of the water oxidation reaction. CO or other fuels are the desired products of a \ce{CO2} reduction reaction. Analyzing reaction probability flows helps identify the most energy-efficient (minimum energy reaction) pathways. In an NQD simulation, the greater the probability density in the product region over time, the higher the reaction rate. The \textbf{reaction rate ($k$)} is intrinsically related to $\rho(\boldsymbol{R}, t)$ and $Q$, and $k$ is calculated by integrating the probability flux ($J(\boldsymbol{R}, t)$) over time, with the probability flux being $J(\boldsymbol{R}, t)= \int_{\text{product}} \chi^{*}(\boldsymbol{R}, t) [\delta(\boldsymbol{R}-\boldsymbol{R_P})\nabla_{\boldsymbol{R}}]\chi(\boldsymbol{R}, t)d \boldsymbol{R}$. The reaction rate is $k = \frac{\langle J \rangle}{\rho_{\text{reactants}}}$ and $
\langle J \rangle = \frac{1}{t_{\text{total}}} \int_0^{t_{\text{total}}} J(\boldsymbol{R}, t) \, dt$. Here, probability flux describes the flow of probability density across a surface that divides reactant and product regions. This workflow offers a way to implement the role of quantum effects on various reaction mechanisms in artificial photosynthesis.\\

\begin{figure}[htbp]
\centering
\includegraphics[width=0.98\linewidth]{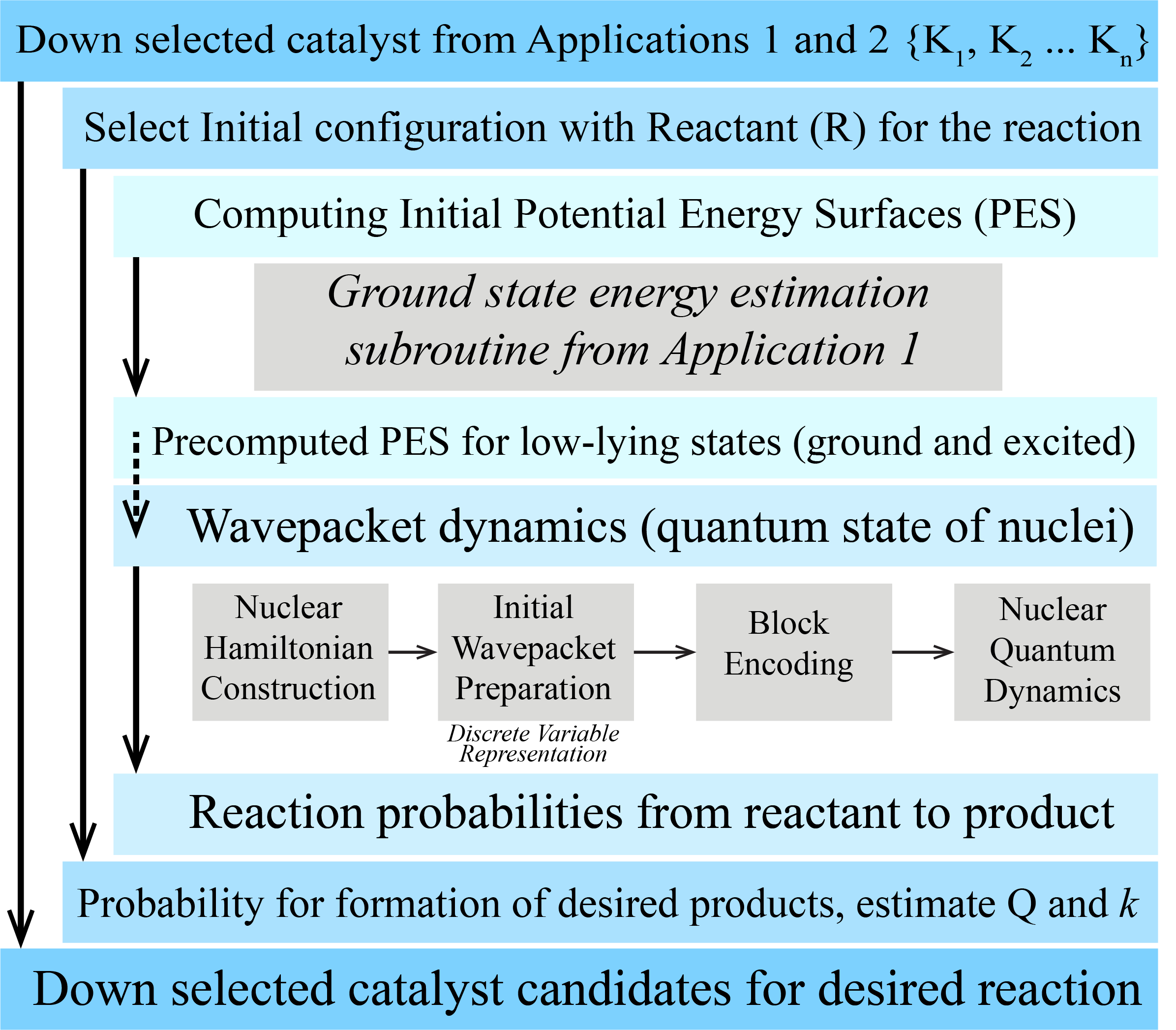}
\caption{\label{flowchart-nuclear} Flowchart of selecting catalysts for different reaction mechanisms of artificial photosynthesis using nuclear quantum dynamics. 
} 
\end{figure}

\noindent\textbf{Hard Computational Modules:}
Similar to \textit{Applications 1 and 2}, the first ``hard computational module" of NQD simulation is to calculate the PES, which involves estimating the ground state energy. Nevertheless, unlike \textit{Application 2}, where ground state energies are calculated at each timestep to estimate forces during system evolution, here, PES for low-lying states are calculated once. Though, as established in earlier applications, quantum computers provide significant advantages in efficiently calculating ground and lowest excited states energies. The next ``hard computational module" of NQD simulation is the computationally challenging task of evolving the nuclear wavepackets on the pre-computed PESs. Each timestep, the nuclear Hamiltonian propagates the wavepacket to show the system’s current state. Therefore, quantum computing can be utilized to solve these challenges (PES calculation and wavepacket dynamics) efficiently with higher accuracy. 

Another challenging step associated with NQD simulations is accurately representing and propagating a nuclear wavepacket in a high-dimensional space. The computational cost of the DVR representation of nuclear coordinates exponentially increases with the number of nuclei. This also increases the memory requirement for saving the nuclear wavefunctions. Furthermore, performing TDSE simulation for nuclear wavepacket dynamics is computationally expensive and challenging using classical computers. Therefore, by leveraging quantum parallelism and entanglement, quantum computing provides a promising solution to overcome the computationally demanding tasks in NQD simulations, enabling substantial progress in quantum chemistry and artificial photosynthesis processes.

\noindent The whole workflow shown in Figure~\ref{flowchart-nuclear} can be summarized as follows:

\begin{enumerate}
    \item \textbf{Select Reaction Mechanism}: Begin by choosing the reaction mechanism involved in artificial photosynthesis, water oxidation reaction, or \ce{CO2} reduction reactions.
        
    \item \textbf{Select Catalyst from Pool of Candidates}: Accordingly choose the potential catalysts, $K_1, K_2, \ldots, K_n$. Note that we only use down-selected catalysts from \textit{Applications 1 and 2}.
    
    \item \textbf{Select Initial Configuration}: Select initial configuration with reactant and catalyst. 

    \item \textbf{Initial Potential Energy Surface (PES)}: The ground state energy estimation subroutine from \textit{Application 1} is used to compute the low-level PES.

    \item \textbf{Wavepacket Dynamics}: Perform nuclear wavepacket dynamics for the system.
        
    \begin{enumerate}
    
    \item \textbf{Nuclear Hamiltonian Construction}: The nuclear kinetic operators and PES are used to construct the nuclear Hamiltonian.

    \item \textbf{Initial Wavepacket Preparation}:  
    There are different ways of preparing initial states. Depending on the physics of the problems, we can prepare the initial state as a trivial product state or a more complicated form (like superpositioned or entangled states)~\cite{Beck:2000wm}. Here, we take the adiabatic state as the initial instate. To this aim, the (nuclear) ground state wave function is first determined by a quantum algorithm, which is used as the initial wave packet on an excited state within the Frank-Condon approximation~\cite{Wu:2020vc}. 
    
    \item \textbf{Block Encoding}: Encode the nuclear wavepacket and Hamiltonian onto a computer. 

    \item \textbf{Nuclear quantum dynamics}: The nuclei are propagated on PES by solving the TDSE.
    \end{enumerate}
    
    \item \textbf{Evolution Over Time}: Estimate the reaction probabilities and identify reactions that result in the formation of diverse products. 
    
    \item \textbf{Reaction rate Calculation}: Estimation of the probability density and flux for the formation of the desired products, facilitating the computation of quantum yield ($Q$) and reaction rate ($k$).
        
    \item \textbf{Catalyst Selection for Reactions}: Select the ideal catalyst that best facilitates the reaction.

    \item \textbf{Selected Catalysts for Artificial Photosynthesis}: List all promising catalysts for reactions, both water oxidation reaction and \ce{CO2} reduction reactions.
\end{enumerate}

\subsubsection{Nuclear quantum dynamics encoding}
\label{sec:nqdtheory}

As shown in Eq.~\ref{eq:ctqd}, the NQD can be readily simulated on quantum computers by the Hamiltonian simulation once the nuclear Hamiltonian is encoded into Pauli strings. Solving the Schr\"odinger equation for nuclei requires the potential energy surface $E_I(\boldsymbol{R})$ at arbitrary position $\boldsymbol{R}=(R_1, R_2, \cdots, R_{3N})$ with a dimension of $M=3N$ (or $3N-6$ and $3N-5$ for nonlinear and linear molecules if rotational and translational invariance are utilized), where $N$ is the number of atoms. This is achieved by precomputing PES on discretized grids. Note, the computation of full-dimensional PESs exactly for even a small molecule is a classically intractable task, as described by the challenge of \textit{Application 1}. 

Nevertheless, after PESs are available, the TDSE for nuclear motions can be numerically solved in conjunction with a certain basis set for nuclear wavefunction. To achieve this, we need to encode the nuclear Hamiltonian in terms of Pauli operators to simulate NQD on quantum computers. Different approaches are used to construct the Second-Quantized Hamiltonian and subsequent mapping into qubit format. Here, we discuss two approaches: 1) grid-based representation and 2) vibrational mode-based representation. The former represents the nuclear wavefunction in terms of basis set functions on a product of grids along each direction. While the latter expresses the nuclear wavefunction in terms of harmonic oscillator eigenstates (which is the superposition of position states). We will start with the former and briefly discuss the latter approach.

{\bf Grid-based encoding.} Discrete variable representation (DVR)~\cite{light1985generalized, colbert1992novel} and finite-basis set are the two widely used representations~\cite{Gatti2014}. Without losing generality, we use the DVR representation to discuss the quantization of nuclear Hamiltonian. The DVR discretizes the nuclear coordinates onto a grid and simplifies the complex NQD problem by ensuring no coupling between grid points and sparse matrix presentation. These grids are defined using orthogonal polynomials, making the basis functions localized. 
With the generalized multidimensional DVR~\cite{light1985generalized,colbert1992novel}, the nuclear Hamiltonian can be rewritten as
\begin{equation}\label{eq:nucH_dvr}
   \hat{H}_I = \sum^{M}_{a}\hat{T}^a + \sum^L_{q_1\cdots q_M} E_I(x_{q_1}, \cdots, x_{q_M})\ket{q_1,\cdots, p_M}\bra{q_1, \cdots, q_M}
\end{equation}
Where $M$ is the number of dimensions. $L$ is the number of grids per dimension. Hence, there are $L^M$ grids in total. The kinetic energy operator of the $a$-th dimension ($T^a$) is calculated using matrix representation (non-diagonal matrix) from orthogonal polynomials,
\begin{equation}
    T^a_{ij} = \sum_{k=1}^{L} \langle \xi_i | \frac{d^2}{dR^2_a} | \xi_j \rangle
\end{equation} 
where $\ket{\xi_i}$ are the DVR functions. The kinetic energy can be estimated by approximating the second derivative of the wavefunction at each point using a grid-based finite-difference method.
For the potential energy operator $E_I(\bR)$, the corresponding matrix element $E_{I, ij}$ between two DVR basis functions $\chi_i$ and $\chi_j$  at grid points $\bR_i$ and $ \bR_j$ is defined as, $E_{I, ij}=E_I(\bR_i)\delta_{ij}$, and 
\begin{equation}
    E_I(\bR)=\text{diag}(E_I(\bR_1), E_I(\bR_2),..., E_I(\bR_n)),
\end{equation} 
i.e., the potential matrix becomes diagonal in DVR. In principle, finer grids improve the numerical accuracy but, at the same time, increase the computational cost. Practical calculations need to balance accuracy and efficiency, in analogy to the truncation in the basis set of electronic structure calculations. Besides, the nuclear Hamiltonian in the DVR representational is generally very sparse. This is because 1) the potential operator is diagonal, and 2) the kinetic operator is the trivial product of operators of each dimension. Such sparsity can be utilized in Hamiltonian simulation to reduce the number of quantum gates.

{\it Direct mapping.}
With the Hamiltonian in the DVR representation, we can directly simulate the nuclear dynamics on quantum computers by mapping the DVR Hamiltonian into Pauli strings either via {\it direct} or {\it binary} mappings. 
Within the direct mapping, each position states $\ket{p_m}$ is encoded into 1 qubit (one-to-one mapping),
\begin{equation}
    \ket{q_m} = \prod^{m-1}_{j=0}\ket{0}_j \ket{1}_m \prod^{L-1}_{j=m+1}\ket{0}_j.
\end{equation}
Therefore, for a given nuclear Hamiltonian with $M$ DOFs and $L$ DVR basis set per dimension, the nuclear wavepacket is represented by $ML$ qubits, leading to significant qubit resource requirements. Hence, an efficient coding technique to represent the large number of position states is essential to reduce the qubits requirements. 

{\it Binary mapping.}
Within the binary mapping, the position operator associated with a grid point $m$ can be encoded as
\begin{equation}
  \ket{p_m}= \otimes^{K-1}_{j=0} \ket{b_j} \equiv \ket{b_{K-1}}\ket{b_{K-2}}\cdots\ket{b_0}
\end{equation}
where $m = \sum^{K-1}_{j=0} b_j 2^j$. And each pair of $\ket{b_K}\bra{b^{'}_K}$ are expressed in terms of the Pauli matrices. With this scheme, the DVR Hamiltonian with $ML$ grid points will be encoded into Pauli Hamiltonian with $M\log_2(L)$ qubits.

{\bf Vibrational mode based encoding.} Alternatively, we can rewrite the nuclear Hamiltonian in the eigenstate of quantum harmonic oscillators (vibrational mode). There are two steps to obtain the nuclear Hamiltonian in the vibrational mode representation. First, the vibrational modes are computed by employing the harmonic approximation to the PES, 
$E_I(\boldsymbol{R}) \simeq \frac{1}{2}\sum_{\alpha} \omega^2_\alpha \boldsymbol{Q}^2_\alpha$, where $\boldsymbol{Q}_\alpha$ is the nuclear vibrational mode coordinates. This procedure is analogous to the Hartree-Fock (HF) calculations used in electronic structure calculations. Like the HF molecular orbitals are a linear combination of atomic basis sets, nuclear vibrational modes are a linear combination of original nuclear coordinates, representing the collective motions of the nuclei. They are not associated with individual nuclei but with the system as a whole.
Second, we rewrite the potential $V(\boldsymbol{R})$ in the vibrational mode representation~\cite{McArdle:2019wf}, which results in the high-order interacting terms among the vibrational mode, $V(\boldsymbol{R})= \sum_{J\geq 2} K_J \boldsymbol{Q}^J$. This step is analogous to the second quantization of the electronic Hamiltonian in the HF molecular orbitals. However, unlike the electronic Hamiltonian, many-body expansion of the original nuclear Hamiltonian in the vibrational mode representation ends up with arbitrarily high coupling terms~\cite{McArdle:2019wf}. If we truncated the many-body interaction to second order (two-body interaction), the quantized nuclear Hamiltonian can be expressed in a similar manner as the electronic ones
\begin{equation}\label{eq:manyH}
    \hH^n = \sum_{pq}\epsilon_{pq}a^\dag_p a_q + \sum_{prqs}V_{pqrs}a^\dag_p a^\dag_q a_r a_s. 
\end{equation}
where $p,q,r,s\in\{M_k\}$ where $M_k$ is the number of basis set for the $k^{\rm th}$ dimension (of the $d$-dimensional system). Even though usually second or third-order expansions are sufficient to obtain high accuracy in the vibrational structure according to previous studies~\cite{Ollitrault:2020wh, McArdle:2019wf}, higher order terms are needed for nuclear quantum dynamics.

With the vibrational mode representation, the original nuclear Hamiltonian (which can be a boson or a mixture of fermions and bosons depending on the spins of nuclei) is rewritten as the vibrational Hamiltonian, which was used to compute the vibrational structures. 
The nuclear Hamiltonian in the vibrational mode representation can be easily mapped to qubits via the boson-to-qubit mapping. Unlike the fermions, where the particles occupy mutually orthogonal orbitals, there is no such restriction on the bosons where multiple bosons can occupy the same orbitals. The number of Fock states for each particle can be infinite. In practice, we will truncate the Fock state to a certain number of $n$. Hence, unlike the fermionic case, where one spin-orbital can be naturally mapped to one qubit, one bosonic orbital requires more than one qubit to encode. Similar to the encoding of DVR Hamiltonian in terms of Pauli strings, the Hamiltonian in the vibrational mode representation can also be encoded into qubit Hamiltonian via either direct~\cite{Somma:2003wn, Dumitrescu:2018vu} or binary~\cite{McArdle:2019wf} mappings. Here, we take the binary mapping as an example. 
In binary mapping, a binary representation is used to encode the Fock state~\cite{McArdle:2019wf}
\begin{equation}
    \ket{\nu}=\otimes^{K-1}_{j=0} \ket{b_j} \equiv \ket{b_{K-1}}\ket{b_{K-2}}\cdots\ket{b_0}
\end{equation}
where $\nu=\sum^{K-1}_{j=0}b_{j}2^j$. As a result, only $K=[\log_2(n)]$ qubits are needed to encode one mode, similar to the binary encoding of one DVR dimension discussed above.

\subsubsection{Why classical methods are not sufficient to perform the hard computational module}

As shown by Eq.~\ref{eq:nucH_dvr}, although the kinetic operator of nuclei is decoupled, the nuclei are correlated via interaction potential $V$, making the total wavefunction not a product of single-particle wavefunctions. Classical algorithms face significant challenges when using DVR for NQD simulation. As mentioned earlier, the number of DVR grid points increases exponentially with the number of nuclear coordinates. For a system with $M$ nuclear coordinates and $L$ grid points per coordinate, the total number of grid points becomes $L^M$. Consequently, the memory requirements for saving wavefunctions, performing complex Hamiltonian matrix operations, and the computational cost of time evolution increase drastically. To overcome these challenges, classical methods often employ dimensionality reduction, basis set truncation, tensor decomposition, and low-rank approximations to the nuclear wavefunction~\cite{Larsson:2024tt}. Dimensionality reduction involves reducing nuclear DOF to the most influential ones that affect reactions, fixing non-influential nuclear coordinates during simulation, and reducing the dimension of PES (reaction pathways). The truncation approach involves limiting DVR basis functions, using sparse and adaptive DVR grids, and approximating wavepacket propagation. The tensor decomposition and the low-rank approximation represent the many-body wavefunction of the low-rank product of single-particle functions and truncate many-body interaction at lower orders~\cite{Beck:2000wm}. These limitations indicate that quantum computers are ideal for NQD simulations, as they can naturally represent the exponentially large nuclear wavefunction with quantum circuits. 

Here, we take the cases in Table~\ref{tab:comparemapping} as examples to demonstrate the limitations of a classical algorithm for the NQD.  We assume 256 grid points per dimension. 
In the case of \ce{Co2O9H12} catalyst for the water oxidization, there are 26 atoms and $72$ dimensions. And for the case of \ce{CO2-CoPc} reactions, the total nuclear DOFs is $174$. Consequently, the all-atom NQD with 256 grid points for each dimensional requires $256^{72}$ grids and $256^{174}$ for \ce{H2O-Co2O9H12} and \ce{CO2-CoPc} cases, respectively. Even if we only consider the atoms in \ce{Co2O9H12} (or \ce{CoPc}) catalyst that have direct interaction with the water molecule (or \ce{CO2}) and freeze other atoms, we still need at least 5 atoms, three water atoms and two \ce{Co} atoms, or 7 atoms, three atoms in $\ce{CO2}$, three atoms in water, one \ce{Co} atoms. Note this is a very crude approximation as other atoms will also change configurations during the reactions). The simulation will still require 15 (or 21) dimensions since now there is no rotational and translation invariance as other atoms are frozen, leading to $256^{15}\simeq 10^{36}$ (or for $256^{21}\simeq 10^{50}$) grid points, which still is an intractable task for classical computers to store the operators. Consequently, such a system is already very challenging for classical methods, like MCTDH, even with additional approximations (truncation at small order). This is also the reason why the current state-of-the-art classical nuclear quantum dynamics methods can only treat a small number of dimensions ($<10$) even after many numerical ticks are applied~\cite{Beck:2000wm}.

\begin{table}[!htb]
  \centering
  \caption{Resource requirements for nuclear quantum dynamics}
  \label{tab:comparemapping}
  \begin{tabular}{c|c|c|c}

  \hline\hline
  Systems   & Atoms ($N$) & Reduced dimensions ($M^*$) & Number of grids per dimension ($L$)   \\ \hline 
  \ce{H2O}-\ce{Co2O9H12} & 26 & 15 & 256$^{\#}$ \\ \hline
  \ce{CO2-CoPc}          & 60 & 21 & 256 \\ \hline\hline
  \end{tabular} \\
  $^*$ Analogous to active space selection in electronic structure calculations\\
  $^{\#}$ A typical number we have used in nuclear quantum dynamics~\cite{Wu:2020vc} though the optimal number is system-dependent.
\end{table}

\subsubsection{Concrete problem instantiations}
Here, we discuss concrete instance requirements for studying the quantum dynamics of $N$ nuclei on a parameterized potential energy surface $V(\boldsymbol{R})$. 

It is essential to recognize that accurately computing PES is a complex and nontrivial task. It's even a subcommittee of many research groups that spend much effort calculating and fitting these PESs for different (usually small) molecules for community uses~\cite{shu_potlib}. 
Hence, to keep the resource requirement reasonable for the test case, we chose a toy model of \ce{H2O} molecule as a sample. There are a total of 3 (i.e., $3N-6$) nuclear DOFs after excluding the rotational and translational DOFs. The qubits Hamiltonians with different encoding methods are shown in the GitHub repository \cite{qc-applications-notebooks}.

In the previous section, we showed that NQD simulation is classically intractable due to the exponential growth in the memory requirement for expressing the nuclear many-body wavefunction. 
In contrast to the huge memory requirement for studying the cases in Table~\ref{tab:comparemapping}, the nuclear function's encoding only requires $15 \times 256=3840$ and $15\times\log_2(256)=120$ qubits on quantum computers with the direction and binary mappings, respectively, leading to an exponential advantage in the expression capability. Even if we consider all the atoms in the \ce{H2O-Co2O9H12} system or half of the atoms in the \ce{Co-Pc} (the ligand outside the $N$ ring in the \ce{Co-Pc} molecules has a small effect on the reaction), the resulting 72 and $\sim90$ nuclear DOFs in the \ce{H2O-Co2O9H12} and \ce{Co-Pc} systems only requires $552$ and 720 qubits, respectively. In contrast to the exponential complexity of classical algorithms, the polynomial scaling of qubits requirements makes it possible to run NQD with a larger number of nuclear DOFs. 
For simulation, we run 1ps quantum trajectories with 100 steps per dynamics to track probabilities along the time evolution.

The table in Sec.~\ref{sec: summary} outlines the quantum resource requirements for NQD simulations on quantum computers. We set a minimum of 15 nuclear DOFs to achieve reasonably realistic NQD simulations, corresponding to an effective system of 5 atoms. Our target, however, is to reach approximately 90 nuclear DOFs. While the total number of DOFs, \(3N - 6\), can exceed 90 for systems with more than 30 atoms, it’s generally unnecessary to include all atoms in NQD simulations; atoms or dimensions with negligible changes during reactions can be frozen, reducing resource requirements, similar to active space selection. If quantum resources allow for more than about 720 qubits, it may be more advantageous to allocate additional qubits to increase grid resolution in the simulations.

\subsubsection{List of candidate systems where a similar process is relevant?}

A similar workflow can be used for cases mentioned in \textit{Application 2}, where dynamical evolution of the system is required, and the quantum effect is important. Some additional examples are

\begin{itemize}
    \item Hydrogenases (enzyme), understanding the hydrogen binding and release mechanisms~\cite{hat1}.
    \item Vibrational energy transfer and mode coupling, cis-trans Isomerization in chemistry and biology~\cite{hat2,hat3}
    \item Isotope effects: Hydrogen-Deuterium exchange reactions~\cite{hat4,hat5} and transmutation in fusion reactors~\cite{hat6}. 
    \item In materials science, hydrogen storage in metal hydrides, such as hydrogen absorption and release in magnesium-based hydrides~\cite{hat7}.
\end{itemize}

\section{Requirements summary}
\label{sec: summary}
This section summarizes the application requirements for all the applications described in this chapter.
\subsection*{Application 1: Energetics of Reaction Pathways for Down Selection of Catalysts for Artificial Photosynthesis}

\begin{small}

\begin{tabular}{ |l l l| }
    \hline
        & & \\
     & Total time limit & 1 week per catalyst candidate \\
     & & (screening of 100 catalysts in 2 years)
     \\ [1ex]
     & & \\[1ex]
    \textbf{Workload:} & Number of subroutine calls required & 100 = 20 (transition states) \\
    & & $\times$ 5 (reaction pathways per catalyst)
    \\ [1ex]
     & Maximum subroutine time limit & 1.6 hours \\ [1ex]
     & Model type & Fermionic Hamiltonian / Electronic Structure \\[1ex]
     & Size & minimum $64$ spin-orbitals\\
     & & target $650$ spin-orbitals \\[1ex]
     \textbf{Problem specifications:} &  Interaction Structure & Sparse irregular \\[1ex]
     & Computational target & ground state \\[1ex]
     & Accuracy requirement & 1 mHartree  \\
     & & \\
     \hline
\end{tabular}

\subsection*{Application 2: Reaction Dynamics through Quantum Molecular Dynamics Simulations}

\begin{tabular}{ |l l l| }
    \hline
        & & \\
     & Total time limit & 1 month per catalyst candidate \\
     & & (most promising screened candidates)
     \\ [1ex]
     & & \\[1ex]
    \textbf{Workload:} & Number of subroutine calls required & 100,000 = 1000 (steps of molecular dynamics) \\
    & & $\times$ 100 (initial conditions per catalyst)
    \\ [1ex]
     & Maximum subroutine time limit & 0.5 minute \\
     & & (progress on the classical side required)
     \\ [1ex]
     & Model type & Fermionic Hamiltonian / Electronic Structure \\[1ex]
     & Size & minimum $64$ spin-orbitals\\
     & & target $650$ spin-orbitals \\[1ex]
     \textbf{Problem specifications:} &  Interaction Structure & Sparse irregular \\[1ex]
     & Computational target & ground state and forces \\[1ex]
     & Accuracy requirement & 1 mHartree  \\
     & & \\
     \hline
\end{tabular}

\subsection*{Application 3: Fine-tuning Reaction Mechanism through Nuclear Quantum Dynamics Simulations}

\begin{tabular}{ |l l l| }
    \hline
        & & \\
     & Total time limit & 1 month per catalyst candidate \\
     & & (most promising screened candidates)
     \\ [1ex]
     & & \\[1ex]
    \textbf{Workload:} & Number of subroutine calls required & 100 \\ [1ex]
     & Maximum subroutine time limit & 7.2 hours \\ [1ex]
     & Model type & Nuclear Hamiltonian simulation\\
     & & and PES computation \\[1ex]
     & Size & minimum $15$ dimensions\\
     & & target $90$ dimensions \\[1ex]
     \textbf{Problem specifications:} &  Interaction Structure & Sparse irregular \\[1ex]
     & Computational target & Reaction probability and rate \\[1ex]
     & Accuracy requirement & 10\% in Reaction probability Rate \\
     & & (related to 1 mHartree accuracy in\\
     & & barrier and PES computations)  \\
     & & \\
     \hline
\end{tabular}

\end{small}

\section{Quantum implementation}
An example of the workflow explained in Application 1 using an example of water oxidation reaction with 8 transition states and an active space of 28 electrons in 19 orbitals (38 spin orbitals) with quantum implementation and resource estimation can be found in \cite{qc-applications-notebooks}.  Additionally, a sample script for encoding the nuclear Hamiltonian using a \ce{H2O} molecule toy model with three nuclear DOF (i.e., $3N - 6$, excluding rotational and translational DOFs) is also provided in \cite{qc-applications-notebooks}.

\newpage
\end{chapter}

\begin{chapter}{Simulations of quantum chromodynamics and nuclear astrophysics} \label{ch:qcd}

This chapter presents applications relevant to nuclear physics. We focus on applications that probe extreme regimes which are experimentally or observationally accessible, but not simulable by known classical methods. Relevant experiments range from the Relativistic Heavy-Ion Collider (an experimental facility at Brookhaven National Laboratory) to IceCube (a neutrino observatory in Antarctica supervised by the University of Wisconsin-Madison, and the largest neutrino telescope in the world). The computational capability required is quantum Hamiltonian simulation: the Hamiltonians used are lattice discretizations of field theory Hamiltonians (QCD and an effective theory of neutrinos). Current classical approaches in all cases suffer from severe systematic errors of unknown magnitude; these systematics come from approximations that must be made in order to render the calculations classically tractable. The potential benefit of quantum computers lies in their ability to perform the simulations with all systematics under control.\\

\noindent \textbf{Hamiltonian Type}:  Neutrino Effective Field Theory and QCD Hamiltonian.\\
\textbf{Quantum Computational Kernel}: Hamiltonian Simulation.

\clearpage

\section{Application area overview}

Nuclear physics concerns itself with the many-body dynamics of the Standard Model of physics. The dynamics of the Standard Model is inherently quantum mechanical with many degrees of freedom, as particles can be freely created and annihilated at high energies. Thus a fully error-corrected large-scale quantum computer will be required to predict the exotic states of matter which emerge from quantum field theory. While essentially every sub-domain of nuclear physics can profit from a quantum computer for at least some range of its problems, two important examples addressed here are the behavior of the dense gas of neutrinos created in a core-collapse supernova and the properties of the quark-gluon plasma, a state of matter that formed and dominated the universe early in cosmological history. Both of these examples connect to world-leading experiments that have been built or are under construction by the United States. In the case of supernovae, the neutrino experiments DUNE and IceCube act as supernova neutrino observatories, capable of measuring the neutrino output of a supernova, while the Facility for Rare Isotope Beams (FRIB) (see~\cite{Horowitz_2019}) will provide important insight into the r-Process of Nucleosynthesis, which in astrophysical environments can radically modify the momentum and flavor distribution of the neutrino gas. In the case of the quark-gluon plasma, the Relativistic Heavy Ion Collider both established the existence of this state of matter, and has led the experimental investigation into its properties. Each of these facilities, while covering a myriad of physics applications (covered in~\cite{NPLRP}), hosts experiments that would benefit profoundly from the quantum calculations we describe in this section.

\section{Problem and computational workflows}

\subsection{Application 1: Supernova neutrinos}

Core-collapse supernovae are among the most catastrophic events in the universe, and play an important role in cosmological history, providing an engine by which elements heavier than iron can be synthesized. Supernovae simulations require understanding physics at many different scales~\cite{RevModPhys.62.801,2007PhR...442...38J}. Although supernovae are ultimately driven by the microscopic mechanisms in nuclear physics, their simulation naively appears to be a problem of classical physics of transport and fluid dynamics. However, these supernovae generate a tremendous number of neutrinos, with the bulk of the energy of the supernova being carried away by the neutrinos. While generally a weakly interacting particle, the neutrino densities in a supernova are such that the flavor and momentum distributions can be significantly impacted by their quantum mechanical self-interactions, and new dynamical phenomena may occur beyond the well-understood vacuum oscillations of neutrino flavor (see review~\cite{Volpe:2023met}).  For a complete treatment of the physics of a supernova, a critical question to answer is the exact momentum and flavor distribution of the neutrinos during the course of the supernova. At such large densities, traditional classical computational methods of kinetic theory become impractical, as it is necessary to determine the time-evolution of the full many-body density matrix, and recently, much effort in the field has been devoted to using quantum computation and quantum information science as a way forward~\cite{Balantekin:2023qvm}.

\subsubsection*{Specific background of the application}

Tracing out a static nuclear and electron background that the neutrinos are embedded in, the hamiltonian for the neutrinos is given by~\cite{SIGL1993423} (see also~\cite{Volpe:2023met} and further references therein):
\begin{align}\label{eq:H_nunu}
\hat{H}&=\hat{H}_{1}+\hat{H}_{2}\,,\\
\hat{H}_1&=\sum_{f,f'}\sum_{s,s'}\int d^3\vec{p} B_{ff',ss'}(t,\vec{p})\hat{a}^{\dagger}_{f,s}(\vec{p})\hat{a}_{f',s'\vec{p}}\,,\\
\hat{H}_2&=\sum_{f_1,f_2,f_3,f_4}\sum_{s_{1},s_{2},s_{3},s_{4}}\int d^3\vec{p}_{1}d^3\vec{p}_{2}d^3\vec{p}_{3}d^3\vec{p}_{4}C(\vec{p}_1^{f_1s_1},\vec{p}_2^{f_2s_2},\vec{p}_3^{f_3s_3},\vec{p}_4^{f_4s_4})\delta^{(3)}\Big(\sum_{i}\vec{p}_i\Big)\nonumber\\
&\qquad\qquad\qquad\qquad\qquad\qquad\qquad\qquad\qquad\qquad\times\hat{a}^{\dagger}_{f_3,s_3}(\vec{p}_3)\hat{a}_{f_1,s_1}(\vec{p}_1)\hat{a}^{\dagger}_{f_4,s_4}(\vec{p}_4)\hat{a}_{f_2,s_2}(\vec{p}_2)\,.
\end{align}
The operators $\hat{a}^{\dagger}_{f,s,\vec{p}},\hat{a}_{f,s,\vec{p}}$ are fermionic creation and annihilation operators with anti-commutation relations that create and annihilate states of flavor $f$, spin $s$ and momentum $\vec{p}$ (above $f_i$ and $s_i$ refer to flavor and spin labels of the operators):
\begin{align}[\hat{a}_{f,s}(\vec{p}),\hat{a}_{f',s'}(\vec{q})]_{+}&=0\,,\\
[\hat{a}^{\dagger}_{f,s}(\vec{p}),\hat{a}^{\dagger}_{f',s'}(\vec{q})]_{+}&=0\,,\\
[\hat{a}^{\dagger}_{f,s}(\vec{p}),\hat{a}_{f',s'}(\vec{q})]_{+}&=2(2\pi)^3 E_f(\vec{k})\delta_{ff'}\delta_{ss'}\delta^{(3)}(\vec{p}-\vec{q})\,,\\
E_f(\vec{k})&=\sqrt{\vec{k}^2+m_f^2}\,.
\end{align}
In the notation of Ref. \cite{Dixon:1996wi}, the coefficient $C$ for relativistic neutrinos is easily expressed in terms of helicity projected 2-spinors, as helicity and chirality dependence become identical, and is given as:
\begin{align}  C(\vec{p}_1^{f},\vec{p}_2^{f'},\vec{p}_3^{f},\vec{p}_4^{f'})=\frac{G_F}{\sqrt{2}} \langle p_1|\gamma^{\mu}|p_3]\langle p_2|\gamma_{\mu}|p_4]\,.
\end{align}
For neutrino-anti-neutrino annihilation we have:
\begin{align}  C(\vec{p}_1^{f},\vec{p}_2^{\bar{f}},\vec{p}_3^{f'},\vec{p}_4^{\bar{f}'})=\frac{G_F}{\sqrt{2}} \langle p_1|\gamma^{\mu}|p_2]\langle p_3|\gamma_{\mu}|p_4]\,.
\end{align}
The hamiltonians $\hat{H}_1$ and $\hat{H}_2$ represent the one and two-body neutrino hamiltonians. $G_F$ is the low energy coupling constant from the 4-Fermi effective theory of the standard model that results from integrating out the Z-boson.
The one body hamiltonian includes the neutrino kinetic energy terms, the mass mixing matrix, and the interactions of neutrinos against any fixed background of nuclear or electron matter.


Another simplifying assumption we can impose is that the neutrinos are relativistic in momenta. Given that typical nuclear reactions producing the neutrinos have an energy scale on the order of a few MeV, and the neutrino mass is less than $0.1eV$, this is a safe assumption for the neutrinos as they travel through the supernova. Treating the neutrinos as massless then forces all neutrinos to have left-handed chirality, and anti-neutrino's right-handed chirality, and we need not keep track of the spin degree of freedom. Eventually vacuum oscillations will prove important at length scales on the order of a few kilometers, but for neutrino-neutrino scattering, this is a higher order effect.

The primary goal of the theorist analyzing the problem is to calculate the final distribution of momenta and flavors of the neutrinos given a specific initial distribution. If $\hat{\rho}$ is the initial density matrix of all the neutrinos, this is the asymptotic flavor-dependent momentum spectrum:
\begin{align}\label{eq:momentum_spectrum}
N_f(t,\vec{p})&=\text{tr}[\hat{a}_{f}^{\dagger}(\vec{p})\hat{a}_{f}(\vec{p})e^{it\hat{H}}\hat{\rho}e^{-it\hat{H}}], \text{ as } t\gg T_{sn}\,.
\end{align}

We note that vacuum oscillations prohibit the neutrinos from have a fixed momentum distribution per flavor asymptotically. Physically, of course, infinite time evolution is not needed for the full hamiltonian. On a certain time scale, which we call $T_{\rm sn}$,\footnote{This is expected to be around 10s~\cite{2007PhR...442...38J}.} we can expect the neutrinos will escape the supernova, and the neutrino-neutrino ($\hat{H}_2$ above) and neutrino-matter (included in $\hat{H}_1$ above) interactions will become irrelevant, as the matter and neutrino densities plummet. From this time forward, only vacuum oscillations of the neutrinos would matter, included in the one-body term above, and are straightforward to calculate. 

\subsubsection*{Overview of the value of the application}

Scientifically, supernova challenge our capabilities to model complex thermonuclear explosions, requiring the incorporation of microscopic nuclear physics reactions, neutrino dynamics, and fluid modelling into a physically coherent and predictive package. Given that neutrinos carry $\sim 90\%$ of the energy released in such an event, proper handling of their dynamics is critical to correct modeling of the supernova. Moreover, these neutrinos can be directly observed in dedicated science experiments that are either on-going or under constructions~\cite{2012ARNPS..62...81S}.

While not the sole reason for these experiments, some of the science goals for the U.S. led and funded neutrino experiments IceCube (via the National Science Foundation, $\sim \$242$ million) and the Deep Underground Neutrino Experiment (DUNE, funded via the Department of Energy, Office of Science, High Energy Physics, projected total $\sim \$5$ billion), see~\cite{2018JPhCS1029a2001K,2021EPJC...81..423A}, is the detection of neutrinos from a supernova occurring in the Milky-Way galaxy. These can be expected to occur one to three times every hundred years, with the 1987a supernova event being the last nearby core-collapse supernova. The observation of neutrinos from this event has been a critical source of information used to constrain both stellar collapse models as well as beyond the Standard Model physics. 

DUNE will be able to provide both directionality information on the neutrinos origin, count the number of neutrinos, as well as measure their energy. Given accurate neutrino cross-sections for the DUNE detector, all this information will directly constrain the predicted spectrum of neutrinos from the core-collapse supernova, and direct experimental measurements for Eq.~\eqref{eq:momentum_spectrum} can be performed for $f=e$, the electron neutrino. 


\subsubsection{Concrete utility estimation}
Within Los Alamos National Laboratory, the supernova neutrino problem has been worked on for well over a decade, with an expenditure varying between \$400K to \$1 million annually. Moreover, supernova simulation which requires the correct characterization of the neutrino background has been a long-standing problem of interest to the Laboratory as a test bed for fluid dynamics simulations and multi-scale, multi-physics modeling. It has received an average investment of \$2-5 million, though the exact number is hard to estimate due to its spread over many programs and the underlying multi-use functionality of the developed code and models. From a physics point of view, these simulations are deficient without a proper understanding of the underlying neutrino background, and will not be fully capable of claiming a definitive understanding of core-collapse supernova without a solution to the neutrino scattering problem.   

\subsubsection*{Objectives}

The primary objective would be to produce a quantitative prediction for the time-scale when Eq.~\eqref{eq:momentum_spectrum} reaches an equilibrium value, given physically reasonable initial density matricies and matter profiles. More generally, one would like to answer the question whether the density matrix thermalizes in an appropriate sense before $T_{sn}$, the time it takes for the supernova to end. An example of such a calculation of the thermalization time for a small system in a simplified model of neutrino-neutrino scattering is given in figure~\ref{fig:sigz_v_t}; see~\cite{Martin:2023gbo} for details.


\begin{figure}
    \centering
    \includegraphics[scale=0.3]{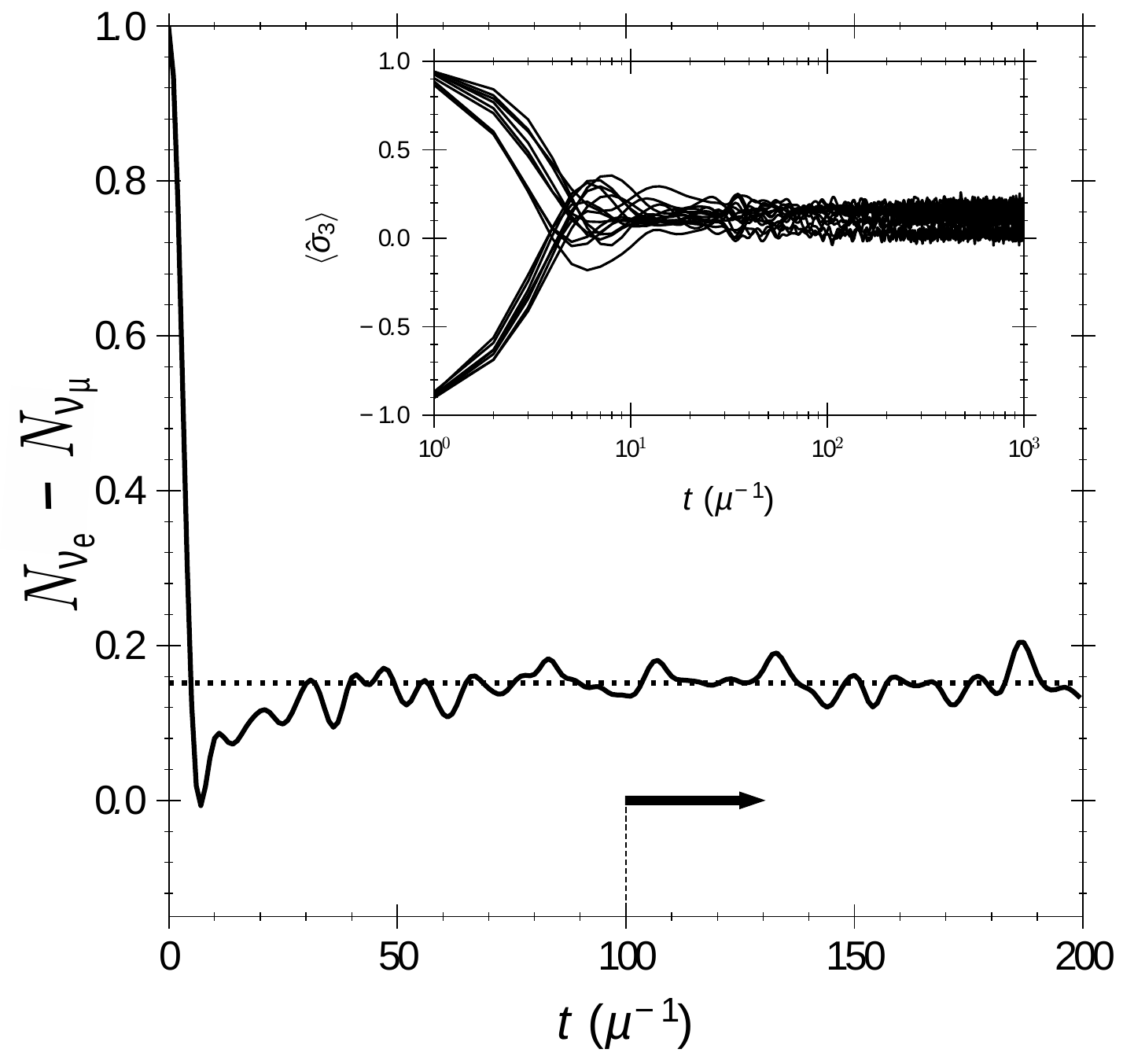}
    \caption{We plot the evolution of the difference in flavor occupation, $N_{\nu_e}-N_{\nu_\mu}$, for neutrino in momentum mode $i=1$ compared to the prediction utilizing the grand-canonical partition function (black dotted line), which represents the equilibrated long-time answer. The calculation is performed for a toy model of the full neutrino problem where the number of momentum modes equals the number of neutrinos, with an unphysically small number of neutrinos. The momentum modes are not allowed to change, only the flavor occupation in the simplified model. Thus each neutrino is labeled by its specific fixed momentum mode $i=1,...,16$, and $N_{\nu_e}(i)-N_{\nu_\mu}(i)$ captures the flavor distribution of that mode. The line and arrow indicate when the time-averaging begins for comparison against the grand canonical prediction.  
   \textit{Inset:} $N_{e}(i)-N_{\mu}(i)$ as a function of time for every neutrino on the entire considered time domain. Adapted from~\cite{Martin:2023gbo}.}
    \label{fig:sigz_v_t}
\end{figure}

\subsubsection*{End-to-end computational workflow}

Direct time evolution of the second-quantized hamiltonian for the initial $\sim 10^{38}$ per cubic meter of neutrinos would be hopeless to calculate, and unnecessary. One simply would need to show that Eq.~\eqref{eq:momentum_spectrum} reaches a stable distribution as the number of neutrinos grows large. The desired work-flow would proceed as follows, with the steps requiring quantum computational methods being explicitly indicated, and all other steps can be assumed to be a classical workload.
\begin{enumerate}
    \item Discretize the number of momentum modes into a fixed set $\mathcal{P}=\{\vec{p}_{i}\}_{i=1}^{K}$, and we can associate each momentum mode with a specific quantum state specifying whether or not the momentum mode is occupied and by what flavor of (anti)neutrino. More explicitly, for two flavors of neutrinos+anti-neutrinos, we have a set of labels denoting the occupation of a state:
    \begin{align}
    \mathcal{F}=\{0,\nu_{e},\nu_{\mu},\nu_{e}+\nu_{\mu},\bar\nu_{e},\bar\nu_{\mu},\bar\nu_{e}+\bar\nu_{\mu},\nu_{e}+\bar\nu_{e},\nu_{\mu}+\bar\nu_{\mu},\nu_{e}+\bar\nu_{\mu},\bar\nu_{e}+\nu_{\mu}\}\,.
    \end{align}
    Where $0$ denotes no occupation. Then we write any state of the discretized system as:
    \begin{align}
    |\psi\rangle =\sum_{\phi_1\in\mathcal{F},\phi_2\in\mathcal{F},...} c(\phi_1,...,\phi_K)\otimes_{i=1}^{K} |\phi_i\rangle_{\vec{p}_{i}}\,.
    \end{align}
    Thus for a two flavor example, we need 4 qubits per momentum mode to represent the occupation of the system. We note that this first quantized formalism can neglect the pair creation of neutrinos, as this process is not density enhanced, and is suppressed by the weak coupling constant $G_F$. Two-body scattering processes that preserve the total number of neutrinos minus anti-neutrinos will be most relevant. However, nothing precludes the inclusion of pair creation, though the number of momentum modes needed for an accurate description of the dynamics may grow so that artificial discretization effects are sufficiently suppressed. An example of this implementation, albeit on a small, unphysical scale, can be found in~\cite{Yukari}.
    \item Fix number of momentum modes $K$, and total number of neutrinos $N$ and anti-neutrinos $\bar N$. $K$ will specify the qubit requirements of the quantum computer.
    \item Initiate a physically reasonable set of initial states: these are product states, which for each neutrino is Gaussian distributed over momenta, with the momentum peaked at the nuclear scales at which the neutrinos are generated, i.e., $1-10$ MeV. Product states suffice due to the uncorrelated production of the neutrinos, and the gaussian distribution localizes the neutrinos to specific regions of the supernova. The simulation will be performed with no matter background or vacuum oscillations at all, neglecting $\hat{H}_1$. 
    \item Perform quantum evolution until the required momentum spectra reaches an equilibrium value. The equilibrium value can be deduced as a steady state value in time of the momentum spectrum, where moment to moment fluctuations are small relative to the long-time averaged observable~\cite{Srednicki_1996,2009PhRvE..79f1103L}. For an example of such an plot, see Fig.~\ref{fig:sigz_v_t}, where the flavor distribution of a fixed momentum mode is plotted as a function of time. 
    \item Repeat steps 2-4, increasing $K, N$, and $\bar{N}$ until discretization effects are negligible, and one can estimate the momenta spectra for the large $N$/$\bar{N}$ limit. 
    \item Now repeat the calculation (steps 1-5) for more sophisticated matter backgrounds informed by classical supernova simulations (such effects are reviewed in~\cite{radice2018turbulence}) and with vacuum oscillations induced by masses, which will result in the inclusion of the $\hat{H}_1$ term.
\end{enumerate}

\subsubsection*{Why classical methods are not sufficient to perform the hard computational module}

In general, calculating time-dependent quantities with classical methods is exponentially hard in the time-frame needed to be resolved, see~\cite{2023PRXQ....4b0340W}. Since the question we wish to resolve is the needed thermalization time for the observables of the system, barring special insights into the specific problem, full exploration of the thermalization time would be classically expensive. Aside from the classically tractable solutions to the classical limit of the equations of motion for the forward scattering neutrino interactions (itself a simplified model of the full neutrino scattering problem), no attempt has been made to classically calculate the quantum hamiltonian for neutrino scattering for any sort of realistic system sizes, as it would be impossible to do so on current hardware.

\subsubsection*{Notes on quantum implementation and resource estimation}

We can easily give a lower bound to the computational quantum gate complexity of our hamiltonian by considering a simplified case of the neutrino scattering hamiltonian known as the coherent forward scattering model of~\cite{Balantekin:2006tg}. In the integrals over all possible momenta, one only considers the cases where $\vec{p}_1=-\vec{p}_3$ and $\vec{p}_2=-\vec{p}_4$, or $\vec{p}_1=-\vec{p}_4$ and $\vec{p}_2=-\vec{p}_3$, which represent the forward/back-scattering scenarios. Further, one neglects anti-neutrinos, and restricts to a two-flavor model. From the point of view of the first-quantized formalism, un-occupied momentum modes cannot become occupied, and all that can happen is that the flavor swaps between the momentum modes. It is straightforward to see then that the forward scattering model fixes the number of momentum modes to be the number of neutrinos, $N=K$, and the hamiltonian will take a form that is analyzed in~\cite{2018PhRvL.120k0501K}. Further, the forward scattering model is a strict subset of the full problem, and thus any gate count and qubit requirements for the forward scattering problem lower bounds the difficulty of the full problem.  

Within the forward scattering model, then, we can write the hamiltonian as an all-to-all Heisenberg model, yielding:
\begin{align}
    \hat{H} &= \sum_{i=1}^N\vec{b}_i\cdot \hat{\vec{\sigma}}_i+\frac{G_F \mu}{\sqrt{2}N}\sum_{1
    \leq i<j \leq N}p_i\cdot p_j\hat{\vec{\sigma}}_i\cdot \hat{\vec{\sigma}}_j\,.
    \label{eq:forward_scattering}
\end{align}
Where now $N$ is the number of neutrinos or equivalently momentum modes, and $\vec{b}_i$ is the one-body coupling matrix, $\vec{p}_i$ is the momentum of the i-th neutrino, $\mu$ is density of neutrinos, which we can take to be $1$ with a redefinition of time after setting the one-body terms to zero, and $\hat{\vec{\sigma}}_i$ is a Pauli matrix operating on the flavor space of the $i$-th momentum mode. Ignoring the one-body terms in the hamiltonian, and focusing on the two-body interactions, the Heisenberg interactions can easily be written in terms a swap gate, and the exponentiation of a term in the Hamiltonian is again a sum of a swap gate with the identity. With an all-to-all connected architecture, the simplest Trotterization of this Hamiltonian yields a circuit with $N$ depth for a single Trotter step. Following~\cite{2021PhRvD.104f3009H,2018PhRvL.120k0501K} we can estimate the computational gate complexity assuming a linear connectivity for the qubit architecture, which now yields an order $N^2$ circuit for a single Trotter step. 

\subsubsection*{Concrete problem instantiations}

The simplest case that can prove to be of scientific use will be to consider a gas of momentum plane-wave neutrinos, undergoing two-body interactions without a complicating matter background, in the relativistic limit. Then the one-body hamiltonian reduces to the identity, being the conserved total kinetic energy of the system. Then the goal will be to see how equilibration is reached for flavor distribution of single momentum mode in Eq.~\eqref{eq:momentum_spectrum} for a given distribution of initial neutrino momenta and flavors. Actual equilibration of single-body observables can happen with relatively small system sizes in simplified models~\cite{Martin:2023gbo}. Thus it is reasonable to start with system sizes with $\sim 100$ momentum modes. Multiple questions must be explored, whose answer is not known a priori:
\begin{enumerate}
\item Fixing $K$, $N$ and $\bar{N}$, what is the time scale $t_{\rm eq.}(N,\bar{N},K)$ for equilibration? If the time-scale is power-law dependent on these quantities, this can be deduced from only a few runs of the quantum computer with a linear progression in the number of neutrinos and momentum modes. Constant or logarithmic dependence would require scanning an order of magnitude. 
\item Fixing $N$ and $\bar{N}$, what are the number of momentum modes $K$ needed before discretization effects are negligible, i.e., order $1\%$, for the late time value of Eq.~\eqref{eq:momentum_spectrum}? We denote this number as $K(N,\bar{N})$, and now the time scale $t_{\rm eq.}$ will simply be a function of $N \,\&\, \bar{N}$.
\item Deduce the behavior of $t_{\rm eq.}$ as $N \,\&\, \bar{N}$ grow, to order $1\%$ accuracy.
\end{enumerate}
As we do not expect to be close to any phase transitions or critical points of the system for \emph{generic} initial conditions, it is reasonable to expect that we do not need large system sizes to "extrapolate to infinity."\footnote{Being at a phase transition is very sensitive to system size, as a true phase transition can only occur at infinite system size. Finite size effects will smooth out the expected singularities associated to the phase transition.} Of course, this would easily be verified in the course of the calculation. Depending on the behavior of the time-scale with the system size, then, starting with system sizes of order $\sim 100$ momentum modes (or 400 qubits), and scaling to $\sim 1000$ modes (or 4000 qubits), would be sufficient to determine if anything \emph{unusual} is happening for order $100$ neutrinos. Thus the largest quantum computer we would need should have order $4000$ logical qubits, capable of running circuits with order $4000$ gates for multiple Trotter time-steps. Note that for finding the equilibrium value, we can expect a single shot should be sufficient, as the system can be expected to be self-averaging. Again, this is the expectation that the system will equilibrate to an effective thermal system at high temperatures for generic initial conditions~\cite{Srednicki1994,PhysRevA.43.2046,Srednicki_1996}.

\subsection{Application 2: Hydrodynamics of hot and dense matter}

Theoretical predictions of the behavior of the quark-gluon plasma require the calculation of a small number of \emph{transport coefficients}, of which the most significant is the shear viscosity of the plasma. This computation is not possible with current classical algorithms, nor with those that are foreseeable in the near term; however a straightforward quantum algorithm is known.

\subsubsection*{Specific background of the application}

The Relativistic Heavy-Ion Collider (RHIC) at Brookhaven National Laboratory performs high-energy collisions of gold nuclei. (CERN's Large Hadron Collider spends around one-tenth of its time performing similar experiments, albeit with lead instead of gold.) The immediate result of each such collision is a rapidly expanding ball of hot, dense matter known as the quark-gluon plasma (QGP). Investigating the QGP's properties has been a major goal of nuclear physics in the U.S.~since its discovery in the early 2000s. 

The quark-gluon plasma is a state of matter obtained at temperatures in excess of $\sim 10^{12}$ Kelvin. It is well described by quantum chromodynamics (QCD); however simulating a full heavy-ion collision directly from QCD is prohibitively expensive, even with a quantum computer. Modern modeling of the QGP is centered around the belief that at large scales (larger than $10^{-15}$ meters) it behaves as a fluid, and can therefore be treated with (relativistic) hydrodynamics. This belief comes from both theoretical work (see~\cite{Romatschke:2017ejr}) and evidence for hydrodynamic behavior in experiments~\cite{Kolb:2003dz}.

The central object of relativistic hydrodynamics~\cite{Romatschke:2017ejr} is the \emph{energy-momentum tensor} $T^{\mu\nu}$. This is a conserved quantity, obeying
\begin{equation}
    \partial_\mu T^{\mu\nu} = 0\text.
\end{equation}
This equation has no content without the definition of $T$. This tensor is obtained from a gradient expansion and expressed in terms of the local velocity and density fields. Certain components of $T$ have simple physical interpretation: in particular $T^{00}$ is the energy density, and $T^{0i} = p^i$ is the momentum density.

Relativistic hydrodynamics is a somewhat exotic theory. It has a far more familiar non-relativistic limit: the Navier-Stokes equation. This is a partial differential equation describing the behavior of a fluid with pressure $p$, density $\rho$, and velocity $u$:
\begin{equation}
    \rho \left(\frac{\partial u_i}{\partial t} + u_j \nabla_j u_i\right) + \nabla p
    =
    \eta \partial_j \left(\partial_i u_j + \partial_j u_i\right)
    +
    \left(\zeta- \frac 2 3 \eta\right) \partial_i \partial_j u_j
    \text.
\end{equation}
The coefficients $\eta$ and $\zeta$, referred to respectively as the shear and bulk viscosities, are free parameters of the model. They appear as free parameters in this equation because they are free parameters in the mapping from the density and velocity fields to the energy-momentum tensor. They must be determined either by a fit to data, or by microscopic (i.e.~non-hydrodynamic) calculations. Physically, these terms represent dissipation---they cause macroscopic disturbances to decay, with the energy of those disturbances converted into heat.

Full relativistic hydrodynamics has more free parameters, but it remains the case that shear viscosity is expected to be the dominant dissipative term (see e.g.~\cite{Rath:2020beo} for a comparison of shear and bulk viscosities, and~\cite{Romatschke:2017ejr} for a discussion of higher-order coefficients). The simulation of relativistic hydrodynamics is a well studied problem; a modern simulation code used for heavy-ion collisions is MUSIC~\cite{Schenke:2010nt,Schenke:2010rr,Paquet:2015lta}.

Such simulators require the viscosities as inputs. In the case of the QGP, no reliable microscopic calculation of the viscosities is available---they can be determined only by fits to data, and then somewhat imprecisely. A large-scale quantum computer would rectify this, allowing first-principles determination of the transport coefficients. This calculation proceeds by simulating the time-evolution of hot and dense matter under a lattice discretization of the QCD Hamiltonian. The simplest such discretization is~\cite{Kogut:1974ag}
\begin{equation}\label{eq:lattice-qcd-hamiltonian}
    H_{\mathrm{QCD}} = -\frac{1}{g^2 a} \sum_P \mathop{\mathrm{Re}} \mathop{\mathrm{Tr}} \left[
    \prod_{\langle i j \rangle \in P} U_{ij}
    \right]
    + \frac{4 g^2}{a} \sum_{\langle i j\rangle} \pi^2_{ij}
    + m \sum_i \bar\psi_i \psi_i + \sum_{\langle i j} \bar\psi_i U_{ij} \psi_j
    \text.
\end{equation}
This Hamiltonian addresses continuous objects (the gauge fields $U$) not suitable for direct implementation on a qubit-based quantum computer. A large and rapidly growing literature seeks the most efficient ``qubitization'' of the Hamiltonian (of which a small sample is~\cite{Ji:2020kjk,Ji:2022qvr,Alexandru:2019nsa,Ciavarella:2021nmj}); this question will not be addressed further here, but rather in the spirit of~\cite{Lamm:2019bik} we will treat the qubitization as a subroutine, any implementation of which can be used by the higher-level algorithm for studying hydrodynamics.

Our focus is on the shear viscosity, which is believed to be the dominant dissipative force in the QGP. The shear viscosity may be understood as a constant determining the rate of decay of a shear wave. A shear wave is a particular (low-energy) excitation of a fluid in which one component of the velocity, say $v_1(x)$, depends on a separate coefficient of the position. As a simplest example we might have
\begin{equation}
    v_1(x,t) = A(t) \sin k x_3
\end{equation}
with $v_2 = v_3 = 0$. The coefficient $A$ is the amplitude of the shear wave, and at late times has an exponential decay proportional to $\eta$.

In order to measure the shear viscosity we will need to measure correlation functions of $T^{0i}$. This operator is defined in the continuum, not on the lattice. We must find a sequence of lattice operators, constructed in terms of the fundamental operators used to construct the Hamiltonian of Eq.~(\ref{eq:lattice-qcd-hamiltonian}), that converge to $T^{0i}$ in the continuum. One such calculation is described in~\cite{Cohen:2021imf}; a more careful nonperturbative calculation (done classically via lattice QCD) may improve the convergence of the quantum calculation.

\subsubsection*{Overview of the value of the application}
The QGP is the central object of study at BNL's Relativistic Heavy-Ion Collider, see for example~\cite{rhic-new-physics}, which highlights the study of the low-viscosity quark-gluon-plasma at RHIC. RHIC's original construction cost was around \$617 million~\cite{IG-0543}, and in 2006 its yearly operational budget was estimated at the level of \$115.5 million~\cite{wiki:RHIC}.

The ability to \emph{reliably} calculate transport coefficients---particularly the shear viscosity---is of theoretical interest independent from the behavior of the standard model and QCD. A long-standing conjecture holds that the shear viscosity of a fluid can never fall below the \emph{KSS bound} $\eta \ge \frac{1}{4\pi} s$, where $s$ is the entropy density~\cite{Kovtun:2004de}. The bound itself is now widely believed invalid in general~\cite{Cremonini:2011iq}; however comparatively little is known about field theories exhibiting low ratios $\frac\eta s$, and it is not known whether or not QCD in particular violates the bound.

\subsubsection*{Concrete utility estimation}
To further emphasize the importance of QGP studies at RHIC, and the importance of viscosity to the QGP, we offer the following statistics, gleaned from theoretical physics articles posted to the arxiv through January 20, 2024:
\begin{itemize}
    \item Of $5582$ articles mentioning RHIC or LHC in the title or abstract, a total of $503$ also mention the QGP ($9\%$).
    \item Of $5582$ articles mentioning RHIC or LHC in the title or abstract, a total of $259$ also mention viscosity ($5\%$).
    \item Specifically with regards to RHIC: of $2976$ articles mentioning RHIC is the title or abstract, a total of $215$ ($7\%$) mention viscosity.
    \item Of $1302$ articles mentioning the QGP in the title or abstract, a total of $182$ also mention viscosity ($14\%$).
\end{itemize}
Taking these crude measures at face value, it appears that viscosity is deeply relevant to at least $7\%$ of the scientific work of RHIC, corresponding to \$43 million of the construction cost or to \$8 million of yearly operational cost (based on the 2006 estimation~\cite{wiki:RHIC}). 

\subsubsection*{Objectives}

To predict the angle-dependent distribution of energy produced in heavy-ion collisions at RHIC and LHC.

\subsubsection*{End-to-end computational workflow}

There are three large-scale steps to this calculation: a classical pre-processing step for determining lattice parameters, the core step utilizing the quantum processor, and a classical post-processing step in which hydrodynamic simulations are performed:
\begin{enumerate}
    \item Determine a suitable set of lattice parameters of a qubitized Hamiltonian. This determines the free parameters to be used in the quantum circuit for the second step.
    \item Estimate transport coefficients (and other thermodynamic properties) of QCD matter, using this Hamiltonian as an approximation to continuum QCD, across a range of chemical potentials and temperatures.
    \item Using these coefficients, perform hydrodynamics simulations of the sort being performed today, to predict the distribution of energy yields in a collision at RHIC.
\end{enumerate}

In what follows we will, for concreteness, assume that the gauge fields are qubitized as described in~\cite{Alexandru:2019nsa}. The quark fields are unaffected by this qubitization, and no additional free parameters are introduced in the fermionic part of the Hamiltonian $H_f$. The bosonic fields are significantly restricted: each link is now associated to a 1080-dimensional Hilbert space formally spanned by the elements of the Valentiner subgroup of $SU(3)$. To permit a closer approach to the continuum limit, the Hamiltonian to be used has an additional free parameter $\gamma$:
\begin{equation}\label{eq:valentiner-hamiltonian}
    H_{\mathrm{QCD}} = -\frac{1}{g^2 a} \sum_P \mathop{\mathrm{Re}} \mathop{\mathrm{Tr}} \left[
    \prod_{\langle i j \rangle \in P} U_{ij}
    +
    \gamma \left(\prod_{\langle i j \rangle \in P} U_{ij}\right)^2
    \right]
    + \frac{4 g^2}{a} \sum_{\langle i j\rangle} \pi^2_{ij}
    + H_f
    \text.
\end{equation}
The parameter $\gamma$ has been determined~\cite{Alexandru:2019nsa} and validated~\cite{Alexandru:2021jpm} for pure gauge fields in the action formalism. These values are likely to be approximately correct for gauge fields coupled to fermions in the Hamiltonian formalism, but not exactly. We describe now a procedure for determining these values, simultaneously with the fermion mass parameter, via classical lattice calculations (see~\cite{gattringer2009quantum} for an introductory reference). The entire procedure takes place at a fixed desired lattice spacing $a$.
\begin{enumerate}
    \item For a given set of parameters $g^2$,$a$,$\gamma$,$m$, perform HMC sampling (as described in~\cite{gattringer2009quantum} or similar) to collect $\sim 100$ statistically independent samples.\label{step:sample}
    \item Measure the pion mass and string tension.
    \item If the error bars on the above measurements exceed $10\%$, return to step~\ref{step:sample} and collect more samples.
    \item Repeat the above for a grid of parameters $(g^2,\gamma,m)$ to identify those with pion mass and string tension matching the physical values.
    \item Via the transfer matrix formalism~\cite{Creutz:1976ch}, convert these parameters in the action to Hamiltonian parameters.
\end{enumerate}
With this procedure complete, we have determined a Hamiltonian system that is a good approximation to continuum QCD. The approximation can be improved by various means; the most important (here elided) is to repeat the above procedure for a somewhat finer lattice spacing (a difference of $25\%$ is reasonable), and then perform the same simulations on both lattices. The difference between the two simulations provides an estimate of the error due to lattice artifacts, and the two can be used to extrapolate to the physical ($a=0$) point.

We now turn to the second step above; that is, the computation of the shear viscosity. The outline of this procedure is simple, crudely resembling a physical experiment. We must first prepare a sample of the material to be studied (hot, dense QCD), and then we inspect the behavior of shear fluctuations, brought into existence by the finite temperature, over time. This procedure is repeated for each sample of interest---that is, each number and energy density.

The preparation of the sample is the least obvious step---unlike Hamiltonian evolution (which is more-or-less a one-size-fits-all algorithm, at least for local lattice theories), state preparation is awash with system-specific complications. Happily the states we are interested in are at relatively high temperature, where state preparation is much easier. A variety of algorithms for this sort of state preparation have been discussed in the literature~\cite{Cohen:2023dll,Cohen:2023rhd,Cohen:2021imf}; here we make use of a heat bath, which leads to an algorithm with a particular conceptual simplicity. We need one addition to the QCD Hamiltonian above:
\begin{equation}\label{eq:H_QCD_coupled}
    H_{\mathrm{coupled}} = H_{\mathrm{QCD}}
    + \sum_{\langle i j \rangle} \frac 1 2 (\phi_i - \phi_j)^2
    + \sum_i \left(\frac 1 2 \pi_i^2 \lambda \phi_i^4\right)
    + g' \sum_i \bar\psi_i \psi_i \phi_i
\end{equation}
The scalar fields $\phi$ constitute the heat bath, coupled to the fermions by a Yukawa coupling. The single-index $\pi_i$ operator is canonically conjugate to $\phi_i$, not to be confused with the two-index (and conceptually similar) $\pi_{ij}$ of the QCD Hamiltonian. The simulation of strongly coupled scalar fields has been carefully described in~\cite{Jordan:2012xnu}, including the preparation of the vacuum. Therefore we are able to cool the QCD system by preparing a zero-temperature heat bath (with strong self-coupling $\lambda$) and turning on a weak Yukawa coupling to drain energy. Note that bringing the heat bath into the continuum limit is unimportant, so we can simply choose $\lambda=0.1$ from the outset and disregard the hypothetical $\phi^2$ term.

In detail, the computation of the shear viscosity proceeds as follows~\cite{Cohen:2021imf}:
\begin{enumerate}
    \item Prepare (on a quantum processor, with appropriate lattice couplings) a QCD state with the desired fermion density $n$. The state prepared does not matter: for concreteness let it be a product state, with all gauge fields in their strong-coupling ground state. Preparation of this sort of state is detailed in~\cite{Lawrence:2020irw}. Note that this is not the ground state of any Hamiltonian we are interested in---it is in fact a very high-temperature state of the Hamiltonian under which we will evolve.
    \label{qgp-loop-start}
    \item On a separate set of qubits prepare, following~\cite{Jordan:2012xnu}, the ground state of the strongly coupled scalar field. From this point on the two sets of qubits are to be considered one system.
    \item \label{qcd-cooling} Perform one step of time-evolution under the coupled Hamiltonian of Eq.~(\ref{eq:H_QCD_coupled}), with Yukawa coupling $g' \sim 0.01$. The coupling is chosen to not substantially change the thermodynamics of the QCD system, while allowing equilibration to take place in a reasonable number of time-steps. 
    \item Measure the energy in the QCD fields ($\langle H_{\mathrm{QCD}} \rangle$). Return to step \ref{qcd-cooling} and repeat as long as the energy density $\epsilon$ is above the desired value.
    \item Turn off the Yukawa coupling. Evolve for a short time additional ($\sim 10\,\mathrm{fm}\,c^{-1}$ will suffice) to equilibrate.
    \item The simulated system is now in equilibrium, with the desired energy and fermion density. We now measure the amplitude of the shear wave; that is, the expectation value of the operator
    \begin{equation}
        \hat A_k = \int dx\, \sin k x_3 \,\hat T^{01}(x)
        \text,
    \end{equation}
    where the integral is taken over spatial positions $x$ (after lattice discretization~\cite{Cohen:2021imf} this is a sum).
    The most interesting wavevector $k$ to consider is the largest one available in the simulated system. The operator $\hat T^{ij}$ is the energy-momentum tensor; the component $T^{01}$ represents momentum density along the $x_1$ direction, and an expression for it in terms of fundamental lattice operators is given in~\cite{Cohen:2021imf}.
    \item \label{begin-qcd-te}Continue to time-evolve according to an appropriate qubitization of Eq.~(\ref{eq:lattice-qcd-hamiltonian}), while periodically measuring the amplitude of the shear wave. Due to thermal fluctuations this is not a constant, but rather a random walk with some autocorrelation time. Evolve for $\sim 5\,\mathrm{fm}\,c^{-1}$.
    \item Inspect the autocorrelator $\langle A(0)A(t)\rangle$ and fit to an exponential decay at $t = 4\,\mathrm{fm}\,c^{-1}$. The decay coefficient yields an estimate of the shear viscosity.
    \item The error bars on this extraction are likely too large. Return to step~\ref{begin-qcd-te} and repeat until relative error on $\eta$ is within $20\%$.
    \label{qgp-loop-end}
    \item Repeat \ref{qgp-loop-start}-\ref{qgp-loop-end} for each pair $(\epsilon,n)$ of energy and fermion density of interest.
\end{enumerate}
The result of this procedure is a first-principles estimate of the rate of decay of shear waves in hot, dense QCD. Several approximations have been made (finite lattice size and so on), but all are in principle controllable by following the same procedure with different coefficients. In short: the calculation above is novel and of substantial physical interest.

\paragraph{A note on limits.} In ``concrete problem instantiation'' below, a reasonable set of parameters is listed for obtaining a first estimate of the shear viscosity. Moreover, byproducts of this calculation (particularly the measurement of the shear correlator as a function of time) are of interest even without a high-precision estimate of the shear viscosity.

Nevertheless it is worth mentioning what is necessary in order to attain a higher precision calculation of the shear viscosity itself, beyond this first calculation. Several extrapolations must be performed. For the purpose of illustration (that is, with no guarantee of completeness), the limits which are to be considered are:
\begin{itemize}
    \item The continuum limit: the lattice spacing is taken to zero, and the bare parameters in the Hamiltonian are tuned to achieve the physical meson and baryon masses.
    \item The infinite-volume limit: finite-volume effects are expected to be algebraic in $mL$, where $L$ is the physical side-length of the lattice and $m$ is a characteristic mass scale.
    \item The long-wavelength limit: the shear viscosity described the decay of a mode with asymptotically low wavenumber $k$.
    \item The long-time limit: the shear viscosity is defined from the \emph{asymptotic} behavior of the correlator discussed above.
    \item The high-statistics limit: the determination of shear viscosity can only be done when the measured correlator is sufficiently free from statistical noise.
\end{itemize}
Not all of these limits commute, and where they do not, physical considerations select the correct order of limits. Schematically this order is:
\begin{equation}
    \lim_{k \rightarrow 0}
    \lim_{t \rightarrow 0}
    \lim_{L \rightarrow \infty}
    \lim_{a \rightarrow 0}
    \lim_{\mathrm{statistics}}
    \text.
\end{equation}

In practice these limits are accomplished by performing calculations close enough to the asymptotic regime that the scaling (and therefore systematic errors) can be estimated from one or two data points. For example, one might perform one calculation at a lattice spacing of $0.15\,\mathrm{fm}$, and another at a lattice spacing of $0.12\,\mathrm{fm}$; the difference between the two offers a reasonable estimate of the systematic error induced by the finite lattice spacing.

\subsubsection*{Why classical methods are not sufficient to perform the hard computational module}

The hard computational piece is the determination of shear viscosity from QCD. Shear viscosities have been computed via perturbation theory for weakly-coupled fields theories~\cite{Jeon:1994if}. However, at the energy scales probed by RHIC, QCD is strongly coupled---by definition, perturbation theory is a very poor approximation. (This is a large part of what makes RHIC physics so interesting.) Indeed, perturbative estimates of the shear viscosity of the quark-gluon plasma have historically varied over multiple orders of magnitude.

This leaves nonperturbative methods, which in the context of quantum chromodynamics means lattice QCD.  Classical computational methods for quantum systems are most successful when computing thermodynamic (static) properties of matter at vanishing chemical potential---that is, when there is an equal number of fermions and antifermions. The resource scaling of these methods is a low-order polynomial for such systems. When dynamical quantities are desired, the resource scaling becomes exponential in the size of the system being simulated. Similarly, when there is a strong fermion-antifermion imbalance, the resource scaling of known algorithms is again exponential due to the famous ``fermion sign problem''. This latter problem is exceptionally well-studied, and has been the target of intense research for three decades. A solution for QCD remains elusive.

\subsubsection*{Concrete problem instantiation}

To concretely instantiate the above algorithm we need to pick precise physical parameters to be simulated. There are three sets of physical parameters: those defining the Hamiltonian, those defining the matter to be simulated, and those defining the estimation of shear viscosity. Our recommendation for a valuable initial calculation is:

\vspace{8pt}
\centerline{
\begin{tabular}{|c|c|c|c|}
\hline
    \textbf{Category} & \textbf{Parameter} & \textbf{Value} & \textbf{Justification}\\
    \hline
    \multirow{2}{*}{Hamiltonian} & $a$ & $0.2\,\mathrm{fm}$ & $aT \ll 1$\\
    & $L$ & $2\,\mathrm{fm}$ & $LT \gg 1$\\
    \hline
    \multirow{2}{*}{Matter} & $T$ & $160\,\mathrm{MeV}$& Pseudocritical temperature~\cite{Philipsen:2012nu}\\
    & $n$ & $0$ & Previous estimates available\\
    \hline
    \multirow{2}{*}{Fit} & $k$ & $2\pi/L$ & Fits in box\\
    & $t$ & $4\,\mathrm{fm}\,c^{-1}$ & Heavy-ion timescale\\
    \hline
\end{tabular}
}
\vspace{8pt}
The final column in the table above summarizes the primary consideration justifying the choice of parameter. Note also that the above table reports a temperature rather than an energy density---it is readily converted to an energy density via the equation of state (see~\cite{Philipsen:2012nu}).

The above calculation is physically informative and would on its own represent a major advance in our understanding of the hydrodynamics of QCD matter. However, to perform a hydrodynamic simulation we need not just one value of the shear viscosity, but measurements across a range of energy and number densities. 
Happily, at least away from the conjectured QCD critical point, the transport coefficients may be expected to vary smoothly with temperature and chemical potential. Therefore, a sensible initial task is to compute the shear viscosity as described above on a coarse grid of temperatures and chemical potentials. A $10\times 10$ grid with dimensions set by the largest energy and number density occurring in a current hydrodynamic calculation should be sufficient. For concreteness, take the peak energy density to be $T_{\mathrm{max}} = 3\,\mathrm{GeV}\,\mathrm{fm}^{-3}$~\cite{Bjorken:1982qr} and the peak number density to be $n_{\mathrm{max}} = 3 \rho_0 = 0.45\,\mathrm{fm}^{-3}$.

\section{Requirements summary}
This section summarizes the application requirements for all the applications described in this chapter.
\subsection*{Application 1: Supernova neutrinos}

\begin{small}

\begin{tabular}{ |l l l| }
    \hline
        & & \\
     & Total time limit & 5 years (order of magnitude related to the \\
     & & lifespan of the corresponding experiments)
     \\ [1ex]
     & & \\[1ex]
    \textbf{Workload:} & Number of subroutine calls required & 1000 = 10 (coupling to supernova sims.) \\
    & & $\times$ 100 (iterations over $K,N,\bar{N}$)
    \\ [1ex]
     & Maximum subroutine time limit & N/A \\ [1ex]
     & Model type & Fermionic Hamiltonian \\[1ex]
     & Size & $1000$ momenta modes \\[1ex]
     \textbf{Problem specifications:} &  Interaction Structure & Fully connected \\[1ex]
     & Computational target & Flavor occupation numbers \\[1ex]
     & Accuracy requirement & 1\% relative on flavor occupation numbers \\
     & & \\
     \hline
\end{tabular}

\subsection*{Application 2: Hydrodynamics of hot and dense matter}

\begin{tabular}{ |l l l| }
    \hline
        & & \\
     & Total time limit & 5 years (order of magnitude related to the \\
     & & lifespan of the corresponding experiments)
     \\ [1ex]
     & & \\[1ex]
    \textbf{Workload:} & Number of subroutine calls required & 200 = 10 (energies) \\
    & & $\times$ 10 (densities) $\times$ 2 (lattice spacings)
    \\ [1ex]
     & Maximum subroutine time limit & N/A \\ [1ex]
     & Model type & Qubitized QCD Hamiltonian \\[1ex]
     & Size & 36,000 = $10 \times 10 \times 10$ (lattice sites) \\
    & & $\times 11$ (Valentiner qubitization)\\
    & & $\times 3$ (links per site) + 3000 (fermions)
     \\[1ex]
     \textbf{Problem specifications:} &  Interaction Structure & Sparse regular (3D lattice) \\[1ex]
     & Computational target & Shear viscosity \\[1ex]
     & Accuracy requirement & 20\% relative on shear viscosity  \\
     & & \\
     \hline
\end{tabular}

\end{small}

\section{Quantum implementation}
The fully specified Hamiltonian \eqref{eq:forward_scattering}, quantum implementation and resource estimation can be found in \cite{qc-applications-notebooks}.

\newpage
\end{chapter}

\begin{chapter}{End Notes}
    
\begin{section}{Author Contributions}
The design and oversight of this document was contributed by Carleton Coffrin, Stephan Eidenbenz, Andrey Lokhov, Sidhant Misra and Marc Vuffray.
Contributions on quantum algorithms were provided by Andreas B\"artschi and Abhijith Jayakumar.
Implementations of quantum algorithms and related supplementary materials were prepared by Jonhas Colina, Zachary Morrell and Zain Mughal.
The MAGLAB user facility application chapter was lead by Minseong Lee and Sidhant Misra.
The high-temperature super conductivity application chapter was lead by Francesco Caravelli and Marc Vuffray.
The driven-dissipative Dicke model application chapter was lead by Andrei Piryatinski and Marc Vuffray.
The exotic phases of magnetic materials application chapter was lead by Allen Scheie and Sidhant Misra.
The catalysis for artificial photosynthesis application chapter was lead by Avanish Mishra, Yu Zhang and Andrey Lokhov.
The nuclear physics and astrophysics applications chapter was lead by Duff Neill, Scott Lawrence and Andrey Lokhov.
\end{section}

\begin{section}{Acknowledgements}
The contributors of this manuscript kindly thank:
Dr. Joe Altepeter for making this research possible;
Dr. Dave Clader, Dr. Rachael Al-Saadon, Dr. Yeuan-Ming Sheu and Ms. Eliana Krakovskyfor technical feedback on drafts of this document;
Dr. Jian-Xin Zhu for helpful discussions on modeling superconductivity in materials;
Ms. Anna Trujillo for project management support of this research;
Los Alamos National Laboratory's High Performance Computing Division for institutional computing workload data and discussions.
This research was developed with funding from the Defense Advanced Research Projects Agency (DARPA).
The views, opinions and/or findings expressed are those of the authors and should not be interpreted as representing the official views or policies of the Department of Defense or the U.S. Government.
\end{section}

\begin{section}{Change Log}

\begin{itemize}
    \item v0.1.0 - First document version. Application chapters on experimental magnetic materials, Fermi-Hubbard materials, dissipative Dicke model, magnetic material phases, artificial photosynthesis, and nuclear physics and astrophysics.
    \item v0.2.0 - Added references to quantum resource estimates; Minor technical revisions to Fermi-Hubbard materials chapter; Minor revisions to artificial photosynthesis chapter to improve readability and added a new subsection on nuclear dynamics.
    \item v0.3.0 - Minor updates to quantum computational kernels to reflect recent publications; Minor technical corrections to Fermi-Hubbard materials chapter and addition of subsection on Statics vs Dynamics in superconductivity; Minor revisions to the MAGLAB chapter to include additional application requirements needed for resource estimation.
\end{itemize}

\end{section}
\newpage
\end{chapter}


    





\ShowBibliography


@book{feynmanLectures,
  title={The Feynman lectures on physics},
  author={Feynman, Richard P and Leighton, Robert B and Sands, Matthew},
  volume={2},
  year={1964},
  publisher={California Institute of Technology}
}

@article{bethe1931theorie,
  title={Zur theorie der metalle: I. Eigenwerte und eigenfunktionen der linearen atomkette},
  author={Bethe, Hans},
  journal={Zeitschrift f{\"u}r Physik},
  volume={71},
  number={3-4},
  pages={205--226},
  year={1931},
  publisher={Springer}
}

@article{onsager1944crystal,
  title={Crystal statistics. I. A two-dimensional model with an order-disorder transition},
  author={Onsager, Lars},
  journal={Physical Review},
  volume={65},
  number={3-4},
  pages={117},
  year={1944},
  publisher={APS}
}

@article{anderson1952approximate,
  title={An approximate quantum theory of the antiferromagnetic ground state},
  author={Anderson, Philip W},
  journal={Physical Review},
  volume={86},
  number={5},
  pages={694},
  year={1952},
  publisher={APS}
}

@article{haldane1983continuum,
  title={Continuum dynamics of the 1-D Heisenberg antiferromagnet: Identification with the O (3) nonlinear sigma model},
  author={Haldane, F Duncan M},
  journal={Physics letters a},
  volume={93},
  number={9},
  pages={464--468},
  year={1983},
  publisher={Elsevier}
}

@article{sachdev2011quantum,
  title={Quantum criticality},
  author={Sachdev, Subir and Keimer, Bernhard},
  journal={Physics Today},
  volume={64},
  number={2},
  pages={29--35},
  year={2011},
  publisher={AIP Publishing}
}

@article{grunberg2008nobel,
  title={Nobel Lecture: From spin waves to giant magnetoresistance and beyond},
  author={Gr{\"u}nberg, Peter A},
  journal={Reviews of Modern Physics},
  volume={80},
  number={4},
  pages={1531},
  year={2008},
  publisher={APS}
}

@book{fishman2018spin,
  title={Spin-wave theory and its applications to neutron scattering and THz spectroscopy},
  author={Fishman, Randy S and Fernandez-Baca, Jaime A and R{\~o}{\~o}m, Toomas},
  year={2018},
  publisher={Morgan \& Claypool Publishers}
}

@article{dieny2020opportunities,
  title={Opportunities and challenges for spintronics in the microelectronics industry},
  author={Dieny, Bernard and Prejbeanu, Ioan Lucian and Garello, Kevin and Gambardella, Pietro and Freitas, Paulo and Lehndorff, Ronald and Raberg, Wolfgang and Ebels, Ursula and Demokritov, Sergej O and Akerman, Johan and others},
  journal={Nature Electronics},
  volume={3},
  number={8},
  pages={446--459},
  year={2020},
  publisher={Nature Publishing Group UK London}
}

@article{PhysRevB.92.041105,
  title = {Spin liquid phase of the $S=\frac{1}{2}/{J}_{1}-{J}_{2}$ Heisenberg model on the triangular lattice},
  author = {Zhu, Zhenyue and White, Steven R.},
  journal = {Phys. Rev. B},
  volume = {92},
  issue = {4},
  pages = {041105},
  numpages = {4},
  year = {2015},
  month = {7},
  publisher = {American Physical Society},
  doi = {10.1103/PhysRevB.92.041105},
  url = {https://link.aps.org/doi/10.1103/PhysRevB.92.041105}
}

@article{PhysRevB.92.140403,
  title = {Competing spin-liquid states in the spin-$\frac{1}{2}$ Heisenberg model on the triangular lattice},
  author = {Hu, Wen-Jun and Gong, Shou-Shu and Zhu, Wei and Sheng, D. N.},
  journal = {Phys. Rev. B},
  volume = {92},
  issue = {14},
  pages = {140403},
  numpages = {6},
  year = {2015},
  month = {10},
  publisher = {American Physical Society},
  doi = {10.1103/PhysRevB.92.140403},
  url = {https://link.aps.org/doi/10.1103/PhysRevB.92.140403}
}

@article{PhysRevB.93.144411,
  title = {Spin liquid nature in the Heisenberg ${J}_{1}\ensuremath{-}{J}_{2}$ triangular antiferromagnet},
  author = {Iqbal, Yasir and Hu, Wen-Jun and Thomale, Ronny and Poilblanc, Didier and Becca, Federico},
  journal = {Phys. Rev. B},
  volume = {93},
  issue = {14},
  pages = {144411},
  numpages = {14},
  year = {2016},
  month = {4},
  publisher = {American Physical Society},
  doi = {10.1103/PhysRevB.93.144411},
  url = {https://link.aps.org/doi/10.1103/PhysRevB.93.144411}
}

@article{PhysRevB.94.121111,
  title = {Symmetry fractionalization in the topological phase of the spin-$\frac{1}{2}$ $J_1-J_2$ triangular Heisenberg model},
  author = {Saadatmand, S. N. and McCulloch, I. P.},
  journal = {Phys. Rev. B},
  volume = {94},
  issue = {12},
  pages = {121111},
  numpages = {6},
  year = {2016},
  month = {9},
  publisher = {American Physical Society},
  doi = {10.1103/PhysRevB.94.121111},
  url = {https://link.aps.org/doi/10.1103/PhysRevB.94.121111}
}

@article{PhysRevB.95.035141,
  title = {Chiral spin liquid and quantum criticality in extended $S=\frac{1}{2}$ Heisenberg models on the triangular lattice},
  author = {Wietek, Alexander and L\"auchli, Andreas M.},
  journal = {Phys. Rev. B},
  volume = {95},
  issue = {3},
  pages = {035141},
  numpages = {6},
  year = {2017},
  month = {1},
  publisher = {American Physical Society},
  doi = {10.1103/PhysRevB.95.035141},
  url = {https://link.aps.org/doi/10.1103/PhysRevB.95.035141}
}

@article{PhysRevB.96.075116,
  title = {Global phase diagram and quantum spin liquids in a spin-$\frac{1}{2}$ triangular antiferromagnet},
  author = {Gong, Shou-Shu and Zhu, W. and Zhu, J.-X. and Sheng, D. N. and Yang, Kun},
  journal = {Phys. Rev. B},
  volume = {96},
  issue = {7},
  pages = {075116},
  numpages = {10},
  year = {2017},
  month = {8},
  publisher = {American Physical Society},
  doi = {10.1103/PhysRevB.96.075116},
  url = {https://link.aps.org/doi/10.1103/PhysRevB.96.075116}
}

@article{PhysRevLett.123.207203,
  title = {Dirac Spin Liquid on the Spin-$1/2$ Triangular Heisenberg Antiferromagnet},
  author = {Hu, Shijie and Zhu, W. and Eggert, Sebastian and He, Yin-Chen},
  journal = {Phys. Rev. Lett.},
  volume = {123},
  issue = {20},
  pages = {207203},
  numpages = {6},
  year = {2019},
  month = {11},
  publisher = {American Physical Society},
  doi = {10.1103/PhysRevLett.123.207203},
  url = {https://link.aps.org/doi/10.1103/PhysRevLett.123.207203}
}

@article{paddison2017continuous,
  title={Continuous excitations of the triangular-lattice quantum spin liquid YbMgGaO4},
  author={Paddison, Joseph AM and Daum, Marcus and Dun, Zhiling and Ehlers, Georg and Liu, Yaohua and Stone, Matthew B and Zhou, Haidong and Mourigal, Martin},
  journal={Nature Physics},
  volume={13},
  number={2},
  pages={117--122},
  year={2017},
  publisher={Nature Publishing Group UK London}
}

@article{li2015gapless,
  title={Gapless quantum spin liquid ground state in the two-dimensional spin-1/2 triangular antiferromagnet YbMgGaO4},
  author={Li, Yuesheng and Liao, Haijun and Zhang, Zhen and Li, Shiyan and Jin, Feng and Ling, Langsheng and Zhang, Lei and Zou, Youming and Pi, Li and Yang, Zhaorong and others},
  journal={Scientific reports},
  volume={5},
  number={1},
  pages={16419},
  year={2015},
  publisher={Nature Publishing Group UK London}
}

@article{shen2016evidence,
  title={Evidence for a spinon Fermi surface in a triangular-lattice quantum-spin-liquid candidate},
  author={Shen, Yao and Li, Yao-Dong and Wo, Hongliang and Li, Yuesheng and Shen, Shoudong and Pan, Bingying and Wang, Qisi and Walker, HC and Steffens, Paul and Boehm, Martin and others},
  journal={Nature},
  volume={540},
  number={7634},
  pages={559--562},
  year={2016},
  publisher={Nature Publishing Group UK London}
}

@article{PhysRevX.8.031028,
  title = {Valence Bonds in Random Quantum Magnets: Theory and Application to ${\mathrm{YbMgGaO}}_{4}$},
  author = {Kimchi, Itamar and Nahum, Adam and Senthil, T.},
  journal = {Phys. Rev. X},
  volume = {8},
  issue = {3},
  pages = {031028},
  numpages = {34},
  year = {2018},
  month = {7},
  publisher = {American Physical Society},
  doi = {10.1103/PhysRevX.8.031028},
  url = {https://link.aps.org/doi/10.1103/PhysRevX.8.031028}
}

@article{rao2021survival,
  title={Survival of itinerant excitations and quantum spin state transitions in YbMgGaO4 with chemical disorder},
  author={Rao, X and Hussain, G and Huang, Q and Chu, WJ and Li, N and Zhao, X and Dun, Z and Choi, ES and Asaba, T and Chen, L and others},
  journal={Nature communications},
  volume={12},
  number={1},
  pages={4949},
  year={2021},
  publisher={Nature Publishing Group UK London}
}

@article{PhysRevLett.119.057203,
  title = {Quasiparticle Breakdown and Spin Hamiltonian of the Frustrated Quantum Pyrochlore ${\mathrm{Yb}}_{2}{\mathrm{Ti}}_{2}{\mathrm{O}}_{7}$ in a Magnetic Field},
  author = {Thompson, J. D. and McClarty, P. A. and Prabhakaran, D. and Cabrera, I. and Guidi, T. and Coldea, R.},
  journal = {Phys. Rev. Lett.},
  volume = {119},
  issue = {5},
  pages = {057203},
  numpages = {6},
  year = {2017},
  month = {8},
  publisher = {American Physical Society},
  doi = {10.1103/PhysRevLett.119.057203},
  url = {https://link.aps.org/doi/10.1103/PhysRevLett.119.057203}
}

@article{scheie2020multiphase,
  title={Multiphase magnetism in Yb2Ti2O7},
  author={Scheie, Allen and Kindervater, Jonas and Zhang, Shu and Changlani, Hitesh J and Sala, Gabriele and Ehlers, Georg and Heinemann, Andre and Tucker, Gregory S and Koohpayeh, Seyed M and Broholm, Collin},
  journal={Proceedings of the National Academy of Sciences},
  volume={117},
  number={44},
  pages={27245--27254},
  year={2020},
  publisher={National Acad Sciences}
}

@article{banerjee2016proximate,
  title={Proximate Kitaev quantum spin liquid behaviour in a honeycomb magnet},
  author={Banerjee, A and Bridges, CA and Yan, J-Q and Aczel, AA and Li, L and Stone, MB and Granroth, GE and Lumsden, MD and Yiu, Y and Knolle, Johannes and others},
  journal={Nature materials},
  volume={15},
  number={7},
  pages={733--740},
  year={2016},
  publisher={Nature Publishing Group UK London}
}

@article{Banerjee2017Science,
author = {Arnab Banerjee  and Jiaqiang Yan  and Johannes Knolle  and Craig A. Bridges  and Matthew B. Stone  and Mark D. Lumsden  and David G. Mandrus  and David A. Tennant  and Roderich Moessner  and Stephen E. Nagler },
title = {Neutron scattering in the proximate quantum spin liquid \&\#x3b1;-RuCl<sub>3</sub>},
journal = {Science},
volume = {356},
number = {6342},
pages = {1055-1059},
year = {2017},
doi = {10.1126/science.aah6015},
URL = {https://www.science.org/doi/abs/10.1126/science.aah6015},
eprint = {https://www.science.org/doi/pdf/10.1126/science.aah6015}
}

@article{kitaev2006anyons,
  title={Anyons in an exactly solved model and beyond},
  author={Kitaev, Alexei},
  journal={Annals of Physics},
  volume={321},
  number={1},
  pages={2--111},
  year={2006},
  publisher={Elsevier}
}

@article{Nayak_2008,
  title = {Non-Abelian anyons and topological quantum computation},
  author = {Nayak, Chetan and Simon, Steven H. and Stern, Ady and Freedman, Michael and Das Sarma, Sankar},
  journal = {Rev. Mod. Phys.},
  volume = {80},
  issue = {3},
  pages = {1083--1159},
  numpages = {0},
  year = {2008},
  month = {9},
  publisher = {American Physical Society},
  doi = {10.1103/RevModPhys.80.1083},
  url = {https://link.aps.org/doi/10.1103/RevModPhys.80.1083}
}

@article{tokura2017emergent,
  title={Emergent functions of quantum materials},
  author={Tokura, Yoshinori and Kawasaki, Masashi and Nagaosa, Naoto},
  journal={Nature Physics},
  volume={13},
  number={11},
  pages={1056--1068},
  year={2017},
  publisher={Nature Publishing Group},
  url={https://doi.org/10.1038/nphys4274}
}

@article{Dai_2021,
  title = {Spinon Fermi Surface Spin Liquid in a Triangular Lattice Antiferromagnet ${\mathrm{NaYbSe}}_{2}$},
  author = {Dai, Peng-Ling and Zhang, Gaoning and Xie, Yaofeng and Duan, Chunruo and Gao, Yonghao and Zhu, Zihao and Feng, Erxi and Tao, Zhen and Huang, Chien-Lung and Cao, Huibo and Podlesnyak, Andrey and Granroth, Garrett E. and Everett, Michelle S. and Neuefeind, Joerg C. and Voneshen, David and Wang, Shun and Tan, Guotai and Morosan, Emilia and Wang, Xia and Lin, Hai-Qing and Shu, Lei and Chen, Gang and Guo, Yanfeng and Lu, Xingye and Dai, Pengcheng},
  journal = {Phys. Rev. X},
  volume = {11},
  issue = {2},
  pages = {021044},
  numpages = {10},
  year = {2021},
  month = {5},
  publisher = {American Physical Society},
  doi = {10.1103/PhysRevX.11.021044},
  url = {https://link.aps.org/doi/10.1103/PhysRevX.11.021044}
}

@Article{Xie2023,
author={Xie, Tao and Eberharter, A. A. and Xing, Jie and Nishimoto, S.
and Brando, M. and Khanenko, P. and Sichelschmidt, J. and Turrini, A. A.
and Mazzone, D. G. and Naumov, P. G. and Sanjeewa, L. D. and Harrison, N.
and Sefat, Athena S. and Normand, B. and L{\"a}uchli, A. M. and Podlesnyak, A. and Nikitin, S. E.},
title={Complete field-induced spectral response of the spin-1/2 triangular-lattice antiferromagnet {CsYbSe$_2$}},
journal={npj Quantum Materials},
year={2023},
month={9},
day={23},
volume={8},
number={1},
pages={48},
issn={2397-4648},
doi={10.1038/s41535-023-00580-9},
url={https://doi.org/10.1038/s41535-023-00580-9}
}

@Article{Scheie2024,
author={Scheie, A. O. and Ghioldi, E. A. and Xing, J. and Paddison, J. A. M. and Sherman, N. E. and Dupont, M.
and Sanjeewa, L. D. and Lee, Sangyun and Woods, A. J. and Abernathy, D. and Pajerowski, D. M. and Williams, T. J. 
and Zhang, Shang-Shun and Manuel, L. O. and Trumper, A. E. and Pemmaraju, C. D. and Sefat, A. S. and Parker, D. S. 
and Devereaux, T. P.and Movshovich, R. and Moore, J. E. and Batista, C. D. and Tennant, D. A.},
title={Proximate spin liquid and fractionalization in the triangular antiferromagnet KYbSe$_2$},
journal={Nature Physics},
year={2024},
month={1},
day={01},
volume={20},
number={1},
pages={74-81},
issn={1745-2481},
doi={10.1038/s41567-023-02259-1},
url={https://doi.org/10.1038/s41567-023-02259-1}
}

@article{Scheie_2024_KYS_PRB,
  title = {Nonlinear magnons and exchange Hamiltonians of the delafossite proximate quantum spin liquid candidates ${\text{KYbSe}}_{2}$ and ${\text{NaYbSe}}_{2}$},
  author = {Scheie, A. O. and Kamiya, Y. and Zhang, Hao and Lee, Sangyun and Woods, A. J. and Ajeesh, M. O. and Gonzalez, M. G. and Bernu, B. and Villanova, J. W. and Xing, J. and Huang, Q. and Zhang, Qingming and Ma, Jie and Choi, Eun Sang and Pajerowski, D. M. and Zhou, Haidong and Sefat, A. S. and Okamoto, S. and Berlijn, T. and Messio, L. and Movshovich, R. and Batista, C. D. and Tennant, D. A.},
  journal = {Phys. Rev. B},
  volume = {109},
  issue = {1},
  pages = {014425},
  numpages = {12},
  year = {2024},
  month = {1},
  publisher = {American Physical Society},
  doi = {10.1103/PhysRevB.109.014425},
  url = {https://link.aps.org/doi/10.1103/PhysRevB.109.014425}
}

@article{scheie2022quantum,
  title={Quantum wake dynamics in Heisenberg antiferromagnetic chains},
  author={Scheie, Allen and Laurell, Pontus and Lake, Bella and Nagler, Stephen E and Stone, Matthew B and Caux, Jean-Sebastian and Tennant, D Alan},
  journal={Nature Communications},
  volume={13},
  number={1},
  pages={5796},
  year={2022},
  publisher={Nature Publishing Group UK London},
  url={https://doi.org/10.1038/s41467-022-33571-8},
  doi={10.1038/s41467-022-33571-8}
}

@article{scheie2021witnessing,
  title = {Witnessing entanglement in quantum magnets using neutron scattering},
  author = {Scheie, A. and Laurell, Pontus and Samarakoon, A. M. and Lake, B. and Nagler, S. E. and Granroth, G. E. and Okamoto, S. and Alvarez, G. and Tennant, D. A.},
  journal = {Phys. Rev. B},
  volume = {103},
  issue = {22},
  pages = {224434},
  numpages = {16},
  year = {2021},
  month={6},
  publisher = {American Physical Society},
  doi = {10.1103/PhysRevB.103.224434},
  url = {https://link.aps.org/doi/10.1103/PhysRevB.103.224434}
}

@article{laurell2020dynamics,
  title = {Quantifying and Controlling Entanglement in the Quantum Magnet ${\mathrm{Cs}}_{2}{\mathrm{CoCl}}_{4}$},
  author = {Laurell, Pontus and Scheie, Allen and Mukherjee, Chiron J. and Koza, Michael M. and Enderle, Mechtild and Tylczynski, Zbigniew and Okamoto, Satoshi and Coldea, Radu and Tennant, D. Alan and Alvarez, Gonzalo},
  journal = {Phys. Rev. Lett.},
  volume = {127},
  issue = {3},
  pages = {037201},
  numpages = {7},
  year = {2021},
  month = {7},
  publisher = {American Physical Society},
  doi = {10.1103/PhysRevLett.127.037201},
  url = {https://link.aps.org/doi/10.1103/PhysRevLett.127.037201}
}

@article{PhysRevB.106.085110,
  title = {Magnetic excitations, nonclassicality, and quantum wake spin dynamics in the Hubbard chain},
  author = {Laurell, Pontus and Scheie, Allen and Tennant, D. Alan and Okamoto, Satoshi and Alvarez, Gonzalo and Dagotto, Elbio},
  journal = {Phys. Rev. B},
  volume = {106},
  issue = {8},
  pages = {085110},
  numpages = {15},
  year = {2022},
  month = {8},
  publisher = {American Physical Society},
  doi = {10.1103/PhysRevB.106.085110},
  url = {https://link.aps.org/doi/10.1103/PhysRevB.106.085110}
}

@article{PhysRevB.107.054422,
  title = {Multipartite entanglement in the one-dimensional spin-$\frac{1}{2}$ Heisenberg antiferromagnet},
  author = {Menon, Varun and Sherman, Nicholas E. and Dupont, Maxime and Scheie, Allen O. and Tennant, D. Alan and Moore, Joel E.},
  journal = {Phys. Rev. B},
  volume = {107},
  issue = {5},
  pages = {054422},
  numpages = {10},
  year = {2023},
  month = {2},
  publisher = {American Physical Society},
  doi = {10.1103/PhysRevB.107.054422},
  url = {https://link.aps.org/doi/10.1103/PhysRevB.107.054422}
}

@Article{Hauke2016,
	author={Hauke, Philipp 	and Heyl, Markus
	and Tagliacozzo, Luca 	and Zoller, Peter},
	title={Measuring multipartite entanglement through dynamic susceptibilities},
	journal={Nat. Phys.},
	year={2016},
	volume={12},
	number={8},
	pages={778-782},
	issn={1745-2481},
	doi={10.1038/nphys3700},
	url={https://doi.org/10.1038/nphys3700}
}

@article{PhysRevB.94.075121,
  title = {Quantum variance: A measure of quantum coherence and quantum correlations for many-body systems},
  author = {Fr\'erot, Ir\'en\'ee and Roscilde, Tommaso},
  journal = {Phys. Rev. B},
  volume = {94},
  issue = {7},
  pages = {075121},
  numpages = {15},
  year = {2016},
  month = {8},
  publisher = {American Physical Society},
  doi = {10.1103/PhysRevB.94.075121},
  url = {https://link.aps.org/doi/10.1103/PhysRevB.94.075121}
}

@Article{Frerot2019,
  author   = {Fr\'erot, Ir\'en\'ee and Tommaso Roscilde},
  title    = {Reconstructing the quantum critical fan of strongly correlated systems using quantum correlations},
  journal  = {Nat. Commun.},
  year     = {2019},
  volume   = {10},
  pages    = {577},
  doi      = {10.1038/s41467-019-08324-9},
}

@article{Roscilde_2004,
  title = {Studying Quantum Spin Systems through Entanglement Estimators},
  author = {Roscilde, Tommaso and Verrucchi, Paola and Fubini, Andrea and Haas, Stephan and Tognetti, Valerio},
  journal = {Phys. Rev. Lett.},
  volume = {93},
  issue = {16},
  pages = {167203},
  numpages = {4},
  year = {2004},
  month = {10},
  publisher = {American Physical Society},
  doi = {10.1103/PhysRevLett.93.167203},
  url = {https://link.aps.org/doi/10.1103/PhysRevLett.93.167203}
}

@Article{Hyllus_2012,
  author    = {Hyllus, Philipp and Laskowski, Wies\l{}aw and Krischek, Roland and Schwemmer, Christian and Wieczorek, Witlef and Weinfurter, Harald and Pezz\'e, Luca and Smerzi, Augusto},
  title     = {Fisher information and multiparticle entanglement},
  journal   = {Phys. Rev. A},
  year      = {2012},
  volume    = {85},
  pages     = {022321},
  doi       = {10.1103/PhysRevA.85.022321},
  issue     = {2},
  numpages  = {10},
  publisher = {American Physical Society},
  url       = {https://link.aps.org/doi/10.1103/PhysRevA.85.022321},
}

@article{Zhang_2019,
  title = {Dynamical Structure Factor of the Three-Dimensional Quantum Spin Liquid Candidate ${\mathrm{NaCaNi}}_{2}{\mathrm{F}}_{7}$},
  author = {Zhang, Shu and Changlani, Hitesh J. and Plumb, Kemp W. and Tchernyshyov, Oleg and Moessner, Roderich},
  journal = {Phys. Rev. Lett.},
  volume = {122},
  issue = {16},
  pages = {167203},
  numpages = {6},
  year = {2019},
  month = {4},
  publisher = {American Physical Society},
  doi = {10.1103/PhysRevLett.122.167203},
  url = {https://link.aps.org/doi/10.1103/PhysRevLett.122.167203}
}

@article{plumb2019continuum,
  title={Continuum of quantum fluctuations in a three-dimensional S= 1 Heisenberg magnet},
  author={Plumb, KW and Changlani, Hitesh J and Scheie, A and Zhang, Shu and Krizan, JW and Rodriguez-Rivera, JA and Qiu, Yiming and Winn, B and Cava, RJ and Broholm, Collin L},
  journal={Nature Physics},
  volume={15},
  number={1},
  pages={54--59},
  year={2019},
  publisher={Nature Publishing Group},
  url={https://doi.org/10.1038/s41567-018-0317-3}
}

@article{han2012fractionalized,
  title={Fractionalized excitations in the spin-liquid state of a kagome-lattice antiferromagnet},
  author={Han, Tian-Heng and Helton, Joel S and Chu, Shaoyan and Nocera, Daniel G and Rodriguez-Rivera, Jose A and Broholm, Collin and Lee, Young S},
  journal={Nature},
  volume={492},
  number={7429},
  pages={406--410},
  year={2012},
  publisher={Nature Publishing Group},
  url={https://doi.org/10.1038/nature11659}
}

@article{RevModPhys.88.041002,
  title = {Colloquium: Herbertsmithite and the search for the quantum spin liquid},
  author = {Norman, M. R.},
  journal = {Rev. Mod. Phys.},
  volume = {88},
  issue = {4},
  pages = {041002},
  numpages = {14},
  year = {2016},
  month={12},
  publisher = {American Physical Society},
  doi = {10.1103/RevModPhys.88.041002},
  url = {https://link.aps.org/doi/10.1103/RevModPhys.88.041002}
}

@article{scheie2023reconstructing,
  title={Reconstructing the spatial structure of quantum correlations},
  author={Scheie, Allen and Laurell, Pontus and Dagotto, Elbio and Tennant, D Alan and Roscilde, Tommaso},
  journal={arXiv preprint arXiv:2306.11723},
  year={2023}
}

@article{LIU20221034,
title = {Gapless quantum spin liquid and global phase diagram of the spin-1/2 J1-J2 square antiferromagnetic Heisenberg model},
journal = {Science Bulletin},
volume = {67},
number = {10},
pages = {1034-1041},
year = {2022},
issn = {2095-9273},
doi = {https://doi.org/10.1016/j.scib.2022.03.010},
url = {https://www.sciencedirect.com/science/article/pii/S2095927322001001},
author = {Wen-Yuan Liu and Shou-Shu Gong and Yu-Bin Li and Didier Poilblanc and Wei-Qiang Chen and Zheng-Cheng Gu}
}

@article{PhysRevB.38.9335,
  title = {Possible spin-liquid state at large $S$ for the frustrated square Heisenberg lattice},
  author = {Chandra, P. and Doucot, B.},
  journal = {Phys. Rev. B},
  volume = {38},
  issue = {13},
  pages = {9335--9338},
  numpages = {0},
  year = {1988},
  month = {11},
  publisher = {American Physical Society},
  doi = {10.1103/PhysRevB.38.9335},
  url = {https://link.aps.org/doi/10.1103/PhysRevB.38.9335}
}

@article{PhysRevLett.63.2148,
  title = {Phase diagram of the frustrated spin-1/2 Heisenberg antiferromagnet in 2 dimensions},
  author = {Dagotto, Elbio and Moreo, Adriana},
  journal = {Phys. Rev. Lett.},
  volume = {63},
  issue = {19},
  pages = {2148--2151},
  numpages = {0},
  year = {1989},
  month = {11},
  publisher = {American Physical Society},
  doi = {10.1103/PhysRevLett.63.2148},
  url = {https://link.aps.org/doi/10.1103/PhysRevLett.63.2148}
}

@article{PhysRevB.41.4619,
  title = {Exact diagonalization of finite frustrated spin-(1/2 Heisenberg models},
  author = {Figueirido, F. and Karlhede, A. and Kivelson, S. and Sondhi, S. and Rocek, M. and Rokhsar, D. S.},
  journal = {Phys. Rev. B},
  volume = {41},
  issue = {7},
  pages = {4619--4632},
  numpages = {0},
  year = {1990},
  month={3},
  publisher = {American Physical Society},
  doi = {10.1103/PhysRevB.41.4619},
  url = {https://link.aps.org/doi/10.1103/PhysRevB.41.4619}
}

@article{PhysRevB.41.9323,
  title = {Bond-operator representation of quantum spins: Mean-field theory of frustrated quantum Heisenberg antiferromagnets},
  author = {Sachdev, Subir and Bhatt, R. N.},
  journal = {Phys. Rev. B},
  volume = {41},
  issue = {13},
  pages = {9323--9329},
  numpages = {0},
  year = {1990},
  month = {5},
  publisher = {American Physical Society},
  doi = {10.1103/PhysRevB.41.9323},
  url = {https://link.aps.org/doi/10.1103/PhysRevB.41.9323}
}

@article{PhysRevB.43.10970,
  title = {Static and dynamical correlations in a spin-1/2 frustrated antiferromagnet},
  author = {Poilblanc, Didier and Gagliano, Eduardo and Bacci, Silvia and Dagotto, Elbio},
  journal = {Phys. Rev. B},
  volume = {43},
  issue = {13},
  pages = {10970--10983},
  numpages = {0},
  year = {1991},
  month = {5},
  publisher = {American Physical Society},
  doi = {10.1103/PhysRevB.43.10970},
  url = {https://link.aps.org/doi/10.1103/PhysRevB.43.10970}
}

@article{PhysRevB.44.12050,
  title = {Dimer stability region in a frustrated quantum Heisenberg antiferromagnet},
  author = {Chubukov, Andrey V. and Jolicoeur, Th.},
  journal = {Phys. Rev. B},
  volume = {44},
  issue = {21},
  pages = {12050--12053},
  numpages = {0},
  year = {1991},
  month={12},
  publisher = {American Physical Society},
  doi = {10.1103/PhysRevB.44.12050},
  url = {https://link.aps.org/doi/10.1103/PhysRevB.44.12050}
}

@article{schulz1992finite,
  title={Finite-size scaling for the two-dimensional frustrated quantum Heisenberg antiferromagnet},
  author={Schulz, HJ and Ziman, TAL},
  journal={Europhysics Letters},
  volume={18},
  number={4},
  pages={355},
  year={1992},
  publisher={IOP Publishing}
}

@article{PhysRevB.46.8206,
  title = {Frustrated two-dimensional quantum Heisenberg antiferromagnet at low temperatures},
  author = {Ivanov, N. B. and Ivanov, P. Ch.},
  journal = {Phys. Rev. B},
  volume = {46},
  issue = {13},
  pages = {8206--8213},
  numpages = {0},
  year = {1992},
  month = {10},
  publisher = {American Physical Society},
  doi = {10.1103/PhysRevB.46.8206},
  url = {https://link.aps.org/doi/10.1103/PhysRevB.46.8206}
}

@article{PhysRevB.51.6151,
  title = {Direct calculation of the spin stiffness in the ${\mathit{J}}_{1}$-${\mathit{J}}_{2}$ Heisenberg antiferromagnet},
  author = {Einarsson, T. and Schulz, H. J.},
  journal = {Phys. Rev. B},
  volume = {51},
  issue = {9},
  pages = {6151--6154},
  numpages = {0},
  year = {1995},
  month={3},
  publisher = {American Physical Society},
  doi = {10.1103/PhysRevB.51.6151},
  url = {https://link.aps.org/doi/10.1103/PhysRevB.51.6151}
}

@article{schulz1996magnetic,
  title={Magnetic order and disorder in the frustrated quantum Heisenberg antiferromagnet in two dimensions},
  author={Schulz, HJ and Ziman, TAL and Poilblanc, Didier},
  journal={Journal de Physique I},
  volume={6},
  number={5},
  pages={675--703},
  year={1996},
  publisher={EDP Sciences}
}

@article{PhysRevB.54.9007,
  title = {Valence-bond crystal phase of a frustrated spin-1/2 square-lattice antiferromagnet},
  author = {Zhitomirsky, M. E. and Ueda, Kazuo},
  journal = {Phys. Rev. B},
  volume = {54},
  issue = {13},
  pages = {9007--9010},
  numpages = {0},
  year = {1996},
  month = {10},
  publisher = {American Physical Society},
  doi = {10.1103/PhysRevB.54.9007},
  url = {https://link.aps.org/doi/10.1103/PhysRevB.54.9007}
}

@article{PhysRevB.60.7278,
  title = {Dimer order with striped correlations in the ${J}_{1}{\ensuremath{-}J}_{2}$ Heisenberg model},
  author = {Singh, Rajiv R. P. and Weihong, Zheng and Hamer, C. J. and Oitmaa, J.},
  journal = {Phys. Rev. B},
  volume = {60},
  issue = {10},
  pages = {7278--7283},
  numpages = {0},
  year = {1999},
  month = {9},
  publisher = {American Physical Society},
  doi = {10.1103/PhysRevB.60.7278},
  url = {https://link.aps.org/doi/10.1103/PhysRevB.60.7278}
}

@article{PhysRevLett.84.3173,
  title = {Spontaneous Plaquette Dimerization in the ${\mathit{J}}_{1}--{\mathit{J}}_{2}$ Heisenberg Model},
  author = {Capriotti, Luca and Sorella, Sandro},
  journal = {Phys. Rev. Lett.},
  volume = {84},
  issue = {14},
  pages = {3173--3176},
  numpages = {0},
  year = {2000},
  month = {4},
  publisher = {American Physical Society},
  doi = {10.1103/PhysRevLett.84.3173},
  url = {https://link.aps.org/doi/10.1103/PhysRevLett.84.3173}
}

@article{PhysRevLett.87.097201,
  title = {Resonating Valence Bond Wave Functions for Strongly Frustrated Spin Systems},
  author = {Capriotti, Luca and Becca, Federico and Parola, Alberto and Sorella, Sandro},
  journal = {Phys. Rev. Lett.},
  volume = {87},
  issue = {9},
  pages = {097201},
  numpages = {4},
  year = {2001},
  month = {8},
  publisher = {American Physical Society},
  doi = {10.1103/PhysRevLett.87.097201},
  url = {https://link.aps.org/doi/10.1103/PhysRevLett.87.097201}
}

@article{PhysRevLett.91.067201,
  title = {Valence-Bond Spin-Liquid State in Two-Dimensional Frustrated Spin-$1/2$ Heisenberg Antiferromagnets},
  author = {Zhang, Guang-Ming and Hu, Hui and Yu, Lu},
  journal = {Phys. Rev. Lett.},
  volume = {91},
  issue = {6},
  pages = {067201},
  numpages = {4},
  year = {2003},
  month = {8},
  publisher = {American Physical Society},
  doi = {10.1103/PhysRevLett.91.067201},
  url = {https://link.aps.org/doi/10.1103/PhysRevLett.91.067201}
}

@article{PhysRevLett.91.197202,
  title = {Nonlinear $\ensuremath{\sigma}$ Model Method for the ${J}_{1}-{J}_{2}$ Heisenberg Model: Disordered Ground State with Plaquette Symmetry},
  author = {Takano, Ken'ichi and Kito, Yoshiya and \ifmmode \bar{O}\else \={O}\fi{}no, Yoshiaki and Sano, Kazuhiro},
  journal = {Phys. Rev. Lett.},
  volume = {91},
  issue = {19},
  pages = {197202},
  numpages = {4},
  year = {2003},
  month = {11},
  publisher = {American Physical Society},
  doi = {10.1103/PhysRevLett.91.197202},
  url = {https://link.aps.org/doi/10.1103/PhysRevLett.91.197202}
}

@article{PhysRevB.73.184420,
  title = {$J_1-J_2$ model: First-order phase transition versus deconfinement of spinons},
  author = {Sirker, J. and Weihong, Zheng and Sushkov, O. P. and Oitmaa, J.},
  journal = {Phys. Rev. B},
  volume = {73},
  issue = {18},
  pages = {184420},
  numpages = {7},
  year = {2006},
  month = {5},
  publisher = {American Physical Society},
  doi = {10.1103/PhysRevB.73.184420},
  url = {https://link.aps.org/doi/10.1103/PhysRevB.73.184420}
}

@article{PhysRevLett.97.157201,
  title = {Quantum $J_1-J_2$ Antiferromagnet on a Stacked Square Lattice: Influence of the Interlayer Coupling on the Ground-State Magnetic Ordering},
  author = {Schmalfu\ss{}, D. and Darradi, R. and Richter, J. and Schulenburg, J. and Ihle, D.},
  journal = {Phys. Rev. Lett.},
  volume = {97},
  issue = {15},
  pages = {157201},
  numpages = {4},
  year = {2006},
  month = {10},
  publisher = {American Physical Society},
  doi = {10.1103/PhysRevLett.97.157201},
  url = {https://link.aps.org/doi/10.1103/PhysRevLett.97.157201}
}

@article{PhysRevB.74.144422,
  title = {Plaquette valence-bond crystal in the frustrated Heisenberg quantum antiferromagnet on the square lattice},
  author = {Mambrini, Matthieu and L\"auchli, Andreas and Poilblanc, Didier and Mila, Fr\'ed\'eric},
  journal = {Phys. Rev. B},
  volume = {74},
  issue = {14},
  pages = {144422},
  numpages = {11},
  year = {2006},
  month = {10},
  publisher = {American Physical Society},
  doi = {10.1103/PhysRevB.74.144422},
  url = {https://link.aps.org/doi/10.1103/PhysRevB.74.144422}
}

@article{PhysRevB.78.214415,
  title = {Ground state phases of the spin-1/2 ${J}_{1}--{J}_{2}$ Heisenberg antiferromagnet on the square lattice: A high-order coupled cluster treatment},
  author = {Darradi, R. and Derzhko, O. and Zinke, R. and Schulenburg, J. and Kr\"uger, S. E. and Richter, J.},
  journal = {Phys. Rev. B},
  volume = {78},
  issue = {21},
  pages = {214415},
  numpages = {10},
  year = {2008},
  month={12},
  publisher = {American Physical Society},
  doi = {10.1103/PhysRevB.78.214415},
  url = {https://link.aps.org/doi/10.1103/PhysRevB.78.214415}
}

@article{PhysRevB.78.224415,
  title = {Plaquette order in the ${J}_{1}-{J}_{2}-{J}_{3}$ model: Series expansion analysis},
  author = {Arlego, Marcelo and Brenig, Wolfram},
  journal = {Phys. Rev. B},
  volume = {78},
  issue = {22},
  pages = {224415},
  numpages = {6},
  year = {2008},
  month={12},
  publisher = {American Physical Society},
  doi = {10.1103/PhysRevB.78.224415},
  url = {https://link.aps.org/doi/10.1103/PhysRevB.78.224415}
}

@article{PhysRevB.79.024409,
  title = {Hierarchical mean-field approach to the ${J}_{1}-{J}_{2}$ Heisenberg model on a square lattice},
  author = {Isaev, L. and Ortiz, G. and Dukelsky, J.},
  journal = {Phys. Rev. B},
  volume = {79},
  issue = {2},
  pages = {024409},
  numpages = {14},
  year = {2009},
  month = {1},
  publisher = {American Physical Society},
  doi = {10.1103/PhysRevB.79.024409},
  url = {https://link.aps.org/doi/10.1103/PhysRevB.79.024409}
}

@article{PhysRevB.79.195119,
  title = {Exploring frustrated spin systems using projected entangled pair states},
  author = {Murg, V. and Verstraete, F. and Cirac, J. I.},
  journal = {Phys. Rev. B},
  volume = {79},
  issue = {19},
  pages = {195119},
  numpages = {7},
  year = {2009},
  month = {5},
  publisher = {American Physical Society},
  doi = {10.1103/PhysRevB.79.195119},
  url = {https://link.aps.org/doi/10.1103/PhysRevB.79.195119}
}

@article{PhysRevB.79.224431,
  title = {Master equation approach to computing RVB bond amplitudes},
  author = {Beach, K. S. D.},
  journal = {Phys. Rev. B},
  volume = {79},
  issue = {22},
  pages = {224431},
  numpages = {8},
  year = {2009},
  month={6},
  publisher = {American Physical Society},
  doi = {10.1103/PhysRevB.79.224431},
  url = {https://link.aps.org/doi/10.1103/PhysRevB.79.224431}
}

@article{richter2010spin,
  title={The spin-1/2 J1--J2 Heisenberg antiferromagnet on the square lattice: Exact diagonalization for N= 40 spins},
  author={Richter, J and Schulenburg, J},
  journal={The European Physical Journal B},
  volume={73},
  number={1},
  pages={117--124},
  year={2010},
  publisher={Springer}
}

@article{PhysRevB.85.094407,
  title = {Spin-1/2 ${J}_{1}$-${J}_{2}$ Heisenberg antiferromagnet on a square lattice: A plaquette renormalized tensor network study},
  author = {Yu, Ji-Feng and Kao, Ying-Jer},
  journal = {Phys. Rev. B},
  volume = {85},
  issue = {9},
  pages = {094407},
  numpages = {6},
  year = {2012},
  month={3},
  publisher = {American Physical Society},
  doi = {10.1103/PhysRevB.85.094407},
  url = {https://link.aps.org/doi/10.1103/PhysRevB.85.094407}
}

@article{PhysRevB.86.024424,
  title = {Spin liquid ground state of the spin-$\frac{1}{2}$ square ${J}_{1}$-${J}_{2}$ Heisenberg model},
  author = {Jiang, Hong-Chen and Yao, Hong and Balents, Leon},
  journal = {Phys. Rev. B},
  volume = {86},
  issue = {2},
  pages = {024424},
  numpages = {14},
  year = {2012},
  month = {7},
  publisher = {American Physical Society},
  doi = {10.1103/PhysRevB.86.024424},
  url = {https://link.aps.org/doi/10.1103/PhysRevB.86.024424}
}

@article{PhysRevB.86.045115,
  title = {Ground-state phase diagram of the quantum ${J}_{1}\ensuremath{-}{J}_{2}$ model on the square lattice},
  author = {Mezzacapo, Fabio},
  journal = {Phys. Rev. B},
  volume = {86},
  issue = {4},
  pages = {045115},
  numpages = {5},
  year = {2012},
  month = {7},
  publisher = {American Physical Society},
  doi = {10.1103/PhysRevB.86.045115},
  url = {https://link.aps.org/doi/10.1103/PhysRevB.86.045115}
}

@article{PhysRevLett.111.037202,
  title = {Constructing a Gapless Spin-Liquid State for the Spin-$1/2$ ${J}_{1}\ensuremath{-}{J}_{2}$ Heisenberg Model on a Square Lattice},
  author = {Wang, Ling and Poilblanc, Didier and Gu, Zheng-Cheng and Wen, Xiao-Gang and Verstraete, Frank},
  journal = {Phys. Rev. Lett.},
  volume = {111},
  issue = {3},
  pages = {037202},
  numpages = {5},
  year = {2013},
  month = {7},
  publisher = {American Physical Society},
  doi = {10.1103/PhysRevLett.111.037202},
  url = {https://link.aps.org/doi/10.1103/PhysRevLett.111.037202}
}

@article{PhysRevB.88.060402,
  title = {Direct evidence for a gapless ${Z}_{2}$ spin liquid by frustrating N\'eel antiferromagnetism},
  author = {Hu, Wen-Jun and Becca, Federico and Parola, Alberto and Sorella, Sandro},
  journal = {Phys. Rev. B},
  volume = {88},
  issue = {6},
  pages = {060402},
  numpages = {5},
  year = {2013},
  month = {8},
  publisher = {American Physical Society},
  doi = {10.1103/PhysRevB.88.060402},
  url = {https://link.aps.org/doi/10.1103/PhysRevB.88.060402}
}

@article{PhysRevB.89.104415,
  title = {Plaquette valence-bond solid in the square-lattice ${J}_{1}$-${J}_{2}$ antiferromagnet Heisenberg model: A bond operator approach},
  author = {Doretto, R. L.},
  journal = {Phys. Rev. B},
  volume = {89},
  issue = {10},
  pages = {104415},
  numpages = {15},
  year = {2014},
  month={3},
  publisher = {American Physical Society},
  doi = {10.1103/PhysRevB.89.104415},
  url = {https://link.aps.org/doi/10.1103/PhysRevB.89.104415}
}

@article{PhysRevB.89.235122,
  title = {Continuous phase transition from N\'eel state to ${Z}_{2}$ spin-liquid state on a square lattice},
  author = {Qi, Yang and Gu, Zheng-Cheng},
  journal = {Phys. Rev. B},
  volume = {89},
  issue = {23},
  pages = {235122},
  numpages = {9},
  year = {2014},
  month={6},
  publisher = {American Physical Society},
  doi = {10.1103/PhysRevB.89.235122},
  url = {https://link.aps.org/doi/10.1103/PhysRevB.89.235122}
}

@article{PhysRevLett.113.027201,
  title = {Plaquette Ordered Phase and Quantum Phase Diagram in the Spin-$\frac{1}{2}$ ${J}_{1}-{J}_{2}$ Square Heisenberg Model},
  author = {Gong, Shou-Shu and Zhu, Wei and Sheng, D. N. and Motrunich, Olexei I. and Fisher, Matthew P. A.},
  journal = {Phys. Rev. Lett.},
  volume = {113},
  issue = {2},
  pages = {027201},
  numpages = {5},
  year = {2014},
  month = {7},
  publisher = {American Physical Society},
  doi = {10.1103/PhysRevLett.113.027201},
  url = {https://link.aps.org/doi/10.1103/PhysRevLett.113.027201}
}

@article{PhysRevB.90.041106,
  title = {Simulating a two-dimensional frustrated spin system with fermionic resonating-valence-bond states},
  author = {Chou, Chung-Pin and Chen, Hong-Yi},
  journal = {Phys. Rev. B},
  volume = {90},
  issue = {4},
  pages = {041106},
  numpages = {5},
  year = {2014},
  month = {7},
  publisher = {American Physical Society},
  doi = {10.1103/PhysRevB.90.041106},
  url = {https://link.aps.org/doi/10.1103/PhysRevB.90.041106}
}

@article{morita2015quantum,
  title={Quantum Spin Liquid in Spin 1/2 J 1--J 2 Heisenberg Model on Square Lattice: Many-Variable Variational Monte Carlo Study Combined with Quantum-Number Projections},
  author={Morita, Satoshi and Kaneko, Ryui and Imada, Masatoshi},
  journal={Journal of the Physical Society of Japan},
  volume={84},
  number={2},
  pages={024720},
  year={2015},
  publisher={The Physical Society of Japan}
}

@article{richter2015spin,
  title={The spin-1/2 square-lattice J 1-J 2 model: the spin-gap issue},
  author={Richter, Johannes and Zinke, Ronald and Farnell, Damian JJ},
  journal={The European Physical Journal B},
  volume={88},
  pages={1--6},
  year={2015},
  publisher={Springer}
}

@article{PhysRevB.94.075143,
  title = {Tensor-product state approach to spin-$\frac{1}{2}$ square ${J}_{1}-{J}_{2}$ antiferromagnetic Heisenberg model: Evidence for deconfined quantum criticality},
  author = {Wang, Ling and Gu, Zheng-Cheng and Verstraete, Frank and Wen, Xiao-Gang},
  journal = {Phys. Rev. B},
  volume = {94},
  issue = {7},
  pages = {075143},
  numpages = {6},
  year = {2016},
  month = {8},
  publisher = {American Physical Society},
  doi = {10.1103/PhysRevB.94.075143},
  url = {https://link.aps.org/doi/10.1103/PhysRevB.94.075143}
}

@article{PhysRevB.96.014414,
  title = {Quantum critical phase with infinite projected entangled paired states},
  author = {Poilblanc, Didier and Mambrini, Matthieu},
  journal = {Phys. Rev. B},
  volume = {96},
  issue = {1},
  pages = {014414},
  numpages = {11},
  year = {2017},
  month = {7},
  publisher = {American Physical Society},
  doi = {10.1103/PhysRevB.96.014414},
  url = {https://link.aps.org/doi/10.1103/PhysRevB.96.014414}
}

@article{PhysRevLett.121.107202,
  title = {Critical Level Crossings and Gapless Spin Liquid in the Square-Lattice Spin-$1/2$ ${J}_{1}-{J}_{2}$ Heisenberg Antiferromagnet},
  author = {Wang, Ling and Sandvik, Anders W.},
  journal = {Phys. Rev. Lett.},
  volume = {121},
  issue = {10},
  pages = {107202},
  numpages = {7},
  year = {2018},
  month = {9},
  publisher = {American Physical Society},
  doi = {10.1103/PhysRevLett.121.107202},
  url = {https://link.aps.org/doi/10.1103/PhysRevLett.121.107202}
}

@article{PhysRevB.97.174408,
  title = {$U(1)$-symmetric infinite projected entangled-pair states study of the spin-1/2 square ${J}_{1}-{J}_{2}$ Heisenberg model},
  author = {Haghshenas, R. and Sheng, D. N.},
  journal = {Phys. Rev. B},
  volume = {97},
  issue = {17},
  pages = {174408},
  numpages = {10},
  year = {2018},
  month = {5},
  publisher = {American Physical Society},
  doi = {10.1103/PhysRevB.97.174408},
  url = {https://link.aps.org/doi/10.1103/PhysRevB.97.174408}
}

@article{PhysRevB.98.241109,
  title = {Gapless spin liquid ground state of the spin-$\frac{1}{2}$ ${J}_{1}\ensuremath{-}{J}_{2}$ Heisenberg model on square lattices},
  author = {Liu, Wen-Yuan and Dong, Shaojun and Wang, Chao and Han, Yongjian and An, Hong and Guo, Guang-Can and He, Lixin},
  journal = {Phys. Rev. B},
  volume = {98},
  issue = {24},
  pages = {241109},
  numpages = {5},
  year = {2018},
  month={12},
  publisher = {American Physical Society},
  doi = {10.1103/PhysRevB.98.241109},
  url = {https://link.aps.org/doi/10.1103/PhysRevB.98.241109}
}

@article{poilblanc2019critical,
  title={Critical colored-RVB states in the frustrated quantum Heisenberg model on the square lattice},
  author={Poilblanc, Didier and Mambrini, Matthieu and Capponi, Sylvain},
  journal={SciPost Physics},
  volume={7},
  number={4},
  pages={041},
  year={2019}
}

@article{PhysRevB.102.014417,
  title = {Gapless spin liquid and valence-bond solid in the ${J}_{1}$-${J}_{2}$ Heisenberg model on the square lattice: Insights from singlet and triplet excitations},
  author = {Ferrari, Francesco and Becca, Federico},
  journal = {Phys. Rev. B},
  volume = {102},
  issue = {1},
  pages = {014417},
  numpages = {5},
  year = {2020},
  month = {7},
  publisher = {American Physical Society},
  doi = {10.1103/PhysRevB.102.014417},
  url = {https://link.aps.org/doi/10.1103/PhysRevB.102.014417}
}

@article{PhysRevX.11.031034,
  title = {Dirac-Type Nodal Spin Liquid Revealed by Refined Quantum Many-Body Solver Using Neural-Network Wave Function, Correlation Ratio, and Level Spectroscopy},
  author = {Nomura, Yusuke and Imada, Masatoshi},
  journal = {Phys. Rev. X},
  volume = {11},
  issue = {3},
  pages = {031034},
  numpages = {19},
  year = {2021},
  month = {8},
  publisher = {American Physical Society},
  doi = {10.1103/PhysRevX.11.031034},
  url = {https://link.aps.org/doi/10.1103/PhysRevX.11.031034}
}

@article{qian2023absence,
  title = {Absence of spin liquid phase in the ${J}_{1}\ensuremath{-}{J}_{2}$ Heisenberg model on the square lattice},
  author = {Qian, Xiangjian and Qin, Mingpu},
  journal = {Phys. Rev. B},
  volume = {109},
  issue = {16},
  pages = {L161103},
  numpages = {7},
  year = {2024},
  month = {4},
  publisher = {American Physical Society},
  doi = {10.1103/PhysRevB.109.L161103},
  url = {https://link.aps.org/doi/10.1103/PhysRevB.109.L161103}
}

@article{PhysRevB.105.224404,
  title = {Schwinger boson theory of ordered magnets},
  author = {Zhang, Shang-Shun and Ghioldi, E. A. and Manuel, L. O. and Trumper, A. E. and Batista, Cristian D.},
  journal = {Phys. Rev. B},
  volume = {105},
  issue = {22},
  pages = {224404},
  numpages = {24},
  year = {2022},
  month={6},
  publisher = {American Physical Society},
  doi = {10.1103/PhysRevB.105.224404},
  url = {https://link.aps.org/doi/10.1103/PhysRevB.105.224404}
}

@article{Yan_2017,
  title = {Theory of multiple-phase competition in pyrochlore magnets with anisotropic exchange with application to ${\mathrm{Yb}}_{2}{\mathrm{Ti}}_{2}{\mathrm{O}}_{7}, {\mathrm{Er}}_{2}{\mathrm{Ti}}_{2}{\mathrm{O}}_{7}$, and ${\mathrm{Er}}_{2}{\mathrm{Sn}}_{2}{\mathrm{O}}_{7}$},
  author = {Yan, Han and Benton, Owen and Jaubert, Ludovic and Shannon, Nic},
  journal = {Phys. Rev. B},
  volume = {95},
  issue = {9},
  pages = {094422},
  numpages = {39},
  year = {2017},
  month={3},
  publisher = {American Physical Society},
  doi = {10.1103/PhysRevB.95.094422},
  url = {https://link.aps.org/doi/10.1103/PhysRevB.95.094422}
}

@article{benton16-pinchline,
    author =       {Benton, Owen and Jaubert, L. D. C. and Yan, Han and Shannon, Nic},
    title =        {A spin-liquid with pinch-line singularities on the pyrochlore lattice},
    journal =      {Nature Communications},
    volume = {7},
    pages = {11572},
    year =         {2016},
    doi =          {10.1038/ncomms11572}
}

@misc{liu2023deconfined,
      title={Deconfined quantum criticality with emergent symmetry in the extended Shastry-Sutherland model}, 
      author={Wen-Yuan Liu and Xiao-Tian Zhang and Zhe Wang and Shou-Shu Gong and Wei-Qiang Chen and Zheng-Cheng Gu},
      year={2023},
      eprint={2309.10955},
      archivePrefix={arXiv},
      primaryClass={cond-mat.str-el},
      url={https://doi.org/10.48550/arXiv.2309.10955}
}

@article{PhysRevX.9.011005,
  title = {Quantum and Classical Phases of the Pyrochlore Heisenberg Model with Competing Interactions},
  author = {Iqbal, Yasir and M\"uller, Tobias and Ghosh, Pratyay and Gingras, Michel J. P. and Jeschke, Harald O. and Rachel, Stephan and Reuther, Johannes and Thomale, Ronny},
  journal = {Phys. Rev. X},
  volume = {9},
  issue = {1},
  pages = {011005},
  numpages = {34},
  year = {2019},
  month = {1},
  publisher = {American Physical Society},
  doi = {10.1103/PhysRevX.9.011005},
  url = {https://link.aps.org/doi/10.1103/PhysRevX.9.011005}
}

@misc{pohle2023ground,
    title={Ground state of the $S$=1/2 pyrochlore Heisenberg antiferromagnet: A quantum spin liquid emergent from dimensional reduction},
    author={Rico Pohle and Youhei Yamaji and Masatoshi Imada},
    year={2023},
    eprint={2311.11561},
    archivePrefix={arXiv},
    primaryClass={cond-mat.str-el},
    url={arxiv.org/abs/2311.11561}
}

@article{PhysRevLett.126.117204,
  title = {Possible Inversion Symmetry Breaking in the $S=1/2$ Pyrochlore Heisenberg Magnet},
  author = {Hagym\'asi, Imre and Sch\"afer, Robin and Moessner, Roderich and Luitz, David J.},
  journal = {Phys. Rev. Lett.},
  volume = {126},
  issue = {11},
  pages = {117204},
  numpages = {6},
  year = {2021},
  month={3},
  publisher = {American Physical Society},
  doi = {10.1103/PhysRevLett.126.117204},
  url = {https://link.aps.org/doi/10.1103/PhysRevLett.126.117204}
}

@article{PhysRevX.11.041021,
  title = {Broken-Symmetry Ground States of the Heisenberg Model on the Pyrochlore Lattice},
  author = {Astrakhantsev, Nikita and Westerhout, Tom and Tiwari, Apoorv and Choo, Kenny and Chen, Ao and Fischer, Mark H. and Carleo, Giuseppe and Neupert, Titus},
  journal = {Phys. Rev. X},
  volume = {11},
  issue = {4},
  pages = {041021},
  numpages = {20},
  year = {2021},
  month = {10},
  publisher = {American Physical Society},
  doi = {10.1103/PhysRevX.11.041021},
  url = {https://link.aps.org/doi/10.1103/PhysRevX.11.041021}
}

@article{PhysRevLett.131.096702,
  title = {Abundance of Hard-Hexagon Crystals in the Quantum Pyrochlore Antiferromagnet},
  author = {Sch\"afer, Robin and Placke, Benedikt and Benton, Owen and Moessner, Roderich},
  journal = {Phys. Rev. Lett.},
  volume = {131},
  issue = {9},
  pages = {096702},
  numpages = {6},
  year = {2023},
  month = {8},
  publisher = {American Physical Society},
  doi = {10.1103/PhysRevLett.131.096702},
  url = {https://link.aps.org/doi/10.1103/PhysRevLett.131.096702}
}

@Article{Luo2021,
author={Luo, Qiang and Zhao, Jize and Kee, Hae-Young and Wang, Xiaoqun},
title={Gapless quantum spin liquid in a honeycomb $\Gamma$ magnet},
journal={npj Quantum Materials},
year={2021},
month={6},
day={04},
volume={6},
number={1},
pages={57},
issn={2397-4648},
doi={10.1038/s41535-021-00356-z},
url={https://doi.org/10.1038/s41535-021-00356-z}
}

@article{WenJun_2019,
  title = {Nematic spin liquid phase in a frustrated spin-1 system on the square lattice},
  author = {Hu, Wen-Jun and Gong, Shou-Shu and Lai, Hsin-Hua and Hu, Haoyu and Si, Qimiao and Nevidomskyy, Andriy H.},
  journal = {Phys. Rev. B},
  volume = {100},
  issue = {16},
  pages = {165142},
  numpages = {14},
  year = {2019},
  month = {10},
  publisher = {American Physical Society},
  doi = {10.1103/PhysRevB.100.165142},
  url = {https://link.aps.org/doi/10.1103/PhysRevB.100.165142}
}

@article{Pohle_2023,
  title = {Spin nematics meet spin liquids: Exotic quantum phases in the spin-1 bilinear-biquadratic model with Kitaev interactions},
  author = {Pohle, Rico and Shannon, Nic and Motome, Yukitoshi},
  journal = {Phys. Rev. B},
  volume = {107},
  issue = {14},
  pages = {L140403},
  numpages = {6},
  year = {2023},
  month = {4},
  publisher = {American Physical Society},
  doi = {10.1103/PhysRevB.107.L140403},
  url = {https://link.aps.org/doi/10.1103/PhysRevB.107.L140403}
}

@article{PhysRevLett.101.197201,
  title = {${\mathrm{Na}}_{4}{\mathrm{Ir}}_{3}{\mathrm{O}}_{8}$ as a 3D Spin Liquid with Fermionic Spinons},
  author = {Zhou, Yi and Lee, Patrick A. and Ng, Tai-Kai and Zhang, Fu-Chun},
  journal = {Phys. Rev. Lett.},
  volume = {101},
  issue = {19},
  pages = {197201},
  numpages = {4},
  year = {2008},
  month = {11},
  publisher = {American Physical Society},
  doi = {10.1103/PhysRevLett.101.197201},
  url = {https://link.aps.org/doi/10.1103/PhysRevLett.101.197201}
}

@article{PhysRevB.94.235138,
  title = {Competing magnetic orders and spin liquids in two- and three-dimensional kagome systems: Pseudofermion functional renormalization group perspective},
  author = {Buessen, Finn Lasse and Trebst, Simon},
  journal = {Phys. Rev. B},
  volume = {94},
  issue = {23},
  pages = {235138},
  numpages = {18},
  year = {2016},
  month={12},
  publisher = {American Physical Society},
  doi = {10.1103/PhysRevB.94.235138},
  url = {https://link.aps.org/doi/10.1103/PhysRevB.94.235138}
}

@article{PhysRevB.104.094413,
  title = {Theoretical study of quantum spin liquids in $S=\frac{1}{2}$ hyper-hyperkagome magnets: Classification, heat capacity, and dynamical spin structure factor},
  author = {Chern, Li Ern and Kim, Yong Baek},
  journal = {Phys. Rev. B},
  volume = {104},
  issue = {9},
  pages = {094413},
  numpages = {14},
  year = {2021},
  month = {9},
  publisher = {American Physical Society},
  doi = {10.1103/PhysRevB.104.094413},
  url = {https://link.aps.org/doi/10.1103/PhysRevB.104.094413}
}

@article{Zhang_2021_SUN,
  title = {Classical spin dynamics based on $\mathrm{SU}(N)$ coherent states},
  author = {Zhang, Hao and Batista, Cristian D.},
  journal = {Phys. Rev. B},
  volume = {104},
  issue = {10},
  pages = {104409},
  numpages = {12},
  year = {2021},
  month = {9},
  publisher = {American Physical Society},
  doi = {10.1103/PhysRevB.104.104409},
  url = {https://link.aps.org/doi/10.1103/PhysRevB.104.104409}
}

@article{Schollwock_2005,
  title = {The density-matrix renormalization group},
  author = {Schollw\"ock, U.},
  journal = {Rev. Mod. Phys.},
  volume = {77},
  issue = {1},
  pages = {259--315},
  numpages = {0},
  year = {2005},
  month = {4},
  publisher = {American Physical Society},
  doi = {10.1103/RevModPhys.77.259},
  url = {https://link.aps.org/doi/10.1103/RevModPhys.77.259}
}

@article{drescher2022dynamical,
  title = {Dynamical signatures of symmetry-broken and liquid phases in an $S$ = $\frac{1}{2}$ Heisenberg antiferromagnet on the triangular lattice},
  author = {Drescher, Markus and Vanderstraeten, Laurens and Moessner, Roderich and Pollmann, Frank},
  journal = {Phys. Rev. B},
  volume = {108},
  issue = {22},
  pages = {L220401},
  numpages = {6},
  year = {2023},
  month={12},
  publisher = {American Physical Society},
  doi = {10.1103/PhysRevB.108.L220401},
  url = {https://link.aps.org/doi/10.1103/PhysRevB.108.L220401}
}

@article{Sherman_2023,
  title = {Spectral function of the ${J}_{1}\ensuremath{-}{J}_{2}$ Heisenberg model on the triangular lattice},
  author = {Sherman, Nicholas E. and Dupont, Maxime and Moore, Joel E.},
  journal = {Phys. Rev. B},
  volume = {107},
  issue = {16},
  pages = {165146},
  numpages = {19},
  year = {2023},
  month = {4},
  publisher = {American Physical Society},
  doi = {10.1103/PhysRevB.107.165146},
  url = {https://link.aps.org/doi/10.1103/PhysRevB.107.165146}
}

@article{Wu_2021,
  title = {Exact diagonalization study of the anisotropic Heisenberg model related to ${\mathrm{YbMgGaO}}_{4}$},
  author = {Wu, Muwei and Yao, Dao-Xin and Wu, Han-Qing},
  journal = {Phys. Rev. B},
  volume = {103},
  issue = {20},
  pages = {205122},
  numpages = {16},
  year = {2021},
  month = {5},
  publisher = {American Physical Society},
  doi = {10.1103/PhysRevB.103.205122},
  url = {https://link.aps.org/doi/10.1103/PhysRevB.103.205122}
}

@article{PhysRevB.49.5065,
  title = {Lanczos method for the calculation of finite-temperature quantities in correlated systems},
  author = {Jakli\ifmmode \check{c}\else \v{c}\fi{}, J. and Prelov\ifmmode \check{s}\else \v{s}\fi{}ek, P.},
  journal = {Phys. Rev. B},
  volume = {49},
  issue = {7},
  pages = {5065--5068},
  numpages = {0},
  year = {1994},
  month = {2},
  publisher = {American Physical Society},
  doi = {10.1103/PhysRevB.49.5065},
  url = {https://link.aps.org/doi/10.1103/PhysRevB.49.5065}
}

@article{RevModPhys.73.33,
  title = {Quantum Monte Carlo simulations of solids},
  author = {Foulkes, W. M. C. and Mitas, L. and Needs, R. J. and Rajagopal, G.},
  journal = {Rev. Mod. Phys.},
  volume = {73},
  issue = {1},
  pages = {33--83},
  numpages = {0},
  year = {2001},
  month = {1},
  publisher = {American Physical Society},
  doi = {10.1103/RevModPhys.73.33},
  url = {https://link.aps.org/doi/10.1103/RevModPhys.73.33}
}

@article{PhysRevB.92.045110,
  title = {Geometry dependence of the sign problem in quantum Monte Carlo simulations},
  author = {Iglovikov, V. I. and Khatami, E. and Scalettar, R. T.},
  journal = {Phys. Rev. B},
  volume = {92},
  issue = {4},
  pages = {045110},
  numpages = {13},
  year = {2015},
  month = {7},
  publisher = {American Physical Society},
  doi = {10.1103/PhysRevB.92.045110},
  url = {https://link.aps.org/doi/10.1103/PhysRevB.92.045110}
}

@book{boothroyd2020principles,
  title={Principles of neutron scattering from condensed matter},
  author={Boothroyd, Andrew T},
  year={2020},
  publisher={Oxford University Press}
}

@article{syzranov2022eminuscent,
  title={Eminuscent phase in frustrated magnets: a challenge to quantum spin liquids},
  author={Syzranov, SV and Ramirez, AP},
  journal={Nature Communications},
  volume={13},
  number={1},
  pages={2993},
  year={2022},
  publisher={Nature Publishing Group UK London},
  doi={10.1038/s41467-022-30739-0},
  url={https://doi.org/10.1038/s41467-022-30739-0}
}

@article{li2024multipartite,
doi = {10.1088/1367-2630/ad273a},
url = {https://dx.doi.org/10.1088/1367-2630/ad273a},
year = {2024},
month = {2},
publisher = {IOP Publishing},
volume = {26},
number = {2},
pages = {023031},
author = {Yan-Chao Li and Yuan-Hang Zhou and Yuan Zhang and Yan-Kui Bai and Hai-Qing Lin},
title = {Multipartite entanglement serves as a faithful detector for quantum phase transitions},
journal = {New Journal of Physics}
}

@article{Hastings_2007,
doi = {10.1088/1742-5468/2007/05/P05010},
url = {https://dx.doi.org/10.1088/1742-5468/2007/05/P05010},
year = {2007},
publisher = {},
volume = {2007},
number = {05},
pages = {P05010},
author = {M B Hastings},
title = {Quasi-adiabatic continuation in gapped spin and fermion systems: Goldstone’s theorem and
flux periodicity},
journal = {Journal of Statistical Mechanics: Theory and Experiment},
}

@article{Wu:2020vc,
        author = {Wu, Qisheng and Zhou, Linsen and Schatz, George C. and Zhang, Yu and Guo, Hua},
        date = {2020-07-29},
        doi = {10.1021/jacs.0c04491},
        journal = {J. Am. Chem. Soc.},
        journal1 = {J. Am. Chem. Soc.},
        journal2 = {J. Am. Chem. Soc.},
        month = {07},
        number = {30},
        pages = {13090--13101},
        publisher = {American Chemical Society},
        title = {Mechanistic Insights into Photocatalyzed H2 Dissociation on Au Clusters},
        type = {doi: 10.1021/jacs.0c04491},
        url = {https://doi.org/10.1021/jacs.0c04491},
        volume = {142},
        year = {2020},
        year1 = {2020}}

@misc{shu_potlib,
  author       = {Y. Shu and Z. Varga and A. W. Jasper and B. C. Garrett and J. Espinosa-García and J. C. Corchado and R. J. Duchovic and Y. L. Volobuev and G. C. Lynch and K. R. Yang and T. C. Allison and A. F. Wagner and D. G. Truhlar},
  title        = {{POTLIB}},
  howpublished = {\url{http://comp.chem.umn.edu/potlib}},
  note         = {Accessed: 2024-10-15}
}

@book{Gatti2014,
  editor    = {Fabien Gatti},
  title     = {Molecular Quantum Dynamics: From Theory to Applications},
  series    = {Physical Chemistry in Action},
  year      = {2014},
  publisher = {Springer, Berlin, Heidelberg},
  doi       = {10.1007/978-3-642-45290-1},
}

@article{Larsson:2024tt,
        author = {Larsson ,Henrik R. },
        date = {2024-07-17},
        doi = {10.1080/00268976.2024.2306881},
        journal = {Molecular Physics},
        journal1 = {Molecular Physics},
        journal2 = {Molecular Physics},
        month = {07},
        number = {14},
        pages = {e2306881},
        publisher = {Taylor \& Francis},
        title = {A tensor network view of multilayer multiconfiguration time-dependent Hartree methods},
        type = {doi: 10.1080/00268976.2024.2306881},
        url = {https://doi.org/10.1080/00268976.2024.2306881},
        volume = {122},
        year = {2024},
        year1 = {2024}}

@article{Ceriotti:2016wt,
        author = {Ceriotti, Michele and Fang, Wei and Kusalik, Peter G. and McKenzie, Ross H. and Michaelides, Angelos and Morales, Miguel A. and Markland, Thomas E.},
        date = {2016-07-13},
        doi = {10.1021/acs.chemrev.5b00674},
        journal = {Chem. Rev.},
        journal1 = {Chem. Rev.},
        journal2 = {Chem. Rev.},
        month = {07},
        number = {13},
        pages = {7529--7550},
        publisher = {American Chemical Society},
        title = {Nuclear Quantum Effects in Water and Aqueous Systems: Experiment, Theory, and Current Challenges},
        type = {doi: 10.1021/acs.chemrev.5b00674},
        url = {https://doi.org/10.1021/acs.chemrev.5b00674},
        volume = {116},
        year = {2016},
        year1 = {2016}}

@article{Ryabinkin:2017uz,
        author = {Ryabinkin, Ilya G. and Joubert-Doriol, Lo{\"i}c and Izmaylov, Artur F.},
        date = {2017-07-18},
        doi = {10.1021/acs.accounts.7b00220},
        journal = {Accounts of Chemical Research},
        journal1 = {Accounts of Chemical Research},
        journal2 = {Acc. Chem. Res.},
        month = {07},
        number = {7},
        pages = {1785--1793},
        publisher = {American Chemical Society},
        title = {Geometric Phase Effects in Nonadiabatic Dynamics near Conical Intersections},
        type = {doi: 10.1021/acs.accounts.7b00220},
        url = {https://doi.org/10.1021/acs.accounts.7b00220},
        volume = {50},
        year = {2017},
        year1 = {2017}}

@article{hat1,
  title={Hydrogenases},
  author={Lubitz, Wolfgang and Ogata, Hideaki and Rudiger, Olaf and Reijerse, Edward},
  journal={Chem. Rev.},
  volume={114},
  number={8},
  pages={4081--4148},
  year={2014},
  publisher={ACS Publications}
}

@article{hat2,
  title={Possibility of nonequilibrium isomerization of azobenzene triggered by vibrational excitations},
  author={Tanaka, Shigenori and Itoh, Satoshi and Kurita, Noriyuki},
  journal={Chem. Phys. Lett.},
  volume={362},
  number={5-6},
  pages={467--475},
  year={2002},
  publisher={Elsevier}
}

@article{hat3,
    author = {Le, Hung M. and Raff, Lionel M.},
    title = "{Cis-trans, trans-cis isomerizations and N–O bond dissociation of nitrous acid (HONO) on an ab initio potential surface obtained by novelty sampling and feed-forward neural network fitting}",
    journal = {J. Chem. Phys.},
    volume = {128},
    number = {19},
    pages = {194310},
    year = {2008},
    month = {05},
    issn = {0021-9606},
    doi = {10.1063/1.2918503},
    url = {https://doi.org/10.1063/1.2918503},
}

@article{hat4,
  title={Advances in hydrogen/deuterium exchange mass spectrometry and the pursuit of challenging biological systems},
  author={James, Ellie I and Murphree, Taylor A and Vorauer, Clint and Engen, John R and Guttman, Miklos},
  journal={Chem. Rev.},
  volume={122},
  number={8},
  pages={7562--7623},
  year={2021},
  publisher={ACS Publications}
}

@article{hat5,
  title={Hydrogen--deuterium exchange reactions with isobutane over acid zeolites},
  author={Schoofs, Bart and Schuermans, Jochen and Schoonheydt, Robert A},
  journal={Microporous and Mesoporous Materials},
  volume={35},
  pages={99--111},
  year={2000},
  publisher={Elsevier}
}

@article{hat6,
  title={Irradiation effects in tungsten—From surface effects to bulk mechanical properties},
  author={Riesch, J and Feichtmayer, A and Coenen, JW and Curzadd, B and Gietl, H and H{\"o}schen, T and Manhard, A and Schwarz-Selinger, T and Neu, R},
  journal={Nuclear Materials and Energy},
  volume={30},
  pages={101093},
  year={2022},
  publisher={Elsevier}
}

@article{hat7,
  title={Progress in improving hydrogen storage properties of Mg-based materials},
  author={Xinglin, Yang and Xiaohui, Lu and Jiaqi, Zhang and Quanhui, Hou and Junhu, Zou},
  journal={Materials Today Advances},
  volume={19},
  pages={100387},
  year={2023},
  publisher={Elsevier}
}

@article{light1985generalized,
  title={Generalized discrete variable approximation in quantum mechanics},
  author={Light, JC and Hamilton, IP and Lill, JV},
  journal={J. Chem. Phys.},
  volume={82},
  number={3},
  pages={1400--1409},
  year={1985},
  publisher={American Institute of Physics}
}

@article{colbert1992novel,
  title={A novel discrete variable representation for quantum mechanical reactive scattering via the S-matrix Kohn method},
  author={Colbert, Daniel T and Miller, William H},
  journal={J. Chem. Phys.},
  volume={96},
  number={3},
  pages={1982--1991},
  year={1992},
  publisher={American Institute of Physics}
}

@article{heisenberg1927,
  title={{\"U}ber den anschaulichen Inhalt der quantentheoretischen Kinematik und Mechanik},
  author={Heisenberg, Werner},
  journal={Zeitschrift f{\"u}r Physik},
  volume={43},
  number={3},
  pages={172--198},
  year={1927},
  publisher={Springer}
}

@article{einstein1913einige,
  title={Einige Argumente f{\"u}r die Annahme einer molekularen Agitation beim absoluten Nullpunkt},
  author={Einstein, Albert and Stern, Otto},
  journal={Annalen der Physik},
  volume={345},
  number={3},
  pages={551--560},
  year={1913},
  publisher={WILEY-VCH Verlag Leipzig}
}

@misc{melander1980reaction,
  title={Reaction Rates of Isotopic Molecules},
  author={Melander, L},
  year={1980},
  publisher={Wiley-Interscience}
}

@article{bigeleisen1949relative,
  title={The relative reaction velocities of isotopic molecules},
  author={Bigeleisen, Jacob},
  journal={J. Chem. Phys.},
  volume={17},
  number={8},
  pages={675--678},
  year={1949},
  publisher={AIP Publishing}
}

@article{simon1966isotope,
  title={Isotope effects in organic chemistry and biochemistry},
  author={Simon, H and Palm, D},
  journal={Angewandte Chemie International Edition in English},
  volume={5},
  number={11},
  pages={920--933},
  year={1966},
  publisher={Wiley Online Library}
}

@inproceedings{Somma:2003wn,
        author = {Rolando D. Somma and Gerardo Ortiz and Emanuel H. Knill and James Gubernatis},
        booktitle = {Proc.SPIE},
        date = {2003-08-04},
        doi = {10.1117/12.487249},
        pages = {96--103},
        title = {Quantum simulations of physics problems},
        url = {https://doi.org/10.1117/12.487249},
        volume = {5105},
        year = {2003}}

@article{Dumitrescu:2018vu,
        author = {Dumitrescu, E. F. and McCaskey, A. J. and Hagen, G. and Jansen, G. R. and Morris, T. D. and Papenbrock, T. and Pooser, R. C. and Dean, D. J. and Lougovski, P.},
        date = {2018-05-23/},
        day = {23},
        doi = {10.1103/PhysRevLett.120.210501},
        id = {10.1103/PhysRevLett.120.210501},
        j1 = {PRL},
        journal = {Physical Review Letters},
        journal1 = {Phys. Rev. Lett.},
        month = {05},
        number = {21},
        pages = {210501--},
        publisher = {American Physical Society},
        title = {Cloud Quantum Computing of an Atomic Nucleus},
        url = {https://link.aps.org/doi/10.1103/PhysRevLett.120.210501},
        volume = {120},
        year = {2018}}

@article{Ollitrault:2020wh,
        author = {Ollitrault, Pauline J. and Baiardi, Alberto and Reiher, Markus and Tavernelli, Ivano},
        doi = {10.1039/D0SC01908A},
        journal = {Chemical Science},
        journal1 = {Chem. Sci.},
        number = {26},
        pages = {6842--6855},
        publisher = {The Royal Society of Chemistry},
        title = {Hardware efficient quantum algorithms for vibrational structure calculations},
        type = {10.1039/D0SC01908A},
        url = {http://dx.doi.org/10.1039/D0SC01908A},
        volume = {11},
        year = {2020}}

@article{McArdle:2019wf,
        author = {McArdle, Sam and Mayorov, Alexander and Shan, Xiao and Benjamin, Simon and Yuan, Xiao},
        doi = {10.1039/C9SC01313J},
        journal = {Chemical Science},
        journal1 = {Chem. Sci.},
        number = {22},
        pages = {5725--5735},
        publisher = {The Royal Society of Chemistry},
        title = {Digital quantum simulation of molecular vibrations},
        type = {10.1039/C9SC01313J},
        url = {http://dx.doi.org/10.1039/C9SC01313J},
        volume = {10},
        year = {2019}}

@article{Burghardt:2008ut,
        author = {Burghardt, I. and Giri, K. and Worth, G. A.},
        doi = {10.1063/1.2996349},
        journal = {J. Chem. Phys.},
        journal1 = {J. Chem. Phys.},
        number = {17},
        pages = {174104},
        title = {Multimode quantum dynamics using Gaussian wavepackets: The Gaussian-based multiconfiguration time-dependent Hartree (G-MCTDH) method applied to the absorption spectrum of pyrazine},
        url = {https://doi.org/10.1063/1.2996349},
        volume = {129},
        year = {2008},
        year1 = {2008/11/05}
}

@article{Beck:2000wm,
        author = {Beck, M. H. and J{\"a}ckle, A. and Worth, G. A. and Meyer, H. -D.},
        date = {2000-01-01/},
        doi = {https://doi.org/10.1016/S0370-1573(99)00047-2},
        journal = {Physics Reports},
        number = {1},
        pages = {1--105},
        title = {The multiconfiguration time-dependent Hartree (MCTDH) method: a highly efficient algorithm for propagating wavepackets},
        url = {https://www.sciencedirect.com/science/article/pii/S0370157399000472},
        volume = {324},
        year = {2000},
}

@article{Dammak:2009tb,
        author = {Dammak, Hichem and Chalopin, Yann and Laroche, Marine and Hayoun, Marc and Greffet, Jean-Jacques},
        date = {2009-11-05/},
        day = {05},
        doi = {10.1103/PhysRevLett.103.190601},
        id = {10.1103/PhysRevLett.103.190601},
        j1 = {PRL},
        journal = {Physical Review Letters},
        journal1 = {Phys. Rev. Lett.},
        month = {11},
        number = {19},
        pages = {190601--},
        publisher = {American Physical Society},
        title = {Quantum Thermal Bath for Molecular Dynamics Simulation},
        url = {https://link.aps.org/doi/10.1103/PhysRevLett.103.190601},
        volume = {103},
        year = {2009}}

@article{Mora:2018tq,
        author = {Mora, S. Jimena and Odella, Emmanuel and Moore, Gary F. and Gust, Devens and Moore, Thomas A. and Moore, Ana L.},
        date = {2018-02-20},
        doi = {10.1021/acs.accounts.7b00491},
        journal = {Accounts of Chemical Research},
        journal1 = {Accounts of Chemical Research},
        journal2 = {Acc. Chem. Res.},
        month = {02},
        number = {2},
        pages = {445--453},
        publisher = {American Chemical Society},
        title = {Proton-Coupled Electron Transfer in Artificial Photosynthetic Systems},
        type = {doi: 10.1021/acs.accounts.7b00491},
        url = {https://doi.org/10.1021/acs.accounts.7b00491},
        volume = {51},
        year = {2018},
        year1 = {2018}}

@article{Nocera:2022up,
        author = {Nocera, Daniel G. },
        date = {2022-01-26},
        doi = {10.1021/jacs.1c10444},
        journal = {J. Am. Chem. Soc.},
        journal1 = {J. Am. Chem. Soc.},
        journal2 = {J. Am. Chem. Soc.},
        month = {01},
        number = {3},
        pages = {1069--1081},
        publisher = {American Chemical Society},
        title = {Proton-Coupled Electron Transfer: The Engine of Energy Conversion and Storage},
        type = {doi: 10.1021/jacs.1c10444},
        url = {https://doi.org/10.1021/jacs.1c10444},
        volume = {144},
        year = {2022},
        year1 = {2022}}

@article{Garrido-Barros:2022us,
        author = {Garrido-Barros, Pablo and Derosa, Joseph and Chalkley, Matthew J. and Peters, Jonas C.},
        date = {2022-09-01},
        doi = {10.1038/s41586-022-05011-6},
        id = {Garrido-Barros2022},
        journal = {Nature},
        number = {7925},
        pages = {71--76},
        title = {Tandem electrocatalytic N2 fixation via proton-coupled electron transfer},
        url = {https://doi.org/10.1038/s41586-022-05011-6},
        volume = {609},
        year = {2022}}

@article{Horvath:2012vn,
        author = {Horvath, Samantha and Fernandez, Laura E. and Soudackov, Alexander V. and Hammes-Schiffer, Sharon},
        date = {2012-09-25},
        doi = {10.1073/pnas.1118333109},
        journal = {Proceedings of the National Academy of Sciences},
        journal1 = {Proceedings of the National Academy of Sciences},
        journal2 = {Proceedings of the National Academy of Sciences},
        number = {39},
        pages = {15663--15668},
        publisher = {Proceedings of the National Academy of Sciences},
        title = {Insights into proton-coupled electron transfer mechanisms of electrocatalytic H2 oxidation and production},
        type = {doi: 10.1073/pnas.1118333109},
        url = {https://doi.org/10.1073/pnas.1118333109},
        volume = {109},
        year = {2012},
        year1 = {2012}}

@misc{qc-applications-notebooks,

  author = {Morrell, Zachary and Misra, Sidhant and Lokhov, Andrey and Vuffray, Marc and Lee, Minseong and Zhang, Yu and Caravelli, Francesco and Coffrin, Carleton},

  title = {Quantum Computing Applications Specifications},

  year = {2023},

  publisher = {GitHub},

  journal = {GitHub repository},

  howpublished = {\url{https://github.com/lanl-ansi/qc-applications}},

}

@article{Ding:2023wl,
        author = {Ding, Lexin and Knecht, Stefan and Schilling, Christian},
        date = {2023-12-14},
        doi = {10.1021/acs.jpclett.3c02536},
        journal = {J. Phys. Chem. Lett.},
        journal1 = {J. Phys. Chem. Lett.},
        journal2 = {J. Phys. Chem. Lett.},
        month = {12},
        number = {49},
        pages = {11022--11029},
        publisher = {American Chemical Society},
        title = {Quantum Information-Assisted Complete Active Space Optimization (QICAS)},
        type = {doi: 10.1021/acs.jpclett.3c02536},
        url = {https://doi.org/10.1021/acs.jpclett.3c02536},
        volume = {14},
        year = {2023},
        year1 = {2023}}

@article{Handy:1984tq,
        author = {Handy, Nicholas C. and Schaefer, Henry F. , III},
        doi = {10.1063/1.447489},
        journal = {J. Chem. Phys.},
        journal1 = {J. Chem. Phys.},
        number = {11},
        pages = {5031--5033},
        title = {On the evaluation of analytic energy derivatives for correlated wave functions},
        url = {https://doi.org/10.1063/1.447489},
        volume = {81},
        year = {1984},
        year1 = {1984/12/01}}

@article{o2019calculating,
  title={Calculating energy derivatives for quantum chemistry on a quantum computer},
  author={O’Brien, Thomas E and Senjean, Bruno and Sagastizabal, Ramiro and Bonet-Monroig, Xavier and Dutkiewicz, Alicja and Buda, Francesco and DiCarlo, Leonardo and Visscher, Lucas},
  journal={npj Quantum Information},
  volume={5},
  number={1},
  pages={113},
  year={2019},
  publisher={Nature Publishing Group UK London}
}

@article{o2022efficient,
  title={Efficient quantum computation of molecular forces and other energy gradients},
  author={O'Brien, Thomas E and Streif, Michael and Rubin, Nicholas C and Santagati, Raffaele and Su, Yuan and Huggins, William J and Goings, Joshua J and Moll, Nikolaj and Kyoseva, Elica and Degroote, Matthias and others},
  journal={Physical Review Research},
  volume={4},
  number={4},
  pages={043210},
  year={2022},
  publisher={APS}
}

@article{Stein:2016ve,
        author = {Stein, Christopher J. and Reiher, Markus},
        date = {2016-04-12},
        doi = {10.1021/acs.jctc.6b00156},
        journal = {Journal of Chemical Theory and Computation},
        journal1 = {Journal of Chemical Theory and Computation},
        journal2 = {J. Chem. Theory Comput.},
        month = {04},
        number = {4},
        pages = {1760--1771},
        publisher = {American Chemical Society},
        title = {Automated Selection of Active Orbital Spaces},
        type = {doi: 10.1021/acs.jctc.6b00156},
        url = {https://doi.org/10.1021/acs.jctc.6b00156},
        volume = {12},
        year = {2016},
        year1 = {2016}}

@article{Pineda:2022vg,
        author = {Pineda, M. and Stamatakis, M.},
        doi = {10.1063/5.0083251},
        journal = {J. Chem. Phys.},
        journal1 = {J. Chem. Phys.},
        number = {12},
        pages = {120902},
        title = {Kinetic Monte Carlo simulations for heterogeneous catalysis: Fundamentals, current status, and challenges},
        url = {https://doi.org/10.1063/5.0083251},
        volume = {156},
        year = {2022},
        year1 = {2022/03/22}}

@article{Manos:2021cr,
author = {Chen, Benjamin W. J. and Xu, Lang and Mavrikakis, Manos},
title = {Computational Methods in Heterogeneous Catalysis},
journal = {Chem. Rev.},
volume = {121},
number = {2},
pages = {1007-1048},
year = {2021},
doi = {10.1021/acs.chemrev.0c01060},
}

@article{Rachel:2018cr,
author = {Crespo-Otero, Rachel and Barbatti, Mario},
title = {Recent Advances and Perspectives on Nonadiabatic Mixed Quantum–Classical Dynamics},
journal = {Chem. Rev.},
volume = {118},
number = {15},
pages = {7026-7068},
year = {2018},
doi = {10.1021/acs.chemrev.7b00577},
}

@article{Chen:2021cr,
author = {Chen, Benjamin W. J. and Xu, Lang and Mavrikakis, Manos},
title = {Computational Methods in Heterogeneous Catalysis},
journal = {Chem. Rev.},
volume = {121},
number = {2},
pages = {1007-1048},
year = {2021},
doi = {10.1021/acs.chemrev.0c01060},
}

@article{Helgaker:1982tb,
        author = {Helgaker, Trygve Ulf},
        date = {1982-05-01},
        doi = {https://doi.org/10.1002/qua.560210520},
        journal = {International Journal of Quantum Chemistry},
        journal1 = {International Journal of Quantum Chemistry},
        journal2 = {International Journal of Quantum Chemistry},
        journal3 = {Int. J. Quantum Chem.},
        number = {5},
        pages = {939--940},
        publisher = {John Wiley \& Sons, Ltd},
        title = {Simple derivation of the potential energy gradient for an arbitrary electronic wave function},
        url = {https://doi.org/10.1002/qua.560210520},
        volume = {21},
        year = {1982},
        year1 = {1982}}

@article{Parrish:2021arxiv,
  title = {Analytical Ground- and Excited-State Gradients for Molecular Electronic Structure Theory from Hybrid Quantum/Classical Methods},
  author = {Parrish, Robert M. and Anselmetti, Gian-Luca R. and Gogolin, Christian},
  journal = {arXiv:2110.05040 [quant-ph]},
  year = {2021},
  doi = {10.48550/arXiv.2110.05040},
  url = {https://doi.org/10.48550/arXiv.2110.05040},
  comments = {23 pages, 5 figures},
  archivePrefix = {arXiv},
  eprint = {2110.05040},
  primaryClass = {quant-ph},
  version = {1}
}

@article{Henrik:1990jcp,
    author = {Koch, Henrik and Jensen, Hans Jo/rgen Aa. and Jo/rgensen, Poul and Helgaker, Trygve and Scuseria, Gustavo E. and Schaefer, Henry F., III},
    title = "{Coupled cluster energy derivatives. Analytic Hessian for the closed‐shell coupled cluster singles and doubles wave function: Theory and applications}",
    journal = {J. Chem. Phys.},
    volume = {92},
    number = {8},
    pages = {4924-4940},
    year = {1990},
    month = {04},
    issn = {0021-9606},
    doi = {10.1063/1.457710},
    url = {https://doi.org/10.1063/1.457710},
    eprint = {https://pubs.aip.org/aip/jcp/article-pdf/92/8/4924/18986340/4924\_1\_online.pdf},
}

@article{Becke:2014jcp,
    author = {Becke, Axel D.},
    title = "{Perspective: Fifty years of density-functional theory in chemical physics}",
    journal = {J. Chem. Phys.},
    volume = {140},
    number = {18},
    pages = {18A301},
    year = {2014},
    month = {04},
    issn = {0021-9606},
    doi = {10.1063/1.4869598},
    url = {https://doi.org/10.1063/1.4869598},
    eprint = {https://pubs.aip.org/aip/jcp/article-pdf/doi/10.1063/1.4869598/15480507/18a301\_1\_online.pdf},
}

@article{kitaev1995,
  title={Quantum measurements and the Abelian stabilizer problem},
  author={Kitaev, A Yu},
  journal={arXiv preprint quant-ph/9511026},
  year={1995}
}

@article{Rocca2024,
  title = {Reducing the Runtime of Fault-Tolerant Quantum Simulations in Chemistry through Symmetry-Compressed Double Factorization},
  author = {Rocca, Dario and Cortes, Cristian L. and Gonthier, Jerome and Ollitrault, Pauline J. and Parrish, Robert M. and Anselmetti, Gian-Luca and Degroote, Matthias and Moll, Nikolaj and Santagati, Raffaele and Streif, Michael},
  journal = {arXiv:2403.03502 [quant-ph]},
  url = {https://doi.org/10.48550/arXiv.2403.03502},
  year = {2024},
  subjects = {Quantum Physics (quant-ph)}
}

@article{Back:2014tb,
        author = {Back, B. B. and Esbensen, H. and Jiang, C. L. and Rehm, K. E.},
        date = {2014-03-28},
        day = {28},
        doi = {10.1103/RevModPhys.86.317},
        id = {10.1103/RevModPhys.86.317},
        j1 = {RMP},
        journal = {Rev. Mod. Phys.},
        journal1 = {Rev. Mod. Phys.},
        month = {03},
        number = {1},
        pages = {317--360},
        publisher = {American Physical Society},
        title = {Recent developments in heavy-ion fusion reactions},
        url = {https://link.aps.org/doi/10.1103/RevModPhys.86.317},
        volume = {86},
        year = {2014}}

@article{feynman:1982qc,
        author = {Feynman, Richard P.}, 
        doi = {10.1007/BF02650179},
        issn = {1572-9575},
        journal = {Int. J. Theor. Phys.},
        number = {6},
        pages = {467-488},
        title = {Simulating physics with computers},
        type = {Journal Article},
        url = {https://doi.org/10.1007/BF02650179}, 
        volume = {21},
        year = {1982}}

@article{Ching:21lc,
  title = {Cartan Subalgebra Approach to Efficient Measurements of Quantum Observables},
  author = {Yen, Tzu-Ching and Izmaylov, Artur F.},
  journal = {PRX Quantum},
  volume = {2},
  issue = {4},
  pages = {040320},
  numpages = {12},
  year = {2021},
  publisher = {American Physical Society},
  doi = {10.1103/PRXQuantum.2.040320},
  url = {https://link.aps.org/doi/10.1103/PRXQuantum.2.040320}
}

@article{Huang:2020wo,
        author = {Huang, Hsin-Yuan and Kueng, Richard and Preskill, John},
        date = {2020-10-01},
        doi = {10.1038/s41567-020-0932-7},
        id = {Huang2020},
        journal = {Nat. Phys.},
        number = {10},
        pages = {1050--1057},
        title = {Predicting many properties of a quantum system from very few measurements},
        url = {https://doi.org/10.1038/s41567-020-0932-7},
        volume = {16},
        year = {2020}}

@article{Xavier:2020prx,
  title = {Nearly Optimal Measurement Scheduling for Partial Tomography of Quantum States},
  author = {Bonet-Monroig, Xavier and Babbush, Ryan and O'Brien, Thomas E.},
  journal = {Phys. Rev. X},
  volume = {10},
  issue = {3},
  pages = {031064},
  numpages = {12},
  year = {2020},
  publisher = {American Physical Society},
  doi = {10.1103/PhysRevX.10.031064},
  url = {https://link.aps.org/doi/10.1103/PhysRevX.10.031064}
}

@misc{linkedinCatalystMarket,
	author = {},
	title = {{C}atalyst {M}arket {S}ize {R}esearch {R}eport [2023-2030] | 116 {P}ages  --- linkedin.com},
	howpublished = {\url{https://www.linkedin.com/pulse/catalyst-market-size-research-report-gxaef/}},
	year = {},
	note = {[Accessed 08-04-2024]},
}

@article{Simon:2024prxq,
  title = {Improved Precision Scaling for Simulating Coupled Quantum-Classical Dynamics},
  author = {Simon, Sophia and Santagati, Raffaele and Degroote, Matthias and Moll, Nikolaj and Streif, Michael and Wiebe, Nathan},
  journal = {PRX Quantum},
  volume = {5},
  issue = {1},
  pages = {010343},
  numpages = {58},
  year = {2024},
  publisher = {American Physical Society},
  doi = {10.1103/PRXQuantum.5.010343},
  url = {https://link.aps.org/doi/10.1103/PRXQuantum.5.010343}
}

@article{exfo1,
  title={Mechanistic insight into the chemical exfoliation and functionalization of Ti3C2 MXene},
  author={Srivastava, Pooja and Mishra, Avanish and Mizuseki, Hiroshi and Lee, Kwang-Ryeol and Singh, Abhishek K},
  journal={ACS applied materials \& interfaces},
  volume={8},
  number={36},
  pages={24256--24264},
  year={2016},
  publisher={ACS Publications}
}

@article{exfo2,
  title={Liquid-phase exfoliation of phosphorene: design rules from molecular dynamics simulations},
  author={Sresht, Vishnu and Padua, Agilio AH and Blankschtein, Daniel},
  journal={ACS nano},
  volume={9},
  number={8},
  pages={8255--8268},
  year={2015},
  publisher={ACS Publications}
}

@article{exfo3,
  title={Deciphering the role of solvents in the liquid phase exfoliation of hexagonal boron nitride: A molecular dynamics simulation study},
  author={Mukhopadhyay, Titas Kumar and Datta, Ayan},
  journal={J. Phys. Chem. C},
  volume={121},
  number={1},
  pages={811--822},
  year={2017},
  publisher={ACS Publications}
}

@article{bat1,
  title={Li ion diffusion mechanisms in LiFePO4: an ab initio molecular dynamics study},
  author={Yang, Jianjun and Tse, John S},
  journal={J. Phys. Chem. A},
  volume={115},
  number={45},
  pages={13045--13049},
  year={2011},
  publisher={ACS Publications}
}

@article{bat2,
  title={Two-dimensional infrared spectroscopy and molecular dynamics simulation studies of nonaqueous lithium ion battery electrolytes},
  author={Lim, Joonhyung and Lee, Kyung-Koo and Liang, Chungwen and Park, Kwang-Hee and Kim, Minjoo and Kwak, Kyungwon and Cho, Minhaeng},
  journal={J. Phys. Chem. B},
  volume={123},
  number={31},
  pages={6651--6663},
  year={2019},
  publisher={ACS Publications}
}

@article{mechno1,
  title={First principles dynamics and minimum energy pathways for mechanochemical ring opening of cyclobutene},
  author={Ong, Mitchell T and Leiding, Jeff and Tao, Hongli and Virshup, Aaron M and Mart{\'\i}nez, Todd J},
  journal={J. Am. Chem. Soc.},
  volume={131},
  number={18},
  pages={6377--6379},
  year={2009},
  publisher={ACS Publications}
}

@article{mechno2,
  title={Mechanochemistry for ammonia synthesis under mild conditions},
  author={Han, Gao-Feng and Li, Feng and Chen, Zhi-Wen and Coppex, Claude and Kim, Seok-Jin and Noh, Hyuk-Jun and Fu, Zhengping and Lu, Yalin and Singh, Chandra Veer and Siahrostami, Samira and others},
  journal={Nat. Nano.},
  volume={16},
  number={3},
  pages={325--330},
  year={2021},
  publisher={Nature Publishing Group UK London}
}

@article{mechno3,
  title={Extemporaneous mechanochemistry: Shock-wave-induced ultrafast chemical reactions due to intramolecular strain energy},
  author={Hamilton, Brenden W and Kroonblawd, Matthew P and Strachan, Alejandro},
  journal={J. Phys. Chem. Lett.},
  volume={13},
  number={29},
  pages={6657--6663},
  year={2022},
  publisher={ACS Publications}
}

@article{cat1,
  title={Evolution of metastable structures at bimetallic surfaces from microscopy and machine-learning molecular dynamics},
  author={Lim, Jin Soo and Vandermause, Jonathan and Van Spronsen, Matthijs A and Musaelian, Albert and Xie, Yu and Sun, Lixin and O\'Connor, Christopher R and Egle, Tobias and Molinari, Nicola and Florian, Jacob and others},
  journal={J. Am. Chem. Soc.},
  volume={142},
  number={37},
  pages={15907--15916},
  year={2020},
  publisher={ACS Publications}
}

@article{cat2,
  title={Diffusion mechanisms of metal atoms in PdAu bimetallic catalyst under CO atmosphere based on ab initio molecular dynamics},
  author={Yang, Yongpeng and Shen, Xiangjian and Han, Yi-Fan},
  journal={Applied Surf. Sci.},
  volume={483},
  pages={991--1005},
  year={2019},
  publisher={Elsevier}
}

@article{cat3,
  title={Structural rearrangement of Au--Pd nanoparticles under reaction conditions: An ab initio molecular dynamics study},
  author={Xu, Cong-Qiao and Lee, Mal-Soon and Wang, Yang-Gang and Cantu, David C and Li, Jun and Glezakou, Vassiliki-Alexandra and Rousseau, Roger},
  journal={ACS nano},
  volume={11},
  number={2},
  pages={1649--1658},
  year={2017},
  publisher={ACS Publications}
}

@article{Weinberg:2012cr,
author = {Weinberg, David R. and Gagliardi, Christopher J. and Hull, Jonathan F. and Murphy, Christine Fecenko and Kent, Caleb A. and Westlake, Brittany C.  and Paul, Amit and Ess, Daniel H. and McCafferty, Dewey Granville and Meyer, Thomas J.},
title = {Proton-Coupled Electron Transfer},
journal = {Chem. Rev.},
volume = {112},
number = {7},
pages = {4016-4093},
year = {2012},
doi = {10.1021/cr200177j},
}

@article{collins2002molecular,
  title={Molecular potential-energy surfaces for chemical reaction dynamics},
  author={Collins, Michael A},
  journal={Theoretical Chemistry Accounts},
  volume={108},
  pages={313--324},
  year={2002},
  publisher={Springer}
}

@article{Schiffer:2015jacs,
author = {Hammes-Schiffer, Sharon},
title = {Proton-Coupled Electron Transfer: Moving Together and Charging Forward},
journal = {J. Am. Chem. Soc.},
volume = {137},
number = {28},
pages = {8860-8871},
year = {2015},
doi = {10.1021/jacs.5b04087},
}

@article{hynes1985chemical,
  title={Chemical reaction dynamics in solution},
  author={Hynes, James T},
  journal={Annual Review of Physical Chemistry},
  volume={36},
  number={1},
  pages={573--597},
  year={1985},
  publisher={Annual Reviews 4139 El Camino Way, PO Box 10139, Palo Alto, CA 94303-0139, USA}
}

@article{Curchod2018CR,
        author = {Curchod, Basile F. E. and Mart{\'{i}}nez, Todd J.},
        creationdate = {2023-03-15T23:35:00},
        doi = {10.1021/acs.chemrev.7b00423},
        issn = {0009-2665},
        journal = {Chem. Rev.},
        modificationdate = {2023-03-15T23:35:04},
        number = {7},
        pages = {3305--3336},
        title = {Ab Initio Nonadiabatic Quantum Molecular Dynamics},
        urldate = {2020-10-17},
        volume = {118},
        year = {2018}}

@article{Nelson:2018cr,
        author = {Nelson, Tammie R. and White, Alexander J. and Bjorgaard, Josiah A. and Sifain, Andrew E. and Zhang, Yu and Nebgen, Benjamin and Fernandez-Alberti, Sebastian and Mozyrsky, Dmitry and Roitberg, Adrian E. and Tretiak, Sergei},
        doi = {10.1021/acs.chemrev.9b00447},
        journal = {Chem. Rev.},
        number = {4},
        pages = {2215-2287},
        title = {Non-adiabatic Excited-State Molecular Dynamics: Theory and Applications for Modeling Photophysics in Extended Molecular Materials},
        volume = {120},
        year = {2020}}

@article{Bernhard:2003ep,
author = {Schlegel, H. Bernhard},
title = {Exploring potential energy surfaces for chemical reactions: An overview of some practical methods},
journal = {Journal of Computational Chemistry},
volume = {24},
number = {12},
pages = {1514-1527},
doi = {https://doi.org/10.1002/jcc.10231},
url = {https://onlinelibrary.wiley.com/doi/abs/10.1002/jcc.10231},
eprint = {https://onlinelibrary.wiley.com/doi/pdf/10.1002/jcc.10231},
year = {2003}
}

@article{marx2000ab,
  title={Ab initio molecular dynamics: Theory and implementation},
  author={Marx, Dominik and Hutter, Jurg},
  journal={Modern methods and algorithms of quantum chemistry},
  volume={1},
  number={301-449},
  pages={141},
  year={2000},
  publisher={John von Neumann Institute for Computing J{\"u}lich}
}

@article{tuckerman2002ab,
  title={Ab initio molecular dynamics: basic concepts, current trends and novel applications},
  author={Tuckerman, Mark E},
  journal={Journal of Physics: Condensed Matter},
  volume={14},
  number={50},
  pages={R1297},
  year={2002},
  publisher={IOP Publishing}
}

@article{PhysRevLett.55.2471,
  title = {Unified Approach for Molecular Dynamics and Density-Functional Theory},
  author = {Car, R. and Parrinello, M.},
  journal = {Phys. Rev. Lett.},
  volume = {55},
  issue = {22},
  pages = {2471--2474},
  numpages = {0},
  year = {1985},
  publisher = {American Physical Society},
  doi = {10.1103/PhysRevLett.55.2471},
  url = {https://link.aps.org/doi/10.1103/PhysRevLett.55.2471}
}

@article{RevModPhys.64.1045,
  title = {Iterative minimization techniques for ab initio total-energy calculations: molecular dynamics and conjugate gradients},
  author = {Payne, M. C. and Teter, M. P. and Allan, D. C. and Arias, T. A. and Joannopoulos, J. D.},
  journal = {Rev. Mod. Phys.},
  volume = {64},
  issue = {4},
  pages = {1045--1097},
  numpages = {0},
  year = {1992},
  publisher = {American Physical Society},
  doi = {10.1103/RevModPhys.64.1045},
  url = {https://link.aps.org/doi/10.1103/RevModPhys.64.1045}
}

@article{andp.19273892002,
author = {Born, M. and Oppenheimer, R.},
title = {Zur Quantentheorie der Molekeln},
journal = {Annalen der Physik},
volume = {389},
number = {20},
pages = {457-484},
doi = {https://doi.org/10.1002/andp.19273892002},
url = {https://onlinelibrary.wiley.com/doi/abs/10.1002/andp.19273892002},
eprint = {https://onlinelibrary.wiley.com/doi/pdf/10.1002/andp.19273892002},
year = {1927}
}

@article{ap2,
  title={Artificial photosynthesis: Beyond mimicking nature},
  author={Dau, Holger and Fujita, Etsuko and Sun, Licheng},
  journal={ChemSusChem},
  volume={10},
  number={22},
  pages={4228--4235},
  year={2017},
  publisher={Wiley Online Library}
}

@article{ap1,
  title={Water-splitting chemistry of photosystem II},
  author={McEvoy, James P and Brudvig, Gary W},
  journal={Chem. Rev.},
  volume={106},
  number={11},
  pages={4455--4483},
  year={2006},
  publisher={ACS Publications}
}

@article{McArdle:2020we,
	author = {McArdle, Sam and Endo, Suguru and Aspuru-Guzik, Al{\'a}n and Benjamin, Simon C. and Yuan, Xiao},
	date = {2020-03-30},
	day = {30},
	doi = {10.1103/RevModPhys.92.015003},
	id = {10.1103/RevModPhys.92.015003},
	j1 = {RMP},
	journal = {Rev. Mod. Phys.},
	month = {03},
	number = {1},
	pages = {015003--},
	publisher = {American Physical Society},
	title = {Quantum computational chemistry},
	url = {https://link.aps.org/doi/10.1103/RevModPhys.92.015003},
	volume = {92},
	year = {2020}}

@article{Bauer:2020ud,
	author = {Bauer, Bela and Bravyi, Sergey and Motta, Mario and Chan, Garnet Kin-Lic},
	date = {2020-11-25},
	doi = {10.1021/acs.chemrev.9b00829},
	journal = {Chem. Rev.},
	month = {11},
	number = {22},
	pages = {12685--12717},
	publisher = {American Chemical Society},
	title = {Quantum Algorithms for Quantum Chemistry and Quantum Materials Science},
	type = {doi: 10.1021/acs.chemrev.9b00829},
	url = {https://doi.org/10.1021/acs.chemrev.9b00829},
	volume = {120},
	year = {2020},
	year1 = {2020}}

@article{Chan:2011vs,
	author = {Chan, Garnet Kin-Lic and Sharma, Sandeep},
	date = {2011-05-05},
	doi = {10.1146/annurev-physchem-032210-103338},
	journal = {Annual Review of Physical Chemistry},
	journal1 = {Annual Review of Physical Chemistry},
	journal2 = {Annu. Rev. Phys. Chem.},
	n2 = {The density matrix renormalization group is a method that is useful for describing molecules that have strongly correlated electrons. Here we provide a pedagogical overview of the basic challenges of strong correlation, how the density matrix renormalization group works, a survey of its existing applications to molecular problems, and some thoughts on the future of the method.},
	number = {1},
	pages = {465--481},
	publisher = {Annual Reviews},
	title = {The Density Matrix Renormalization Group in Quantum Chemistry},
	type = {doi: 10.1146/annurev-physchem-032210-103338},
	url = {https://doi.org/10.1146/annurev-physchem-032210-103338},
	volume = {62},
	year = {2011},
	year1 = {2011}}

@article{Bartlett:2007wj,
	author = {Bartlett, Rodney J. and Musia{\l}, Monika},
	date = {2007-02-22},
	day = {22},
	doi = {10.1103/RevModPhys.79.291},
	id = {10.1103/RevModPhys.79.291},
	j1 = {RMP},
	journal = {Rev. Mod. Phys.},
	month = {02},
	number = {1},
	pages = {291--352},
	publisher = {American Physical Society},
	title = {Coupled-cluster theory in quantum chemistry},
	url = {https://link.aps.org/doi/10.1103/RevModPhys.79.291},
	volume = {79},
	year = {2007}}

@article{Kim:2015uj,
	author = {Kim, Dohyung and Sakimoto, Kelsey K. and Hong, Dachao and Yang, Peidong},
	date = {2015-03-09},
	doi = {https://doi.org/10.1002/anie.201409116},
	journal = {Angew. Chem., Int. Ed.},
	journal3 = {Angew. Chem. Int. Ed.},
	n2 = {Abstract The apparent incongruity between the increasing consumption of fuels and chemicals and the finite amount of resources has led us to seek means to maintain the sustainability of our society. Artificial photosynthesis, which utilizes sunlight to create high-value chemicals from abundant resources, is considered as the most promising and viable method. This Minireview describes the progress and challenges in the field of artificial photosynthesis in terms of its key components: developments in photoelectrochemical water splitting and recent progress in electrochemical CO2 reduction. Advances in catalysis, concerning the use of renewable hydrogen as a feedstock for major chemical production, are outlined to shed light on the ultimate role of artificial photosynthesis in achieving sustainable chemistry.},
	number = {11},
	pages = {3259--3266},
	publisher = {John Wiley \& Sons, Ltd},
	title = {Artificial Photosynthesis for Sustainable Fuel and Chemical Production},
	url = {https://doi.org/10.1002/anie.201409116},
	volume = {54},
	year = {2015},
	year1 = {2015}}

@article{Ye:2019wj,
	author = {Ye, Sheng and Ding, Chunmei and Liu, Mingyao and Wang, Aoqi and Huang, Qinge and Li, Can},
	date = {2019-12-01},
	doi = {https://doi.org/10.1002/adma.201902069},
	journal = {Adv. Mat.},
	n2 = {Abstract Water oxidation is the primary reaction of both natural and artificial photosynthesis. Developing active and robust water oxidation catalysts (WOCs) is the key to constructing efficient artificial photosynthesis systems, but it is still facing enormous challenges in both fundamental and applied aspects. Here, the recent developments in molecular catalysts and heterogeneous nanoparticle catalysts are reviewed with special emphasis on biomimetic catalysts and the integration of WOCs into artificial photosystems. The highly efficient artificial photosynthesis depends largely on active WOCs integrated into light harvesting materials via rational interface engineering based on in-depth understanding of charge dynamics and the reaction mechanism.},
	number = {50},
	pages = {1902069},
	publisher = {John Wiley \& Sons, Ltd},
	title = {Water Oxidation Catalysts for Artificial Photosynthesis},
	url = {https://doi.org/10.1002/adma.201902069},
	volume = {31},
	year = {2019},
	year1 = {2019}}

@article{Karkas:2014wv,
	author = {K{\"a}rk{\"a}s, Markus D. and Verho, Oscar and Johnston, Eric V. and {\AA}kermark, Bj{\"o}rn},
	date = {2014-12-24},
	doi = {10.1021/cr400572f},
	journal = {Chem. Rev.},
	month = {12},
	number = {24},
	pages = {11863--12001},
	publisher = {American Chemical Society},
	title = {Artificial Photosynthesis: Molecular Systems for Catalytic Water Oxidation},
	type = {doi: 10.1021/cr400572f},
	url = {https://doi.org/10.1021/cr400572f},
	volume = {114},
	year = {2014},
	year1 = {2014}}

@article{von2021quantum,
	author = {von Burg, Vera and Low, Guang Hao and H{\"a}ner, Thomas and Steiger, Damian S and Reiher, Markus and Roetteler, Martin and Troyer, Matthias},
	journal = {Phys. Rev. Res.},
	number = {3},
	pages = {033055},
	publisher = {APS},
	title = {Quantum computing enhanced computational catalysis},
	volume = {3},
	year = {2021}}

@article{chemrev.8b00803,
	author = {Cao, Yudong and Romero, Jonathan and Olson, Jonathan P. and Degroote, Matthias and Johnson, Peter D. and Kieferov{\'a}, M{\'a}ria and Kivlichan, Ian D. and Menke, Tim and Peropadre, Borja and Sawaya, Nicolas P. D. and Sim, Sukin and Veis, Libor and Aspuru-Guzik, Al{\'a}n},
	doi = {10.1021/acs.chemrev.8b00803},
	journal = {Chem. Rev.},
	number = {19},
	pages = {10856-10915},
	title = {Quantum Chemistry in the Age of Quantum Computing},
	volume = {119},
	year = {2019}}

@article{parity2012,
	author = {Seeley,Jacob T. and Richard,Martin J. and Love,Peter J.},
	doi = {10.1063/1.4768229},
	journal = {J. Chem. Phys.},
	number = {22},
	pages = {224109},
	title = {The Bravyi-Kitaev transformation for quantum computation of electronic structure},
	url = {https://doi.org/10.1063/1.4768229},
	volume = {137},
	year = {2012}}

@article{BRAVYI2002210,
	author = {Sergey B. Bravyi and Alexei Yu. Kitaev},
	doi = {https://doi.org/10.1006/aphy.2002.6254},
	issn = {0003-4916},
	journal = {Ann. Phys.},
	number = {1},
	pages = {210 - 226},
	title = {Fermionic Quantum Computation},
	url = {http://www.sciencedirect.com/science/article/pii/S0003491602962548},
	volume = {298},
	year = {2002}}

@article{JW:1928,
	author = {Jordan, P. and Wigner, E.},
	doi = {10.1007/BF01331938},
	issn = {0044-3328},
	journal = {Zeitschrift f{\"u}r Physik},
	number = {9},
	pages = {631-651},
	title = {{\"U}ber das Paulische {\"A}quivalenzverbot},
	type = {Journal Article},
	url = {https://doi.org/10.1007/BF01331938},
	volume = {47},
	year = {1928}}

@article{Rissler2006,
	author = {J{\"o}rg Rissler and Reinhard M. Noack and Steven R. White},
	journal = {Chem. Phys},
	pages = {519-531},
	title = {Measuring orbital interaction using quantum information theory},
	url = {https://www.sciencedirect.com/science/article/pii/S0301010405005069},
	volume = {323},
	year = {2006}}

@article{tkachenko2021correlation,
	author = {Tkachenko, Nikolay V and Sud, James and Yu Zhang and Tretiak, Sergei and Anisimov, Petr M and Arrasmith, Andrew T and Coles, Patrick J and Cincio, Lukasz and Dub, Pavel A},
	journal = {PRX Quantum},
	number = {2},
	pages = {020337},
	publisher = {APS},
	title = {Correlation-Informed Permutation of Qubits for Reducing Ansatz Depth in the Variational Quantum Eigensolver},
	volume = {2},
	year = {2021}}

@article{clusterVQE,
	author = {Zhang, Yu and Cincio, Lukasz and Negre, Christian FA and Czarnik, Piotr and Coles, Patrick and Anisimov, Petr M and Mniszewski, Susan M and Tretiak, Sergei and Dub, Pavel A},
	issue = {1},
	journal = {Npj Quantum Inf.},
	pages = {96},
	title = {Variational Quantum Eigensolver with Reduced Circuit Complexity},
	volume = {8},
	year = {2022}}

@article{Mills:1005sc,
	author = {Gregory Mills and Hannes J{\'o}nsson and Gregory K. Schenter},
	doi = {https://doi.org/10.1016/0039-6028(94)00731-4},
	issn = {0039-6028},
	journal = {Surf. Sci.},
	number = {2},
	pages = {305-337},
	title = {Reversible work transition state theory: application to dissociative adsorption of hydrogen},
	url = {https://www.sciencedirect.com/science/article/pii/0039602894007314},
	volume = {324},
	year = {1995}}

@article{tkachenko_quantum_2022,
        author = {Tkachenko, Nikolay V and Cincio, Lukasz and Boldyrev, Alexander I and Tretiak, Sergei and Dub, Pavel A and Zhang, Yu},
        date = {2024-04-22},
        doi = {10.1088/2058-9565/ad3a97},
        journal = {Quantum Sci. Technol.},
        number = {3},
        pages = {035012},
        publisher = {IOP Publishing},
        title = {Quantum Davidson algorithm for excited states},
        url = {https://dx.doi.org/10.1088/2058-9565/ad3a97},
        volume = {9},
        year = {2024}}

@article{lee_evaluating_2023,
	author = {Lee, Seunghoon and Lee, Joonho and Zhai, Huanchen and Tong, Yu and Dalzell, Alexander M. and Kumar, Ashutosh and Helms, Phillip and Gray, Johnnie and Cui, Zhi-Hao and Liu, Wenyuan and Kastoryano, Michael and Babbush, Ryan and Preskill, John and Reichman, David R. and Campbell, Earl T. and Valeev, Edward F. and Lin, Lin and Chan, Garnet Kin-Lic},
	doi = {10.1038/s41467-023-37587-6},
	issn = {2041-1723},
	journal = {Nat. Commun.},
	language = {en},
	month = apr,
	number = {1},
	pages = {1952},
	title = {Evaluating the evidence for exponential quantum advantage in ground-state quantum chemistry},
	url = {https://www.nature.com/articles/s41467-023-37587-6},
	urldate = {2023-12-19},
	volume = {14},
	year = {2023}}

@misc{tubman_postponing_2018,
	author = {Tubman, Norm M. and Mejuto-Zaera, Carlos and Epstein, Jeffrey M. and Hait, Diptarka and Levine, Daniel S. and Huggins, William and Jiang, Zhang and McClean, Jarrod R. and Babbush, Ryan and Head-Gordon, Martin and Whaley, K. Birgitta},
	month = sep,
	note = {arXiv:1809.05523 [cond-mat, physics:physics, physics:quant-ph]},
	publisher = {arXiv},
	shorttitle = {Postponing the orthogonality catastrophe},
	title = {Postponing the orthogonality catastrophe: efficient state preparation for electronic structure simulations on quantum devices},
	url = {http://arxiv.org/abs/1809.05523},
	urldate = {2023-12-19},
	year = {2018}}

@article{atwater2023artificial,
	author = {Atwater, Harry A},
	journal = {Physics Today},
	number = {12},
	pages = {32--39},
	publisher = {AIP Publishing},
	title = {Artificial photosynthesis: A pathway to solar fuels},
	volume = {76},
	year = {2023}}

@article{yang2023landscape,
	author = {Yang, Ke R and Kyro, Gregory W and Batista, Victor S},
	journal = {Nat. Comput. Sci.},
	number = {6},
	pages = {504--513},
	publisher = {Nature Publishing Group US New York},
	title = {The landscape of computational approaches for artificial photosynthesis},
	volume = {3},
	year = {2023}}

@article{lyu2023time,
	author = {Lyu, Danya and Xu, Jinchang and Wang, Zhenyou},
	journal = {Front. Chem.},
	publisher = {Frontiers Media SA},
	title = {Time-resolved in situ vibrational spectroscopy for electrocatalysis: challenge and opportunity},
	volume = {11},
	year = {2023}}

@article{mishra2017atomistic,
	author = {Mishra, Avanish and Srivastava, Pooja and Carreras, Abel and Tanaka, Isao and Mizuseki, Hiroshi and Lee, Kwang-Ryeol and Singh, Abhishek K},
	journal = {J. Phys. Chem. C},
	number = {34},
	pages = {18947--18953},
	publisher = {ACS Publications},
	title = {Atomistic origin of phase stability in oxygen-functionalized MXene: a comparative study},
	volume = {121},
	year = {2017}}

@article{parida2020vertically,
	author = {Parida, Shayani and Mishra, Avanish and Chen, Jie and Wang, Jin and Dobley, Arthur and Carter, C Barry and Dongare, Avinash M},
	journal = {Journal of the American Ceramic Society},
	number = {11},
	pages = {6603--6614},
	publisher = {Wiley Online Library},
	title = {Vertically stacked 2H-1T dual-phase MoS2 microstructures during lithium intercalation: a first principles study},
	volume = {103},
	year = {2020}}

@article{johansen2015aluminum,
	author = {Johansen, KM and Vines, Lasse and Bj{\o}rheim, Tor Svendsen and Schifano, Ramon and Svensson, Bengt Gunnar},
	journal = {Phys. Rev. Applied},
	number = {2},
	pages = {024003},
	publisher = {APS},
	title = {Aluminum migration and intrinsic defect interaction in single-crystal zinc oxide},
	volume = {3},
	year = {2015}}

@article{roubelakis2012proton,
	author = {Roubelakis, Manolis M and Bediako, D Kwabena and Dogutan, Dilek K and Nocera, Daniel G},
	journal = {Energy \& Environmental Science},
	number = {7},
	pages = {7737--7740},
	publisher = {Royal Society of Chemistry},
	title = {Proton-coupled electron transfer kinetics for the hydrogen evolution reaction of hangman porphyrins},
	volume = {5},
	year = {2012}}

@article{pei2015defective,
	author = {Pei, Dan-Ni and Gong, Li and Zhang, Ai-Yong and Zhang, Xing and Chen, Jie-Jie and Mu, Yang and Yu, Han-Qing},
	journal = {Nat. Commun.},
	number = {1},
	pages = {8696},
	publisher = {Nature Publishing Group UK London},
	title = {Defective titanium dioxide single crystals exposed by high-energy $\{$001$\}$ facets for efficient oxygen reduction},
	volume = {6},
	year = {2015}}

@article{koza2012electrodeposition,
	author = {Koza, Jakub A and He, Zhen and Miller, Andrew S and Switzer, Jay A},
	journal = {Chemistry of Materials},
	number = {18},
	pages = {3567--3573},
	publisher = {ACS Publications},
	title = {Electrodeposition of Crystalline Co$_3$O$_4$ A Catalyst for the Oxygen Evolution Reaction},
	volume = {24},
	year = {2012}}

@article{Zhang:2017tn,
	author = {Zhang, Xing and Wu, Zishan and Zhang, Xiao and Li, Liewu and Li, Yanyan and Xu, Haomin and Li, Xiaoxiao and Yu, Xiaolu and Zhang, Zisheng and Liang, Yongye and Wang, Hailiang},
	date = {2017-03-08},
	doi = {10.1038/ncomms14675},
	id = {Zhang2017},
	journal = {Nat. Commun.},
	number = {1},
	pages = {14675},
	title = {Highly selective and active CO2 reduction electrocatalysts based on cobalt phthalocyanine/carbon nanotube hybrid structures},
	url = {https://doi.org/10.1038/ncomms14675},
	volume = {8},
	year = {2017}}

@article{Li:2020wr,
	author = {Li, Dandan and Kassymova, Meruyert and Cai, Xuechao and Zang, Shuang-Quan and Jiang, Hai-Long},
	date = {2020-06-01},
	doi = {https://doi.org/10.1016/j.ccr.2020.213262},
	journal = {Coordination Chemistry Reviews},
	pages = {213262},
	title = {Photocatalytic CO2 reduction over metal-organic framework-based materials},
	url = {https://www.sciencedirect.com/science/article/pii/S0010854520300461},
	volume = {412},
	year = {2020}}

@article{Fang:2023tf,
	author = {Fang, Siyuan and Rahaman, Motiar and Bharti, Jaya and Reisner, Erwin and Robert, Marc and Ozin, Geoffrey A. and Hu, Yun Hang},
	date = {2023-08-10},
	doi = {10.1038/s43586-023-00243-w},
	id = {Fang2023},
	journal = {Nature Reviews Methods Primers},
	number = {1},
	pages = {61},
	title = {Photocatalytic CO2 reduction},
	url = {https://doi.org/10.1038/s43586-023-00243-w},
	volume = {3},
	year = {2023}}

@article{Liu:2019nc,
	author = {Liu, Yingshuo and McCrory, Charles C. L.},
	date = {2019-04-11},
	doi = {10.1038/s41467-019-09626-8},
	id = {Liu2019},
	journal = {Nat. Commun.},
	number = {1},
	pages = {1683},
	title = {Modulating the mechanism of electrocatalytic CO2 reduction by cobalt phthalocyanine through polymer coordination and encapsulation},
	url = {https://doi.org/10.1038/s41467-019-09626-8},
	volume = {10},
	year = {2019}}

@article{birkenheuer2005model,
	author = {Birkenheuer, Uwe and Gordienko, Aleksei B and Nasluzov, Vladimir A and Fuchs-Rohr, Monika K and R{\"o}sch, Notker},
	journal = {Int. Quantum Chem.},
	number = {5},
	pages = {743--761},
	publisher = {Wiley Online Library},
	title = {Model density approach to the Kohn--Sham problem: Efficient extension of the density fitting technique},
	volume = {102},
	year = {2005}}

@article{wasielewski1992photoinduced,
	author = {Wasielewski, Michael R},
	journal = {Chem. Rev.},
	number = {3},
	pages = {435--461},
	publisher = {ACS Publications},
	title = {Photoinduced electron transfer in supramolecular systems for artificial photosynthesis},
	volume = {92},
	year = {1992}}

@article{meyer1989chemical,
	author = {Meyer, Thomas J},
	journal = {Acc. Chem. Res.},
	number = {5},
	pages = {163--170},
	publisher = {ACS Publications},
	title = {Chemical approaches to artificial photosynthesis},
	volume = {22},
	year = {1989}}

@article{benniston2008artificial,
	author = {Benniston, Andrew C and Harriman, Anthony},
	journal = {Materials Today},
	number = {12},
	pages = {26--34},
	publisher = {Elsevier},
	title = {Artificial photosynthesis},
	volume = {11},
	year = {2008}}

@article{barber2013natural,
	author = {Barber, James and Tran, Phong D},
	journal = {Journal of The Royal Society Interface},
	number = {81},
	pages = {20120984},
	publisher = {The Royal Society},
	title = {From natural to artificial photosynthesis},
	volume = {10},
	year = {2013}}

@book{hall1999photosynthesis,
	author = {Hall, David Oakley and Rao, Krishna},
	publisher = {Cambridge University Press},
	title = {Photosynthesis},
	year = {1999}}

@article{costentin2013catalysis,
	author = {Costentin, Cyrille and Robert, Marc and Sav{\'e}ant, Jean-Michel},
	journal = {Chem. Soc. Rev.},
	number = {6},
	pages = {2423--2436},
	publisher = {Royal Society of Chemistry},
	title = {Catalysis of the electrochemical reduction of carbon dioxide},
	volume = {42},
	year = {2013}}

@article{reyes2022molecular,
	author = {Reyes Cruz, Edgar A and Nishiori, Daiki and Wadsworth, Brian L and Nguyen, Nghi P and Hensleigh, Lillian K and Khusnutdinova, Diana and Beiler, Anna M and Moore, GF},
	journal = {Chem. Rev.},
	number = {21},
	pages = {16051--16109},
	publisher = {ACS Publications},
	title = {Molecular-Modified Photocathodes for Applications in Artificial Photosynthesis and Solar-to-Fuel Technologies},
	volume = {122},
	year = {2022}}

@article{C3CS60405E,
	author = {Berardi, Serena and Drouet, Samuel and Franc{\`a}s, Laia and Gimbert-Suri{\~n}ach, Carolina and Guttentag, Miguel and Richmond, Craig and Stoll, Thibaut and Llobet, Antoni},
	doi = {10.1039/C3CS60405E},
	issue = {22},
	journal = {Chem. Soc. Rev.},
	pages = {7501-7519},
	publisher = {The Royal Society of Chemistry},
	title = {Molecular artificial photosynthesis},
	url = {http://dx.doi.org/10.1039/C3CS60405E},
	volume = {43},
	year = {2014}}

@article{yang2023,
	author = {Yang, Ke R and Kyro, Gregory W and Batista, Victor S},
	journal = {Nat. Comput. Sci.},
	number = {6},
	pages = {504--513},
	publisher = {Nature Publishing Group US New York},
	title = {The landscape of computational approaches for artificial photosynthesis},
	volume = {3},
	year = {2023}}

@article{matheu2015,
	author = {Matheu, Roc and Ertem, Mehmed Z and Benet-Buchholz, Jordi and Coronado, Eugenio and Batista, Victor S and Sala, Xavier and Llobet, Antoni},
	journal = {J. Am. Chem. Soc.},
	number = {33},
	pages = {10786--10795},
	publisher = {ACS Publications},
	title = {Intramolecular proton transfer boosts water oxidation catalyzed by a Ru complex},
	volume = {137},
	year = {2015}}

@article{liao2015,
	author = {Liao, Kuo and Askerka, Mikhail and Zeitler, Elizabeth L and Bocarsly, Andrew B and Batista, Victor S},
	journal = {Topics in Catalysis},
	pages = {23--29},
	publisher = {Springer},
	title = {Electrochemical reduction of aqueous imidazolium on Pt (111) by proton coupled electron transfer},
	volume = {58},
	year = {2015}}

@article{Dau2010,
	author = {Dau, Holger and Limberg, Christian and Reier, Tobias and Risch, Marcel and Roggan, Stefan and Strasser, Peter},
	doi = {https://doi.org/10.1002/cctc.201000126},
	eprint = {https://chemistry-europe.onlinelibrary.wiley.com/doi/pdf/10.1002/cctc.201000126},
	journal = {ChemCatChem},
	number = {7},
	pages = {724-761},
	title = {The Mechanism of Water Oxidation: From Electrolysis via Homogeneous to Biological Catalysis},
	url = {https://chemistry-europe.onlinelibrary.wiley.com/doi/abs/10.1002/cctc.201000126},
	volume = {2},
	year = {2010}}

@article{Kanan2008sci,
	author = {Matthew W. Kanan and Daniel G. Nocera},
	doi = {10.1126/science.1162018},
	eprint = {https://www.science.org/doi/pdf/10.1126/science.1162018},
	journal = {Science},
	number = {5892},
	pages = {1072-1075},
	title = {In Situ Formation of an Oxygen-Evolving Catalyst in Neutral Water Containing Phosphate and $\text{C}o^{2+}$},
	url = {https://www.science.org/doi/abs/10.1126/science.1162018},
	volume = {321},
	year = {2008}}

@article{jp511805x,
	author = {Fernando, Amendra and Aikens, Christine M.},
	doi = {10.1021/jp511805x},
	journal = {J. Phys. Chem. C},
	number = {20},
	pages = {11072-11085},
	title = {Reaction Pathways for Water Oxidation to Molecular Oxygen Mediated by Model Cobalt Oxide Dimer and Cubane Catalysts},
	volume = {119},
	year = {2015}}

@article{cao2019quantum,
	author = {Cao, Yudong and Romero, Jonathan and Olson, Jonathan P and Degroote, Matthias and Johnson, Peter D and Kieferov{\'a}, M{\'a}ria and Kivlichan, Ian D and Menke, Tim and Peropadre, Borja and Sawaya, Nicolas PD and others},
	journal = {Chem. Rev.},
	number = {19},
	pages = {10856--10915},
	publisher = {ACS Publications},
	title = {Quantum chemistry in the age of quantum computing},
	volume = {119},
	year = {2019}}

@article{reiher2017elucidating,
	author = {Reiher, Markus and Wiebe, Nathan and Svore, Krysta M and Wecker, Dave and Troyer, Matthias},
	journal = {Proc. Natl. Acad. Sci.},
	number = {29},
	pages = {7555--7560},
	publisher = {National Acad Sciences},
	title = {Elucidating reaction mechanisms on quantum computers},
	volume = {114},
	year = {2017}}

@article{list1b,
  title={Recent advances in catalyst design and activity enhancement induced by a magnetic field for electrocatalysis},
  author={Wang, Kun and Yang, Qian and Zhang, Haowen and Zhang, Meiling and Jiang, Hunan and Zheng, Chen and Li, Jinyang},
  journal={Journal of Materials Chemistry A},
  volume={11},
  number={15},
  pages={7802--7832},
  year={2023},
  publisher={Royal Society of Chemistry}
}

@article{list1a,
  title={Computational optimization of electric fields for better catalysis design},
  author={Welborn, Valerie Vaissier and Ruiz Pestana, Luis and Head-Gordon, Teresa},
  journal={Nature Catalysis},
  volume={1},
  number={9},
  pages={649--655},
  year={2018},
  publisher={Nature Publishing Group UK London}
}

@article{list1c,
  title={Enhancement of electrocatalysis through magnetic field effects on mass transport},
  author={Vensaus, Priscila and Liang, Yunchang and Ansermet, Jean-Philippe and Soler-Illia, Galo JAA and Lingenfelder, Magal{\'\i}},
  journal={Nature Communications},
  volume={15},
  number={1},
  pages={2867},
  year={2024},
  publisher={Nature Publishing Group UK London}
}

@article{list2a,
  title={Non-adiabatic molecular dynamics of molecules in the presence of strong light-matter interactions},
  author={Zhang, Yu and Nelson, Tammie and Tretiak, Sergei},
  journal={J. Chem. Phys.},
  volume={151},
  number={15},
  year={2019},
  publisher={AIP Publishing}
}

@article{list2b,
  title={Non-adiabatic excited-state molecular dynamics: Theory and applications for modeling photophysics in extended molecular materials},
  author={Nelson, Tammie R and White, Alexander J and Bjorgaard, Josiah A and Sifain, Andrew E and Zhang, Yu and Nebgen, Benjamin and Fernandez-Alberti, Sebastian and Mozyrsky, Dmitry and Roitberg, Adrian E and Tretiak, Sergei},
  journal={Chem. Rev.},
  volume={120},
  number={4},
  pages={2215--2287},
  year={2020},
  publisher={ACS Publications}
}

@article{list5a,
  title={Surface and interface control in nanoparticle catalysis},
  author={Xie, Chenlu and Niu, Zhiqiang and Kim, Dohyung and Li, Mufan and Yang, Peidong},
  journal={Chem. Rev.},
  volume={120},
  number={2},
  pages={1184--1249},
  year={2019},
  publisher={ACS Publications}
}

@article{list5b,
  title={Chemical reactions at surfaces and interfaces from first principles: Theory and application},
  author={Liu, Z-P},
  journal={Pure and applied chemistry},
  volume={76},
  number={12},
  pages={2069--2083},
  year={2004},
  publisher={De Gruyter}
}

@Inbook{list5c,
editor={Wood, John
and Lindqvist, Oliver
and Helgesson, Claes
and Vannerberg, Nils-Gosta},
bookTitle={Reactivity of Solids},
year={1977},
publisher={Springer US},
pages={15--42},
doi={10.1007/978-1-4684-2340-2_2},
}

@article{list6a,
  title={Fuel cells--proton-exchange membrane fuel cells| catalysts: life-limiting considerations},
  author={Zhang, J and Carter, RN and Yu, PT and Gu, W and Wagner, FT and Gasteiger, HA},
  year={2009},
  publisher={Elsevier}
}

@article{list6b,
  title={Advancing the Rigor and Reproducibility of Electrocatalyst Stability Benchmarking and Intrinsic Material Degradation Analysis for Water Oxidation},
  author={Edgington, Jane and Seitz, Linsey C},
  journal={ACS Catalysis},
  volume={13},
  number={5},
  pages={3379--3394},
  year={2023},
  publisher={ACS Publications}
}

@misc{maglab_budget,

  

  title = {Private communication with Vivien Zapf and Minseong Lee},

}

@article{trebst2022kitaev,
  title={Kitaev materials},
  author={Trebst, Simon and Hickey, Ciar{\'a}n},
  journal={Physics Reports},
  volume={950},
  pages={1--37},
  year={2022},
  publisher={Elsevier}
}

@article{zhou2017quantum,
  title={Quantum spin liquid states},
  author={Zhou, Yi and Kanoda, Kazushi and Ng, Tai-Kai},
  journal={Reviews of Modern Physics},
  volume={89},
  number={2},
  pages={025003},
  year={2017},
  publisher={APS}
}

@article{savary2016quantum,
  title={Quantum spin liquids: a review},
  author={Savary, Lucile and Balents, Leon},
  journal={Reports on Progress in Physics},
  volume={80},
  number={1},
  pages={016502},
  year={2016},
  publisher={IOP Publishing}
}

@article{broholm2020quantum,
  title={Quantum spin liquids},
  author={Broholm, C and Cava, RJ and Kivelson, SA and Nocera, DG and Norman, MR and Senthil, T},
  journal={Science},
  volume={367},
  number={6475},
  pages={eaay0668},
  year={2020},
  publisher={American Association for the Advancement of Science}
}

@article{loidl2021proximate,
  title={On the proximate Kitaev quantum-spin liquid $\alpha$-RuCl3: thermodynamics, excitations and continua},
  author={Loidl, Alois and Lunkenheimer, Peter and Tsurkan, Vladimir},
  journal={Journal of Physics: Condensed Matter},
  volume={33},
  number={44},
  pages={443004},
  year={2021},
  publisher={IOP Publishing}
}

@inproceedings{gao2021review,
  title={Review of magnetoelectric sensors},
  author={Gao, Junqi and Jiang, Zekun and Zhang, Shuangjie and Mao, Zhineng and Shen, Ying and Chu, Zhaoqiang},
  booktitle={Actuators},
  volume={10},
  number={6},
  pages={109},
  year={2021},
  organization={MDPI}
}

@article{cheong2007multiferroics,
  title={Multiferroics: a magnetic twist for ferroelectricity},
  author={Cheong, Sang-Wook and Mostovoy, Maxim},
  journal={Nature materials},
  volume={6},
  number={1},
  pages={13--20},
  year={2007},
  publisher={Nature Publishing Group UK London}
}

@article{eerenstein2006multiferroic,
  title={Multiferroic and magnetoelectric materials},
  author={Eerenstein, Wilma and Mathur, ND and Scott, James F},
  journal={nature},
  volume={442},
  number={7104},
  pages={759--765},
  year={2006},
  publisher={Nature Publishing Group UK London}
}

@article{maglab1,
  title={https://nationalmaglab.org},
  author={},
  journal={},
  volume={},
  number={},
  pages={},
  year={},
  publisher={}
}

@article{maglab2,
  title={https://nationalmaglab.org/about-the-maglab/facts-figures/world-records/},
  author={},
  journal={},
  volume={},
  number={},
  pages={},
  year={},
  publisher={}
}

@article{kitaev2003fault,
  title={Fault-tolerant quantum computation by anyons},
  author={Kitaev, A Yu},
  journal={Annals of physics},
  volume={303},
  number={1},
  pages={2--30},
  year={2003},
  publisher={Elsevier}
}

@article{lahtinen2017short,
  title={A short introduction to topological quantum computation},
  author={Lahtinen, Ville and Pachos, Jiannis},
  journal={SciPost Physics},
  volume={3},
  number={3},
  pages={021},
  year={2017}
}

@article{klimov2017magnetoelectric,
  title={Magnetoelectric write and read operations in a stress-mediated multiferroic memory cell},
  author={Klimov, Alexey and Tiercelin, Nicolas and Dusch, Yannick and Giordano, Stefano and Mathurin, Th{\'e}o and Pernod, Philippe and Preobrazhensky, Vladimir and Churbanov, Anton and Nikitov, Sergei},
  journal={Applied Physics Letters},
  volume={110},
  number={22},
  year={2017},
  publisher={AIP Publishing}
}

@article{shalf2020future,
  title={The future of computing beyond Moore’s Law},
  author={Shalf, John},
  journal={Philosophical Transactions of the Royal Society A},
  volume={378},
  number={2166},
  pages={20190061},
  year={2020},
  publisher={The Royal Society Publishing}
}

@article{wen2002quantum,
  title={Quantum orders and symmetric spin liquids},
  author={Wen, Xiao-Gang},
  journal={Physical Review B},
  volume={65},
  number={16},
  pages={165113},
  year={2002},
  publisher={APS}
}

@article{jackeli2009mott,
  title={Mott insulators in the strong spin-orbit coupling limit: from Heisenberg to a quantum compass and Kitaev models},
  author={Jackeli, George and Khaliullin, Giniyat},
  journal={Physical review letters},
  volume={102},
  number={1},
  pages={017205},
  year={2009},
  publisher={APS}
}

@article{rau2016spin,
  title={Spin-orbit physics giving rise to novel phases in correlated systems: Iridates and related materials},
  author={Rau, Jeffrey G and Lee, Eric Kin-Ho and Kee, Hae-Young},
  journal={Annual Review of Condensed Matter Physics},
  volume={7},
  pages={195--221},
  year={2016},
  publisher={Annual Reviews}
}

@article{rau2014generic,
  title={Generic spin model for the honeycomb iridates beyond the Kitaev limit},
  author={Rau, Jeffrey G and Lee, Eric Kin-Ho and Kee, Hae-Young},
  journal={Physical review letters},
  volume={112},
  number={7},
  pages={077204},
  year={2014},
  publisher={APS}
}

@article{kim2015kitaev,
  title={Kitaev magnetism in honeycomb RuCl 3 with intermediate spin-orbit coupling},
  author={Kim, Heung-Sik and Catuneanu, Andrei and Kee, Hae-Young and others},
  journal={Physical Review B},
  volume={91},
  number={24},
  pages={241110},
  year={2015},
  publisher={APS}
}

@article{lee2014heisenberg,
  title={Heisenberg-Kitaev model on the hyperhoneycomb lattice},
  author={Lee, Eric Kin-Ho and Schaffer, Robert and Bhattacharjee, Subhro and Kim, Yong Baek},
  journal={Physical Review B},
  volume={89},
  number={4},
  pages={045117},
  year={2014},
  publisher={APS}
}

@article{kasahara2018majorana,
  title={Majorana quantization and half-integer thermal quantum Hall effect in a Kitaev spin liquid},
  author={Kasahara, Y and Ohnishi, T and Mizukami, Y and Tanaka, O and Ma, Sixiao and Sugii, K and Kurita, N and Tanaka, H and Nasu, J and Motome, Y and others},
  journal={Nature},
  volume={559},
  number={7713},
  pages={227--231},
  year={2018},
  publisher={Nature Publishing Group UK London}
}

@article{nasu2015thermal,
  title={Thermal fractionalization of quantum spins in a Kitaev model: Temperature-linear specific heat and coherent transport of Majorana fermions},
  author={Nasu, Joji and Udagawa, Masafumi and Motome, Yukitoshi},
  journal={Physical Review B},
  volume={92},
  number={11},
  pages={115122},
  year={2015},
  publisher={APS}
}

@article{yamaji2016clues,
  title={Clues and criteria for designing a Kitaev spin liquid revealed by thermal and spin excitations of the honeycomb iridate Na 2 IrO 3},
  author={Yamaji, Youhei and Suzuki, Takafumi and Yamada, Takuto and Suga, Sei-ichiro and Kawashima, Naoki and Imada, Masatoshi},
  journal={Physical Review B},
  volume={93},
  number={17},
  pages={174425},
  year={2016},
  publisher={APS}
}

@article{tanaka2022thermodynamic,
  title={Thermodynamic evidence for a field-angle-dependent Majorana gap in a Kitaev spin liquid},
  author={Tanaka, O and Mizukami, Y and Harasawa, R and Hashimoto, Kenichiro and Hwang, K and Kurita, N and Tanaka, H and Fujimoto, S and Matsuda, Y and Moon, E-G and others},
  journal={Nature Physics},
  volume={18},
  number={4},
  pages={429--435},
  year={2022},
  publisher={Nature Publishing Group UK London}
}

@article{reschke2018sub,
  title={Sub-gap optical response in the Kitaev spin-liquid candidate $\alpha$-RuCl3},
  author={Reschke, Stephan and Mayr, Franz and Widmann, Sebastian and Von Nidda, Hans-Albrecht Krug and Tsurkan, Vladimir and Eremin, Mikhail V and Do, Seung-Hwan and Choi, Kwang-Yong and Wang, Zhe and Loidl, Alois},
  journal={Journal of Physics: Condensed Matter},
  volume={30},
  number={47},
  pages={475604},
  year={2018},
  publisher={IOP Publishing}
}

@article{johnson2015monoclinic,
  title={Monoclinic crystal structure of $\alpha$- RuCl 3 and the zigzag antiferromagnetic ground state},
  author={Johnson, Roger D and Williams, SC and Haghighirad, AA and Singleton, J and Zapf, V and Manuel, P and Mazin, II and Li, Y and Jeschke, Harald Olaf and Valent{\'\i}, R and others},
  journal={Physical Review B},
  volume={92},
  number={23},
  pages={235119},
  year={2015},
  publisher={APS}
}

@article{yadav2016kitaev,
  title={Kitaev exchange and field-induced quantum spin-liquid states in honeycomb $\alpha$-RuCl3},
  author={Yadav, Ravi and Bogdanov, Nikolay A and Katukuri, Vamshi M and Nishimoto, Satoshi and Van Den Brink, Jeroen and Hozoi, Liviu},
  journal={Scientific reports},
  volume={6},
  number={1},
  pages={37925},
  year={2016},
  publisher={Nature Publishing Group UK London}
}

@article{zhang2023electronic,
  title={Electronic and magnetic phase diagrams of the Kitaev quantum spin liquid candidate Na 2 Co 2 TeO 6},
  author={Zhang, Shengzhi and Lee, Sangyun and Woods, Andrew J and Peria, William K and Thomas, Sean M and Movshovich, Roman and Brosha, Eric and Huang, Qing and Zhou, Haidong and Zapf, Vivien S and others},
  journal={Physical Review B},
  volume={108},
  number={6},
  pages={064421},
  year={2023},
  publisher={APS}
}

@article{zhang2023magnetic,
  title={A magnetic continuum in the cobalt-based honeycomb magnet BaCo2 (AsO4) 2},
  author={Zhang, Xinshu and Xu, Yuanyuan and Halloran, T and Zhong, Ruidan and Broholm, C and Cava, RJ and Drichko, N and Armitage, NP},
  journal={Nature Materials},
  volume={22},
  number={1},
  pages={58--63},
  year={2023},
  publisher={Nature Publishing Group UK London}
}

@article{roberts2022fidelity,
  title={Fidelity of the Kitaev honeycomb model under a quench},
  author={Roberts, Wesley and Vogl, Michael and Fiete, Gregory A},
  journal={arXiv preprint arXiv:2208.07732},
  year={2022}
}

@misc{maglab2022funding,
	author = {National Science Foundation},
	title = {{N}{S}{F} to sustain the world's most powerful magnet lab through 2027},
	howpublished = {\url{https://new.nsf.gov/news/nsf-sustain-worlds-most-powerful-magnet-lab}},
	year = {2022},
	note = {[Accessed 13-03-2024]},
}

@article{gu2021fast,
  title={Fast-forwarding quantum evolution},
  author={Gu, Shouzhen and Somma, Rolando D and {\c{S}}ahino{\u{g}}lu, Burak},
  journal={Quantum},
  volume={5},
  pages={577},
  year={2021},
  publisher={Verein zur F{\"o}rderung des Open Access Publizierens in den Quantenwissenschaften}
}

@article{feynman1982simulating,
  title={Simulating Physics with Computers},
  author={Feynman, Richard P},
  journal={International Journal of Theoretical Physics},
  volume={21},
  number={6/7},
  year={1982}
}

@article{chen2023quantum,
  title={Quantum thermal state preparation},
  author={Chen, Chi-Fang and Kastoryano, Michael J and Brand{\~a}o, Fernando GSL and Gily{\'e}n, Andr{\'a}s},
  journal={arXiv preprint arXiv:2303.18224},
  year={2023}
}

@article{rall2023thermal,
  title={Thermal state preparation via rounding promises},
  author={Rall, Patrick and Wang, Chunhao and Wocjan, Pawel},
  journal={Quantum},
  volume={7},
  pages={1132},
  year={2023},
  publisher={Verein zur F{\"o}rderung des Open Access Publizierens in den Quantenwissenschaften}
}

@article{terhal2000problem,
  title={Problem of equilibration and the computation of correlation functions on a quantum computer},
  author={Terhal, Barbara M and DiVincenzo, David P},
  journal={Physical Review A},
  volume={61},
  number={2},
  pages={022301},
  year={2000},
  publisher={APS}
}

@article{manzano2020short,
  title={A short introduction to the Lindblad master equation},
  author={Manzano, Daniel},
  journal={Aip Advances},
  volume={10},
  number={2},
  year={2020},
  publisher={AIP Publishing}
}

@article{cerezo2021variational,
  title={Variational quantum algorithms},
  author={Cerezo, Marco and Arrasmith, Andrew and Babbush, Ryan and Benjamin, Simon C and Endo, Suguru and Fujii, Keisuke and McClean, Jarrod R and Mitarai, Kosuke and Yuan, Xiao and Cincio, Lukasz and others},
  journal={Nature Reviews Physics},
  volume={3},
  number={9},
  pages={625--644},
  year={2021},
  publisher={Nature Publishing Group UK London}
}

@article{tilly2022variational,
  title={The variational quantum eigensolver: a review of methods and best practices},
  author={Tilly, Jules and Chen, Hongxiang and Cao, Shuxiang and Picozzi, Dario and Setia, Kanav and Li, Ying and Grant, Edward and Wossnig, Leonard and Rungger, Ivan and Booth, George H and others},
  journal={Physics Reports},
  volume={986},
  pages={1--128},
  year={2022},
  publisher={Elsevier}
}

@article{roland2002quantum,
  title={Quantum search by local adiabatic evolution},
  author={Roland, J{\'e}r{\'e}mie and Cerf, Nicolas J},
  journal={Physical Review A},
  volume={65},
  number={4},
  pages={042308},
  year={2002},
  publisher={APS}
}

@article{albash2018adiabatic,
  title={Adiabatic quantum computation},
  author={Albash, Tameem and Lidar, Daniel A},
  journal={Reviews of Modern Physics},
  volume={90},
  number={1},
  pages={015002},
  year={2018},
  publisher={APS}
}
\end{document}